\documentclass[journal]{IEEEtran} 

\makeatletter
\def\thickhline{%
	\noalign{\ifnum0=`}\fi\hrule \@height \thickarrayrulewidth \futurelet
	\reserved@a\@xthickhline}
\def\@xthickhline{\ifx\reserved@a\thickhline
	\vskip\doublerulesep
	\vskip-\thickarrayrulewidth
	\fi
	\ifnum0=`{\fi}}
\makeatother

\newlength{\thickarrayrulewidth}
\usepackage{epigraph}
\usepackage{longtable}

\usepackage{textcomp}

\usepackage{amssymb}

\usepackage[hidelinks]{hyperref}
\hypersetup{
	colorlinks=true,
	linkcolor=black,
	filecolor=black,
	citecolor=black,
	urlcolor=blue,
}

\usepackage{booktabs}

\usepackage{multirow}

\usepackage{color}
\usepackage[table, dvipsnames]{xcolor}
\definecolor{PyOrange}{RGB}{255, 201, 14}
\definecolor{PyBlue}{RGB}{112, 146, 190}

\definecolor{WordGreen}{RGB}{100, 136, 40}
\definecolor{WordDarkGrey}{RGB}{82, 82, 82}
\definecolor{WordRed}{RGB}{192, 80, 77}
\definecolor{WordBlue}{RGB}{0, 122, 192}

\definecolor{WordLightBlue}{RGB}{218, 238, 243}
\definecolor{WordLightGreen}{RGB}{234, 241, 221}

\definecolor{WordFillGreen}{RGB}{194, 214, 155}
\definecolor{WordFillRed}{RGB}{252, 214, 182}
\definecolor{WordFillGray}{RGB}{217, 217, 217}

\newcommand{\MyPaperTitle}{Internet of Underwater Things and Big Marine\\ Data Analytics -- A Comprehensive Survey}
\newcommand{\MyPaperTitlePlain}{Internet of Underwater Things and Big Marine Data Analytics -- A Comprehensive Survey}
\newcommand{\MyT}{\textcolor{white}{.}\includegraphics[width=0.012\textwidth,clip,keepaspectratio]{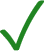}}

\ifCLASSOPTIONcompsoc
  \usepackage[caption=false,font=normalsize,labelfont=sf,textfont=sf]{subfig}
\else
  \usepackage[caption=false,font=footnotesize]{subfig}
\fi

\usepackage{cite}

\usepackage{graphicx}
\graphicspath{{./Graphics/}}
\DeclareGraphicsExtensions{.pdf,.jpeg,.png,.jpg}

\usepackage{amsmath}
\DeclareMathOperator*{\myArgMax}{arg\,max}

\usepackage{array}

\usepackage{url}

\begin{document}
\title{\MyPaperTitle}

\author{
	Mohammad~Jahanbakht\textsuperscript{\href{https://orcid.org/0000-0002-3609-9677}{\includegraphics[scale=0.035]{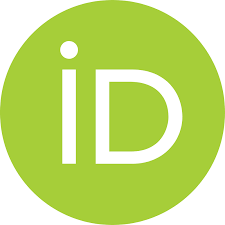}}},~\IEEEmembership{Student Member,~IEEE,}
	Wei~Xiang\textsuperscript{\href{https://orcid.org/0000-0002-0608-065X}{\includegraphics[scale=0.035]{Graphics/Logo_ORCID.png}}},~\IEEEmembership{Senior Member,~IEEE,}
    Lajos~Hanzo\textsuperscript{\href{https://orcid.org/0000-0002-2636-5214}{\includegraphics[scale=0.035]{Graphics/Logo_ORCID.png}}},~\IEEEmembership{Fellow,~IEEE,}
    and~Mostafa~Rahimi~Azghadi\textsuperscript{\href{https://orcid.org/0000-0001-7975-3985}{\includegraphics[scale=0.035]{Graphics/Logo_ORCID.png}}},~\IEEEmembership{Senior Member,~IEEE}%
    \thanks{This work is funded by the Australian Government Research Training Program Scholarship, and in part by Beijing Natural Science Foundation under Grant L182032 (Corresponding authors: Wei~Xiang and M. Rahimi~Azghadi).}
    \thanks{M. Jahanbakh and M. Rahimi Azghadi are with the College of Science and Engineering, James Cook University, Queensland, Australia (e-mail: \href{mailto:mohammad.jahanbakht@my.jcu.edu.au}{mohammad.jahanbakht@my.jcu.edu.au}; \href{mailto:mostafa.rahimiazghadi@jcu.edu.au}{mostafa.rahimiazghadi@jcu.edu.au}).}
    \thanks{Wei Xiang is with the School of Engineering and Mathematical Sciences, La Trobe University, Melbourne, VIC 3086, Australia (e-mail: \href{mailto:w.xiang@latrobe.edu.au}{w.xiang@latrobe.edu.au}).}
    \thanks{L. Hanzo is with the School of Electronics and Computer Science, University of Southampton, Southampton SO17 1BJ, U.K. (e-mail: \href{mailto:lh@ecs.soton.ac.uk}{lh@ecs.soton.ac.uk}).}%
    \thanks{L. Hanzo would like to acknowledge the financial support of the Engineering and Physical Sciences Research Council projects EP/N004558/1, EP/P034284/1, EP/P034284/1, EP/P003990/1 (COALESCE), of the Royal Society's Global Challenges Research Fund Grant as well as of the European Research Council's Advanced Fellow Grant QuantCom.}
%
}

\ifCLASSOPTIONpeerreview
	\markboth{ACCEPTED BY IEEE COMMUNICATION SURVEYS \& TUTORIALS, 2021}%
	{\MyPaperTitlePlain}
\else
	\markboth{ACCEPTED BY IEEE COMMUNICATION SURVEYS \& TUTORIALS, 2021}%
	{Jahanbakht, \MakeLowercase{\textit{et al.}}: \MyPaperTitlePlain}
\fi

\maketitle

\begin{abstract}
	The Internet of Underwater Things (IoUT) is an emerging communication ecosystem developed for connecting underwater objects in maritime and underwater environments. The IoUT technology is intricately linked with intelligent boats and ships, smart shores and oceans, automatic marine transportations, positioning and navigation, underwater exploration, disaster prediction and prevention, as well as with intelligent monitoring and security.
	The IoUT has an influence at various scales ranging from a small scientific observatory, to a mid-sized harbor, and to covering global oceanic trade. The network architecture of IoUT is intrinsically heterogeneous and should be sufficiently resilient to operate in harsh environments. This creates major challenges in terms of underwater communications, whilst relying on limited energy resources.
	Additionally, the volume, velocity, and variety of data produced by sensors, hydrophones, and cameras in IoUT is enormous, giving rise to the concept of Big Marine Data (BMD), which has its own processing challenges.
	Hence, conventional data processing techniques will falter, and bespoke Machine Learning (ML) solutions have to be employed for automatically learning the specific BMD behavior and features facilitating knowledge extraction and decision support. The motivation of this paper is to comprehensively survey the IoUT, BMD, and their synthesis. It also aims for exploring the nexus of BMD with ML. We set out from underwater data collection and then discuss the family of IoUT data communication techniques with an emphasis on the state-of-the-art research challenges. We then review the suite of ML solutions suitable for BMD handling and analytics. We treat the subject deductively from an educational perspective, critically appraising the material surveyed. Accordingly, the reader will become familiar with the pivotal issues of IoUT and BMD processing, whilst gaining an insight into the state-of-the-art applications, tools, and techniques. Finally, we analyze the architectural challenges of the IoUT, followed by proposing a range of promising direction for research and innovation in the broad areas of IoUT and BMD. Our hope is to inspire researchers, engineers, data scientists, and governmental bodies to further progress the field, to develop new tools and techniques, as well as to make informed decisions and set regulations related to the maritime and underwater environments around the world.
\end{abstract}

\ifCLASSOPTIONpeerreview
\else
\fi

\IEEEpeerreviewmaketitle

\section*{Nomenclature}
	\begin{tabbing}
		ASIC~~~~~~~~~ \= Application-Specific Integrated Circuit\\
		AUV \> Autonomous Underwater Vehicles\\
		BER \> Bit Error Rate\\
		BMD \> Big Marine Data\\
		CNN \> Convolutional Neural Networks\\
		DBN \> Deep Belief Network\\
		DCS \> Distributed Computing System\\
		DL \> Deep Learning\\
		FPGA \> Field-Programmable Gate Array\\
		GIS \> Geographic Information System\\
		GPU \> Graphics Processing Unit\\
		IaaS \> Infrastructure as a Service\\
		IoT \> Internet of Things\\
		IoUT \> Internet of Underwater Things\\
		IP \> Internet Protocol\\
		MAC \> Media Access Control\\
		MEC \> Mobile Edge Computing\\
		ML \> Machine Learning\\
		NN \> Neural Network\\
		PaaS \> Platform as a Service\\
		QoS \> Quality of Service\\
		RBM \> Restricted Boltzmann Machine\\
		RFID \> Radio Frequency Identification\\
		RNN \> Recurrent Neural Network\\
		RNTN \> Recursive Neural Tensor Network\\
		ROV \> Remotely Operated Vehicles\\
		SaaS \> Software as a Service\\
		SDN \> Software-Defined Network\\
		SLAM \> Simultaneous Localization and Mapping\\
		SNR \> Signal-to-Noise Ratio\\
		SVM \> Support Vector Machine\\
		TCP \> Transmission Control Protocol\\
		ToF \> Time of Flight\\
		UWSN \> Underwater Wireless Sensor Network\\
	\end{tabbing}

\begin{figure*}[!t]
	\centering
	\includegraphics[width=\textwidth, clip, trim={50 515 50 65}]{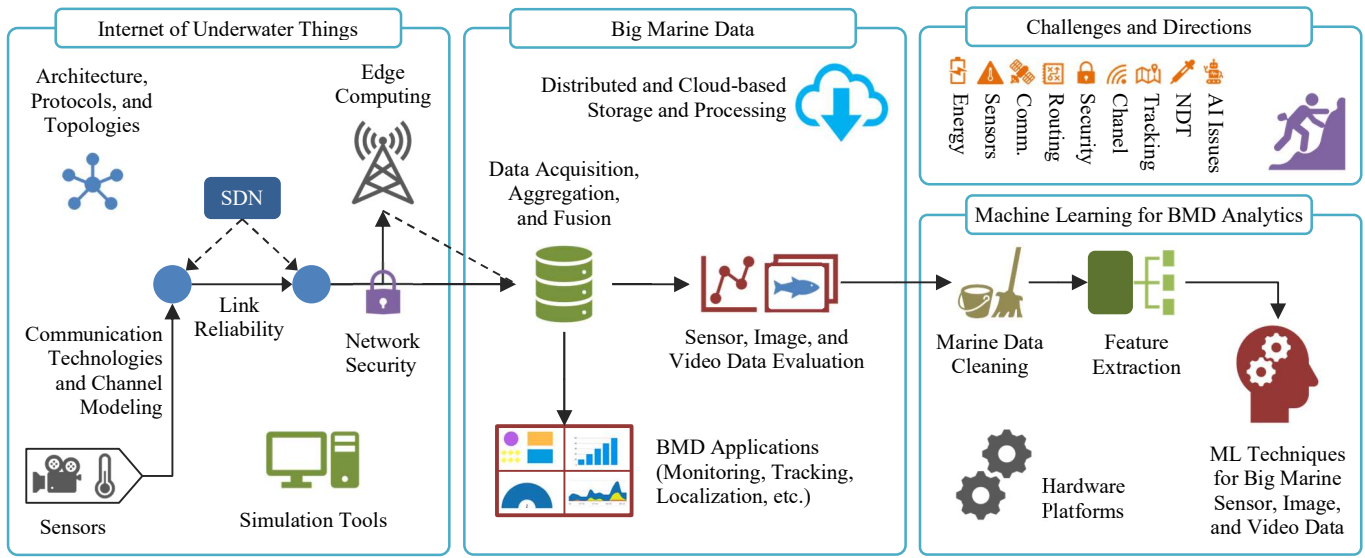}
	\caption{Qualitative relationship between the diverse system components of the IoUT and BMD analytics, starting from underwater sensors and ending up to ML solutions and future directions.}
	\label{Fig_Subjects_Amalgam}
\end{figure*}

\begin{figure}[!t]
	\centering
	\includegraphics[width=0.95\columnwidth, clip, trim={50 83 317 50}]{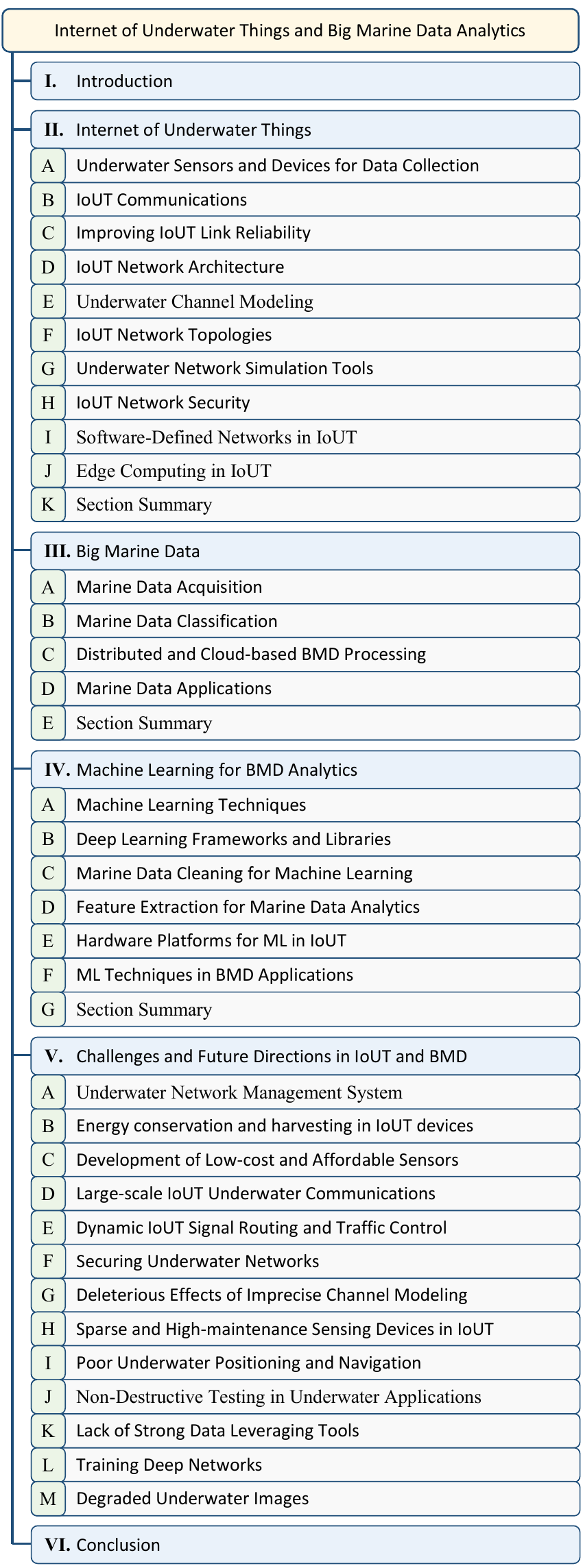}
	\caption{Outline at a glance.}
	\label{Fig_Tree_Paper_Outline}
\end{figure}

\section{Introduction}\label{Section_Intro}
	\IEEEPARstart{T}{HE INTERNET OF THINGS} (IoT) augmented with machine intelligence and big data analytics is expected to transform and revolutionize the way we live in almost every technological area. Broadly speaking, IoT can be defined as an infrastructure of the information society that connects equipment/devices (things) to the Internet and to one another. By this means, the IoT could connect devices in any place on earth to help us have better interaction with our living environment \cite{Al-Fuqaha2015}.
	
	To date, the existing networks in terrestrial and urban areas have been the domain of influence for the IoT and have been researched extensively. This has made a fairly strong foundation for the industrial IoT developments, which are emerging with an astonishing pace at the time of writing \cite{Eldefrawy2019}. However, the underwater section of IoT, i.e. IoUT has not attracted as much attention as it deserves and it is a rather unexplored research area. This is mainly because underwater applications are still in their infancy and the new era of scientific endeavor to better understand, control, and interact with the oceans and seas through underwater technologies is yet to flourish.
	
	Although $44\%$ of the earth's population lives within 150~km of the sea, $95\%$ of sea area remains unexplored by the humankind \cite{Kao2017}. Oceans cover more than $70\%$ of the earth's surface and $90\%$ of international trades are through nautical transportations \cite{Domingo2012}. Astonishingly, 12 people have spent 300 hours on the surface of the moon, while only 3 people have spent about 3 hours at 6 km depth of the ocean. In addition, about 90 million tons of salt-water fish are caught worldwide each year, and the coral reefs are estimated to provide food for almost 500 million people \cite{WHOIKnowYourOcean}. Hence, underwater research and development could have a significant impact on many aspects of human's life by establishing and rolling out the IoUT.
	
	\epigraph{"On the surface of the ocean, men wage war and destroy each other; but down here, just a few feet beneath the surface, there is a calm and peace, unmolested by man."}{--- \textup{Jules Verne}}
	
	Although the IoUT has many technical similarities with its ground-based counterpart (IoT) such as its structure and function, it has many technical differences arising from its different communication/telecommunication environments, computational limitations, and constrained energy resources. To address these gaps between the IoT and IoUT, technical concepts in the field of IoUT will be extensively discussed. These include both the underwater communications \cite{LiY2018, Xiao2017, Zeng2017}, as well as the underwater sensors and devices \cite{Perez2017}.
	
	By connecting an increasing number of devices and machines to the Internet, the IoT and IoUT ecosystems produce enormous amounts of data. This high volume of data is referred to in parlance as \textit{big data}. Big data is currently being generated by various technological ecosystems and perhaps the most ubiquitous data types in today's world is the data produced throughout the IoT. This is also set to increase, since the number of Internet-connected devices is projected to increase from the current 30 billion to over 50 billion by 2020 \cite{Javed2018}.
	
	In the age of sparse data production, analytical mathematics and statistical techniques, widely known as data mining, were employed to infer knowledge from data. However, in the current era of data proliferation, when the produced data volume in the last five years exceeds the whole amount of data generated before that, conventional data processing techniques will soon fall short \cite{Cao2018, Verma2017}. These traditional big data handling methods relying on statistical descriptive, predictive, and prescriptive analytics usually suffer from the lack of generalization. That is, they cannot automatically learn the behavior and features in smaller datasets and use them in big data scenarios.
	
	To address this significant problem, machine learning has risen as one of the practical solutions. ML has been created to facilitate an automatic approach to learning and extracting knowledge from data. This could revolutionize various aspects of our lives, ranging from treating formidable diseases, to boosting the economies, to understanding the universe, to defense and military decisions \cite{Maimo2018, Meng2018, Sun2018}. As listed bellow, ML has also been used in a variety of sparse and big underwater data applications, including:
	
	\begin{itemize}
		\item \textit{Evaluation and discoveries}: Examples of these include the evaluation of corals and their inhabitants \cite{AniBrownMary2017}, seabed analysis and mapping (photo-mosaicking) \cite{Reggiannini2017}, object classification and discovery \cite{Jian2018, Schoening2016}, plant identification \cite{Bonin-Font2017}, the automatic recognition of fish \cite{Meng2018, Chuang2017}, lobster \cite{Lau2012, Osterloff2016}, plankton \cite{Schmid2016}, and other species, as well as tracking and direction finding \cite{Faillettaz2016}.
		
		\item \textit{Monitoring and management}: Examples of these include environmental monitoring (e.g. water quality and pollution) \cite{Pearlman2016}, fish farming \cite{Domingo2012}, pipeline monitoring and corrosion investigation (e.g. in oil and gas industry) \cite{Foresti2001, Khalifeh2016}, harbor security and military surveillance \cite{Kao2017}, navigation assistance \cite{Paull2014}, marine forecast and warning systems (e.g. tsunami, red-tide, flood) \cite{Miyoshi2016}, and maritime geographic information systems \cite{Kalyvas2017}.
	\end{itemize}
	
	ML as an indispensable tool in the IoUT, offers intelligent solutions for analyzing BMD, and thus it will be thoroughly investigated in this paper.
	
	
	This paper is motivated by the fact that the IoUT, BMD, and machine/deep learning are salient topics emerging in the scientific literature. Nevertheless, there is no comprehensive survey to cover the joint applications of these three. In other words, many previous articles can be found in the literature that cover IoT \cite{Javed2018, Al-Fuqaha2015}, big data \cite{Cao2018}, the joint aspects of IoT and big data \cite{Ahmed2017}, and even big data analytics in IoT \cite{Mohammadi2018, Verma2017}. In a clear contrast, the amount of research published on the IoUT and BMD is very limited. Of these limited publications, some cover the IoUT \cite{Kao2017, Domingo2012}, while others cover BMD \cite{LiY2018, HuangD2015}. However, to the best of our knowledge, this paper is the first survey article that provides a comprehensive overview of the IoUT and BMD analytics relying on the most recent radical machine/deep learning approaches. This makes the present article beneficial for data scientists, ML engineers, data analyst, big data engineers, and policy makers in the marine-related disciplines.
	
	To boldly and explicitly illustrate the contributions of this paper in the fields of IoUT and BMD analytics, Table~\ref{Table_Compare_To_Other_Publications} contrasts our unique contributions to other published treatises in the area. For all other articles in this table, a tick mark ({\color{PineGreen}\checkmark}) is granted, even if those dedicated only a few relevant sentences to the given subject. However, none of those surveys are similar to ours in terms of their coverage. Putting the depth of each discussion in other papers aside, there are many topics that are only covered in this paper on IoUT and BMD analytics.

	\begin{table}[!t]
		\renewcommand{\arraystretch}{1.3}
		\caption{The Contribution of This Paper in the Field of IoUT and BMD Analytics, Compared to Other Previously Published Works}
		\label{Table_Compare_To_Other_Publications}
		\centering
		\begin{tabular}{@{}m{0.55\columnwidth}@{} @{}m{0.045\columnwidth}@{} @{}m{0.045\columnwidth}@{} @{}m{0.045\columnwidth}@{} @{}m{0.045\columnwidth}@{} @{}m{0.045\columnwidth}@{} @{}m{0.045\columnwidth}@{} @{}m{0.045\columnwidth}@{} @{}m{0.045\columnwidth}@{} @{}m{0.045\columnwidth}@{} @{}m{0.045\columnwidth}@{} @{}m{0\columnwidth}@{}}
				\toprule
			\centering \textbf{\newline\newline\newline\newline Discussed Subjects} & \centering\rotatebox[origin=c]{90}{\parbox[c]{3.2cm}{\textcolor{white}{..}Domingo \cite{Domingo2012}, 2012}} & \centering\rotatebox[origin=c]{90}{\parbox[c]{3.2cm}{\textcolor{white}{..}Huang \textit{et al.} \cite{HuangD2015}, 2015}} & \centering\rotatebox[origin=c]{90}{\parbox[c]{3.2cm}{\textcolor{white}{..}Kao \textit{et al.} \cite{Kao2017}, 2017}} & \centering\rotatebox[origin=c]{90}{\parbox[c]{3.2cm}{\textcolor{white}{..}Li \textit{et al.} \cite{LiY2018}, 2018}} & \centering\rotatebox[origin=c]{90}{\parbox[c]{3.2cm}{\textcolor{white}{..}Xu \textit{et al.} \cite{XuG2019}, 2019}} &
			\centering\rotatebox[origin=c]{90}{\parbox[c]{3.2cm}{\textcolor{white}{..}Jouhari \textit{et al.} \cite{Jouhari2019}, 2019}} &
			\centering\rotatebox[origin=c]{90}{\parbox[c]{3.2cm}{\textcolor{white}{..}Qiu \textit{et al.} \cite{QiuT2020}, 2020}} &
			\centering\rotatebox[origin=c]{90}{\parbox[c]{3.2cm}{\textcolor{white}{..}Raj \textit{et al.} \cite{RajD2020}, 2020}} &
			\centering\rotatebox[origin=c]{90}{\parbox[c]{3.2cm}{\textcolor{white}{..}Khalil \textit{et al.} \cite{KhalilR2020}, 2020}} &
			
			\centering\rotatebox[origin=c]{90}{\parbox[c]{3.2cm}{\textcolor{white}{..}This Work}}&\tabularnewline
			\hline
			\multicolumn{11}{c}{{\cellcolor[gray]{0.95}Internet of Underwater Things}}&\tabularnewline
			\hline\\\\[-3.5\medskipamount]
			Sensors for Marine Data Collection &\MyT&&\MyT&&\MyT&&&&\MyT&\MyT&\tabularnewline
			Undersea Non-Destructive Testing &&&&&&&&&&\MyT&\tabularnewline
			Energy Consumption and Harvesting &\MyT&&\MyT&&\MyT&\MyT&&\MyT&\MyT&\MyT&\tabularnewline
			Communication Technologies &\MyT&&\MyT&\MyT&&\MyT&\MyT&\MyT&\MyT&\MyT&\tabularnewline
			Link Reliability and Routing Improvement &\MyT&&\MyT&\MyT&\MyT&&\MyT&\MyT&&\MyT&\tabularnewline
			Network Architecture &\MyT&&\MyT&\MyT&\MyT&&\MyT&\MyT&\MyT&\MyT&\tabularnewline
			Communication Protocols &\MyT&&\MyT&&\MyT&\MyT&\MyT&\MyT&&\MyT&\tabularnewline
			Wired and Wireless Channel Modeling &&&\MyT&\MyT&&\MyT&&&&\MyT&\tabularnewline
			Underwater Network Simulation Tools &&&&&&&&&&\MyT&\tabularnewline
			Security and Cryptography in IoUT &\MyT&\MyT&\MyT&&&&\MyT&\MyT&&\MyT&\tabularnewline
			Software-Defined Networks (SDN) &&&&&&&&\MyT&&\MyT&\tabularnewline
			Edge Computing in IoUT &&&&&&&\MyT&&&\MyT&\tabularnewline
			\hline
			\multicolumn{11}{c}{{\cellcolor[gray]{0.95}Big Marine Data}}&\tabularnewline
			\hline\\\\[-3.5\medskipamount]
			Data Acquisition, Aggregation, and Fusion &\MyT&\MyT&&&&&\MyT&&&\MyT&\tabularnewline
			Sensor, Image, and Video Data Evaluation &&&&&&&&&&\MyT&\tabularnewline
			Open Access Databases &&&&&\MyT&&&&&\MyT&\tabularnewline
			Distributed and Cloud-based Data Proc. &&\MyT&&&&&\MyT&&&\MyT&\tabularnewline
			Applications (Monitoring, Tracking, etc.) &\MyT&\MyT&\MyT&\MyT&\MyT&\MyT&\MyT&\MyT&\MyT&\MyT&\tabularnewline
			\hline
			\multicolumn{11}{c}{{\cellcolor[gray]{0.95}Machine Learning and Deep Learning for BMD Analytics}}&\tabularnewline
			\hline\\\\[-3.5\medskipamount]
			Machine Learning Techniques Briefing &&&&&&&&&&\MyT&\tabularnewline
			Sensor, Image, and Video Data Cleaning &&&&&&&&&&\MyT&\tabularnewline
			Feature Extraction from BMD &&&&\MyT&&&&&&\MyT&\tabularnewline
			Hardware Platforms in BMD Analytics &&&&&&&&&&\MyT&\tabularnewline
			ML for Sensor, Image, and Video BMD &&\MyT&&\MyT&\MyT&&\MyT&&&\MyT&\tabularnewline
			
			\bottomrule
		\end{tabular}
	\end{table}
	
	The synthesis of research directions in this paper is handled in a smooth progression, starting from IoUT data collection, followed by networking, through to big data analytics and BMD processing. This flow of logic is illustrated in Fig.~\ref{Fig_Subjects_Amalgam}, which evolves from the distinct research directions to their amalgamation into big data analytics for IoUT.
	The four blue blocks labeled in this figure correspond to the remaining Sections~\ref{Section_IoUT}, \ref{Section_Big_Data}, \ref{Section_Data_Proc}, and \ref{Section_Challenges}. Additionally, any text or symbol inside these blocks represents a dedicated subsection.
	Accordingly, the reader will gradually become familiar with state-of-the-art tools and techniques, whilst gaining an insight into the challenges and opportunities in the broad areas of IoUT, BMD, and processing BMD in IoUT.
	
	This treatise is organized as illustrated in Fig.~\ref{Fig_Tree_Paper_Outline}. Section~\ref{Section_IoUT} presents an overview of the IoUT by associating its ecosystem to the concepts and methodologies defined for IoT.
	In Section~\ref{Section_Big_Data}, we discuss the usual challenges in the field of BMD and provide insights concerning oceanic sensors, image, and video data, which are widely available through several databases.
	In order to advance our knowledge and coordinate efforts in the field of BMD, sophisticated data analytic techniques and methodologies are required. As mentioned, one of the main approaches to meet this demand is to use ML techniques. Section~\ref{Section_Data_Proc} reviews several ML techniques conceived for automatic data leveraging from growing big marine databases. Finally, in Section~\ref{Section_Challenges}, the challenges and opportunities in the emerging fields of IoUT, BMD, and underwater data processing are discussed, whilst offering further insights into the opportunities and potential solutions to the challenges. The paper is concluded in Section~\ref{Section_conclusion}.
	
\section{Internet of Underwater Things}\label{Section_IoUT}
	The concept of networks is broadly defined as a collection of independent machines, which exchange meaningful data through pre-arranged technologies (e.g. Ethernet, Wi-Fi, Bluetooth). Accordingly, the Internet (Worldwide Interconnected Networks) can be considered as a distributed network, or simply a network of networks. This network has an open standard and constitutes a widely accessible ecosystem with lots of users and a variety of applications. Within the Internet, IoT is the largest sub-ecosystem, which connects devices in any place on earth to the World Wide Web.
	
	Similar to the definition of IoT, IoUT may also be defined as worldwide interconnected networks of digitally identified underwater objects, which all obey the communication protocols of a pre-specified reference model such as TCP/IP or OSI \cite{Al-Fuqaha2015}. Based on this definition, a detailed discussion of the IoUT objects (e.g. sensors) and underwater communications will be provided. Additionally, the family of IoUT network standard models and protocols will be surveyed.
	
	\subsection{Underwater Sensors and Devices for Data Collection}\label{SubSection_Sensors_Intro}
		Our knowledge of the underwater environment is rather limited. This is a consequence of having underdeveloped monitoring technologies for this environment. In addition, due to the large operational areas, sea and coastal monitoring tend to suffer from sparse sensor deployment \cite{Zhou2016}. To overcome these shortcomings, low-energy sensors that are capable of working in the vast, hostile, and dynamic underwater conditions are required.
		
		As mentioned in the introduction section, the perception achieved by understanding the data collected using sensors in oceanic areas are essential both to human life and to environmental sustainability. These sensors, for instance, can evaluate the impact of human activities on resources in marine ecosystems and also make us aware of the amount of pollution dumped into the sea \cite{Adamo2015}.
		At the time of writing small-scale Underwater Wireless Sensor Networks (UWSN) and hydrographic research vessels that contain a variety of marine sensors are deployed locally to assess environmental pollution and also to evaluate the seawater quality. However, the main disadvantage of these UWSNs is their small coverage area, which cannot cover the seas in scales of thousands of square kilometers \cite{Al-Zaidi2018}.
		
		Accordingly, it is important to connect all the existing sensor networks to the Internet, giving birth to the IoUT, in order to create an infrastructure for monitoring marine life on a global scale. This internationally accessible IoUT that measures essential chemical and physical parameters at sea, provides both historical and real-time measurements from myriads of marine locations worldwide. The collated oceanic sensor data will help experts predict future phenomena and also help policy makers ratify informed decisions \cite{Hughes2017}.
		
		Therefore, the IoUT infrastructure should consist of sensing objects and communication components in its underwater layers of the architectural model. These objects and components are known as nodes and sinks \cite{Kao2017}. To elaborate \cite{XuG2019},

		\begin{itemize}
			\item Underwater endpoint nodes are the end devices at the underwater side of the network, including various types of sensors, cameras, hydrophones, data storage micro-chips, actuators, acoustic tags, radio frequency tags, tag-readers, etc.
			
			\item Underwater mid-layer nodes are deployed above the underwater endpoint nodes, and are composed of data redistribution points, modems, gateways, repeaters, relays, etc.
			
			\item Sink nodes in IoUT terminology are the overwater nodes along with the land-side facilities, like buoys, exploration platforms, ships, satellites, onshore stations, etc.
		\end{itemize}
		
		Perhaps, the most important data collection components among all the different IoUT endpoint nodes are sensors, which not only collect data, but help activate other underwater components such as cameras, hydrophones, data storage micro-chips, and actuators. Some of the most popular environmental parameters in underwater applications are listed in Table~\ref{Table_Primary_Data_Source_Sensor}. Typical industrial sensors that offer accurate measurement of these parameters are also listed. The BMD generated from the continuous operation of these sensors is transferred, stored and processed in the IoUT ecosystem.

		\begin{table*}[!t]
			\caption{IoUT Sensors to Measure Underwater \colorbox{WordFillGreen!40}{Physical}, \colorbox{WordFillGray!50}{Optical}, \colorbox{WordLightBlue!60}{Fluid}, and \colorbox{WordFillRed!40}{Chemical} Parameters \cite{XuG2019, Fuentes-Perez2018, Al-Zaidi2018, Perez2017, Khalifeh2016}}
			
			
			\label{Table_Primary_Data_Source_Sensor}
			\centering
			\begin{tabular}{@{}m{0.27\columnwidth}@{} @{}m{0.07\columnwidth}@{} @{}m{0.64\columnwidth}@{} @{}m{0.07\columnwidth}@{} @{}m{0.99\columnwidth}@{}}
				
				\toprule
				\textbf{Environmental \newline Parameters} &&
				\multicolumn{1}{c}{\textbf{Typical Underwater Sensor Products}} &&
				\multicolumn{1}{c}{\textbf{Brief Description}}\tabularnewline
				
				\midrule
				
				{\cellcolor{WordFillGreen!25}Temperature $(^{\circ}C)$}&&
				$\bullet$ RBRCoda T by \href{https://rbr-global.com/products/sensors}{RBR} \newline $\bullet$ SBE series by \href{https://www.seabird.com/modular/family?productCategoryId=54627473796}{Sea-Bird} \newline $\bullet$ T Xchange by \href{https://amloceanographic.com/solutions/oem-sensors/}{AML Oceanographic} &&
				Digital (RS232) underwater temperature readings in the range of $-5\sim45\,^{\circ}C$ with an accuracy of $\pm0.001\,^{\circ}C$ and up to $10,000\,m$ depth rating \newline\tabularnewline
				
				\midrule
				
				{\cellcolor{WordFillGreen!25}Conductivity $(S/m)$ \newline and Salinity $(\textrm{ppt})$}&&
				$\bullet$ TTurb by \href{https://www.trios.de/en/tcon.html}{TriOS} \newline $\bullet$ 4319 and 4419 by \href{https://www.aanderaa.com/productsdetail.php?Conductivity-sensor-9}{Aanderaa} \newline $\bullet$ C Xchange by \href{https://amloceanographic.com/solutions/oem-sensors/}{AML Oceanographic} \newline $\bullet$ SBE 4 series by \href{https://www.seabird.com/modular/sbe-4-conductivity-sensor/family?productCategoryId=54627473797}{Sea-Bird} &&
				Digital (RS232 and LAN) conductivity measurement in the range of $0\sim200\,mS/cm$ with an accuracy of more than $\pm0.01\,mS/cm$ and up to $6000\,m$ depth rating (conductivity value is used in some products to indirectly derive salinity)\tabularnewline
				
				\midrule
				
				{\cellcolor{WordFillGreen!25}Depth $(m)$ and \newline Pressure $(\textrm{Bar})$ \newline for Bathymetry}&&
				$\bullet$ PTM, PR36, and 2600 Series by \href{http://www.omniinstruments.co.uk/level-distance-sensors/submersible-depth-sensors-water-level-sensor.html}{Omni} \newline
				$\bullet$ 8000 series by \href{http://www.paroscientific.com/pdf/D50_Series_8000.pdf}{Paroscientific} \newline
				$\bullet$ P Xchange by \href{https://amloceanographic.com/solutions/oem-sensors/}{AML Oceanographic} \newline
				$\bullet$ LMK and LMP Series by \href{https://www.bdsensors.de/en/level/submersible-probes/}{BD Sensors} \newline
				$\bullet$ Nortek Scour Monitor by \href{https://osil.com/Products/OtherMarineInstruments/tabid/56/agentType/View/PropertyID/392/Default.aspx}{OSIL} &&
				Pressure-based Analog (current and voltage) and digital (RS232) sensors can measure water depth ($0\sim7000\,m$), underwater absolute pressure ($0\sim700\,Bar$), and underwater differential pressure ($0\sim6000\,Bar$) with an accuracy of more than $\pm0.1\%$. Some other acoustic sensors (e.g. Nortek Scour Monitor) can measure depth by using echo sounders.\tabularnewline
				
				\midrule
				
				{\cellcolor{WordFillGreen!25}Hydrophone \newline $(\textrm{dBV}/\mu Pa)$}&&
				$\bullet$ Variety of products by \href{https://www.benthowave.com/products/hydrophone.html}{Benthowave Instrument} \newline
				$\bullet$ DH-4 and TH-2 by \href{https://www.sonotronics.com/?page_id=1077}{Sonotronics} \newline
				$\bullet$ C and CR series by \href{http://www.cetaceanresearch.com/hydrophones/index.html}{Cetacean Technology} \newline
				$\bullet$ TC series by \href{http://www.etec.dk/hydrophones.html}{etec electronic engineering} \newline
				$\bullet$ icListen and icTalk series by \href{http://oceansonics.com/iclisten-smart-hydrophones/}{Ocean Sonics} &&
				Analog (voltage) and digital (LAN) $10\,mHz\sim2\,MHz$ active and passive hydrophones with $1^{\circ}$ to Omnidirectional beamwidth for applications up to $2000\,m$ underwater and versatile beam shapes (e.g. Single and Multibeam) with $-230\sim-110\,dBV/\mu Pa$ voltage sensitivity \newline\tabularnewline
				
				\midrule
				
				{\cellcolor{WordFillGray!35}Turbidity and \newline Visibility $(\textrm{NTU})$}&&
				$\bullet$ OBS series by \href{https://www.campbellsci.com.au/turbidity}{Campbell Scientific} \newline
				$\bullet$ InPro 8000 series by \href{https://www.mt.com/au/en/home/products/Process-Analytics/turbidity-meter.html}{Mettler Toledo} \newline
				$\bullet$ WQ730 by \href{https://www.xylem-analytics.com.au/productsdetail.php?Global-Water-WQ730-Turbidity-Sensor-WQ770-B-Turbidity-Meter-106}{xylem} \newline
				$\bullet$ 4112 by \href{https://www.aanderaa.com/productsdetail.php?Turbidity-Sensor-20}{Aanderaa} &&
				Analog (current and voltage) and digital (RS232) $0\sim4000\,NTU$  turbidity measurement (using forward scatter, side scatter light, and backscatter light) with an accuracy of $\pm2\%$ in up to $6000\,m$ depth rating \newline\tabularnewline
				
				\midrule
				
				{\cellcolor{WordFillGray!35}Optical Attenuation \newline (Absorption) $(m^{-1})$}&&
				$\bullet$ FAS series by \href{https://www.seabird.com/transmissometers/ac-s-spectral-absoption-and-attenuation-sensor/family?productCategoryId=54627869911}{Sea-Bird} \newline
				$\bullet$ OLAS by \href{https://www.werne-thiel.de/en/OLAS-optical-lightabsorption-sensor.php}{Werne \& Thiel} &&
				Scanning of the light spectrum between $400\sim730\,nm$ with $\pm0.01\,m^{-1}$ accuracy in up to $5000\,m$ depth rating with $10\sim25\,cm$ light path length\tabularnewline
				
				\midrule
				
				{\cellcolor{WordFillGray!35}Optical Backscatter \newline or Volume Scattering \newline $(sr^{-1} m^{-1})$}&&
				$\bullet$ HydroScat series by \href{https://www.hobilabs.com/cms/index.cfm/37/152/1253/1254/index.html}{HOBI Labs} \newline
				$\bullet$ Hyperion by \href{https://www.valeport.co.uk/Products/Optical-Sensors}{Valeport} \newline
				$\bullet$ ECO series by \href{https://www.seabird.com/scattering-sensors/eco-scattering-sensor/family?productCategoryId=54627869916}{Sea-Bird} &&
				Digitally (RS232) sensing of $0\sim5\,sr^{-1} m^{-1}$ backscatter in $420\sim880\,nm$ with $0.003\,sr^{-1} m^{-1}$ scattering sensitivity and $140^{\circ}$ nominal backscattering angle at up to $6000\,m$ depth rating\tabularnewline
				
				\midrule
				
				{\cellcolor{WordFillGray!35}Photosynthetic PAR \newline $(\mu mol \, m^{-2} s^{-1})$}&&
				$\bullet$ ECO and RXA series by \href{https://www.seabird.com/multispectral-radiometers/family?productCategoryId=54627869937}{Sea-Bird} \newline
				$\bullet$ LI-192 by \href{https://www.licor.com/env/products/light/quantum_underwater.html}{LI-COR} \newline
				$\bullet$ RBRCoda PAR by \href{https://rbr-global.com/products/sensors}{RBR} &&
				Analog (voltage) and digital (USB) sensors in $389\sim700\,nm$ spectral range for linear PAR measurement with an accuracy of $\pm2\%$ in up to $7000\,m$ depth rating\tabularnewline
				
				\midrule
				
				{\cellcolor{WordFillGray!35}Spectral Irradiance \newline (e.g. Fluorometers, \newline Radiometers, etc.) \newline (application- \newline specific \; physical \newline units)}&&
				$\bullet$ Cyclops series by \href{https://www.turnerdesigns.com/fluorometers-and-sensors}{Turner Designs} \newline
				$\bullet$ \href{https://www.seabird.com/fluorometers/family?productCategoryId=54627869904}{ECO, SeaOWL, and WETStar} Fluorometers and \newline \textcolor{white}{.}\; \href{https://www.seabird.com/multispectral-radiometers/multispectral-radiometers/family?sortBy=sequence&productCategoryId=54627869938&secondPageNumber=1&hideObsolete=true&pimContext=SeabirdUSen&erpSystem=SITE_ADMIN&focusResults=true}{Multispectral} and \href{https://www.seabird.com/hyperspectral-radiometers/hyperocr-radiometer/family?productCategoryId=54627869935}{HyperOCR} Radiometers by \newline \textcolor{white}{.}\; Sea-Bird \newline
				$\bullet$ Photo-, Fluoro-, and Radiometers by \href{https://www.trios.de/en/sensors.html}{TriOS} \newline
				$\bullet$ AP Series by \href{https://www.aquaread.com/need-help/what-are-you-measuring/depth/}{Aquaread} &&
				Continuous scanning of the entire ultraviolet, visible, and infrared spectrum between $190\sim950\,nm$ with an accuracy of $\pm0.2\,nm$ in up to $6000\,m$ depth rating for versatile applications such as chlorophyll, phycocyanin (Freshwater Algae), Phycoerythrin (Marine Algae), CDOM/FDOM, dye tracing (fluorescein, PTSA, and rhodamine), hydrocarbons (oil and fuel), nitrate, turbidity, wastewater monitoring, etc.\tabularnewline
				
				\midrule
				
				{\cellcolor{WordLightBlue!45}Water Flow and \newline Current Velocity \newline $(m/s)$}&&
				$\bullet$ Current meters by \href{https://www.valeport.co.uk/Products/Current-Meters}{Valeport} \newline
				$\bullet$ Water flow meters by \href{https://www.ott.com/products/water-flow-127/}{OTT HydroMet} \newline
				$\bullet$ DCPS and ZPulse DCS by \href{https://www.aanderaa.com/productsdetail.php?Current-Sensors-10}{Aanreraa} \newline
				$\bullet$ ISM series by \href{http://www.dr-schlueter-vdi.de/englisch/produkte.htm}{HS Engineers} &&
				Analog (voltage) and digital (RS232) $0\sim15\,m/s$ current flow sensors, using mechanical (impeller), magnetic induction, acoustic Doppler, and K-band radar Doppler technologies with an accuracy of $\pm0.5\%$ and up to $6000\,m$ depth rating\tabularnewline
				
				\midrule
				
				{\cellcolor{WordLightBlue!45}Tide and Wave \newline Elevation $(m)$ \newline and Direction $(^{\circ})$}&&
				$\bullet$ 4648, 5218, and 44xx series by \href{https://www.aanderaa.com/productsdetail.php?Wave-and-Tide-Sensor-13}{Aanderaa} \newline
				$\bullet$ S500 by \href{http://www.nexsens.com/knowledge-base/nexsens-sensors/s500-wave-sensor/s500-inertial-wave-sensor.htm}{NexSens Technology} \newline
				$\bullet$ SVS-603 by \href{https://www.seaviewsystems.com/products/data-buoy-instruments/svs-603-inertial-wave-sensor/}{Seaview Systems} \newline
				$\bullet$ ACM-PLUS series by \href{http://www.falmouth.com/sensors.html}{Falmouth Scientific} \newline &&
				Digital (RS232) tide height, wave height, and wave direction read with an accuracy of $\pm0.5\,cm$ in height and $\pm2^{\circ}$ in direction and $\pm1\%$ in period, plus $0\sim30^{\circ}$ wave tilt angle read with an accuracy of $\pm0.5^{\circ}$ for versatile applications (e.g. energy period, steepness, irregularity, wave Fourier spectrum, moment spectral, tide pressure, etc.)\tabularnewline
				
				\midrule
				
				{\cellcolor{WordFillRed!25}Nutrients (e.g. \newline Nitrate (NO\textsubscript{2}, NO\textsubscript{3}), \newline Phosphates (PO\textsubscript{4}), \newline etc.) (ppm)}&&
				$\bullet$ AP Series by \href{https://www.aquaread.com/need-help/what-are-you-measuring/depth/}{Aquaread} \newline
				$\bullet$ Ion-Selective Electrodes by \newline \textcolor{white}{.}\; \href{https://www.waterprobes.com/sensors-for-sondes-and-monitoring}{Eureka Water Probes} \newline
				$\bullet$ SUNA and HydroCycle series by \href{https://www.seabird.com/nutrient-sensors/family?productCategoryId=54627869921}{Sea-Bird} \newline
				$\bullet$ Marine Nutrient Analyzer by \href{http://subctech.com/sensor-systems/various-sensors/}{SubCtech}
				&&
				Nutrient sensors can be made by solid-state, liquid membrane, or UV spectrum technologies to measure $0\sim30,000\,ppm$ nitrates with an accuracy of $\pm2\,\mu Mol/L$ in up to $2000\,m$ depth rating and also to measure phosphates in up to $200\,m$ depth rating. Other included ion-selective electrodes are ammonium, ammonia, chloride, fluoride, calcium, sodium, silica, etc.
				\tabularnewline
				
				\midrule
				
				{\cellcolor{WordFillRed!25}Potential of \newline Hydrogen (pH)} &&
				$\bullet$ pH Sensor by \href{http://subctech.com/sensor-systems/various-sensors/}{SubCtech} \newline
				$\bullet$ DP-HP5 by \href{https://www.presens.de/products/detail/ph-dipping-probe-dp-hp5.html}{PreSens Precision Sensing} \newline
				$\bullet$ SP series by \href{http://www.sensorlab.es/products}{SensorLab} \newline
				$\bullet$ TpH and TpH-D sensors by \href{https://www.trios.de/en/echem.html}{TriOS} &&
				Analog (current and voltage) and digital (RS232 and USB) conventional  and differential pH sensors in the range of $0\sim14\,pH$ with an accuracy of more than $\pm0.01\,pH$ in up to $2000\,m$ depth ratio \newline\tabularnewline
				
				\midrule
				
				{\cellcolor{WordFillRed!25}Dissolved Oxygen \newline $(\mu Mol/L)$} &&
				$\bullet$ \href{https://www.seabird.com/oxygen-sensors/family?productCategoryId=54627869931}{SBE} and \href{https://www.seabird.com/moored/hydrocat-ep-conductivity-temperature-depth-optical-dissolved-oxygen-ph-turbidity-and-chlorophyll/family?productCategoryId=54627473777}{Hydro-CAT} series by Sea-Bird \newline
				$\bullet$ Oxygen Optode series by \href{https://www.aanderaa.com/productsdetail.php?Oxygen-Optodes-2}{Aanderaa} \newline
				$\bullet$ AP Series by \href{https://www.aquaread.com/need-help/what-are-you-measuring/depth/}{Aquaread} \newline
				$\bullet$ Oxygen sensors by \href{http://www.amt-gmbh.com/}{AMT} &&
				Analog (current and voltage) and digital (RS232) dissolved oxygen sensor in the range of $0\sim1600\,\mu Mol/L$, which is equal to $500\%$ of surface saturation in all natural waters (fresh and salt) with an accuracy of $\pm1\%$ and up to $12,000\,m$ depth rating\tabularnewline
				
				\midrule
				
				{\cellcolor{WordFillRed!25}CO\textsubscript{2} and pCO\textsubscript{2} \newline (ppm)} &&
				$\bullet$ C-sense by \href{https://www.turnerdesigns.com/c-sense-in-situ-pco2-sensor}{Turner Designs} \newline
				$\bullet$ pCO\textsubscript{2} series by \href{http://subctech.com/pco2-monitoring/}{SubCtech} \newline
				$\bullet$ CO\textsubscript{2} series by \href{http://www.sunburstsensors.com/}{Sunburst Sensors} &&
				Analog (current and voltage) and digital (RS232) CO\textsubscript{2} and pCO\textsubscript{2} sensors in the range of $0\sim10,000\,ppm$ with an accuracy of $\pm1\%$ in up to $6000\,m$ depth rating \tabularnewline
				
				\bottomrule
			\end{tabular}
		\end{table*}
		
		In Sections~\ref{Section_Big_Data} and \ref{Section_Data_Proc}, further discussions on marine sensors, as well as on sensor data storing and processing, will be provided. In addition, a number of IoUT-based observatory systems will be introduced. Before any sensor can be deployed in an underwater environment, efficient and reliable underwater data communications should be realized. This poses one of the greatest challenges for the pervasive sensor deployment in IoUT ecosystems due to the extremely low acoustic and electromagnetic channel capacities and high signal attenuations over long maritime distances, which affect reliable underwater data communications.
		The next section will discuss and survey data transportation as well as communication methods and protocols of the IoUT. 
		
	\subsection{IoUT Communications}\label{SubSection_UW_Comm}
		Today's marine vehicular communication systems utilize the Very High Frequency (VHF) automatic identification system to provide essential shipping information \cite{Al-Zaidi2018} (e.g. vessel name, position, speed, destination, etc.). In addition, high-speed satellite communications are available as an expensive alternative to existing VHF systems.
		However, none of these systems are capable of supporting long-range underwater applications, where the acoustic waves are the dominant communication media. Nonetheless, these acoustic carriers also suffer from high propagation delay, fading, narrow-bandwidth, and high-attenuation. Therefore, in underwater applications, it is sometimes inevitable to use a combination of different technologies (i.e. acoustic, electromagnetic, and optical) to overcome the communication challenges such as signal attenuation. 
		
		There is always a high level of signal attenuation, when passing through water. This attenuation affects every telecommunication technology in a different way. The signal attenuation is directly related to the main design constraints such as the maximum reliable data-rate and the maximum possible communications distance. In the following subsections, the signal attenuation is discussed in detail.
		
		\subsubsection{Electromagnetic signal attenuation}
			Electromagnetic waves and radio frequency signals do not propagate well underwater. This is mainly due to the high conductivity of seawater. The penetration depth of electromagnetic waves is inversely proportional both to the conductivity ($\sigma\,[S/m]$) and to the frequency ($f\,[Hz]$) \cite{BalanisAnt2016}:
			\begin{equation}\label{Eq_Penetr_Depth}
				\delta\approx\dfrac{1}{\sqrt{\pi\mu f\sigma}}\,,
			\end{equation}
			where $\mu$ is the water permeability and the penetration depth expressed in $\delta\,[m]$ is defined as the distance that an electromagnetic wave travels before becoming attenuated to $e^{-1}$ of its initial amplitude. Based on this formula, only low frequency signals (with small channel capacity) can travel long distances in seawater, before they completely fade out. Therefore, the antenna size ($L\,[m]$) increases, as the frequency decreases \cite{BalanisAnt2016}:
			\begin{equation}\label{Eq_Ant_Length}
				L\propto\dfrac{v}{f}\,,
			\end{equation}
			where $v$ is the speed of the electromagnetic wave in water that is almost equal to the speed of light in free space.
			
		\subsubsection{Acoustic signal attenuation}
			Due to the hostile electromagnetic underwater environment, most communications in the IoUT are based on acoustic links, which also suffer from a narrow frequency bandwidth. However, the attenuation of low frequency acoustic waves ($\alpha\,[dB/km]$) is lower than that of electromagnetic waves \cite{Ainslie1998}:
			\begin{equation}\label{Eq_Acoustic_Att}
				\alpha\approx F_{1}\left( f, pH \right) + F_{2}\left( f, T, S, z \right)  \,,
			\end{equation}
			where $f$ is the frequency in $[KHz]$, pH is the water acidity (i.e. almost $6.0 \sim 8.5$ for both freshwater and seawater), $T$ is the water temperature in centigrade, $S$ is the water salinity in parts-per-thousand (ppt), and $z$ is depth in kilometer.
			
		\subsubsection{Optical signal attenuation}
			Another possible communication technology for underwater environments relies on an optical channel. However, similar to the previous pair of channel types, optical channels also suffer from signal attenuation.
			
			Absorption and scattering are the two main causes of optical signal attenuation under water. Several previous studies have performed numerical simulations to estimate the attenuation \cite{Zeng2017, Gabriel2013, IlliE2019}. These include solving the radiative transfer equation, which is time-consuming and complex, but precise; or using simplified models (e.g. Monte Carlo), which are typically fast but imprecise \cite{Zeng2017}.
			
			One of the most successful examples of these simplified Monte Carlo-based approaches was proposed by Gabriel \textit{et al.} \cite{Gabriel2013}, where the spectral beam attenuation coefficient is calculated as,
			\begin{equation}\label{Eq_Optical_Attenuation}
				c(\lambda)=a(\lambda)+b(\lambda)\,,
			\end{equation}
			where $a(\lambda)$ and $b(\lambda)$ are the spectral absorption and spectral scattering coefficients, respectively, and $\lambda\,[m]$ is the wavelength. Both of these coefficients constitute intrinsic optical properties and are calculated by simple volume integration over a solid angle ($\Psi$).
			
			Alternatively, the relationship between the received light intensity ($L(t,r,\theta,\varphi)$) and the transmitted optical power ($S(t)$) expressed in spherical coordinates $(r,\theta,\varphi)$ can be defined by the well-known radiative transfer differential equation \cite{Jaffe2015}.
			One of the most accurate solvers for this differential equation in underwater optical communications was proposed by Illi \textit{et al.} \cite{IlliE2019}, who have formulated their time-domain approach as,
			\begin{equation}\label{Eq_RTE}
				\begin{split}
					\left[ \frac{1}{v}\frac{\partial}{\partial t} + \vec{n}\cdot\nabla \right] L \, = \, & -c(\lambda) \, L \, + \, S \\
					& + \iint VSF \times L \, sin(\varphi^\prime) d\theta^\prime d\varphi^\prime
					\,,
				\end{split}
			\end{equation}
			where $v$ is the speed of light, $\vec{n}$ denotes the direction of propagation, and $c(\lambda)$ accounts for the same absorption and scattering coefficients of \eqref{Eq_Optical_Attenuation}. The Volume Scattering Function ($VSF$) will be discussed in Section~\ref{SubSection_UW_Image_Data}.
			
			The improvement in solving \eqref{Eq_RTE} was achieved by enhancing the finite-difference method as well as by proposing a better approximation for the definite integral.
			Again, this is an accurate, but computationally expensive method of optical signal attenuation characterization.
			
			The signal attenuation encountered by different telecommunication technologies discussed above directly affect the bit error rate and the overall underwater link reliability in the IoUT, which is discussed in the following section.
			
	\subsection{Improving IoUT Link Reliability}\label{SubSection_Link_Reliability}

		The underwater channel quality is significantly affected by several dynamic factors, including tidal-waves, pressure gradients, temperature gradients, floating sediments, and changes in water ingredients (chemical compounds). These channel dynamics affect the signal amplitude (distortion), frequency (dispersion), and speed (refraction) \cite{WangJ2018}. They also result in different delays, corresponding to delay jitter. Due to these challenges, all underwater channels are considered as unreliable links, which requires us to define a reliability metric and then, use this metric for quantifying and optimizing the quality of service in our IoUT network \cite{SunW2018}.
		
		Previous contributions in this field tend to assess the link reliability either by software-based or hardware-based metrics \cite{ZhangH2017, SunW2018, Gomes2019, Murali2019}.
		Hardware-based metrics tend to measure Signal-to-Noise Ratio (SNR) and signal-to-interference ratio as their reliability metrics \cite{ZhangH2017}. The signal strength values can be directly read from the hardware transceiver. On the other hand, software-based metrics predominantly rely on comparing the overall end-to-end data delivery of the communication systems \cite{SunW2018}. The channel or link reliability metric is defined as the ratio of bits as well as packets that are delivered successfully through the link. This metric can be evaluated either at the bit- or the packet-level. At the bit-level, the Bit Reception Ratio (BRR) is the reciprocal of the Bit Error Rate (BER):
		
		\begin{equation}\label{Eq_BER}
		BER = \dfrac{N_{Errorous \; Receptions} \: [bits]}{N_{Total \; Transmissions} \: [bits]}\,,
		\end{equation}
		\begin{equation}\label{Eq_BRR}
		BRR = 1 - BER\,.
		\end{equation}
		
		By contrast, for a packet of $m$ bytes (i.e. $8 \times m$ bits), successful reception of a packet means that all the $m$ bytes were received correctly. Thus, the link reliability metric at the packet-level can be defined by the Packet Reception Ratio (PRR) as \cite{SunW2018}:
		
		\begin{equation}\label{Eq_PRR}
		PRR = \left( 1 - BER \right) ^ {8m}\,.
		\end{equation}

		Another commonly used alternative to the BER of \eqref{Eq_BER} for defining a link reliability metric is the Required Number of Packet transmissions (RNP), defined as \cite{Gomes2019}:
		
		\begin{equation}\label{Eq_RNP}
		RNP = \dfrac{N_{Total \; Transmissions} \: [packets]}{N_{Correct \; Receptions} \: [packets]}\,.
		\end{equation}

		In contrast to the BER that is measured at the receiver side, the RNP metric is designed to be measured at the transmitter side.
		
		Another software-based link reliability metric, namely the Expected Transmission Count (ETX), is defined as the number of expected transmissions that a node requires for successful delivery of a packet \cite{Murali2019}, which can be directly calculated as the reciprocal of the PRR value of:
		
		\begin{equation}\label{Eq_ETX}
		ETX = \dfrac{1}{PRR}\,.
		\end{equation}

		Based on the band-limited nature of both the overwater and underwater telecommunication channels, the BER, RNP, and ETX values that have been respectively calculated in \eqref{Eq_BER}, \eqref{Eq_RNP}, and \eqref{Eq_ETX} are related to the originally transmitted data-rate, initially transmitted power, as well as to the distance between the consecutive transceiver pairs \cite{BalanisAnt2016}. In this regard, reliable communication in the presence of random background noise, requires a certain minimum received power. This guarantees reliable data flow throughout the IoUT infrastructure for the ensuring big marine data processing.
		
		The data-rate in bits-per-second (bps) and the transmission range, alongside the other characteristics of communication technologies are shown in Table~\ref{Table_bps_and_Att}. The simplified simulation plot of attenuation vs. transmission range in this table is calculated using the same conditions as in the studies discussed in Section~\ref{SubSection_UW_Comm}.
		Here, the electromagnetic conductivity of the seawater and freshwater are considered to be $44,000$ and $100\,\mu S/cm$, respectively \cite{Fink2005}. The salinity values of the seawater and freshwater are also considered to be $35$ and $0.4\,ppt$. The water depth only has a minor effect on the results and it is considered to be $1\,km$ and $1\,m$ for seawater and freshwater, respectively. The water temperature and its pH value are assumed to be $10^{\circ}C$ and $7.7$.
		
		\begin{table*}[!t]
			\renewcommand{\arraystretch}{1.5}
			\caption{Data-Rates and other Characteristics of Communication Technologies in Underwater Applications \cite{Jouhari2019, Zeng2017, Kaushal2016}}
			\label{Table_bps_and_Att}
			\centering
			\includegraphics[width=\textwidth, clip, trim={40 353 40 60}]{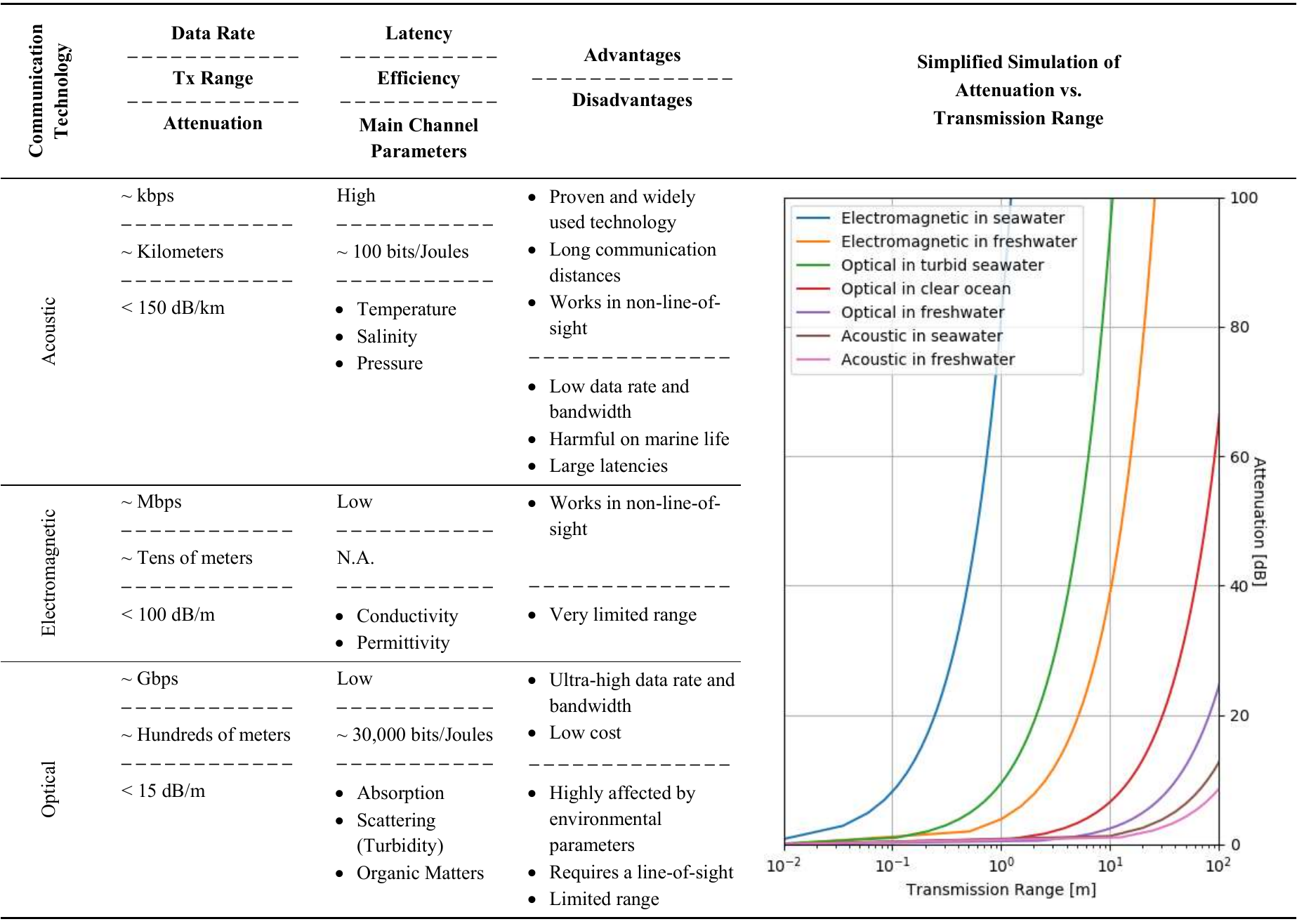}
		\end{table*}
		
		By comparing the details provided in Table~\ref{Table_bps_and_Att}, one can readily conclude that reliable underwater wireless communications are restricted to low data-rate acoustic waves for long distances or high data-rate optical rays for short distances. Short-distance and low data-rate electromagnetic waves are substantially outperformed by the other two technologies.
		
		Despite the restrictive nature of the above-mentioned underwater communication technologies, innovative techniques can be developed to boost both the software-based and the hardware-based reliability metrics of underwater communication links. These techniques include \textit{ad hoc} routing improvements and hop-count optimization, which are discussed in the following subsections.
		
		\subsubsection{Routing improvement}\label{SubSection_Routing}
\begin{figure}[!t]
				\centering
				\includegraphics[width=.87\columnwidth, clip, trim={45 560 300 57}]{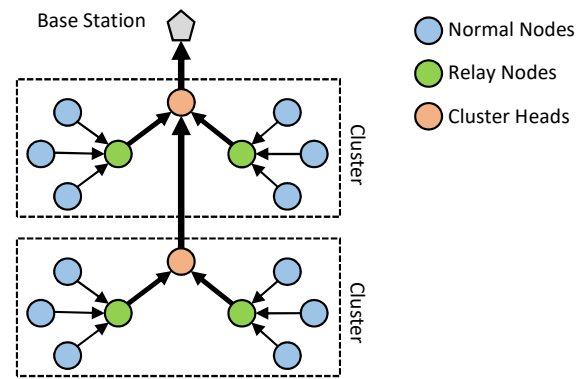}
				\caption{Reliable and energy-efficient multi-cluster network topology \cite{Rani2017}.}
				\label{Fig_Sch_Routing_Improvm}
			\end{figure}

			The specific choice of the link reliability metrics to be optimized has a substantial influence on the routing design of \textit{ad hoc} networks \cite{LiZ2017}. Therefore, the design and implementation of a reliable network under the above-mentioned restrictive conditions of underwater channels requires an efficient data routing scheme \cite{Zhou2016}.
			One such scheme has been proposed by Rani \textit{et al.} \cite{Rani2017} for UWSN, which can be adapted to IoUT nodes as well. As shown in Fig.~\ref{Fig_Sch_Routing_Improvm}, the whole underwater wireless network in \cite{Rani2017} is divided into multiple sub-regions or clusters. In every cluster, relay nodes cooperate with the local normal nodes to forward data to the cluster-heads. The cluster-heads of each cluster are responsible for routing and transmission of data to the next cluster-head situated in the upper sub-region, termed as \textit{cluster coordinator}.
			
			The scheme presented in \cite{Rani2017} relies on a pair of efficient algorithms. The first algorithm constitutes a location-free and energy-based policy that is devised to group individual nodes into cluster sets, aiming to increase SNR reliability metric. The second algorithm involves allowing cluster-heads and relay nodes to store data before routing and transmission. This helps the whole system to increase the RNP metric in \eqref{Eq_RNP}, by avoiding retransmission of the same data packets. By implementing these ideas, the protocol tends to transmit fewer packets of data with a higher reliability and also carries out load balancing in packet routing throughout the network.
			
			Another noteworthy data routing scheme was proposed by Tran-Dang \textit{et al.} \cite{TranDang2019} for underwater acoustic sensor networks. Their work follows the same structure as \cite{Rani2017} in Fig.~\ref{Fig_Sch_Routing_Improvm}. By contrast, their cooperative routing algorithm does not divide the UWSN into multiple clusters. Instead, every normal node independently selects its own relay node and cluster-head. This selection is based on the SNR link reliability metric of Section~\ref{SubSection_Link_Reliability}, on the physical distances represented by hop-count in Section~\ref{SubSection_Hop_Count_optim}, and the time of arrival, which is extracted from the timestamps of the packages.
			Moreover, this algorithm constantly monitors the environment in UWSN to estimate the SNR, hop-count, and time of arrival throughout the network. These parameters are frequently updated to adapt to the dynamic underwater environment. However, the simulation results reported in this paper require experimental fieldwork for validation.
			
			More discussions on the effects of network topology on the software-based link reliability metrics (e.g. BER) are provided in Section~\ref{Subsection_Network_Topology}. In the next section, a hardware-based reliability metric (i.e. SNR) will be used for the evaluation and optimization of the hop-counts in IoUT.
			
		\subsubsection{Hop-count optimization}\label{SubSection_Hop_Count_optim}
			Another subcategory of methods capable of increasing the underwater link reliability is to optimize the hop-count, which is defined as the number of intermediate hardware devices (Modem, Gateway, Switch, Router, HUB, and Repeater) conveying the data between successive sources and destinations \cite{ZhaoR2019}. For example we could minimize the number of hops, provided that the transceiver is capable of reliably communicating over higher distances with the aid of higher transmit power and or more sophisticated receivers.
			
			To characterize the effect of the number of hops on the system's reliability, we have to consider the fact that almost all of the wireless underwater endpoint and mid-layer nodes are battery-limited and the adequate operation of the entire network depends on the charge of their batteries. On the other hand, increasing the transmission distance, which is required for most underwater applications, decreases the effective bandwidth. This in turn increases the power consumption of delivering the payload at a minimum SNR.
			
			To address the challenges imposed by battery constrained underwater nodes on the link's SNR reliability metric, Li \textit{et al.} \cite{LiY2018} suggest reducing the hop-distance by deploying relay nodes along the underwater link to improve the overall transmission performance. This relay-aided transmission scheme is based on power-bandwidth-range dependency, and at the time of writing it offers an energy-efficient method supporting sustainable high data-rate delivery in underwater scenarios.
			
			To establish reliable communication in any IoUT system, a suitable network architecture should be developed. This architecture has to define the model layers, connection topologies, communication protocols, and related equipment sets. All these establishments have a direct influence on our underwater communication paradigm and its subsequent link reliability metrics. This architecture and its influence on all network aspects will be discussed in the next section.
			
	\subsection{IoUT Network Architecture}\label{SubSection_Network_Architect}
		Despite the fact that every standalone network can have its own architecture, the Transmission Control Protocol/Internet Protocol (TCP/IP) is the most ubiquitous network architecture. This reference architecture consists of 5 distinct layers, namely the application, transport, network, data link, and physical layers \cite{Fu2014}. The network layer is also known as the internetwork or Internet layer in some literature. Similarly, the data link and physical layers are sometimes combined for forming a 4-layer TCP/IP model, in which the combined data link and physical layer is referred to as the network access or network interface layer \cite{Dye2007}.

		The TCP/IP layers can be arranged as shown in Table~\ref{Table_Network_Archit} to be matched to the IoUT network architecture. This table presents the IoUT network architecture in a very general form. Here, both the land-side computers and the underwater sensors are considered as network endpoints, located at the highest TCP/IP layer (i.e. the application layer). This architecture covers the entire network, from the underwater application layer up to the overwater application layer. The four rows in this table, from top to bottom, represent the sink nodes, physical channel, underwater mid-layer nodes, and underwater endpoint nodes, based on the definition of those nodes in Section~\ref{SubSection_Sensors_Intro}.
		It should be noted that the standard TCP/IP layer numbers (i.e. 1 for physical, 2 for data link, 3 for network, 4 for transport, and 5 for application) are not used anywhere in this paper. Instead, the layer names are used to specify the corresponding TCP/IP layers.
		
		\begin{table}[!t]
			\renewcommand{\arraystretch}{1.5}
			\caption{Matching IoUT Architectures to the 5-layer TCP/IP Model \cite{Fu2014}}
			\label{Table_Network_Archit}
			\centering
			\includegraphics[width=\columnwidth, clip, trim={40 504 305 65}]{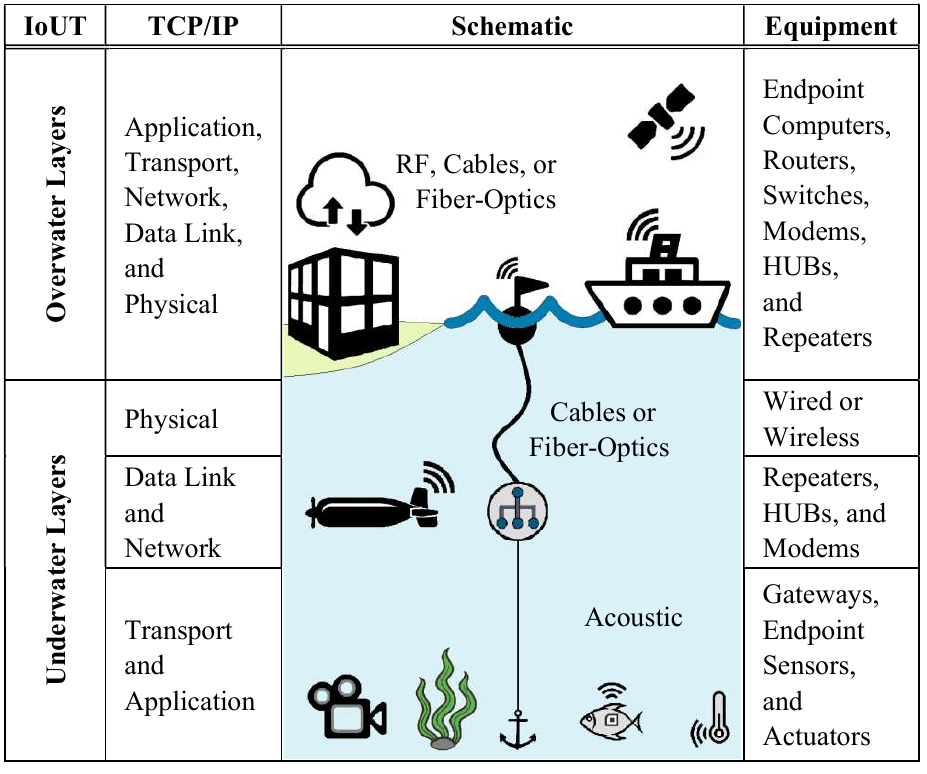}
		\end{table}
		
		Every TCP/IP layer shown in Table~\ref{Table_Network_Archit} relies on some protocols to govern the communications between the IoUT devices, regardless of their underlying structure and design.
		Based on these protocols, every single entity in the network will be aware of the data formats, communication syntax, synchronization methods, security concerns, and error control schemes utilized.
		A layer-wised collection of these protocols is referred to in parlance as \textit{protocol stack}.
		Similar to the IoT \cite{Ahmed2017}, the IoUT ecosystem does not require a universal protocol stack. It can rely on multiple protocols that co-exist in the model layers of a single project, as a result of diverse demands and requirements. These layer-wise stack of protocols will be discussed in the following subsections, along with other layer-wise security protocols in Section~\ref{SubSection_Secure_Comm}. These protocols are required for supporting different aspects of an IoUT infrastructure to meet various challenges, including \cite{Cui2006}:
		
		\begin{itemize}
			\item Connecting to the existing IoT and WEB protocols;
			\item Providing an acceptable latency;
			\item Harmonizing the packet lengths with the channel's coherence times;
			\item Overcoming the significant signal attenuations;
			\item Matching the bit error rate to the application;
			\item Accommodating the low underwater bandwidth;
			\item Resisting the everyday security threats.
		\end{itemize}
		
		In the following subsections, the layers shown in Table~\ref{Table_Network_Archit} along with their relevant protocols are discussed, starting from the application and transport layers.
		
		\subsubsection{Underwater application layer} \label{SubSection_UW_App_Layer}
			The underwater application layer in IoUT is mainly responsible for identifying each individual object (i.e. sensor's id, sensor type, sensor location, etc.) and then, gathering data, processing information, and delivering commands. Gathering data in application layer includes sensing, tracking, recording, and streaming of (live) data, which will be discussed in Section~\ref{Section_Big_Data}. Data processing on the other hand, will take the gathered data and process them, which results in the information sought. This topic will be discussed in more detail in Section~\ref{Section_Data_Proc}. Some actuators may also belong to this layer to react to the environment as instructed by the commands of scientists, visitors, policy-makers, educators, owners, or even from a ML model. The action of these actuators is recordable and controllable through the IoUT.
			
			The existing IoT ecosystem is well equipped with application layer protocols, covering all technological demands in the diverse IoT applications.
			These protocols are comprehensively reviewed by Al-Fuqaha \textit{et al.} \cite{Al-Fuqaha2015}. By studying these IoT protocols it is readily seen that not all of them are suitable for IoUT applications. Some of these protocols rely on wideband operation, while others have protocol headers that are redundant in underwater communications, without any significant effect on the link reliability.
			
			For example, while the HyperText Transfer Protocol (\href{https://tools.ietf.org/html/rfc2616}{HTTP}) is widely used in the application layer of the worldwide web to deliver web-pages, its internal separation into header section and body section is not necessary in a simple sensor data transmission in IoUT. By this means, some suitable protocols to be used in the underwater application layer include but are not limited to the Inter-Module Communication (\href{https://www.lsts.pt/toolchain/imc}{IMC}), Serial Line Interface Protocol (\href{https://tools.ietf.org/html/rfc1055}{SLIP}), eXtensible Markup Language (\href{https://www.w3.org/XML/}{XML}), Message Queuing Telemetry Transport (\href{https://mqtt.org/}{MQTT}), Data Distribution Service (\href{https://www.rti.com/products/dds-standard}{DDS}), and \href{http://modbus.org/docs/PI_MBUS_300.pdf}{Modbus} \cite{Al-Fuqaha2015}. Another protocol is the Teletype Network (\href{https://en.wikipedia.org/wiki/Telnet}{Telnet}), which is considered as the origin of modern application layer protocols \cite{MusilovaS2020}. The simple 8-bit and text-oriented data structure of Telnet makes it suitable for IoUT applications.
			
		\subsubsection{Underwater transport layer}
			As already mentioned in Section~\ref{SubSection_Sensors_Intro}, the sensors, cameras, and actuators in the application layer are typically considered as underwater endpoint nodes, which offer a specific service in IoUT. Considering the nature of data produced by any specific service, the transport layer will be responsible for splitting the data into packets prior to its transmission through the network gateway. This layer also takes into account the importance of data order and any potential data loss.
			
			In this regard, \href{https://tools.ietf.org/html/std7}{TCP} is a reliable transport layer protocol that has widely been used in the Internet.
			In contrast to the concept of a \textit{reliable link}, which has been studied in Section~\ref{SubSection_Link_Reliability}, a \textit{reliable protocol} (e.g. TCP) is defined as a protocol with handshaking, error detection, and error correction capabilities. Based on this definition, reliable protocols typically bear more overhead than unreliable protocols, and as a result, grow in the required channel capacity. This is not a desirable feature in IoUT, where a few lost sensor data packets can be acceptable, in return of freeing up some channel resources. Therefore, User Datagram Protocol (\href{https://tools.ietf.org/html/rfc768}{UDP}) is more recommended in underwater applications, as it does not use acknowledgments at all.
			
			In addition to the TCP and UDP, the Datagram Congestion Control Protocol (\href{https://tools.ietf.org/html/rfc4340}{DCCP}) is another transport layer protocol, which is message-oriented and reliable. DCCP offers congestion notification and congestion control that is useful in UWSNs.
			
			Finally, nodes and gateways in the transport layer may have wired or wireless modes.
			In the case of wireless modes, the undersea nodes should be as multi-modal as possible for considering the node size, weight, complexity, price, and energy consumption \cite{LiY2018}. Relying on multi-mode operation, access to the IoUT networks is possible by relying on multiple marine telecommunication technologies (i.e. acoustic, electromagnetic, and optical).
			
		\subsubsection{Underwater network layer}
			The major tasks of the network layer include bi-directional data packet handling between individual endpoints as well as protocol translation between adjacent layers (i.e. transport and data link).
			These responsibilities are addressed in the network layer by implementing both the Internet protocol and the data routing.
			
			The Internet Protocol version 4 (\href{https://tools.ietf.org/html/rfc791}{IPv4}) and Internet Protocol version 6 (\href{https://tools.ietf.org/html/rfc8200}{IPv6}) are the main network layer Internet protocols. However, IPv6 is not recommended in IoUT, as it is more verbose than IPv4 owing to its excessive header. Additionally, if DCCP is used in the previous transport layer, using Explicit Congestion Notification (\href{https://tools.ietf.org/html/rfc3168}{ECN}) protocol in the network layer is recommended to allow end-to-end notification of network congestion, without dropping packets.
			
			Alongside the Internet protocols, there are network layer routing protocols that specify how the routers communicate and collaborate with each other. An efficient data routing scheme is an essential part of any UWSN. In addition to the discussed algorithms and protocols in Section~\ref{SubSection_Routing}, some other well-designed routing protocols for underwater communications are E-CARP \cite{Zhou2016} and E-CBCCP \cite{Rani2017}.
			
			The IP and routing protocols in the network layer, along with the scheduling protocols in the next section,
			provide the access of the sink nodes (e.g. surface-floated buoys, ships, satellites, etc.) to the endpoint nodes (e.g. underwater sensors), through the mid-layer nodes \cite{Kao2017}. These mid-layer nodes are not always required, but in the case of long-distance communication and/or in the presence of high signal attenuation, they play an essential role in extending the range.
			
			Given the natural difficulties involved in recharging those endpoint nodes, mid-layer nodes, and sink nodes, 
			underwater networks should have an efficient energy management strategy. Many previous studies have investigated different techniques of minimizing the energy consumption in underwater networks \cite{Cao2018, Rani2017, Xu2017, Koseoglu2017}.
			
			A plausible power conservation strategy was introduced by Koseoglu \textit{et al.} \cite{Koseoglu2017} for underwater networks. This strategy assigns more network-layer resources (i.e. route accesses and scheduling) to the nodes and sinks that have to transmit over a longer distance and/or have less efficient physical layers. By involving cross-layer optimization of the channel access rate instead of the separate optimization of the layers, a $66\%$ less energy consumption per successfully transmitted bit could be achieved. However, this study only considered a single-hop scenario, where the nodes do not perform multi-hopping. Whilst the holistic optimization has to resolve the design-dilemma of having more short hops at a higher delay or less hops at a higher power. Therefore, this solution needs further investigation of multi-hopping and to strike a trade-off in terms of the above dilemma.
			
		\subsubsection{Underwater data link layer}
			As already mentioned in the previous section, data link layer supports the access of underwater sensors to the surface stations by appropriately sending/receiving data from the physical layer, as well as coordinating the data departures. This data exchange with the physical medium requires a conversion between data frames and electronic signals, depending on the underlying communication technology (i.e. acoustic, electromagnetic, and optical).
			
			The nodes and sinks that use any of these technologies, usually share the physical telecommunication channel, and therefore, scheduled access to the shared channel becomes a mandatory action in every UWSN \cite{Xu2017}. This necessity gives rise to the concept of Media Access Control (MAC) in the data link layer. The scheduled access, which is provided by MAC helps to avoid or manage data collision and to complete reliable data transmission in real-time. Access regulation in MAC becomes even more important, when connecting the existing UWSNs to the IoUT infrastructure.
			
			In this regard, MAC scheduling protocol can be either collision-free \cite{AlfouzanF2019} or contention-based \cite{Feng2018}.
			Due to the unique characteristics of underwater acoustic channels, employing contention-based MAC protocols in IoUT are inefficient and costly. In contrast, collision-free MAC protocols achieve higher performance by consuming lower energy and offering better network throughput \cite{AlfouzanF2019}.
			Time-Division Multiple Access (TDMA), which is also known as a slotted MAC is the mainstream collision-free scheduling protocol in most of the UWSN standards \cite{ChenYJ2007}.
			Based on the network topology, as discussed in Section~\ref{Subsection_Network_Topology}, in the slotted MAC, a cluster-head undertakes the responsibility of data frame coordination. On the other hand, using ALOHA as a contention-based MAC protocol is also common in UWSN applications \cite{Koseoglu2017}, which requires no coordinator node at all.
			
			Additionally, due to the peculiar features of underwater communication channels (as will be discussed in Section~\ref{SubSection_UW_Physical_Layer}), conventional MAC designs and protocols cannot work well under water \cite{Jiang2018}. Hence, in UWSN, designers should minimize the MAC and Ethernet frame size, while satisfying all other protocol requirements.
			Accordingly, research has been conducted to propose compliant underwater MAC protocols including TRMAC \cite{ZhaoR2019}, UWOR-MAC \cite{Chen2014} and EAST \cite{Xu2017}.
			
			In addition to shortening the MAC frame size in the light of the limited underwater bandwidth, we also have to consider limited underwater energy resources.
			The problem of providing a scheduling service for sensory data transmission while conserving energy is addressed in \cite{Xu2017} by designing an energy-aware scheduling protocol, which is a compelling solution that offers a reasonable network throughput at the time of writing. However, this method has a complex probability-based model, which relies on time-consuming initialization. As a further constraint, the probabilistic state transition mechanism has to be regularly updated. Nonetheless, these adjustments can be carried out automatically.
			
			Among all the underwater communication challenges, the lack of continuous network connectivity between nodes and sinks is the main reason of the notorious unreliability of links. In computer-networking terminology, communicating links associated with intermittent connectivity are termed as \textit{delay tolerant networks} \cite{Al-Zaidi2018}. To address this, Li \textit{et al.} \cite{LiC2016} introduce a delay tolerant MAC protocol for underwater wireless networks relying on m-fold repeated transmission of every packet. The value of $m$ is optimized with respect to the successful transmission probability determined by the channel's reliability.
			
			Another important IoUT data link layer protocol termed as \textit{Dolphin} is introduced by Fujihashi \textit{et al.} \cite{Fujihashi2018}, which supports high-quality video transmission, based on the video-on-demand streaming model. This protocol transmits multiview videos using acoustic technology. To handle multiview video transmission in low data-rate acoustic channels, the underwater systems are categorized into quality-sensitive and delay-sensitive applications, which are supported by the so-called S-Dolphin or G-Dolphin mode of operations. Both of these methodologies exploit time-shifted slot assignments between the underwater video communication pairs to avoid data collisions, albeit they have not been implemented in practice.
			
			
		\subsubsection{Physical layer in IoUT}\label{SubSection_UW_Physical_Layer}
			The nature of the underwater physical layer can be completely different in the case of wired or wireless communications. In a wired communication regime, we can use either cables or optical fibers. The use of cables is limited to wired energy transfer \cite{Perez2017} as well as narrow-bandwidth data transmission in underwater wired sensor networks \cite{Mohamed2010}. On the other hand, the use cases of optical fibers in underwater physical layer communications include but are not limited to:
			
			\begin{itemize}
				\item \textit{Wideband image and video transmission}: As it will be discussed in Section~\ref{SubSection_OneD_TimeSeries_Data}, there are many open access observatories worldwide, which provide live image and video data using their IoUT infrastructure. Some of these observatories (e.g. \href{http://www.obsea.es/}{OBSEA}, \href{https://www.mbari.org/at-sea/cabled-observatory/}{MARS}, and \href{http://www.oceannetworks.ca/}{ONC}) rely on wired data transmission through fiber-optics.
				
				\item \textit{Telephone and Internet delivery worldwide}: In 1988, TAT-8, the first transatlantic fiber-optic cable, was laid between United States, Britain, and France \cite{Carpenter2013}. Nowadays, we rely almost entirely on 550,000 miles of underwater optical fibers for inter-continental Internet transactions, as they are faster and cheaper than satellite communications.
			\end{itemize}
			
			Despite the benefits of using wired communications in IoUT, they have some major drawbacks, resulting in avoiding them as much as possible. Some of these disadvantages include slow and tedious installation, higher implementation and maintenance costs, hard to locate and repair faults, less flexibility as a result of fixed mechanical junctions, being subject to rotate and twist failures in high underwater nodes mobility, lack of security in either case of intentional damage or unauthorized access, being subject to physical and chemical environmental impairments, being subject to shark bite related damages, etc. \cite{Aalsalem2018}.
			
\begin{figure}[!t]
				\centering
				\includegraphics[width=\columnwidth, clip, trim={142 487 142 72}]{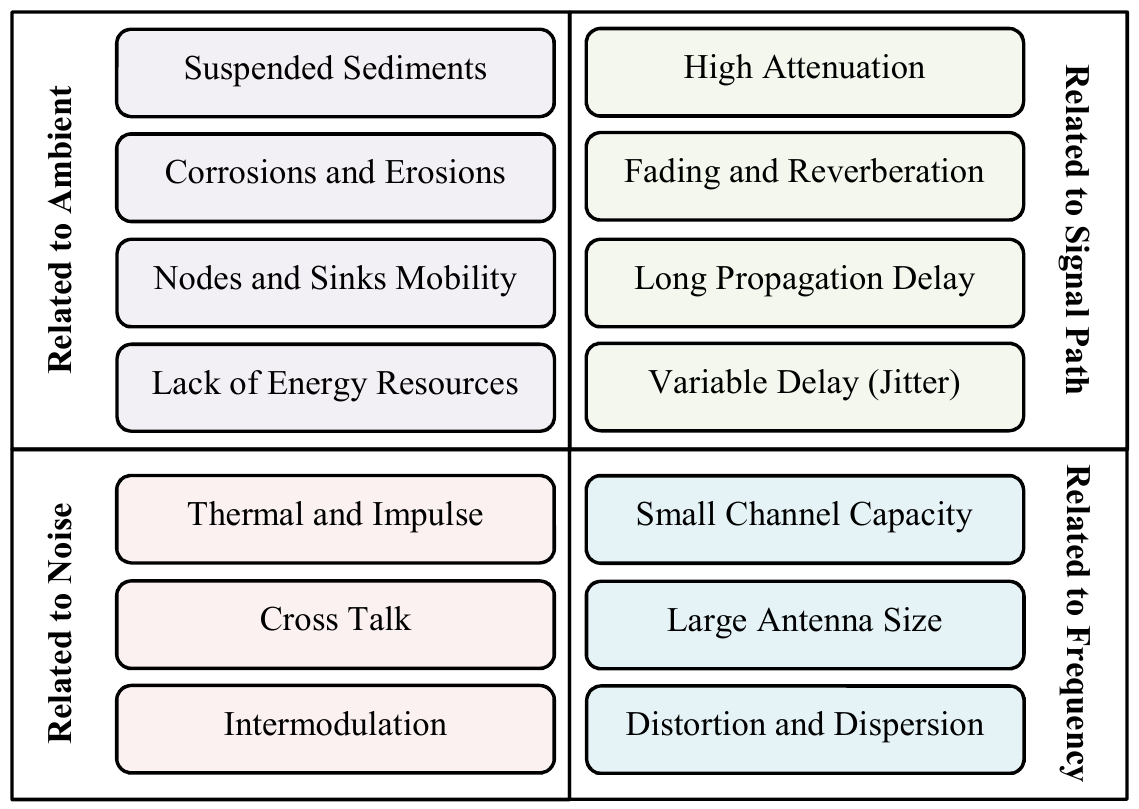}
				\caption{Major signal impairment causes in the underwater physical layer.}
				\label{Fig_InfoG_Phys_Layer_Roughness}
			\end{figure}

			To avoid the above-mentioned disadvantages of wires in the underwater physical layer, wireless communications offer a promising solution. However, underwater wireless signal transmission and wave propagation is more challenging than its aerial/terrestrial counterparts. This is mainly due to the harsh environmental conditions at deep-sea ($>100\,m$), shallow water ($<100\,m$), or even freshwater. Some of the phenomena imposing signal impairments on the underwater physical layer are listed in Fig.~\ref{Fig_InfoG_Phys_Layer_Roughness}.
			
			Each and every challenge shown in Fig.~\ref{Fig_InfoG_Phys_Layer_Roughness} has a different impact on the different underwater applications. For instance, wireless data communication in shallow water is mainly affected by surface noise and multipath fading. By contrast, in the deep-water applications, the attenuation, the narrow-bandwidth, and the high sensitivity to environmental parameters such as temperature are responsible for the most grave signal impairments.
			
			Dut to these challenges, the existing overwater wireless protocols are not readily suitable for undersea applications. Some of these overwater protocols belong to the \href{http://www.ieee802.org/}{IEEE 802} family and to their derivatives (e.g. \href{https://zigbee.org/}{Zigbee}) as well as to the Low-Power Wide Area Networking (LPWAN) protocols (e.g. Long Range (\href{https://www.semtech.com/products/wireless-rf/lora-transceivers}{LoRa}) and Narrowband IoT (\href{ftp://ftp.3gpp.org/tsg_ran/TSG_RAN/TSGR_69/Docs/RP-151621.zip}{NB-IoT})).
			These protocols normally use electromagnetic waves as their transmission medium. Hence, they will have a very short communication range in underwater environments.
			Accordingly, the use cases of these electromagnetic wave based protocols would be limited to a few specific scenarios in IoUT, like:
			\begin{itemize}
				\item \textit{Short-range underwater transmission}: As depicted in the right figure in Table~\ref{Table_bps_and_Att}, electromagnetic waves can only propagate for tens of centimeters in seawater and for couple of meters in freshwater. This propagation distance increases, if the carrier frequency is reduced. Therefore, whenever no long-distance transmission is required, the aforementioned protocols can still be employed in IoUT solutions \cite{BergmannN2014}.
				
				\item \textit{Long-range overwater transmission}:  \href{http://folk.ntnu.no/alfredse/Forslag\%20til\%20prosjektoppgaver\%20hoesten\%202019.htm}{LoRa gateway buoys} and \href{https://internetofbusiness.com/smru-nb-iot-seals-data/}{Underwater NB-IoT} solutions of the sea mammal research unit rely on LPWAN for their overwater data submission.
				Another potential solution is to construct an \textit{ad hoc} point-to-point link, based on the ITU-R M.1842-1 recommendation. This approach uses the VHF radio solution of the Automatic Identification System (AIS) for conveying data packets \cite{Al-Zaidi2018}.
			\end{itemize}
			
			It is worth noting that wired protocols - such as the IEEE 802.3 for LAN (Ethernet) - can still be employed in IoUT projects after modest reconfiguration \cite{Lentz2007}.

			Nevertheless, if long-distance underwater wireless communications are required, acoustic waves constitute the only viable option. An efficient underwater acoustic physical layer IoUT protocol has been conceived by Marchetti and Reggiannini \cite{Marchetti2016}. They achieved a beneficial signal-to-noise ratio gain, by taking advantage of the energy gleaned from all propagation paths, instead of only the strongest path.
			
			As will be discussed in Section~\ref{SubSection_Sim_Tools}, to overcome the challenges of wireless communications portrayed in Fig.~\ref{Fig_InfoG_Phys_Layer_Roughness}, we basically need to undertake an analytical or numerical approach, to computationally evaluate data transmission throughout the network. Taking this mathematical approach will consequently empower us to optimize our communication system in the favor of having more reliable IoUT infrastructure. However, analyzing and optimizing the physical layer of the network requires us to provide the simulation tool with an appropriate underwater physical channel model, which is the topic of the next section.
			
	\subsection{Underwater Channel Modeling}\label{SubSection_Channel_Model}
		As discussed in the previous sections, the underwater physical layer exhibits different behavior in response to both different propagation modes and different channel types. For example, the signal attenuation was found out in Section~\ref{SubSection_UW_Comm} to be calculated differently for electromagnetic, acoustic, and optical carriers. The physical layer investigated in Section~\ref{SubSection_UW_Physical_Layer} also exhibits different behavior in the case of wired and wireless channels. Regardless of the channel types and propagation technologies, we require to have an appropriate channel model to have a better insight on underwater data transmission. This proper model can then be used for predicting the performance of our communication system, designing the optimum underwater location of nodes, and decreasing the overall energy consumption of the system, before its actual deployment \cite{Qarabaqi2013}.
		
		The channel models of wired and wireless networks vary with the choice of the communication technology. In other words, it is not possible to design an accurate channel model for universal employment in every application \cite{WangJ2018}. On the other hand, a feasible model has to undergo some degrees of simplifications, which is strongly correlated with the requirements of the problem itself \cite{ZhouJ2019}. As an example, we might simply neglect the water salinity in an acoustic channel model, while this simplification is not possible in underwater electromagnetic propagation.
		
		Considering the aforementioned factors, channel models are somehow tailored to their own specific use cases. Accordingly, one usually needs to modify and combine the main features of multiple stand-alone models to come up with a dedicated solution to a specific application. For instance, each of the following works considered a specific aspect of the acoustic wave propagation underwater (i.e. layered ocean water, particles in seawater, and the slope in seabed) and neglecting the rest, for simplification. 
		
		\begin{itemize}
			\item It is very common in underwater channel modeling to consider a constant phase velocity throughout the entire medium (isovelocity). But in the contribution of Naderi \textit{et al.} \cite{Naderi2018}, the non-isovelocity acoustic wave propagation in shallow-water environments has been evaluated. They split ocean water into multiple layers of piece-wise linear sound speed profiles. Afterward, they use traditional sound wave propagation techniques to extract the time-variant channel transfer function. The complexity level of their solution can be adjusted by increasing the number of the linear layers. This could easily result in instability or inaccuracy in case of extra-thin or extra-thick layers, respectively.
			
			\item Relatively long wavelength of acoustic waves compared to the floating particles in ocean, makes them less vulnerable to backscatter and forward scatter phenomena. This is not the same for optical waves, as will be studied in Section~\ref{SubSection_UW_Image_Data}. In a relevant study, Zhou \textit{et al.} \cite{ZhouJ2019} have considered the impact of the scattering particles on the propagation characteristics of acoustic waves in underwater environments. They randomly distribute those particles on an assumptive rectangular cross section of ocean. Despite their interesting method, two main drawbacks exist in their statistical approach. First, their evaluation is only a 2D vertical cross section of seawater, neglecting realistic 3D sections. Second, they do not consider the stochastic size of the scattering particles.
			
			\item It is generally considered safe to assume the sea surface as a flat plane in underwater acoustic wave propagation, but this is not the case for the seafloor. In the work done by Naderi \textit{et al.} \cite{Naderi2017}, the up/down slopes in seabed have been taken into account. This important consideration is particularly essential in the case of shallow waters in the presence of coral reefs and plants, and cannot be neglected. 
		\end{itemize}
		
		As mentioned, in order to have a desired customized and simplified underwater channel model, relevant existing models can be combined and unified. The majority of acoustic channel models (including all of the above-mentioned) follow the same mathematical approach. This approach relies on the superposition of a single Euler wave ray, over all the possible propagation paths in the time- or frequency-domain. This superposition relying on resolving the time-variant channel impulse responses can be written as \cite{Naderi2018},
		
		\begin{equation}\label{Eq_AcousticSuperP}
		h(t)=\sum_{n} c_n e^{-j\left( 2\pi f_0 t + \theta_n \right)} \delta\left( t-\Delta t_n \right)\,,
		\end{equation}
		\begin{equation}
		\theta_n=\dfrac{2 \pi f_0}{v_0} \Delta L_n\,,
		\end{equation}
		where $f_0$ and $v_0$ denote the frequency and the acoustic wave speed in water, $\Delta t_n$ and $\Delta L_n$ stand for the propagation delay and the difference in propagation path length, $\theta_n$ is the propagation phase shift, and $\delta(\cdot)$ is the Dirac delta function.
		Here in \eqref{Eq_AcousticSuperP}, $c_n$ encapsulates both the initial gain and the subsequent losses by incorporating the concept of statistical random variables to model the stochastic environmental parameters (i.e. effects of scattering particles, motion-induced Doppler shifting, location uncertainty of nodes, changes in received power, dynamic seabed topology, and other natural variables). Selection of these environmental parameters for encapsulating in $c_n$ differs from one use case to another, depending on the scale of these parameters' effect in every case \cite{Qarabaqi2013}.
		
		The same concept of acoustic superposition in \eqref{Eq_AcousticSuperP} can be used to model electromagnetic channels. But as discussed in Section~\ref{SubSection_UW_Comm}, high-attenuating electromagnetic waves are not common in underwater data transmission and therefore, they are not evaluated here. In a clear contrast, using line-of-sight optical communications as well as fiber-optic channels (as discussed in Section~\ref{SubSection_UW_Physical_Layer}) are very popular in high data-rate underwater communications \cite{Zeng2017}. In this regard, the subject of optical fibers is out of the scope of this article and the reader is encouraged to follow it up from other resources \cite{LiW2016}. Meanwhile, studying the line-of-sight optical channel properties in underwater applications will be conducted in Section~\ref{SubSection_UW_Image_Data}.
		
		As stated at the beginning of this section, the channel modeling is the backbone of every IoUT network simulation to evaluate its proper operation and to optimize its parameters \cite{LiW2019, Liang2018, WangC2018}. One of the very basic parameters in IoUT networks that relies on channel modeling is the node location and it is referred to in parlance as \textit{network topology}, which will be discussed in the next section. Later on, in Section~\ref{SubSection_Sim_Tools} we will review software tools that offer wires and wireless underwater channel simulation, along with network topology design and optimization.
		
	\subsection{IoUT Network Topologies}\label{Subsection_Network_Topology}
\begin{figure}[!t]
			\centering
			\subfloat[Tree network topology \newline \textcolor{white}{.} \;\; \cite{XuG2019, Domingo2012, Perez2017, Jiang2018, Rani2017}]{\includegraphics[width=0.41\columnwidth, clip, trim={1.7cm 22.7cm 16.3cm 1.8cm}]{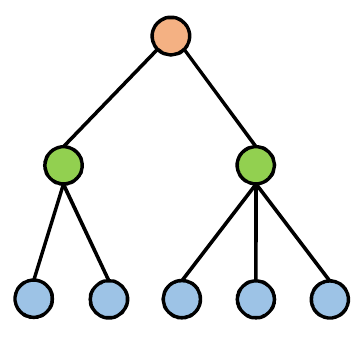}%
				\label{Fig_Sch_Netw_Topology_A}}
			\hfil
			\subfloat[Partially connected mesh \newline \textcolor{white}{.} \;\; \cite{XuG2019, Kao2017, LiY2018, Zeng2017, Jiang2018}]{\includegraphics[width=0.41\columnwidth, clip, trim={6.0cm 22.7cm 12cm 1.8cm}]{./Graphics/Fig_Sch_Netw_Topology.pdf}%
				\label{Fig_Sch_Netw_Topology_B}}
			\caption{Reported network topologies in IoUT applications.}
			\label{Fig_Sch_Netw_Topology}
		\end{figure}

		By utilizing TCP/IP as the reference model for the IoUT network architecture, almost all of the known network topologies can be used in gateway-enabled underwater applications \cite{XuG2019, Jiang2018}. Two of these topologies (i.e. tree and mash), which have high potential and are suitable for implementation in IoUT are shown in Fig.~\ref{Fig_Sch_Netw_Topology}.
		
		As discussed previously in Section~\ref{SubSection_UW_Physical_Layer}, the frequency bandwidth is limited undersea and energy is hard to harvest. Therefore, advanced distributed topologies are rarely used in underwater applications, instead, the conventional tree and mesh topologies are dominant. The tree topology is typically used in small networks relying on one-way protocols \cite{Domingo2012, Perez2017, Jiang2018, Rani2017}. The tier (client-server) negotiation in this topology is based on a request and response process.
		On the other hand, the mesh topology tends to be the option of choice in sophisticated networks to meet the high traffic requirements, while using all available signal routes and frequency bands efficiently. This ensures that the limited channel capacity of underwater environments is efficiently exploited \cite{Kao2017, LiY2018, Zeng2017, Jiang2018}. 
		
		It is worth noting that any signal route of Fig.~\ref{Fig_Sch_Netw_Topology} has its own reliability quantified in terms of its overall bit error rate in any typical IoUT network. This assists us in beneficially choosing a suitable network topology for any application, given its specific bit error rate target.
		
		However, choosing a suitable network topology, designing a multi-layer communication network, and selecting an appropriate protocol for each layer of its architecture is not always a straightforward task. This job requires a tedious work to continuously design, test, debug, integrate, and deploy the newer versions of the network. It will get even more challenging in the harsh underwater conditions. To make this process a lot easier, some simulation tools are available, which can be used in underwater network design as well as layer-wise protocol testing. These software tools will be studied in the next section.
		
	\subsection{Underwater Network Simulation Tools}\label{SubSection_Sim_Tools}
		In IoUT applications, an improperly configured network can cost a lot of time and money \cite{Atyabi2018}. The best way to avoid these costs is to simulate and test every aspect of the network, prior to its first deployment and to any potential re-configuration. To do so, a wide range of open-source and commercial simulation tools are available. Table~\ref{Table_Simulation_Tools} lists a number of the tools that have the capability to model underwater channels and protocols.
		
		All of the tools listed here are discrete-event network simulators, which model a system as a sequence of event triggers, assuming that no changes occur between consecutive events. This highly-repeatable simulation method is easier than analytical models to implement \cite{Naumann2019}. The level of network support in these tools varies from wired and wireless \textit{ad hoc} networks to more advanced cellular and satellite communications.
		
		The first item in this table is the well-known Network Simulator (ns) software and its ns-1, ns-2, ns2-Miracle, and ns-3 upgrades, which have been used by more than 600 academic as well as industrial institutions in more than 50 countries worldwide \cite{Petrioli2015}. Some other tools that rely on ns for their underwater network simulations have also been studied in this table (i.e. AQUASIM, SUNSET, WOSS, and DESERT Underwater). They are all equipped with underwater channel modeling, acoustic transmission, and relevant protocols.
		
		Based on their license level (i.e. commercial, research, education, or military), the simulation tools in Table~\ref{Table_Simulation_Tools} can undertake a verity of analysis and design tasks, including \cite{Dorathy2018}:
		\begin{itemize}
			\item Channel modeling;
			\item Layer-wise protocol design and testing;
			\item Data transmission, traffic, and routing;
			\item Spatially arranging the network topology;
			\item Simulating network elements and commercial devices;
			\item Working with heterogeneous network technologies;
			\item Analyzing the overall network performance.
		\end{itemize}
		
		Additionally, with two exceptions for AQUASIM and OMNeT++, all other software products in this table can perform IoUT network emulation, in conjunction with available hardware as well as software solutions. In a contrast to network simulators that are an abstract mathematical description of a communication system, network emulators are virtual networks that mimic the behavior and also mirror the functionality of a real network, to which the end-systems will physically connect \cite{Veltri2019}.
		
		Network emulation provides a more realistic characterization of the underlying network. Therefore, it is highly recommended in IoUT applications to continue with an emulation step, right after network simulation and just before their actual deployment. By undertaking this simulation middle-step we can assure the quality of the IoUT network, by testing and troubleshooting it in a close to real-life environment.
		
		\begin{table*}[!t]
			\caption{Simulation Tools for Analyzing and Designing Underwater Communication Networks and Protocols}
			\label{Table_Simulation_Tools}
			\centering
			\begin{tabular}{@{}m{0.12\textwidth}@{} @{}m{0.01\textwidth}@{}
					@{}m{0.1\textwidth}@{} @{}m{0.01\textwidth}@{}
					@{}m{0.04\textwidth}@{} @{}m{0.01\textwidth}@{}
					@{}m{0.07\textwidth}@{} @{}m{0.01\textwidth}@{}
					@{}m{0.07\textwidth}@{} @{}m{0.01\textwidth}@{}
					@{}m{0.05\textwidth}@{} @{}m{0.01\textwidth}@{}
					@{}m{0.09\textwidth}@{} @{}m{0.01\textwidth}@{}
					@{}m{0.03\textwidth}@{} @{}m{0.01\textwidth}@{}
					@{}m{0.03\textwidth}@{} @{}m{0.01\textwidth}@{}
					@{}m{0.31\textwidth}@{}}

				\toprule
				\centering\rotatebox[origin=c]{90}{\parbox[c]{3.2cm}{\textcolor{white}{..}\textbf{Software Name}}} &&
				\centering\rotatebox[origin=c]{90}{\parbox[c]{3.2cm}{\textcolor{white}{..}\textbf{Owner}}} &&
				\centering\rotatebox[origin=c]{90}{\parbox[c]{3.2cm}{\textcolor{white}{..}\textbf{Initial Release}}} &&
				\centering\rotatebox[origin=c]{90}{\parbox[c]{3.2cm}{\textcolor{white}{..}\textbf{Based On}}} &&
				\centering\rotatebox[origin=c]{90}{\parbox[c]{3.2cm}{\textcolor{white}{..}\textbf{Operating System(s)}}} &&
				\centering\rotatebox[origin=c]{90}{\parbox[c]{3.2cm}{\textcolor{white}{..}\textbf{Language(s)}}} &&
				\centering\rotatebox[origin=c]{90}{\parbox[c]{3.2cm}{\textcolor{white}{..}\textbf{Main Application(s)}}} &&
				\centering\rotatebox[origin=c]{90}{\parbox[c]{3.2cm}{\textcolor{white}{..}\textbf{Real-Time $\mid$ Emulation}}} &&
				\centering\rotatebox[origin=c]{90}{\parbox[c]{3.2cm}{\textcolor{white}{..}\textbf{Open-Source}}} &&
				\centering \textbf{\newline\newline\newline\newline Other Specifications} \tabularnewline
				
				
				\midrule
				
				Network \newline Simulator \newline (\href{https://www.nsnam.org/}{ns}) \newline\newline\newline\newline&&
				U.S. National \newline Science \newline Foundation \newline\newline\newline\newline&&
				1990s \newline\newline\newline\newline\newline\newline&&
				
				\textcolor{white}{.......}$-$ \newline\newline\newline\newline\newline\newline&&
				
				Windows, \newline Linux \newline\newline\newline\newline\newline&&
				
				TcL, \newline C++ \newline\newline\newline\newline\newline&&
				Research \newline and \newline Education \newline\newline\newline\newline&&
				\textcolor{white}{.}\includegraphics[width=0.015\textwidth,clip,keepaspectratio]{./Graphics/Image_Tick.png} \newline \textcolor{white}{.}\textthreequartersemdash\smallskip \newline \textcolor{white}{.}\includegraphics[width=0.012\textwidth,clip,keepaspectratio]{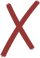} \newline\newline\newline&&
				
				\textcolor{white}{.}\includegraphics[width=0.015\textwidth,clip,keepaspectratio]{./Graphics/Image_Tick.png}\smallskip \newline\newline\newline\newline\newline&&
				$\bullet$ It has had four releases at the time of writing \newline \textcolor{white}{.}\; (i.e. ns-1, ns-2, ns2-Miracle, and ns-3). \newline
				$\bullet$ All releases are object-oriented discrete-event \newline \textcolor{white}{.}\; simulators. \newline
				$\bullet$ ns products are reliable basis for many other \newline \textcolor{white}{.}\; open-source products (e.g. AQUASIM, \newline \textcolor{white}{.}\; SUNSET, WOSS, DESERT Underwater, etc.).
				\tabularnewline
				
				\midrule
				
				Aquatic \newline Simulator \newline (\href{https://www.eawag.ch/en/department/siam/software/}{AQUASIM}) \newline&&
				Swiss Federal \newline Inst. of Aquatic \newline Sci. \& Tech. \newline&&
				2013 \newline\newline\newline&&
				
				ns-2 \newline\newline\newline&&
				
				Windows, \newline Linux, \newline Mac \newline&&
				
				TcL, \newline C++ \newline\newline&&
				Aquatic \newline Research \newline and \newline Education &&
				\textcolor{white}{.}\includegraphics[width=0.015\textwidth,clip,keepaspectratio]{./Graphics/Image_Tick.png} \newline \textcolor{white}{.}\textthreequartersemdash\smallskip \newline \textcolor{white}{.}\includegraphics[width=0.012\textwidth,clip,keepaspectratio]{./Graphics/Image_Cross.png} &&
				
				\textcolor{white}{.}\includegraphics[width=0.015\textwidth,clip,keepaspectratio]{./Graphics/Image_Tick.png}\smallskip \newline\newline&&
				$\bullet$ End-users can define the spatial topology of \newline \textcolor{white}{.}\; the communication network.
				\newline\newline\tabularnewline
				
				\midrule
				
				Sapienza Univ. \newline Net. Framework \newline for Underwater \newline Sim., Emu., and \newline Testing (\href{http://senseslab.di.uniroma1.it/greencastalia/48-sunset-sapienza-university-networking-framework-for-underwater-simulation-emulation-and-real-life-testing}{SUNSET})&&
				Sapienza \newline University of \newline Rome, Italy \newline\newline&&
				2012 \newline\newline\newline\newline&&
				
				ns2-Miracle \newline\newline\newline&&
				
				Windows, \newline Linux \newline\newline\newline&&
				
				TcL, \newline C++ \newline\newline\newline&&
				Underwater \newline Acoustic \newline Sensor \newline Networks \newline&&
				\textcolor{white}{.}\includegraphics[width=0.015\textwidth,clip,keepaspectratio]{./Graphics/Image_Tick.png} \newline \textcolor{white}{.}\textthreequartersemdash\smallskip \newline \textcolor{white}{.}\includegraphics[width=0.015\textwidth,clip,keepaspectratio]{./Graphics/Image_Tick.png} \newline&&
				
				\textcolor{white}{.}\includegraphics[width=0.015\textwidth,clip,keepaspectratio]{./Graphics/Image_Tick.png}\smallskip \newline\newline\newline&&
				$\bullet$ It provides a complete toolset of pre-\newline \textcolor{white}{.}\; deployment (i.e. simulation and emulation) \newline \textcolor{white}{.}\; and deployment-time (i.e. real-time on sea) \newline \textcolor{white}{.}\; testing of the communication protocols \cite{Petrioli2015}.
				\newline\tabularnewline
				
				\midrule
				
				World Ocean \newline Simulation Sys. \newline (\href{http://telecom.dei.unipd.it/ns/woss/}{WOSS}) \newline&&
				University of \newline Padova, Italy \newline\newline&&
				2009 \newline\newline\newline&&
				
				\textcolor{white}{.......}$-$ \newline\newline\newline&&
				
				Windows, \newline Linux, \newline Mac \newline&&
				
				C++ \newline\newline\newline&&
				Underwater \newline Acoustic \newline Networks \newline&&
				\textcolor{white}{.}\includegraphics[width=0.015\textwidth,clip,keepaspectratio]{./Graphics/Image_Tick.png} \newline \textcolor{white}{.}\textthreequartersemdash\smallskip \newline \textcolor{white}{.}\includegraphics[width=0.012\textwidth,clip,keepaspectratio]{./Graphics/Image_Cross.png} &&
				
				\textcolor{white}{.}\includegraphics[width=0.015\textwidth,clip,keepaspectratio]{./Graphics/Image_Tick.png}\smallskip \newline\newline&&
				$\bullet$ It integrates with the existing underwater \newline \textcolor{white}{.}\; channel simulators and improves their \newline \textcolor{white}{.}\; simulation results.
				\newline\tabularnewline
				
				\midrule
				
				\href{http://desert-underwater.dei.unipd.it/}{DESERT \newline Underwater} \newline\newline\newline&&
				University of \newline Padova, Italy \newline\newline\newline&&
				2012 \newline\newline\newline\newline&&
				
				ns2-Miracle \newline\newline\newline&&
				
				Windows, \newline Linux, \newline Mac \newline\newline&&
				
				TcL, \newline C++ \newline\newline\newline&&
				Underwater \newline Networks \newline\newline\newline&&
				\textcolor{white}{.}\includegraphics[width=0.015\textwidth,clip,keepaspectratio]{./Graphics/Image_Tick.png} \newline \textcolor{white}{.}\textthreequartersemdash\smallskip \newline \textcolor{white}{.}\includegraphics[width=0.015\textwidth,clip,keepaspectratio]{./Graphics/Image_Tick.png} \newline&&
				
				\textcolor{white}{.}\includegraphics[width=0.015\textwidth,clip,keepaspectratio]{./Graphics/Image_Tick.png}\smallskip \newline\newline\newline&&
				$\bullet$ The full name is Design, Simulate, \newline \textcolor{white}{.}\; Emulate and Realize Test-beds for Underwater \newline \textcolor{white}{.}\; network protocols. \newline
				$\bullet$ The provided libraries extend the ns2-miracle \newline \textcolor{white}{.}\; simulator.
				\tabularnewline
				
				\midrule
				
				\href{https://www.riverbed.com/au/products/steelcentral/steelcentral-riverbed-modeler.html}{Riverbed \newline Modeler} \newline\newline\newline&&
				Riverbed \newline Technology \newline\newline\newline&&
				1986 \newline\newline\newline\newline&&
				
				\textcolor{white}{.......}$-$ \newline\newline\newline\newline&&
				
				Windows, \newline Linux \newline\newline\newline&&
				
				C++ \newline\newline\newline\newline&&
				Commercial \newline Networks \newline\newline\newline&&
				\textcolor{white}{.}\includegraphics[width=0.015\textwidth,clip,keepaspectratio]{./Graphics/Image_Tick.png} \newline \textcolor{white}{.}\textthreequartersemdash\smallskip \newline \textcolor{white}{.}\includegraphics[width=0.012\textwidth,clip,keepaspectratio]{./Graphics/Image_Cross.png} \newline&&
				
				\textcolor{white}{.}\includegraphics[width=0.012\textwidth,clip,keepaspectratio]{./Graphics/Image_Cross.png}\smallskip \newline\newline\newline&&
				$\bullet$ Riverbed Modeler was formerly known as \newline \textcolor{white}{.}\; \href{www.opnet.com}{OPNET} Modeler Suite. \newline
				$\bullet$ It is an object-oriented discrete-event simulator. \newline
				$\bullet$ Huge libraries of accurate network hardware \newline \textcolor{white}{.}\; models are available \cite{Ghaleb2017}.
				\tabularnewline
				
				\midrule
				
				\href{https://www.scalable-networks.com/qualnet-network-simulation}{QualNet} \newline\newline\newline&&
				SCALABLE \newline Network \newline Technologies \newline&&
				2000 \newline\newline\newline&&
				
				\href{https://www.scalable-networks.com/history}{GloMoSim} \newline\newline\newline&&
				
				Windows, \newline Linux \newline\newline&&
				
				C++ \newline\newline\newline&&
				Commercial \newline Networks \newline\newline&&
				\textcolor{white}{.}\includegraphics[width=0.015\textwidth,clip,keepaspectratio]{./Graphics/Image_Tick.png} \newline \textcolor{white}{.}\textthreequartersemdash\smallskip \newline \textcolor{white}{.}\includegraphics[width=0.012\textwidth,clip,keepaspectratio]{./Graphics/Image_Cross.png} &&
				
				\textcolor{white}{.}\includegraphics[width=0.012\textwidth,clip,keepaspectratio]{./Graphics/Image_Cross.png}\smallskip \newline\newline&&
				$\bullet$ The statistical graphing tools in QualNet can \newline \textcolor{white}{.}\; display hundreds of metrics collected during \newline \textcolor{white}{.}\; the network simulation process.
				\newline\tabularnewline
				
				\midrule
				
				\href{https://www.tetcos.com/}{NetSim} \newline\newline\newline&&
				TETCOS \newline\newline\newline&&
				2002 \newline\newline\newline&&
				
				\textcolor{white}{.......}$-$ \newline\newline\newline&&
				
				Windows \newline\newline\newline&&
				
				Java \newline\newline\newline&&
				Research \newline and \newline Military \newline&&
				\textcolor{white}{.}\includegraphics[width=0.015\textwidth,clip,keepaspectratio]{./Graphics/Image_Tick.png} \newline \textcolor{white}{.}\textthreequartersemdash\smallskip \newline \textcolor{white}{.}\includegraphics[width=0.012\textwidth,clip,keepaspectratio]{./Graphics/Image_Cross.png} &&
				
				\textcolor{white}{.}\includegraphics[width=0.012\textwidth,clip,keepaspectratio]{./Graphics/Image_Cross.png}\smallskip \newline\newline&&
				$\bullet$ NetSim is an end-to-end packet level network \newline \textcolor{white}{.}\; simulator.
				\newline\newline\tabularnewline
				
				\midrule
				
				Objective \newline Modular Net. \newline Testbed in C++ \newline (\href{https://omnetpp.org/}{OMNeT++}) \newline\newline\newline\newline\newline\newline&&
				OpenSim \newline\newline\newline\newline\newline\newline\newline\newline\newline&&
				1997 \newline\newline\newline\newline\newline\newline\newline\newline\newline&&
				
				\href{https://omnest.com/}{OMNEST} \newline\newline\newline\newline\newline\newline\newline\newline\newline&&
				
				Windows, \newline Linux, \newline Mac \newline\newline\newline\newline\newline\newline\newline&&
				
				C++ \newline\newline\newline\newline\newline\newline\newline\newline\newline&&
				Kernel \newline of Other \newline Simulators \newline\newline\newline\newline\newline\newline\newline&&
				\textcolor{white}{.}\includegraphics[width=0.012\textwidth,clip,keepaspectratio]{./Graphics/Image_Cross.png} \newline \textcolor{white}{.}\textthreequartersemdash\smallskip \newline \textcolor{white}{.}\includegraphics[width=0.012\textwidth,clip,keepaspectratio]{./Graphics/Image_Cross.png} \newline\newline\newline\newline\newline\newline&&
				
				\textcolor{white}{.}\includegraphics[width=0.015\textwidth,clip,keepaspectratio]{./Graphics/Image_Tick.png}\smallskip \newline\newline\newline\newline\newline\newline\newline\newline&&
				$\bullet$ Objective Modular Network Simulation Testbed \newline \textcolor{white}{.}\; (\href{https://omnest.com/}{OMNEST}) is the commercially distributed \newline \textcolor{white}{.}\; product of the OpenSim. \newline
				$\bullet$ OMNeT++ does not contain protocol models \newline \textcolor{white}{.}\; and for this, relies on external frameworks, \newline \textcolor{white}{.}\; such as \href{https://inet.omnetpp.org/}{INET}. \newline
				$\bullet$ There are other simulators like \href{https://github.com/boulis/Castalia}{Castalia}, which \newline \textcolor{white}{.}\; are based on OMNeT++ and are useful in \newline \textcolor{white}{.}\; simulating low-power underwater wireless \newline \textcolor{white}{.}\; networks.
				\tabularnewline
				
				\midrule
				
				Java \newline Simulator \newline (\href{https://sites.google.com/site/jsimofficial/}{J-Sim}) \newline\newline\newline\newline&&
				Several \newline Contributors \newline\newline\newline\newline\newline&&
				2001 \newline\newline\newline\newline\newline\newline&&
				
				\textcolor{white}{.......}$-$ \newline\newline\newline\newline\newline\newline&&
				
				Windows, \newline Linux, \newline Mac \newline\newline\newline\newline&&
				
				TcL, \newline Java \newline\newline\newline\newline\newline&&
				Research \newline and \newline Education \newline\newline\newline\newline&&
				\textcolor{white}{.}\includegraphics[width=0.015\textwidth,clip,keepaspectratio]{./Graphics/Image_Tick.png} \newline \textcolor{white}{.}\textthreequartersemdash\smallskip \newline \textcolor{white}{.}\includegraphics[width=0.012\textwidth,clip,keepaspectratio]{./Graphics/Image_Cross.png} \newline\newline\newline&&
				
				\textcolor{white}{.}\includegraphics[width=0.015\textwidth,clip,keepaspectratio]{./Graphics/Image_Tick.png}\smallskip \newline\newline\newline\newline\newline&&
				$\bullet$ J-Sim was formerly known as JavaSim. \newline
				$\bullet$ It is an object-oriented discrete-event simulator. \newline
				$\bullet$ J-Sim uses much smaller memory footprint to \newline \textcolor{white}{.}\; carry out a similar simulation, compared to \newline \textcolor{white}{.}\; ns-2 \cite{Ghaleb2017}. \newline
				$\bullet$ Visualization of the network operations in \newline \textcolor{white}{.}\; J-Sim is relies on the ns-2 network animator.
				\tabularnewline
				
				\bottomrule
			\end{tabular}
		\end{table*}
		
		It is worth mentioning that the accuracy of the simulation tools discussed in this section, as well as of the numerical channel models in Section~\ref{SubSection_Channel_Model} depend on the level of details provided in their inputs. In addition, relying on mathematical estimations, approximations, and statistics to model various aspects of a given natural environment, usually results in limited accuracy. For example, the number of natural parameters that can affect a typical underwater acoustic or optical channel can be huge. Some of these unpredictable parameters include noise, distortion, uncertain electronic specifications, seabed topography, water currents, microscopic organisms, distribution and movement of species, etc. Therefore, the simulation and modeling results should always be supported by experimental fieldworks to discover their weaknesses and to validate the minimum acceptable accuracy of their outputs \cite{KisacikR2019}.
		
		Finally, in addition to the general-purpose software packages and frameworks seen in Table~\ref{Table_Simulation_Tools}, there are some other simulation tools, like Network Security Simulator (\href{http://www.nessi2.de}{$NeSSi^2$}), which are particularly designed to research and evaluate network's resistance against various security threats. These cybersecurity tests in IoUT are important, because underwater networks are often deployed in remote locations with difficult access that makes them vulnerable to cyberattacks.
		
		In this regard, a safe underwater network with a secure communication protocol along with a strong data encryption algorithm can protect us against security threats, as discussed in the next section.
		
	\subsection{IoUT Network Security}\label{SubSection_Network_Security}
		Underwater networks are usually left unattended for long time intervals, exposing them to diverse potential threats. Due to the wide data broadcasting range of these networks, their physical channel (both wired or wireless) is prone to eavesdropping, hence cannot be kept secure. On the other hand, secure communication is crucial in many IoUT infrastructures \cite{Kao2017} (e.g. harbor security, coastal defense, etc.), where unauthorized access can have serious consequences.
		
		The security threats in underwater communications can be divided into passive and active attacks, in accordance to the actions taken by the intruders. In a passive attack, intruders simply try to obtain data, while in an active attack, they will attempt to inject, alter, or delete data by introducing malicious nodes \cite{YangG2019}. In either case, the concept of authenticated data access offers a high-level solution, which can prevent IoUT network both from passive data-extraction and from active data-injection. A correct implementation of this concept will assure
		data confidentiality, data integrity, data freshness, user authentication, and non-repudiation of origin.
		
		However, correctly implementing the concept of authenticated data access can be a challenging task in IoUT, when considering the mobility of network nodes, resource constraints in terms of communication as well as computation capacity, and shortage in energy supplies. This challenging task can be addressed by using cryptographic primitives as well as secure communications, when designing the network \cite{Editorial2019}, which will be discussed in the following subsections.
		
		\subsubsection{Cryptography}
			Cryptography in network security is the science of encrypting data with a symmetric or asymmetric key, prior to its submission, and in a way that only the authorized receiver can decrypt it \cite{YangG2019}.
%
%
			Despite the fact that asymmetric cryptography is more suitable in high-security applications, using symmetric encryption, wherever possible, has substantial benefits for resource-constrained underwater environments. This is because, at a given security level, a symmetric key will be significantly shorter than an asymmetric key \cite{JiangS2019}, which consequently eliminates the need for public key broadcasting as well as sophisticated algorithm deployment and execution.
%
%
			
			Some symmetric cryptography algorithms that are suitable in underwater applications are \href{https://www.schneier.com/academic/blowfish/}{Blowfish} and the associated fish products Twofish and Threefish, as well as the Advanced Encryption Standard (\href{https://nvlpubs.nist.gov/nistpubs/FIPS/NIST.FIPS.197.pdf}{AES}), Rivest Cipher 5 (\href{http://people.csail.mit.edu/rivest/Rivest-rc5rev.pdf}{RC5}), and Rivest Cipher 6 (\href{https://people.csail.mit.edu/rivest/pubs/RRSY98.pdf}{RC6}).
			
			For asymmetric cryptography on the other hand, using Elliptic-Curve Cryptography (\href{https://tools.ietf.org/html/rfc6090}{ECC}) is recommended, as it is energy-efficient with less computation complexity than other popular algorithms, e.g. Rivest–Shamir–Adleman (\href{https://www.di-mgt.com.au/rsa_alg.html}{RSA}) \cite{JiangS2019}. ECC algorithm has also been shown to require smaller keys to provide equivalent security.
			
			In both symmetric and asymmetric cryptography, a common key has to be shared between the communicating pairs. While the classical key-sharing protocols are common in overwater communications, they cannot be readily adopted by IoUT because of their network infrastructure requirements, complexity, and high resource demands \cite{XuM2020}. By contrast, the physical layer key generation seems to offer a promising solution in IoUT. This method relies on the local generation of keys between two legitimate nodes, without broadcasting it in the channel. For example, a multi-party secret key generation scheme is proposed for underwater acoustic channels in \cite{XuM2020}. Their highly confidential key extraction scheme is designed for the circumstances of underwater multipath and Doppler effects. Additionally, the simulation results of \cite{XuM2020} demonstrate a secure communication against the key leakage in active attacks.
			
		\subsubsection{Secure communications}\label{SubSection_Secure_Comm}
			In contrast to cryptographic primitives, which secure an IoUT network by encrypting any data prior to its submission, covert communication techniques as well as secure protocols may also be applied \cite{JiangS2019}. The idea behind covert communication in IoUT is to reduce the average SNR in the propagation medium, which consequently reduces the chances of eavesdropping. Covert communication in underwater applications may rely on the following techniques \cite{YangT2008}:
			
			\begin{itemize}
				\item Using an acoustic phased-arrays to direct data propagation beam toward a desired receiver only;
				\item Using frequency-hopping to repeatedly change the carrier frequency in an unpredictable fashion;
				\item Using spread spectrum and code division multiplexing to spread the transmitted energy over the frequency band;
				\item Using analog network coding with concurrent signal transmissions to have intentionally interfering signals.
			\end{itemize}
			
			In addition to the above methods, camouflaged data transmission is another creative technique of covert communications. Generally speaking, the eavesdropping systems classify engine sounds, environmental noises, and biological voices as ocean noise and try to filter them \cite{JiajiaJ2020}. Here, the underwater acoustic communications can benefit from shaping their output signals in accordance to the time and frequency features of a targeted chaotic signal. Recent studies have verified the efficiency and feasibility of this scheme in IoUT, when the transmitted signal is camouflaged with marine mammal voices (e.g. killer whales) \cite{JiajiaJ2020}, ship-radiated noise \cite{HuangS2020}, etc.
			
			On the other hand, secure protocols can also be employed by IoUT architectural layers for ensuring network-threat-resilient underwater communications. These layer-specific as well as application-oriented secure protocols can be summarized as \cite{JiangS2019}:
			
			\begin{itemize}
				\item \textit{Application layer protocols}: Remote Authentication Dial-in-User Service (\href{https://tools.ietf.org/html/rfc2865}{RADIUS}) and \href{https://tools.ietf.org/html/rfc6733}{Diameter};
				
				\item \textit{Transport layer protocols}: Secure Socket Layer (\href{https://tools.ietf.org/html/rfc6101}{SSL}) and Transport Layer Security (\href{https://tools.ietf.org/html/rfc8446}{TLS});
				
				\item \textit{Network layer protocols}: IP Security (\href{https://tools.ietf.org/html/rfc4301}{IPSec}), Secure FLOOD (SeFLOOD) \cite{Dini2011}, Resilient Pressure Routing (RPR) \cite{Zuba2014}, and Reputation-based Channel Aware Routing Protocol (R-CARP) \cite{Capossele2015};
				
				\item \textit{Data Link and physical layers protocols}: Extensible Authentication Protocol (\href{https://tools.ietf.org/html/rfc3748}{EAP}), IEEE 802.1X EAP over LAN (\href{https://1.ieee802.org/security/802-1x/}{EAPOL}), and IEEE 802.1AE MAC Security (\href{https://1.ieee802.org/security/802-1ae/}{MACsec}).
			\end{itemize}
			All the aforementioned secure protocols, along with the network protocols studied in Section~\ref{SubSection_Network_Architect}, can be categorized under the distributed network management methodologies. In contrast to these distributed controls, a novel centralized approach will be introduced in the next section, which can boost the security of IoUT networks.
			
	\subsection{Software-Defined Networks in IoUT}\label{SubSection_SDN}
\begin{figure}[!t]
			\centering
			\subfloat[Traditional distributed network]{\includegraphics[width=0.45\columnwidth, clip, trim={1.57cm 19.65cm 15.77cm 2.44cm}]{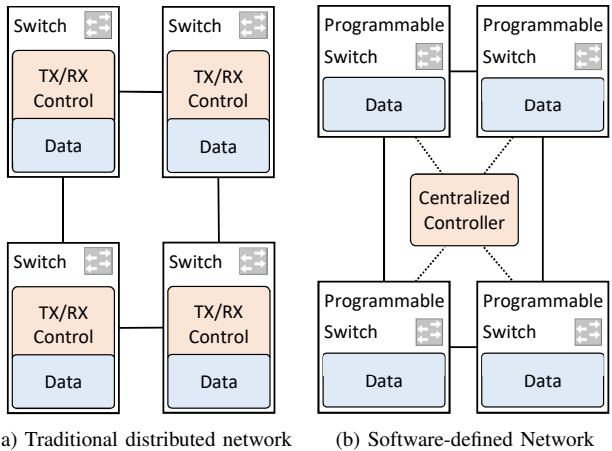}%
			}
			\hfil
			\subfloat[Software-defined Network]{\includegraphics[width=0.45\columnwidth, clip, trim={6.25cm 19.65cm 11.1cm 2.44cm}]{./Graphics/Fig_Sch_SDN.pdf}%
			}
			\caption{Traditional vs. software-defined network management.}
			\label{Fig_Sch_SDN}
		\end{figure}

		As seen in Fig.~\ref{Fig_Sch_SDN}, there is a novel centralized management technique, termed as \textit{Software-Defined Network (SDN)}, which is practically beneficial for IoUT \cite{LuoH2018}. To elaborate, in contrast to distributed networks, where every network node makes its own decisions based on the locally configured routing tables, in SDNs, centralized control nodes continuously monitor the network's traffic flow and manage the network's configuration in order to improve its overall performance.
		
		In contrast to the isolated UWSNs having inflexible hardware-dependent architectures, SDNs tend to be more suitable for the IoUT infrastructure. Some of the beneficial features of SDNs are:
		
		\begin{itemize}
			\item Programmable and hardware-independent;
			\item Automatically manageable and reconfigurable;
			\item Multi-modal (i.e. multi-technology) operation;
			\item Scalable and adaptable in terms of size and topology;
			\item Service-oriented (instead of application-oriented) design to offer multi-application functionality;
			\item Resource-sharing capability;
			\item Convenient troubleshooting.
		\end{itemize}
		
		In addition to the above-mentioned attributes, SDNs also support spectrum-aware communications, which is an outstanding benefit in underwater networks \cite{LuoH2018}. Dynamic spectrum management in SDNs is based on its cognitive radio procedure, which includes informed spectrum sensing, intelligent decision-making, and dynamic spectrum sharing. According to these features, a variety of underwater use cases are reported, including software-defined acoustic modems \cite{Dol2017}, underwater cognitive networks \cite{LuoY2014}, UWSN virtualization \cite{Akyildiz2016}, and underwater sensor cloud \cite{Srimathi2013}.
		
		SDNs are also proved to have a better protection against malicious attacks, as a benefit of their global view over the entire system \cite{JiangS2019}. Basically, the centralized controller node can monitor and detect the irregular behavior of other nodes as well as suspicious data transmissions. It can then mitigate these abnormalities by instructing the network to fix the problem, or simply by ignoring the untrusted nodes.
		
		SDNs have been also conceptualized to provide a multi-mode opto-acoustic technology in IoUT \cite{CelikA2020}. Celik \textit{et al.} have used optical base stations as well as acoustic access points in their infrastructural underwater network. These opto-acoustic mid-layer nodes are used to provide coverage to other endpoint nodes in their vicinity, and to serve as a gateway to the rest of the network \cite{CelikA2020}. They have used SDN as a management enabling technique, both for 'hybridizing' optics and acoustics and for adapting their proposed architecture to the dynamically changing underwater environment. However, this envisioned contribution requires further practical verification, before being accepted as the basis for the future \textit{ad hoc} or infrastructural IoUT designs. Such an experimental study can also reveal the usefulness of this architectural paradigm in support of the IoT and IoUT integration.
		
		Another conceptual hybrid network architecture is proposed by
		Lal \textit{et al.} \cite{Lal2017} for enhancing the security of underwater acoustic networks.
		This hybrid architecture incorporates SDN, context-awareness, and cognition to boost the physical layer security. The efficiency of their proposed architecture in tackling underwater security situations was evaluated for various attacks including jamming, wormhole and ID-spoofing, blackhole and sinkhole, as well as replay and resource exhaustion attacks. Similar to the contribution in \cite{CelikA2020}, this conceptual hybrid SDN can inspire further practical investigations to secure a network, while maintaining reliable communications.
		
		All the approaches discussed above are introduced to improve the IoUT network architecture and protocols to consequently enhance underwater communication and to address its challenges. However, if the devices and nodes in IoUT are capable of local computations at the edge of the network, the underwater network traffic will be significantly reduced and further improvements can be made to the IoUT ecosystem. The concept of these edge computing nodes and their beneficial effects on underwater data traffic will be discussed in the next section.

	\subsection{Edge Computing in IoUT}\label{SubSection_Edge_Computing}
		The concept of edge computing was originally introduced in IoT as an alternative to cloud computing. As the terminology suggests, in edge computing, the endpoint (edge) devices perform all or part of the required computations, so the need for data transfer and communication becomes less of a challenge. This sparse data transfer is ideal for the IoUT ecosystem, which suffers from the hostile communication medium. Within IoUT, edge computing can be defined as a distributed and elastic computing paradigm, in which computing is predominantly carried out in the edge-devices such as, underwater endpoints, mid-layer nodes, and data acquisition tools. In the absence of edge computing capability, the processing should be performed on local computers, servers, or by the centralized clouds \cite{LiuJ2018}, all of which require shuttling data back and forth, hence rendering it unsuitable for IoUT \cite{WangT2020}.
		In edge computing, devices have to expand their capabilities and in addition to data collection and communication, engage in data processing as well. This feature will shift the services from a single centralized point to numerous distributed nodes and closer to the physical world. The major advantages of using such a decentralized edge computing technology in the IoUT ecosystem are data-rate reduction, latency reduction, and prompt inner-network decisions making \cite{WangT2020, CaiS2019, WangQ2020}.

		However, to turn these advantages into reality, edge computing has to tackle significant challenges. As stated in Section~\ref{SubSection_UW_Physical_Layer}, the energy resources of the edge-devices are limited. Furthermore, sustainable power cannot be readily delivered to the processing units of underwater applications. Even though edge computing has been the subject of extensive research in conventional IoT applications, it has its own specific challenges in IoUT. Below, we briefly review two of the IoT edge computing paradigms presented in \cite{Porambage2018} that are suitable for IoUT.
		
		\begin{itemize}
			\item \textit{Mobile Cloud Computing (MCC)}: It is defined in IoT ecosystem, as the combination of cloud computing and mobile nodes to execute computational- as well as storage-heavy mobile applications (e.g. machine learning) in the cloud. MCC can offer rich computational resources to scarce-resource underwater applications. However, its relevance to IoUT is low because of the long propagation distance from the distributed BMD sources to the remote cloud servers \cite{DiX2018}, as well as the narrow underwater bandwidth and limited access to energy.
			Another drawback of MCC is its service accessibility, which is via Internet connection only. This is in contrast to other edge computing paradigms that can offer a direct access \cite{Porambage2018}.
			
			\item \textit{Mobile Edge Computing (MEC)}: The acronym MEC represents mobile edge computing and multi-access edge computing, with the latter one being more inclusive. According to the descriptions offered by the European Telecommunications Standards Institute (\href{https://www.etsi.org/technologies-clusters/technologies/multi-access-edge-computing}{ETSI}), the MEC technique provides cloud computing capabilities at the edge of the network, with close proximity to the end-users (i.e. underwater endpoint nodes).
			In IoUT, the edge of network represents both the land stations at the edge (e.g. cellular towers, data centers, Wi-Fi routers, etc.) \cite{Safavat2019} as well as the overwater sink nodes (i.e. floating buoys, floating vehicles, exploration platforms, etc.).
			These edge gateways will carry out some preliminary and short-term tasks, before handing their results to the cloud servers for more sophisticated and resource-intensive analysis \cite{ChiangM2016}.
			MEC techniques invoked for the IoUT can offer low latency and reliability. Their other advantages in IoUT include their support of wired communications as well as their facilitated direct access (i.e. no Internet connection is required) \cite{Porambage2018}.
			
			
			
		\end{itemize}
		
		Considering the above, MCC and MEC are two of the IoT edge computation techniques that are practically realizable in IoUT, as shown in Table~\ref{Table_Edge_Location}.
		Among them, MEC requires lower bandwidth, lower network traffic, low-latency, and low-power underwater operation, and more reliable access to underwater sensors and cameras in the IoUT \cite{CaiS2019}. Hence, MEC networks constitute the most attractive edge computing paradigm in IoUT applications \cite{CaiS2019, WangQ2020}. They can perform computations in the sink or mid-layer edge-devices that are placed or have access to above water, where for example solar energy may be harvested.
		
		\begin{table}[!t]
			\renewcommand{\arraystretch}{1.5}
			\caption{Location of The Processing Unit in Different IoUT Edge Computing Paradigms}
			\label{Table_Edge_Location}
			\centering
			\includegraphics[width=\columnwidth, clip, trim={40 528 305 65}]{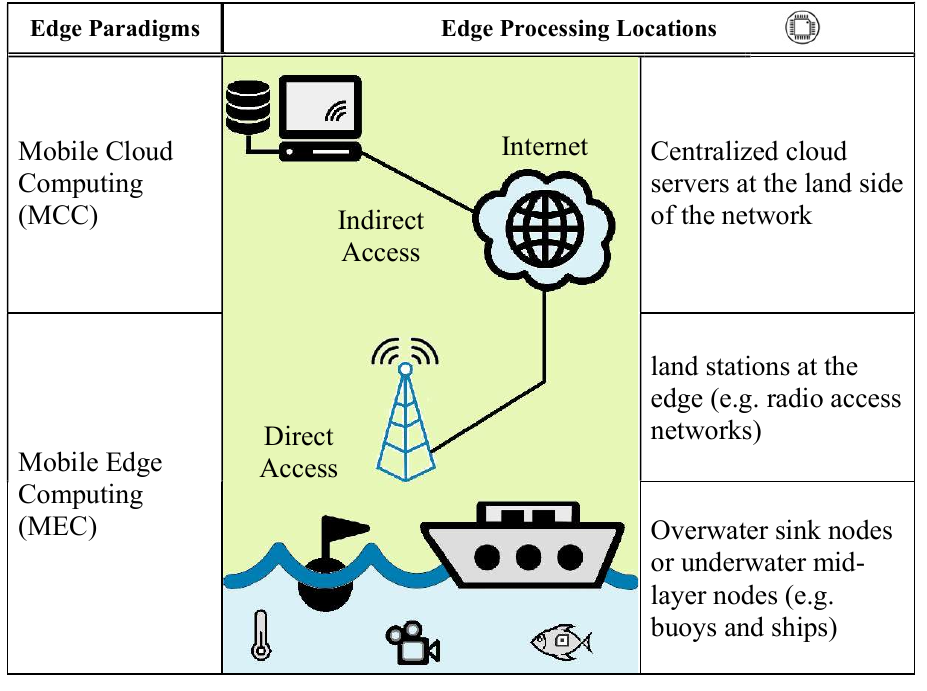}
		\end{table}
		
	\subsection{Section Summary}
		In this section, we studied the essential topics involved in developing the IoUT. We discussed various sensors in the IoUT and learned about their important features and roles in data collection. We then pointed out the major challenges in the IoUT domain and learned that electromagnetic, acoustic and optical signal attenuation significantly affect the IoUT communication and its link reliability. We also learned how to measure and improve the link reliability. Next, we discussed the IoUT architectural model and its layer-wised protocol stack, and learned how the TCP/IP model can be adapted to the IoUT architecture. In order to be able to gain better insight into the network performance and to predict its reliability, network modeling was discussed. The lessons learned were that various channel models result in different levels of complexity and they should be carefully selected, based on the application concerned.
		
		The IoUT network topology was the next topic we covered, where we learned that the tree and partially connected mesh are the most popular topologies in IoUT applications. We also discussed the underwater network simulation tools and how they can be used to facilitate the implementation of the defined protocol stack. We provided a list of these simulation tools, which can be helpful to the community when designing and analyzing underwater communication networks and protocols. Network security in IoUT was another salient topic discussed. We learned that there is a paucity of literature on this important issue, even though it is essential in critical scenarios such as harbor security. Another important topic studied in this section was software-defined networks in IoUT. We learned how it can help with the management of IoUT networks. Finally, we discussed edge computing in IoUT and learned that it may be even more important for IoUT devices to have edge computing capability than for their IoT counterparts, mainly due to the more challenging communications in underwater scenarios.
		
		The IoUT discussed in this section generates a vast amount of data, which may be referred to in parlance as \textit{big marine data}. In the following section, we discuss big marine data and cover its associated data sources, data collection tools, and data processing methods in the oceanic and underwater domain.
		
\section{Big Marine Data}\label{Section_Big_Data}
	This section is dedicated to underwater data types, which are envisioned to have the lion-share in the IoUT data transactions of the near future. We will also cover data acquisition tools, localization as well as tracking, and will introduce some ready-to-use big ocean sensory, imagery, video, and geographic databases, which are useful to researchers and practitioners in the IoUT Domain.
	
	BMD can be succinctly defined as the vast amount of heterogeneous data collected from marine fields. The main characteristics of BMD are temporally long and spatially vast coverage, diverse nature of the data sources (e.g. sensors, cameras, tags, aerial remote sensing), and multi-disciplinary data types (i.e. physical, chemistry, biological, environmental, economical, etc.) \cite{HuangD2015}.
	
	The BMD system components and processing stages are as follows \cite{Ahmed2017}:
	
	\begin{enumerate}
		\item \textit{Acquisition}: Involves collecting raw data.
		
		\item \textit{Transportation and security}: Requires the data to be encrypted and transferred across different communication media to its target storage, while considering its reliability and security.
		
		\item \textit{Storage and privacy}: Deals with policies around data storage (legal concerns and users privacy) and its archival requirements (file formats, retention lifecycle, and replication).
		
		\item \textit{Special-purpose processing}: For complex datasets, bespoke software packages are required for searching, pre-processing by filtering and cleaning, recognition and labeling, post-processing and visualization of results, and updating.
		
		\item \textit{Exploitation and leveraging}: Ensures that users gain benefit in terms of increased revenue from their data. Some benefits of underwater data exploitation are that of monitoring the water's vital cleanliness, help new businesses to grow, support experts by providing informative bespoke data, track worldwide maritime transportations, and protect the environment.
	\end{enumerate}
	
	From the above five BMD system components, \textit{transportation and security} was already discussed in Sections~\ref{Section_IoUT}. The third component, i.e. \textit{storage and privacy} covers legal policies and data handling issues, which are beyond the scope of this paper. Below, we review the \textit{data acquisition} stage and later in sections~\ref{SubSection_Data_Processing_Platforms} and \ref{SubSection_Ocean_Data_Application} we will discuss some aspects of \textit{data leveraging} stage in more details. Additionally, discussions on \textit{data processing} will be provided in Section~\ref{Section_Data_Proc}.
	
	\subsection{Marine Data Acquisition}
		Data acquisition, which is the first component of any BMD solution can be discussed in three different stages including data gathering, data aggregation, and data fusion.
		
		\subsubsection{Data gathering}
		Data gathering in IoUT can be performed using a variety of tools, some of which are listed in Table~\ref{Table_DA_Tools_Usage}. The data acquisition tools in this table are divided into vehicles and primary data sources. Any vehicle in this table can be considered as a data acquisition tool, if and only if it is equipped with one or more primary data sources and tools (e.g. mounted cameras, hydrophones, sensors).
		
		\begin{table*}[!t]
			\setlength{\extrarowheight}{-3pt}%
			\renewcommand{\arraystretch}{1.2}
			\caption{List of Data Acquisition Tools in Underwater Applications with the Percentage of Use of Each Item in Related ML Articles}
			\label{Table_DA_Tools_Usage}
			\centering
			\begin{tabular}{@{}m{0.01\columnwidth}@{} @{}m{0.35\columnwidth}@{} @{}m{0.05\columnwidth}@{} @{}m{0.1\columnwidth}@{} @{}m{0.05\columnwidth}@{} @{}m{0.7\columnwidth}@{} @{}m{0.05\columnwidth}@{} @{}m{0.36\columnwidth}@{}}
			\toprule
			& \textbf{Data Acquisition \newline Tools} && \textbf{Share in ML} && \textbf{Types of Data Acquisition Tools} && \textbf{\quad Used in ML Articles \smallbreak \colorbox{PyBlue}{\textcolor{white}{Before}} and \colorbox{PyOrange}{\textcolor{black}{After}} 2014} \\
			
			\midrule
			
			& \textbf{Vehicles} && $50\%$ &&
			\multirow{2}{*}{\shortstack[l]{Remotely Operated Vehicles (ROV) \\ \quad $\bullet$ Free Swimming \\ \quad $\bullet$ Bottom Crawling \\ \quad $\bullet$ Structurally Reliant \\ \quad $\bullet$ Towed System \\ \quad $\bullet$ Hybrid Remotely Operated Vehicles}} &&
			\includegraphics[width=0.31\columnwidth, clip, trim={67 700 450 70}]{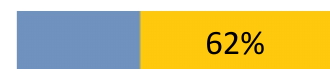} \\
			& \rule{0pt}{40pt} && && && \\
			
			& && $45\%$ &&
			\multirow{2}{*}{\shortstack[l]{Autonomous Underwater Vehicles (AUV) \\ \quad $\bullet$ Underwater AUVs \\ \quad $\bullet$ Underwater Gliders (Seaglider) \\ \quad $\bullet$ Autonomous Submersibles }} &&
			\includegraphics[width=0.31\columnwidth, clip, trim={67 700 450 70}]{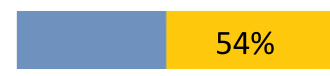} \\
			& \rule{0pt}{19pt} && && && \\
			
			& && $5\%$ &&
			\multirow{2}{*}{\shortstack[l]{Human-Occupied Vehicles \\ \quad $\bullet$ Submersibles (civilian research submarines) \\ \quad $\bullet$ Atmospheric Diving Suits }} &&
			\includegraphics[width=0.31\columnwidth, clip, trim={67 700 450 70}]{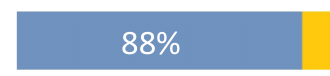} \\
			& \rule{0pt}{15pt} && && && \\
			
			\midrule
			
			& \textbf{\newline Primary \newline Data Sources} && $65\%$ &&
			\multirow{2}{*}{\shortstack[l]{Sensors \\ \quad $\bullet$ Separately Listed in Table~\ref{Table_Primary_Data_Source_Sensor} }} &&
			\includegraphics[width=0.31\columnwidth, clip, trim={67 700 450 70}]{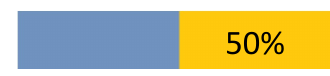} \\
			& && && && \\
			
			& && $35\%$ &&
			\multirow{2}{*}{\shortstack[l]{Cameras \\ \quad $\bullet$ Fixed Cameras (digital still camera) \\ \quad $\bullet$ Baited Remote Videos \\ \quad $\bullet$ Diver Operated Videos \\ \quad $\bullet$ Drifting Underwater Camera System }} &&
			\includegraphics[width=0.31\columnwidth, clip, trim={67 700 450 70}]{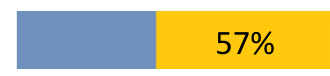} \\
			& \rule{0pt}{33pt} && && && \\
			
			\bottomrule
			\end{tabular}
		\end{table*}
		
		As we will discuss in more details in the forthcoming sections, BMD processing and data leveraging (as the 4\textsuperscript{th} and 5\textsuperscript{th} BMD system components) are not possible without automatic ML approaches. Therefore, a survey was conducted to extract the contribution of data acquisition tools in state-of-the-art ML articles. The survey result is shown in the second and fourth columns of Table~\ref{Table_DA_Tools_Usage}. In the second column, the inter-category contribution of each item is compared to the others. In the vehicular category, the unmanned ROVs and AUVs together were used in about $95\%$ of the relevant ML publications.
		
		In the fourth column of Table~\ref{Table_DA_Tools_Usage}, an item-specific evaluation is performed to quantify the share of each item both in the old and new ML publications (before and after 2014). As can be seen in the table, more than $50\%$ of all underwater research in the area of ML have been carried out after 2014. There is only one exception for human-occupied vehicles. Perhaps, this is due to the recent wide adoption of automated as well as of remote methods and owing to the reduction in academic usage of costly manned vehicles.
		
		\subsubsection{Data aggregation}
		Data aggregation is a statistical data processing stage before higher level calculations and/or before transmission over a band-limited communication channel \cite{Brown2014}. The level of raw data summarization here is to create another set of shortened raw data, by implementing mathematical techniques, such as down-sampling, linear regression, etc.
		
		Data aggregation can be carried out in UWSNs, for example to minimize the bandwidth utilization. It can also reduce the energy consumption in the network-level, by striking a trade-off between low data processing power and high data communication energy. However, using this process in underwater applications may impose some unwanted drawbacks, like \cite{Chhabra2015}:
		
		\begin{itemize}
			\item Increasing the energy consumption and processing requirements in the node-level;
			\item Increasing the overall network latency;
			\item Compromising data accuracy by shortening its volume.
		\end{itemize}
		
		To address these challenges, LEACH, PEGASIS, HEED, and APTEEN, which are some well-known aggregation protocols in wireless sensor networks can be used in UWSNs subject to modest adjustments \cite{Misra2017, Chand2014, Ma2018}. Additionally, heavy computations and high energy demands in data aggregation processes can be mitigated by using edge computing paradigms, as described in Section~\ref{SubSection_Edge_Computing}. 
		
		\subsubsection{Data fusion}\label{SubSection_Data_Fusion}
		The BMD gathered and aggregated may be stored in distinct subsystems or in separate databases. Data fusion is defined as the combination of relevant data from different data sources into an integrated dataset, with the objective of searching for more consistent and more accurate information than that provided by any individual database \cite{Qiu2019}.
		
		Data fusion systems in IoUT mainly rely on edge computing devices as well as on cloud computing servers to handle large amounts of heterogeneous BMD (i.e. sensors, audios, videos, commands, etc.) \cite{QiuT2020}. Three main questions are raised when designing a successful data fusion system in marine applications. The first is the location of data fusing operation, the second is the level of abstraction in the fusion system, and the third is the level of overlapping of the original data.
		
		Considering the first question, data fusion can be performed either in a centralized node or in a distributed network \cite{Qiu2019}. In a centralized scheme, the data fusing node is usually located on overwater edge devices or in an inland facility with good access to energy resources \cite{QiuT2020}. The distributed data fusion will be hierarchically conducted over the entire system, where every cluster-head fuses data from its own cluster nodes, passing them to the next cluster-head. Distributed BMD processing will be discussed in more details in Section~\ref{SubSection_Data_Processing_Platforms}. Considering the second question about abstraction level, the answers offered by \cite{Castanedo2013} are:
		
		\begin{itemize}
			\item \textit{Low level (i.e. sensor fusion)}: If the inputs to our fusion system is the raw data directly gathered from the sensors, cameras, etc.;
			\item \textit{Medium level (i.e. feature fusion)}: If the inputs to our fusion system is the output of feature extracting blocks, as will be discussed in Section~\ref{SubSection_Feature};
			\item \textit{High level (i.e. decision fusion)}: If the inputs to our fusion system is the output of classification or clustering blocks, as will be discussed in Section~\ref{SubSection_ML}.
		\end{itemize}
		
		Low level data fusion is not recommended in BMD, as a result of restricted underwater resources. The only exception is in edge computing, where the raw data is fused together on the edge device, right before extracting their features. On the other hand, medium and high level data fusion can help us with building more sophisticated models or with finding more complex solutions \cite{Nomura2018}.
		
		Considering the third question on the overlapping level, the lessons of \cite{Castanedo2013} are:
		
		\begin{itemize}
			\item \textit{Redundant}: If we have multiple datasets with the same data type, related to the same subject, in the same time interval, by different data acquisition tools or in different sampling time spots;
			\item \textit{Complementary}: If we have multiple datasets of the same data type, related to the same subject, from different angles or in different time intervals;
			\item \textit{Cooperative}: If we have multiple datasets of different data types, related to the same subject.
		\end{itemize}
		
		Fusing redundant or complementary data can increase our confidence over the original datasets. However, based on the expensive data transmission in UWSNs, neither redundant nor complementary levels of overlapping are recommended. In contrast, cooperative data fusion (e.g. couple of sensed parameters from Table~\ref{Table_Primary_Data_Source_Sensor} or associated audio and video data) can obviously increase our knowledge of the subject.
		
		
	\subsection{Marine Data Classification}
		Based on our findings in previous section, which has also been reflected in Table~\ref{Table_DA_Tools_Usage}, cameras and sensors are considered as primary underwater data acquisition tools. Most maritime industrial, research, or observatory project might integrate these data acquisition tools into a wireless or wired sensor network. The sensor networks that are connected to the Internet, all together play a predominant role in the overall IoUT infrastructure. The data collected in the UWSNs within the IoUT may be classified according to its dimension as discussed below. 
		
		\subsubsection{One-dimensional time-series marine data}\label{SubSection_OneD_TimeSeries_Data}
			As described earlier in Sections~\ref{SubSection_Sensors_Intro} and \ref{SubSection_UW_App_Layer}, a sensor is considered as an endpoint node in the application layer, which offers time-series data reception through the Internet according to the end-user preferences. For instance, a biologist might be interested in the water temperature, while an environmentalist might be interested in the water quality.
			
			Conventional marine sensors and marine nano-sensors (as listed in Table~\ref{Table_Primary_Data_Source_Sensor}) measure a variety of oceanic signals and processes within a specific duration. Marine nano-sensors are popular owing to their low power consumption \cite{Zulkifli2018, Kassal2018}. The sequential data provided by sensors and nano-sensors are used in monitoring and surveillance applications to provide long-term and large-scale perception of the environment and to tackle unwanted environmental changes \cite{Hewitt2019}.
			
			Implementation of these sensor nodes in small-scale research-based IoUT structures are repeatedly reported in literature. Some of these contributions are discussed below:
			
			\begin{itemize}
				\item A low-complexity VHF-based IoUT ecosystem is proposed by Al-Zaidi \textit{et al.} \cite{Al-Zaidi2018} for marine data acquisition based on storage devices in the cloud. The proposed structure is equipped with temperature, depth, and wind sensors to produce a near real-time system.
				\item A network of sensor nodes, based on the IEEE 21451 standard is constructed by Adamo \textit{et al.} \cite{Adamo2015} for continuous monitoring of the seawater quality. This system is devised as an IoUT network for making strategic decisions concerning a range of environmental issues.
				\item A low-cost technology for maritime environmental sensing is described by Wright \textit{et al.} \cite{Wright2016}. The technology relies on the IoUT for measuring parameters such as the optical properties of water, ocean temperature, and wave dynamics.
				\item The Great Barrier Reef of Australia is the largest coral system on planet Earth, spanning a distance of over $2300\,km$. Palaniswami \textit{et al.} \cite{Palaniswami2017} constructed an UWSN to capture data from temperature, pressure, and humidity sensors in an IoUT platform, for monitoring the complex ecosystem of the Heron Island in the southern Great Barrier Reef. The big data of sensory time-series collected for their study, has been then analyzed by ML algorithms to detect underwater anomalies. Therefore, \cite{Palaniswami2017} is a notable application of big data analytics in IoUT with prime objective of predicting severe tropical cyclones.
			\end{itemize}
			
			In addition to these small-scale research units, a range of other observatory stations have been established by institutes and organizations, to provide both research-oriented and industrial access to the underwater sensory data. These observatory stations are located all around the planet and they provide a reliable access to BMD through the IoUT.
			
			Here we provide a list of major observatory stations in Table~\ref{Table_Primary_Data_Sources}. The observatories are sorted in an ascending order from the smallest to the largest observatory coverage area. All the entries in this table are open access and indicate active projects with sustainable funding. This guaranties continuous data preparation and update. They are all accessible through a web-form or dedicated software.
			
			\begin{table*}[!t]
				\renewcommand{\arraystretch}{1.5}
				
				
				\caption{Open Access and Active Observatories Worldwide; Providing Up-to-date Primary Data}
				\label{Table_Primary_Data_Sources}
				\centering
				\begin{tabular}{@{}m{0.35\columnwidth}@{} @{}m{0.04\columnwidth}@{} @{}m{0.17\columnwidth}@{} @{}m{0.04\columnwidth}@{} @{}m{1.4\columnwidth}@{}}
				\toprule
				\textbf{Project Name} && \textbf{Funder} && \multicolumn{1}{c}{\textbf{Description}} \\
				\midrule
				
				Expandable Seafloor \newline Observatory (\href{http://www.obsea.es/}{OBSEA}) && University \newline of UPC && A seafloor observatory located in a fishing protected costal area in \textcolor{ForestGreen}{Spain}. It is connected to the coast by a mixture of energy and communication cables to deliver live \textbf{video}, \textbf{acoustic} hydrophone, and \textbf{sensor} data.\\
				
				Woods Hole \newline Oceanographic \newline Institution (\href{http://www.whoi.edu/}{WHOI}) && Non-profit \newline Organization \newline && An independent organization in \textcolor{ForestGreen}{Massachusetts, The USA} that is dedicated to ocean research, exploration, and education. It owns some observatory stations (like \href{http://www.whoi.edu/mvco}{MVCO}) and provides real-time and archived coastal \textbf{sensed} data accessible by the public.\\
				
				Monterey Accelerated \newline Research System (\href{https://www.mbari.org/at-sea/cabled-observatory/}{MARS}) \newline && \href{https://www.mbari.org/}{MBARI} \newline Institution \newline && This cabled observatory provides power and data connections in the deep-sea ($891\,m$ below the surface of Monterey Bay in \textcolor{ForestGreen}{California, The USA}) and provides \textbf{camera} views, \textbf{videos}, \textbf{Acoustics}, and \textbf{sensor} data (O\textsubscript{2}, transmissivity, salinity, etc.) to study the ocean.\\
				
				Lofoten Vester\aa{}len (\href{http://love.statoil.com/}{LoVe}) \newline && Statoil \newline && This station provides a variety of data types, including \textbf{camera} views, active and passive \textbf{acoustics}, and other data gathered from many \textbf{sensors} positioned in the cold-water corals in \textcolor{ForestGreen}{Norway} \cite{Godo2014}.\\
				
				Next Generation of Sensors \newline (\href{http://www.nexosproject.eu/}{NeXOS}) && European \newline Commission && Miscellaneous \textbf{sensor} data from 3 sites around \textcolor{ForestGreen}{Italy and Spain} are gathered (acoustic, carbon, hydrocarbon, fisheries, etc.) \cite{Pearlman2016}.\\
				
				Marine Network \newline (\href{http://www.bsh.de/en/Marine_data/Observations/MARNET_monitoring_network/index.jsp}{MARNET}) && BSH\textsuperscript{\copyright} \newline && A long list of current and historical \textbf{sensor} data (chlorophyll, currents, salinity, turbidity, etc.) for the North sea and Baltic sea of \textcolor{ForestGreen}{Germany} are available to view on-line or to purchase as digital datasets.\\
				
				Channel Coastal \newline Observatory (\href{https://www.channelcoast.org/}{CCO}) && DEFRA \newline Department && The network of regional coastal monitoring program consists of six regional monitoring sites all around the \textcolor{ForestGreen}{UK}. The \textbf{sensory} data gathered in this program include pressure, wave height, tide, etc.\\
				
				Integrated Ocean \newline Observatory System \newline (\href{https://ioos.noaa.gov/}{IOOS}) \newline && NOAA \newline \newline \newline && This project integrates some existing networks of instruments in the western hemisphere (e.g. \href{http://oceanobservatories.org/}{OOI}, \href{http://www.neracoos.org/}{NERACOOS}, \href{https://maracoos.org/}{MARACOOS}) and provides \textbf{videos}, tagged \textbf{camera} views, and sustainable \textbf{sensed} data from regions around \textcolor{ForestGreen}{The USA} including Alaska, Caribbean, California, Gulf of Mexico, Mid-Atlantic, Northeastern-Atlantic, Northwest Pacific, and Southeast Atlantic.\\
				
				Australian Ocean Data \newline Network (\href{https://portal.aodn.org.au/}{AODN}) and \newline Australian Institute of \newline Marine Science (\href{https://www.aims.gov.au/}{AIMS}) && Australian \newline Government \newline \newline && According to AIMS, \textcolor{ForestGreen}{Australia}'s marine territory is the third largest on Earth. AODN and AIMS portals together, provide access to Australian marine and climate \textbf{sensed} data and metadata in coastal areas and the Great Barrier Reef. These databases are collected from miscellaneous sources including \href{http://oceancurrent.imos.org.au/}{IMOS}, NOAA, NASA, ARGO project, etc.\\
				
				Ocean Networks Canada \newline (\href{http://www.oceannetworks.ca/}{ONC}) \newline && University \newline of Victoria \newline && This network continuously delivers data from cabled \textbf{observatories}, \textbf{remote control} systems, and interactive \textbf{sensors} to monitor the west and east coasts of Canada and also the Arctic. This platform merges all individual \textcolor{ForestGreen}{Canadian} observatories.\\
				
				European Multidisciplinary \newline Seafloor and water column \newline Observatory (\href{http://emso.eu/}{EMSO}) && European \newline Union \newline && EMSO consists of 8 observatories and 3 test sites that placed around  \textcolor{ForestGreen}{Europe}, from North East to the Atlantic, through the Mediterranean, to the Black Sea. These observatories are equipped with multiple biological, chemical, and physical \textbf{sensors}, placed along the water column and on the seafloor.\\
				
				EU Marine Observatory \newline and Data Network \newline (\href{http://www.emodnet.eu/}{EMODNet}) && European \newline Union \newline && This \textcolor{ForestGreen}{European} data network includes seven broad disciplinary themes of \href{http://www.emodnet-bathymetry.eu/}{bathymetry}, \href{http://www.emodnet-geology.eu/}{geology}, \href{http://www.emodnet-physics.eu/}{physics}, \href{http://www.emodnet-chemistry.eu/}{chemistry}, \href{http://www.emodnet-biology.eu/}{biology}, \href{http://www.emodnet-seabedhabitats.eu/}{seafloor habitats}, and \href{http://www.emodnet-humanactivities.eu/}{human activities} and provides a wide range of marine \textbf{camera} views, taxonomy, GIS, \textbf{sensed} data, etc.\\
				
				Joint Commission$\ldots$ \newline (\href{https://www.jcomm.info/}{JCOMM}) in \href{http://www.goosocean.org/}{GOOS} && UNESCO \newline && A global scale and \textcolor{ForestGreen}{Intergovernmental} \textbf{sensor} network that integrates many famous observatory programs worldwide, including \href{http://www.argo.net/}{ARGO}, \href{http://www.jcommops.org/dbcp/}{DBCP}, \href{http://www.oceansites.org/}{OceanSITES}, \href{http://www.jcommops.org/sot/}{SOT}, \href{http://www.go-ship.org/}{GO-SHIP}, and \href{http://www.psmsl.org/gloss/}{GLOSS}.\\
				
				\href{https://www.ncdc.noaa.gov/data-access/satellite-data}{NCEI}, Landsat $|$ \newline \href{https://aquarius.nasa.gov/}{Aquarius}, SARAL $|$ \newline \href{https://earth.esa.int/web/guest/data-access/browse-data-products}{CryoSat}, \href{https://www.eumetsat.int/website/home/Data/DataDelivery/index.html}{Jason}, \href{https://www.eumetsat.int/website/home/Data/DataDelivery/index.html}{HY2-A} $|$ \newline IRS && NOAA$|$ \newline NASA$|$ \newline ESA$|$ \newline ISRO && Global \textbf{sensing} of ocean surface by \textcolor{ForestGreen}{USA, European, and Indian} satellites, provides information about sea surface temperature, surface winds and wind stresses, surface Flux analysis, visible and infrared imagery and radiometry, surface salinity, surface topography, etc. Note that \href{https://www.ncdc.noaa.gov/data-access/satellite-data}{NCEI} is a satellite constellation of JPSS, GOES, POES, DMSP.\\
				
				\bottomrule
				\end{tabular}
			\end{table*}
			
			There are some other observatory projects, which are not included in Table~\ref{Table_Primary_Data_Sources}, because they are no longer supported and their databases have become obsolete. For instance, the \href{http://poseidon.hcmr.gr/}{POSEIDON system} in Greece (2008 to 2012) is no longer actively operating, but its atmospheric and marine data are still downloadable through the web. Another example is the \href{http://ijgofs.whoi.edu/}{JGOFS} (Joint Global Ocean Flux Study) project which was funded by the international science council during 1987 to 2003. JGOFS was an international program with participants from more than 20 nations. The rich multi-disciplinary data of this project is also still available to download.
			
			The data provided by any one of the observatory systems in this section, can be affected by environmental noise, outlier records, misread values, and missing quantities. To deal with these measurement errors, we require a series of techniques, which will be provided in Section~\ref{SubSection_Sensor_Preparation}. 
			
		\subsubsection{Two-dimensional underwater image data}\label{SubSection_UW_Image_Data}
			Some observatory stations introduced under Table~\ref{Table_Primary_Data_Sources}, are equipped with cameras to provide ready-to-use 2D image data. In addition to those live IoUT images and videos, there exist a variety of other still-image databases. These new databases are listed in Table~\ref{Table_Image_Data_Sources}, and they are eminently suitable for supervised ML applications (as it will be discussed in Section~\ref{SubSection_ML}), as a benefit of their additional expert labels and annotations.
			These databases are accessible through a web-form or a dedicated software, or even in the form of a downloadable dataset. However, only a few of them are active projects with continuous updates.
			
			\begin{table*}[!t]
				\renewcommand{\arraystretch}{1.5}
				
				
				\caption{Open Access Datasets of Still-images, Taken from Underwater Organisms}
				\label{Table_Image_Data_Sources}
				\centering
				\begin{tabular}{@{}m{0.35\columnwidth}@{} @{}m{0.04\columnwidth}@{} @{}m{0.17\columnwidth}@{} @{}m{0.04\columnwidth}@{} @{}m{1.4\columnwidth}@{}}
				\toprule
				\textbf{Project Name} && \textbf{Funder} && \multicolumn{1}{c}{\textbf{Description}} \\
				\midrule
				Institute for Marine and \newline Antarctic Studies (\href{http://www.imas.utas.edu.au/}{IMAS}) && University \newline of Tasmania && This \textcolor{ForestGreen}{Australian} institute hosts three major image, taxonomy, and atlas database projects including Reef Life Survey (\href{https://reeflifesurvey.com/}{RLS}), Temperate Reef Base (\href{http://temperatereefbase.imas.utas.edu.au/static/landing.html}{TRB}), and \href{http://www.imas.utas.edu.au/zooplankton}{Zooplankton}.\\
				\href{http://www.sealifebase.org/search.php}{Sea Life Base} and \newline \href{https://www.fishbase.de/}{Fish Base} \newline && Non-profit \newline Organization \newline && FishBase is a project of FIN\textsuperscript{\copyright} (\textcolor{ForestGreen}{Philippines}) with around \textbf{59,000 images} of fish. SeaLifeBase is a joint project of the University of British Columbia (\textcolor{ForestGreen}{Canada}) and The FIN\textsuperscript{\copyright} (\textcolor{ForestGreen}{Philippines}), which includes \textbf{12,500 images} of different marine species.\\
				World Register of Marine \newline Species (\href{http://www.marinespecies.org/}{WoRMS}) \newline && Life Watch \newline \newline && Provides an authoritative list of valid and vernacular names for 250,000 global marine organisms. This \textcolor{ForestGreen}{Belgium}-funded infrastructure contains extra information such as \textbf{32,000 images}, literature, biogeographic data, and parent taxon. It is generally used by scientists for classification of marine species.\\
				\href{http://groups.inf.ed.ac.uk/f4k/}{Fish 4 Knowledge} \newline \newline && European \newline Union \newline && Provides \href{http://groups.inf.ed.ac.uk/f4k/GROUNDTRUTH/}{three different datasets} with annotated ground-truth for \href{http://groups.inf.ed.ac.uk/f4k/GROUNDTRUTH/RECOG/}{fish recognition} (\textbf{30,000 images} of 23 species), \href{http://groups.inf.ed.ac.uk/f4k/GROUNDTRUTH/BEHAVIOR/}{fish trajectory} (9000 trajectory of 23 species), and benchmark for \href{http://f4k.dieei.unict.it/datasets/bkg_modeling/}{complex backgrounds} (14 videos of 7 backgrounds). It also offers access to \href{http://groups.inf.ed.ac.uk/f4k/F4KDATASAMPLES/INTERFACE/DATASAMPLES/search.php}{live video feeds} from 10 underwater cameras in reefs of \textcolor{ForestGreen}{Taiwan}.\\
				Encyclopedia of Life \newline (\href{http://eol.org/}{EOL}) \newline && Multiple \newline Organizations \newline && Provides 4,000,000 Images of all life-forms on Earth (e.g. animals, plants, bacteria), including \textbf{15,000 images and videos} from marine environment. It is financially supported by institutions from \textcolor{ForestGreen}{Australia}, \textcolor{ForestGreen}{The UK}, \textcolor{ForestGreen}{The USA}, \textcolor{ForestGreen}{Mexico}, and \textcolor{ForestGreen}{Egypt}.\\
				Shape Queries Using \newline Image Databases (\href{http://www.ee.surrey.ac.uk/CVSSP/demos/css/demo.html}{SQUID}) \newline && University \newline of Surrey \newline && Includes \textbf{1100 images} of marine creatures in a smooth background (one creature in each picture). The ground-truth coordination of boundaries of animal's body in all images is annotated (256 to 1,653 points for each creature). This database is located in \textcolor{ForestGreen}{Tje UK} and will not be updated anymore.\\
				RSMAS, \newline EILAT, EILAT II, \newline and SDMRI \cite{AniBrownMary2017, Shihavuddin2013} && Multiple \newline Organizations \newline && These independent datasets contain \textbf{766, 1123, 303, and 100 images} of corals, which have been categorized by experts into 14, 8, 5, and 20 different classes of coral genera and non-coral, respectively. They are also equipped with color reference pallets for color enhancement.\\
				Moorea Labeled Corals \newline (\href{http://people.eecs.berkeley.edu/~obeijbom/mlc.html}{MLC}) \newline \newline && University \newline of Berkeley \newline \newline && This is a subset of the \href{http://mcr.lternet.edu/}{MCR-LTER} dataset and has \textbf{2055 coral reef images} with almost 200 human expert point-annotations over each one of them. These images are taken from the island of Moorea in \textcolor{ForestGreen}{French} Polynesia and contain 5 coral and 4 non-coral distinct classes. They also contain pallets with known reference colors; making them suitable for image correction and color restoration.\\
				\bottomrule
				\end{tabular}
			\end{table*}

			The multidisciplinary data provided by these image databases can be used in different branches of science. For example, while an ecologist may be interested in counting and tracking of individual inhabitants, a data scientist or computer programmer might need these data to train a ML algorithm and to verify its performance.
			
			When considering the usage of these imagery data, one should be aware of the lighting conditions in the underwater environments. This can be significantly different from overwater photography and it is expected to affect the images taken undersea. To understand the nature of light in undersea imaging systems, we have to discriminate passive and active images, i.e. whether the imaging equipment creates its own light or not. Although the physics of optics are the same in both cases, different sets of parameters have to be considered to provide a better understanding of the context of underwater optical imaging.
			
			Active imaging relies on the explicit usage of artificial light in the process of underwater imaging. This type of photography benefits from a substantial improvement in image quality, especially when the light is appropriately controlled by an optimized hardware configuration. However, active imaging suffers from high underwater energy loss (especially in case of long-term illumination), reduced portability, and unpleasant inhomogeneous intensity and color of the final picture \cite{HanM2018}.
			
\begin{figure}[!t]
				\centering
				\includegraphics[width=.87\columnwidth, clip, trim={75 577 260 68}]{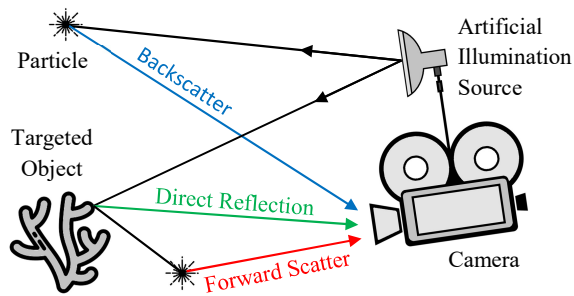}
				\caption{Schematic view of three different optical rays, which are reflected back to the camera, in active underwater imaging.}
				\label{Fig_InfoG_Light_Prop}
			\end{figure}

			As depicted in Fig.~\ref{Fig_InfoG_Light_Prop}, undersea active imaging always encounters 3 sources of light rays into the camera. Direct reflection from the target object is the desired signal, while the two other reflections, termed as \textit{backscatter} (which has not interacted with the target) and \textit{forward scatter} (or blur component) are both undesired.
			
		\textit{Backscatter or Volume Scatter:}
			In active imaging scenarios back scattered light is defined as the one which has never interacted with the target object and usually appears as bright points in the output image. To accurately calculate the magnitude of light that is backscattered to the camera, we have to calculate or measure the light intensity ($I\,[w/m^2]$) first. Light intensity (or irradiance) is the power received by an illuminated surface perpendicular to the direction of propagation, per unit area. This parameter has to be evaluated for the specific volume that the camera is viewing. One of the popular methods to achieve this goal is proposed in \cite{Jaffe2015}, which discretizes the entire volume to small cubic cells and then calculates the Volume Scattering Function ($VSF$)\footnote{$VSF$ is an inherent optical property of water, which does not vary with the incident light field. A number of experimental methods are introduced in \cite{Jaffe2015} to measure the $VSF$.} parameter for each cubic cell.
			
			The $VSF$ number of each unit cell is then used as a weighting function in the next step. Finally, the method of \cite{Jaffe2015} considers the $VSF$, the light intensity $I$, and the angle between the incident and the reflected light for numerically estimating the magnitude of light that is backscattered to the camera.
			Again, this is an efficient and accurate algorithm of estimating the magnitude of the backscattered light. However, it can be used as a fundamental step of any image enhancement procedure, especially in underwater applications, where the presence of backscattered light is indeed a significant challenge. 
			
		\textit{Forward Scatter or Blur:}
			Forward scattered light is defined as a light beam, which interacts with the target and then it is indirectly reflected back to the camera. To approximate the output image of $E_{blur} (x,y)$ in a scattering environment, a common method is to convolve the original image $E_o (x,y)$ with the Point Spread Function ($PSF(r)$) as \cite{Jaffe2015}:
			
			\begin{equation}\label{Eq_Scattered_Light}
			E_{blur}=E_o \otimes PSF(r)\,,
			\end{equation}
			where $r$ is the distance from the camera to the object and $\otimes$ is the convolution operator.
			
			$PSF$ represents the spatial impulse response of a scattering environment between a light source and the point of observation (e.g. camera), as already shown in Fig.~\ref{Fig_InfoG_Light_Prop}. It is used in Fourier optics to calculate the output image of a linear imaging system. The formulation of $PSF$ is simplified to \cite{Jaffe2015}:
			
			\begin{equation}\label{Eq_PSFr}
			PSF(r)=\left[ \left( e^{-Gr}-e^{-cr} \right) F^{-1} \left\lbrace e^{-Brf} \right\rbrace \right]\,,
			\end{equation}
			where $c$ is defined in \eqref{Eq_Optical_Attenuation}, $G$ is an empirical constant ($\left|G\right| \leq c$), $B$ is an empirical damping factor, and $F^{-1}(\cdot)$ is the inverse Fourier transform.
			
			Both $e^{-Gr}$ and $e^{-cr}$ in \eqref{Eq_PSFr} represent the forward scattering amplitude attenuation, while $e^{-Brf}$ stands for the frequency-dependent damping. The use of the convolution operator in \eqref{Eq_Scattered_Light} indicates that this equation is only valid in linear optics, which is the case in the majority of underwater light-based experiments \cite{Jaffe2015}. Although $PSF$ in \eqref{Eq_Scattered_Light} is used for the forward scattering formulation, it can be employed to calculate the $VSF$ in backscattering as well. The $PSF$ formulation as well as its relationship to the $VSF$ has been studied in great detail in \cite{Jaffe2015}.
			
			The forward- and back-scattering process encountered in seawater disperse the light beam, hence resulting in blurred underwater images. Observe from \eqref{Eq_PSFr} that the image gets more blurred, when the distance increases. One can exploit this equation to mitigate the blurring problem with the aid of image deblurring methods, which will be covered in Section~\ref{SubSection_Image_Preparation}.
			
			By contrast to active imaging, in passive imaging, i.e. where no artificial light is generated by the image capturer, even though the power consumption will be significantly lower, other concerns may be present. These concerns include limited visibility, contrast, and color distortion.
			
		\textit{Visibility:}
			The first concern in passive underwater imaging is always the visibility. In clear sunlit water, ambient sunlight provides a clear vision in relatively shallow water. In ordinary line-of-sight underwater applications, the distance of visibility can be modeled as,
			\begin{equation}
			d_{Visible}\approx\dfrac{5}{c-K(\theta, \varphi, z) cos(\theta)}\,,
			\end{equation}
			where $K$ is the diffuse attenuation coefficient, which is an apparent optical parameter, while $\theta$, $\varphi$, and $z$ are the relative spherical coordinates of the subject of interest \cite{Duntley1963}. Based on this formula, underwater visibility decreases rapidly after a few meters. The only known remedy for this physical constraint in deep-waters is to use an artificial light source in active imaging \cite{HanM2018}.
			
		\textit{Contrast:}
			Contrast is defined as the color or gray-level difference between adjacent areas in the presence of light attenuation, optical noise, and vision blurring. If we consider a target object ($O$) against its background ($B$), the contrast of the object and its surrounding areas can be formulated as \cite{Zaneveld2003},
			\begin{equation}
			C_{OB}\overset{\Delta}{=}\dfrac{L_{O}(\theta, \varphi, z)-L_{B}(\theta, \varphi, z)}{L_{B}(\theta, \varphi, z)}\,,
			\end{equation}
			where $L_{O}$ is the light level (radiance) of the object and $L_{B}$ is that of background. The knowledge of $C_{OB}$ will help us design contrast-enhancement algorithms for improving the contrast in underwater environments \cite{Salih2018}.
			
		\textit{Color Distortion:}
			As previously discussed in Section~\ref{SubSection_UW_Comm} and based on \eqref{Eq_Optical_Attenuation}, light absorption (or attenuation) in underwater propagation strongly depends on the wavelength. Accordingly, all of the visible light wavelengths present in sea surface, provide a crystal clear view. By contrast, red light hardly penetrates bellow 10m, which is the reason of the greenish color of seawater. Colors having shorter wavelengths (i.e. blue color family) penetrate even deeper to 20 and up to 50 m in sea water, resulting in low-contrast bluish images.
			
			Color distortion and the visibility are the main reasons why passive imaging is inefficient bellow 10 meters and also why the use of artificial light sources is inevitable in deep-water explorations. To address this important issue, the image enhancement or color restoration methods of Section~\ref{SubSection_Image_Preparation} have to be involved.
			
			Having surveyed some of the 1D signals generated by sensors and the 2D image data types encountered in IoUT, let us now focus our attention on video or 3D data, that forms a large subset of IoUT datasets.
			
		\subsubsection{Three-dimensional underwater video data}\label{SubSection_UW_Video_Data}
			As discussed, some of the observatory stations listed in Table~\ref{Table_Primary_Data_Sources} are capable of delivering live or archival ready-to-use video data. 			
			However, the main challenges in underwater video streaming are related to the limitations of undersea communication, discussed in Section~\ref{SubSection_UW_Comm}, aggravated by the underwater imaging difficulties studied in Section~\ref{SubSection_UW_Image_Data}.
			
			Although the underwater optical parameters and imaging constraints are all the same as in 2D imagery (i.e. contrast, color distortion, visibility, backscatter, and forward scatter), underwater video imaging has an extra intrinsic impediment that has to be addressed. This barrier, is of course, the continuously growing data volume produced by cameras. A typical camera operating at 1 frame/sec may generate 30 million frames (equal to almost 3 Tera bytes of data) per year \cite{Spampinato2010}. This data volume is then multiplied by the number of cameras, mounted in a single observatory system.
			
			As a rule of thumb, in the current version of the semi-automatic marine image-annotation software, every minute of video requires an expert to spend about 15 minutes for manual annotation \cite{Spampinato2010}. Therefore, to analyze the video produced by a single camera in a single month, we approximately need 10,800 man-hour. Hence, there is a growing necessity to develop automatic video processing methods to deal with this excessive data volume. These automatic methods can be used in a variety of underwater video data applications such as, visible light video tracking \cite{Chuang2015}, sonar video tracking \cite{Negahdaripour2013}, photo mosaicing underwater \cite{Bonin-Font2017}, and marine life studies \cite{Meng2018}.
			
			In addition to the above-mentioned potential use cases of automatic video processing, the growing field of depth-based video may also significantly benefit from automatic processing. 
			This technology uses optical multi-camera systems \cite{ChengCh2018, Mazzei2015}, acoustic arrays \cite{Assalih2009}, Time of Flight (ToF) depth sensors \cite{Anwer2016}, and laser beams \cite{WangCh2000} to provide rich 3D information about the scene. This stereoscopic video technology has been used in underwater vehicles (e.g. ROVs, submersibles, etc.) \cite{Ishibashi2009, Negahdaripour2009}, in tethered underwater platforms \cite{Marouchos2017}, towed systems \cite{Chuang2015}, and baited stations \cite{Westling2014}. They are used in a wide variety of applications, including photogrammetric bundle adjustment \cite{Mulsow2014, Trucco2006}, 3D scene and organism reconstruction \cite{Anwer2016, Mazzei2015}, underwater 3D live tracking \cite{Chuang2015, Trucco2006}, quantitative analysis and sizing of targets \cite{Marouchos2017, Chuang2015}, counting and measurement of marine creatures \cite{Westling2014}, as well as improving the segmentation and classification capabilities of traditional 2D algorithms \cite{Westling2014, Johnson-Roberson2006}.
			
			However, the sophisticated depth-based vision technology has its own challenges both in the context of acoustic and optical recording methods. While acoustic depth vision always suffers from variable sound velocity, reverberation, as well as unwanted noise and echoes, optical stereoscopic cameras encounter blur and haziness in turbid water, aggravated by unstable illumination, and light refraction \cite{Chuang2015}.
			Perhaps, the most direct way of addressing this problem is to use improved acoustic and optical channel models for compensating the deficiencies of each technology \cite{Mulsow2014}. Another solution is to combine both technologies into a unified opti-acoustic 3D imaging system. The latter solution is capable of offering improved precision, if both technologies cooperatively calibrate each other~\cite{Negahdaripour2009}.
			
			So far in this section, we have provided a detailed evaluation of the family of IoUT sensors, as well as of image and video data sources. Additionally, some ready-to-use databases were introduced. Once the raw big data has been gathered, a high-performance data processing platform is required for inferring knowledge. Due to the scarcity of resources and owing to the limited power budget at the data collection points, processing cannot be performed locally. Therefore, the system has to rely on cloud-based or on distributed data processing platforms, which will be studied in the next section.
			
	\subsection{Distributed and Cloud-based BMD Processing}\label{SubSection_Data_Processing_Platforms}
		To meet the growing demand for big data processing, many high-quality software and services have been created to offer big data analytics. Some of these platforms are Apache, Amazon EMR, Microsoft Azure HDInsight, Cloudera, Hortonworks (which has recently been merged with Cloudera), SAP-Hana, HP-HAVEn, 1010data, Pivotal data suite, Infobright, etc. \cite{Verma2017, Ahmed2017}. Each of these services responds to the big data processing demands by either providing distributed processing frameworks or cloud computing services.
		
\begin{figure}[!t]
			\centering
			\includegraphics[width=\columnwidth, clip, trim={0 34 0 28}]{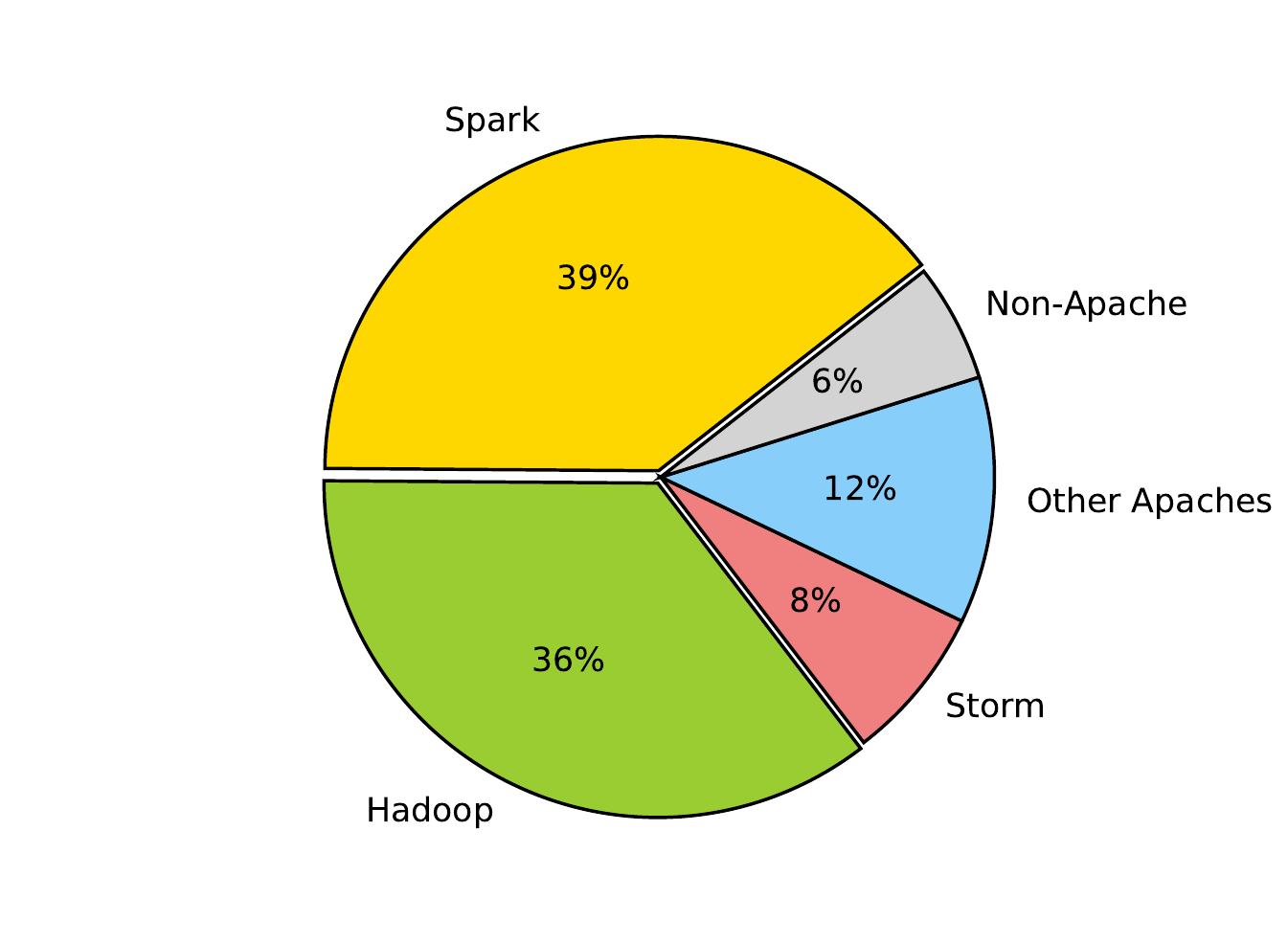}
			\caption{Distributed big data frameworks used in data analytics papers published in IEEE and Web of Science databases.}
			\label{Fig_Pie_Big_Data_Platforms}
		\end{figure}

		Distributed processing systems consist of networked computers, which cooperate with each other to offer high performance data processing \cite{Xingang2019}. As seen in Fig.~\ref{Fig_Pie_Big_Data_Platforms}, among all distributed big data analytics frameworks, the Apache is the dominant platform, which has been used in about $95\%$ of all reported scientific articles. Apache software foundation is a not-for-profit corporation, founded in 1999 to support more than 350 open-source Apache software projects, and 47 of them (about $13\%$) are directly related to big data analytics. More than $80\%$ of the Apache platforms have been developed in only five languages (i.e. Java ($\approx 60\%$), different types of C ($\approx 10\%$), Python ($\approx 5\%$), JavaScript ($\approx 3\%$), and Scala ($\approx 3\%)$) \cite{ASFProjectDirectory}.
		
		Table~\ref{Table_Apache_Platforms} introduces the advantages and disadvantages of the four widely used Apache distributed processing frameworks including Spark, Hadoop, Storm, and Flink. All of these open-source projects are supported by Apache software foundation and they are eminently suitable for high-speed processing of IoUT-generated BMD analytics.
		
		\begin{table*}[!t]
			\renewcommand{\arraystretch}{1.2}
			\caption{Major Apache Frameworks Suggested for Distributed Processing of BMD}
			\label{Table_Apache_Platforms}
			\centering
			\includegraphics[width=\textwidth, clip, trim={48 460 48 60}]{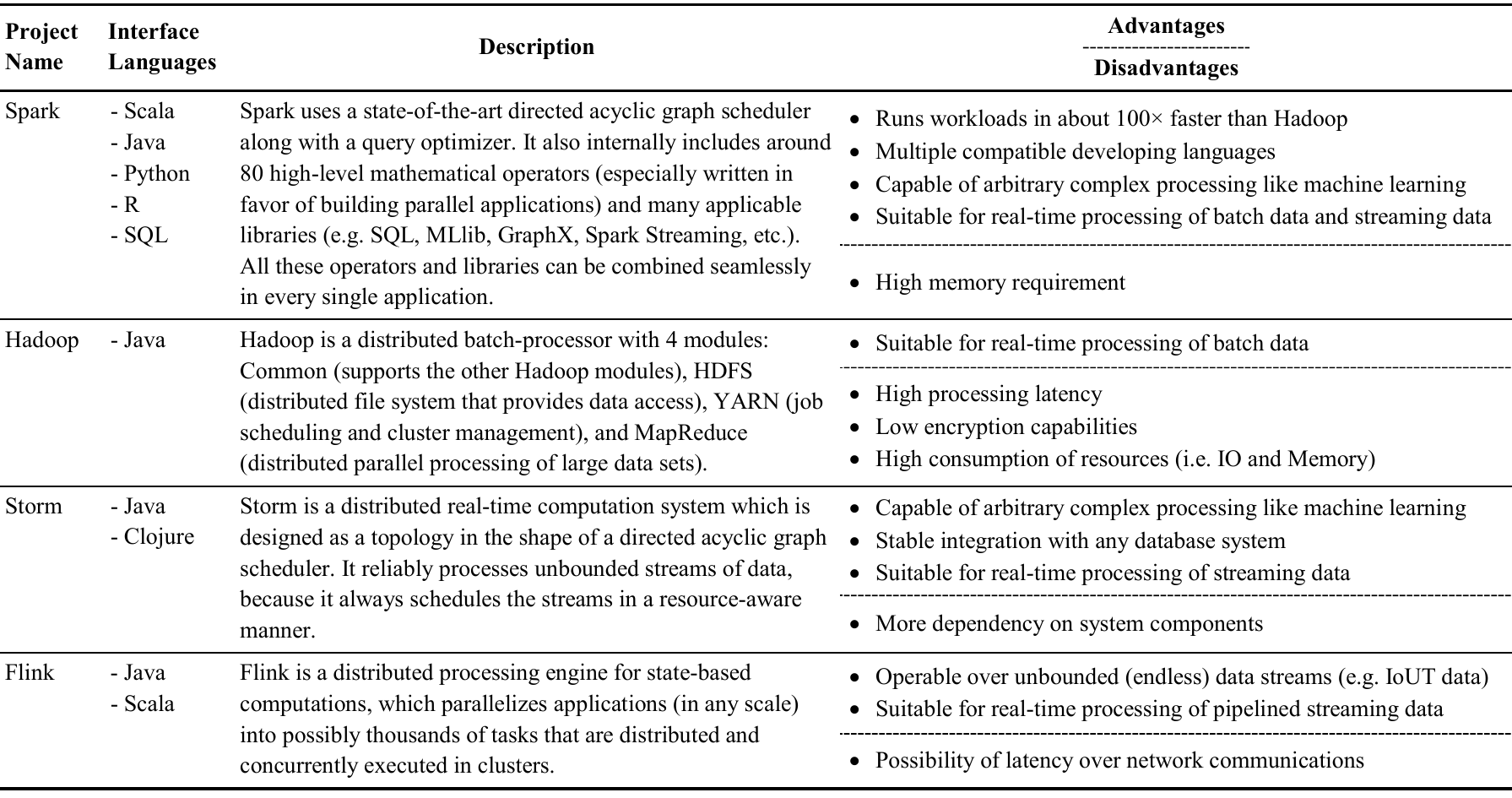}
		\end{table*}
		
		The term \textit{cluster} in Table~\ref{Table_Apache_Platforms} refers to any computer system or embedded system in the parallelized network of computers, each having its own processor, memory, and network IO. Furthermore, the directed acyclic graph scheduler is based on a specific type of directed mathematical graph having no cycles, in which it is impossible for the data to emanate from a vertex (i.e. any standalone computer) and pass over a non-zero number of graph edges (i.e. inter-computer connection cables), and to eventually loop back to the starting vertex again.
		
		In addition to the locally distributed processing systems, there exist companies who offer cloud-based services for all aspects of big data storage, integration, streaming, ML, and \textit{ad hoc} data analysis. According to the \href{https://www.statista.com/statistics/301566/big-data-factory-revenue-by-type/}{Statista}, the share of software and services in overall big data revenue will double in the 6-year period spanning from 2019 to 2025. This means that the organizations will rely more on cloud services to handle their sophisticated use cases \cite{Power2018}. In other words, big data processing is expected to gradually shift from distributed to cloud-based services. This is an ongoing trend, as the number of businesses performing their big data analysis in the cloud has increased from $58\%$ in 2017 to $73\%$ in 2018 \cite{QuboleBDTrends2018}.
		
		As already mentioned in Section~\ref{SubSection_Edge_Computing}, BMD can partly be processed in the IoUT-edge, before being transmitted to land. After transmission, there is no difference between the conventional IoT and IoUT-based distributed or cloud-based data processing. Nevertheless, all available cloud services today are offered in accordance with one of the following models \cite{Zhelev2017}:
		
		\begin{itemize}
			\item \textit{Software as a Service (SaaS)}: This service model relies on centrally hosted software, which delivers specific services to licensed or subscribed clients and usually offers its dedicated functionality through a browser.
			
			\item \textit{Platform as a Service (PaaS)}: This service model targets developers by providing them with operating systems, databases, software packages, application services, etc. It helps developers to focus on the development of their diverse applications, instead of software resource management and hardware maintenance of the underlying infrastructure.
			
			\item \textit{Infrastructure as a Service (IaaS)}: This flexible low-level service model targets both developers and businesses by providing access to the underlying infrastructure (e.g. processors, data partitions, security, backup, etc.). This service tends to rely on high-level Application Programming Interfaces (API) in support of the network operations. While the service provider is responsible for the hardware maintenance, the users are required to configure and maintain both operating system and the required software packages.
		\end{itemize}
		
		According to the aforementioned definitions, cloud computing is capable of processing BMD with the aid of all the above service models. Among them, PaaS strikes an attractive balance between convenience and cost, undertaking the management of the operating system, software packages, and hardware elements. It is worth mentioning that, Big Data as a Service (BDaaS), which has recently appeared in technical glossaries may indeed be classified as a specific form of PaaS, where the statistical analysis software packages are offered alongside the required databases and APIs.
		
		On the other hand, IaaS constitutes a cost-effective solution, where the service providers are responsible for maintaining the underlying hardware. Some of the top Paas and IaaS service providers are Amazon Web Services, Microsoft Azure, Google Compute Engine, DigitalOcean, Oracle Cloud Compute, Rackspace Cloud, IBM Cloud, Linode, HP Enterprise Converged Systems, Green Cloud Technologies, etc.
		
		Despite the fact that SaaS providers tend to be user-oriented and application-specific, some of them also support underwater applications. For instance, marine image-annotation software are ready-to-use software products for semi-automatic annotation of videos and still-images. These software packages are comprehensively reviewed by Gomes-Pereira \textit{et al.} \cite{Gomes-Pereira2016}, where 23 software products from more than 500 publications are summarized.
		
		The semi-automatic solutions offered by marine image-annotation software are rarely applicable in real-time BMD analytics and they tend to rely on data analytic platforms as well as self-developed ML algorithms. As a result, every practical use case of BMD, requires a distributed on-premise platform or cloud-based data processing software for gleaning knowledge from BMD. Next, some of these IoUT data extraction applications are studied.
		
	\subsection{Marine Data Applications}\label{SubSection_Ocean_Data_Application}
		The marine data collected by various sensors and devices of the IoUT ecosystem can be exploited by data processing platforms for compelling applications. These applications offer businesses benefits based on their own data \cite{Ahmed2017}. In the following subsections, we discuss and evaluate some of these applications in the context of IoUT, including maritime applications, underwater localization, and marine life tracking.
		
		\subsubsection{Coastal monitoring and GIS data}
			By appropriately processing the available maritime sensory data, some secondary parameters can be collated with the aid of the IoUT. These include the accurate localization of marine vehicles, the provision of weather and climate data for specific oceanic locations, accessing biogeographic data such as the recognition, counting, and distribution of underwater species, etc.
			Hereafter, we will refer to all these secondary parameters as IoUT Geographic Data, because they are devoted to studying the physical environment, the inhabitants, locations, and things in the particular area of the sea.
			
			Having accurate geographical data is essential in IoUT applications. For this type of IoUT data, there exist a number of open-source on-line geographical databases. These databases are mainly focused on maritime information systems, a particular type of Geographic Information System (GIS), which is indispensable in efficient international transportation \cite{Barale2018}. Fortunately, most of these GISs can be readily merged with the IoUT infrastructure in support of maritime organizations with the aid of tracking and routing information, etc. They can also provide up-to-date access for researchers in order to monitor the global ecosystems.
			
			Kalyvas \textit{et al.} \cite{Kalyvas2017} surveyed more than 180 free real-world GISs having open access databases based on the most trustworthy on-line data sources. They have categorized the GISs into 19 classes, which are distilled into as few as four classes here for simplicity. These classes and their applications include:
			
			\begin{itemize}
				\item \textit{IoUT tracking data}: For vessel monitoring, marine accidents, anti-shipping activities, and navigational aid systems;
				\item \textit{Marine cartographic data}: For essential naval data (like borders), protected and sensitive areas, port status, port locations and facilities, coastline as well as land areas, and bathymetry maps;
				\item \textit{Oceanic climate data}: For nautical weather, natural hazards, tides as well as eddies, and satellite imagery;
				\item \textit{IoUT commerce data}: For shipping companies, flags of convenience, and marine conservation organizations.
			\end{itemize}
			
			An important category that could be added to the above-mentioned classes of Kalyvas \textit{et al.} \cite{Kalyvas2017} is Biogeographical Data for addressing the geospatial distribution of underwater species. In order to define this new category, the open-access \href{http://www.iobis.org/}{OBIS} project (Ocean Biogeographic Information System), funded by UNESCO can be used. OBIS is connected to more than 500 databases in 56 countries and provides observation of 120,000 marine species down to $11,000\,m$ depth, from bacteria to whales.
			
			In addition to the aforementioned need for Biogeographical data, under the Marine Cartographic Data category detailed in \cite{Kalyvas2017}, there is a missing sub-category for Georeferenced Locations. To elaborate a little further, georeferencing in GIS is a subcategory of navigational assistance that aims for precisely associating locations with their equivalent points on the physical maps, which is achieved by the \href{http://marineregions.org/}{Marine Regions} project. This project is composed of a list of 55,000 georeferenced marine places, gathered from the \href{http://www.vliz.be/}{VLIMAR} Gazetteer and \href{http://www.vliz.be/}{MARBOUND} databases.
			
			In addition to the GIS databases, there should also be a directory of applications, which require access to precise undersea locations. This is called underwater positioning. In the following sections, undersea localization of sensors and vehicles, along with the marine life tracking methods will be discussed in more details.
			
		\subsubsection{Underwater localization}\label{SubSection_Geo_Localization}
			A very useful data type that is included in IoUT is related to the positioning of undersea devices, systems, animal species, and data sources. This is even more critical when there is a need to geo-tag IoUT sensory and imagery data.
			
			Underwater positioning is a challenging task, because the otherwise ubiquitous navigation signals of the GPS satellites do not penetrate seawater. Therefore, other underwater navigation methods should be used, including blind positioning relying on miscellaneous inertial sensors, acoustic transponders (with separate transmitters and receivers), ranging sonars (with only a single standalone acoustic transceiver), image-based positioning (using cameras to localize), and Simultaneous Localization and Mapping (SLAM).
			
			Due to the associated challenges including the long latencies, multipath fading, Doppler shifts, frequency limitations, sparse deployment of the nodes, and their high mobility in underwater networks, a single navigation technique will never offer a flawless performance \cite{Liu2018, ZhangT2018}. Therefore, all the vehicles, fixed stations, sinks, and nodes in IoUT applications usually combine some of these localization methods to achieve improved performance in underwater environments.
			Here, we briefly review the aforementioned localization techniques and provide insights concerning their advantages and limitations.
			
		\textit{Blind positioning:}
			which is also known as inertial navigation, is based on the knowledge of a device's relative orientation, acceleration, velocity, and gravity anomaly (i.e. difference between the observed gravity and the predicted value \cite{WangB2019}). In this localization method, the underwater device has to determine its position automatically, i.e., without any positioning support from a ship or transponder. In this method, a combination of sensors is used to estimate the current location.
			
			These sensors include a typical magnetic compass for direction detection, a pressure sensor for underwater depth estimation, a doppler velocity log for velocity measurement, a ring laser gyroscope or Micro Electro-Mechanical Systems (MEMS) gyroscope for angular velocity detection, and a pendulum or MEMS accelerometer for force and acceleration calculations \cite{Paull2014}.
			
			However, these sophisticated blind positioning methods suffer from propagating errors. Even a small positioning error remains in the memory of the system and aggravates the future measurement errors, leading to an unbounded error propagation. Nevertheless, the method's estimation of the exact position can be improved by a variety of integrated sensors. Furthermore, blind positioning is a power-efficient method, compared to other positioning techniques \cite{Chen2016}. Hence these positioning techniques contribute to almost all modern underwater positioning systems \cite{Liu2018, ZhangT2018, Chen2016, WangB2019, Song2018}.
			
\begin{figure}[!t]
				\centering
				\includegraphics[width=.6\columnwidth, clip, trim={70 539 353 68}]{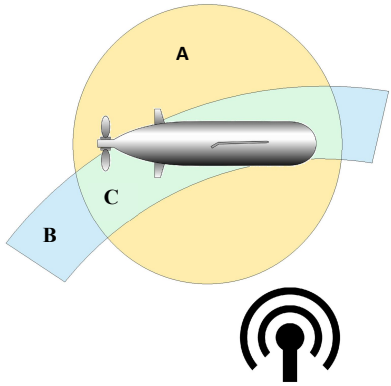}
				\caption{Using a fixed beacon at known location to decrease the location-uncertainty of an AUV from ‘A’ (intersecting with ‘B’) to ‘C’.}
				\label{Fig_Sch_Fixed_Beacon}
			\end{figure}

		\textit{Acoustic transponders:}
			This localization technique uses a transmitter and receiver pair for measuring the ToF of a ping signal in order to perform navigation. ToF $[sec]$ is applicable both to the adjacent transmitter/receiver (i.e. where the transmitter and receiver are closely located) and separate transmitter/receiver (i.e. where transmitter and receiver are distantly located). It is also known as the time of arrival in separate transmitter/receiver scenarios \cite{ZhangT2018}. By measuring the ToF and the acoustic wave speed, one can precisely measure the distance in $[m]$. This positioning method has diverse categories, including:
			
			\begin{itemize}
				\item \textit{Acoustic array}: Similar to the concept of GPS satellites, and by using more than one beacon transponders, the system will be able to determine the position in any underwater location based on the phase-difference of the signals arriving at the transceivers. Short- as well as ultrashort-baseline \cite{Xiao2017} and long-baseline \cite{Batista2015} techniques may be used in this category. In short- and ultrashort-baseline relying on ship-mounted transponders, the undersea system localizes itself relative to the ship floating at the surface; while in long-baseline that uses GPS-intelligence as well as dispersed buoys and beacons, the location can be determined.
				\item \textit{Single fixed transponder}: The idea behind fixed beacon based positioning is shown in Fig.~\ref{Fig_Sch_Fixed_Beacon}. The AUV in this picture has an uncertain prior knowledge about its position (region A). However, it knows its distance from a fixed beacon subject to a degree of uncertainty (region B). With the advent of combining these two pieces of information, the AUV finds its position subject to a reduced uncertainty (region C). Due to the fact that it uses only a single geo-referenced beacon, this positioning system is cheaper and easier to install than multi-beacon long-baseline techniques \cite{HanY2018}.
				\item \textit{Cooperative navigation}: This method is also known as modem-based navigation, because in recent advanced positioning systems, modem transponders are also used to send beacons in support of navigation based on the ToF \cite{Renner2017}. These modems do not have to be stationary, hence they can be installed on a moving vessel having a known geo-position or a swarm of underwater vehicles to communicate and to cooperatively localize each other.
			\end{itemize}
			
		\textit{Range sonars:}
			Sonars are robust, well-established standalone acoustic transceivers that were originally used for imaging and ranging. The main ranging sonars used in underwater navigation and mapping include \cite{Paull2014}:
			
			\begin{itemize}
				\item \textit{Echo sounder}: A single narrow beam is used for determining the distance from obstacles \cite{Sarmiento2018} or from the seafloor.
				\item \textit{Multi-beam}: Measures the ToF for each beam to assemble a bathymetric map \cite{Menna2018}.
			\end{itemize}
			
		\textit{Image-based positioning:}
			This technique uses environmental images, taken either by monocular or stereoscopic cameras (or even by imaging sonars) for navigation. In this positioning method, substantial processing power is required for feature extraction and for processing in order to detect and identify key points, objects, and regions of interest.
			
			The main idea behind the vision-based navigation, also known as visual odometry, is to capture images of the seabed and then to match subsequent images in order to navigate \cite{Al-Kaff2018, Gao2016}. stereoscopic cameras have the advantage of capturing 3D transformations between consecutive image pairs.
			
			It is plausible that errors can propagate and accumulate in the context of this technique. To address this problem and achieve a bounded positioning error, the SLAM tactic can be merged with image-based positioning for improved localization.
			
		\textit{SLAM:}
			Concurrent or Simultaneous Localization and Mapping is defined as the process in which an autonomous vehicle builds a map of a specific area and also localizes itself within it \cite{Paull2014}. The mapping process of SLAM may rely on a variety of devices such as cameras, sonars, or inertial sensors, respectively, leading to vision-based, sonar-based, or sensor-based SLAM techniques \cite{Palomeras2018, Norgren2018}. In all cases, the features of the sensed data or captured image are extracted. Then, based on those features, the position is detected and stored.
			
		\subsubsection{Marine life tracking}\label{SubSection_Marine_Life_Tracking}
			At first glance, underwater animal tracking seems to be nothing more than a memory-aided localization method in IoUT, which was the subject of the previous section. However, undersea animals may be quite small and they cannot carry relatively heavy inertial sensors, transponders, sonars, or cameras. Therefore, sophisticated new tracking methods have to be devised.
			
			Additionally, as discussed in Section~\ref{SubSection_Geo_Localization} and also bearing in mind the penetration depth formula of \eqref{Eq_Penetr_Depth}, the GPS signals having a frequency of $1.2$ and $1.5\,GHz$ cannot penetrate seawater beyond a few meters. Accordingly, alternative marine life tracking methods have to be implemented in IoUT applications, some of which are evaluated in this section. All these methods can make use of a data storage tag to archive data on a memory chip for future retrieval \cite{Fore2017}.
			
		\textit{RFID tags:}
			This tracking technology uses Radio Frequency Identification (RFID) patches and has a limited range of about $10\,m$ in freshwater. However, it does not work well in seawater owing to its high salinity. RFID tags are equipped with a unique Identification code and need an external energy source in the form of a low frequency signal, in order to become activated and to retrieve data. These tags are also available in passive integrated transponder form, which are specific implantable RFID devices \cite{Domingo2012}.
			
		\textit{Acoustic tags:}
			This tracking technology emits signals in the form of acoustic waves and has a reasonable transmission range both in fresh and seawater. In this method, a pinging sound with an embedded ID is periodically transmitted to an array of remote acoustic sensors (hydrophones). This ping is used to locate the animals and the ID is used to identify them \cite{Fore2017}.
			
		\textit{Image-based tracking:}
			Perhaps image-based tracking is one of the most advanced underwater tracking techniques reported to date. It uses images of a moving or fixed camera system to track, recognize, count, measure, and also study the animal's behaviors. In a fixed camera based system, usually a baited or trawled station is used for attracting intended species \cite{Chuang2015}. On the other hand, the tracking of animals in a moving camera based system uses cameras installed on a moving platform. The main challenge in these systems is to promptly process the large variety of images \cite{Chuang2017}.
		
	\subsection{Section Summary}
		
		In this section, we have studied the essential topics of big marine data. At the beginning, we introduced the five BMD system components, followed by further elaborations on the first component, i.e. on data acquisition. Thereat, we reviewed the underwater data gathering tools and techniques, along with the data aggregation protocols for more compactly representing the gathered data. We also studied data fusion methodologies, conceived for the fusion of data from distinct datasets into an integrated database, for the sake of having more generally representative information. Our discussions in this section were then continued by classifying BMD into 1D sensor signals, 2D image data, and 3D video streams. For all these categories, we surveyed the typical IoUT infrastructures and observatory systems, which freely offer their BMD for researchers. The visibility, contrast, color distortion, and light scattering in 2D underwater images as well as 3D video data gathering were also addressed. Then, different distributed as well as cloud-based BMD processing frameworks were introduced and a couple of essential BMD applications were studied. These applications covered the oceanic GIS, underwater localization, and marine life tracking, which all play essential roles in IoUT contexts.
		
		All of the tools and methods that have been discussed in this section, including the sensory and imagery data sources, GIS-based ready-to-use data in IoUT, and underwater object tracking as well as geo-tagging methods will generate BMD, which has to be processed and analysed. In the following section, we will focus our attention on big data processing techniques. Thereby, we will review state-of-the-art technologies in data cleansing and data processing using ML techniques for underwater applications.
		
\section{Machine Learning for BMD Analytics}\label{Section_Data_Proc}
\begin{figure}[!t]
		\centering
		\includegraphics[width=\columnwidth, clip, trim={45 647 310 74}]{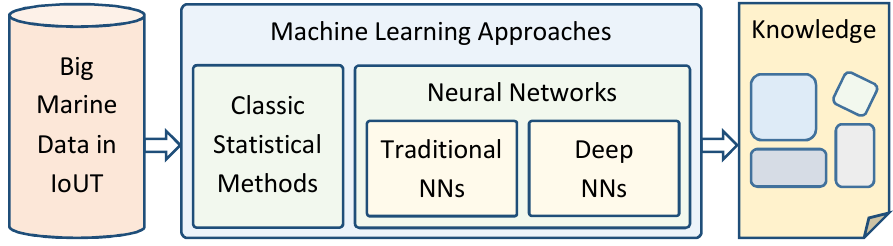}
		\caption{Conceptual categorization of the main topics in Section~\ref{Section_Data_Proc}.}
		\label{Fig_InfoG_ML_Stats_NN_DL}
	\end{figure}

	As mentioned in Section~\ref{Section_Big_Data}, \textit{data processing} is one of the five critical system components in the chain of wealth-creation from data. To address the commercial and industrial demands, ML and its Deep Learning (DL) variant constitute promising solution. In this section, as visualized in Fig.~\ref{Fig_InfoG_ML_Stats_NN_DL}, we focus our attention on the use of ML strategies including the classical ML approaches, as well as the more traditional and emerging NN-based learning methods for BMD processing.  
	
	\subsection{Machine Learning Techniques}\label{SubSection_ML}
		Considering the large amount of data generated both in IoT and IoUT, there is a growing need for powerful tools and techniques capable of distilling and visualizing knowledge gleaned from data. These tools and techniques have evolved over the past century! As a benefit, ML is readily available for the analysis of BMD \cite{Mohammadi2018}. ML can be viewed as a collection of software algorithms that empower intelligent machines to improve their performance on an accomplishing pre-configured specific task. This particular task in turn, may be categorized under descriptive, predictive, and prescriptive models.
		
		
		
		The ML approaches may be divided into supervised and unsupervised methods. The unsupervised methods are mainly used for data clustering, according to the features embedded in the data itself. These methods do not rely on an expert for labeling and for entering inputs. On the other hand, in supervised learning, data will be labeled, prior to its exploitation, by an expert. In this case, the resultant ML solution is referred to as a classifier, which is then involved for the ensuing data classification phase.
		
		Classification methods are devised for categorizing data, representing for example a measurement from a sensor or a pixel in an image into one of the legitimate predefined output classes. When there are only two possible answers to a given question (i.e. yes or no), the classification problem is often termed as \textit{detection}. In cases of three or more legitimate output classes associated with multiple choices, classification is also often referred to as recognition.
		Methodologically, classification algorithms can be divided into statistical methods (also known as conventional data analytics) and Neural Networks (NN). The neural-based methods may themselves be divided into traditional NNs and deep NNs \cite{Schmidhuber2015}.
		
%
			
			All ML approaches are designed to assign a given input data ($X$) to a set of predefined classes ($C_{i}$). This is a well-known classification/clustering problem, which can be solved by either generative or discriminative models. The generative models are those which solve this problem by using the joint probability distribution function of \cite{NohY2018},
			\begin{equation}\label{Eq_Generative_Model}
			i = \myArgMax_C{P(C) P(X|C)} \,.
			\end{equation}
			
			By contrast, the family of discriminative models uses the conditional probability distribution function of \cite{NohY2018},
			\begin{equation}\label{Eq_Discriminative_Model}
			i = \myArgMax_C{P(C|X)} \,.					
			\end{equation}
			
			According to the Bayes' theorem from a theoretical perspective, these methods are identical. In practice however, it is usually easier to calculate $P(C|X)$; especially when we have a large amount of data (i.e. $X$) for training our model. The conventional data analytic methods tend to use generative models, while most NNs rely on discriminative models \cite{AdeliE2019}.
						
			In generative models, the behavior of both $C$ and $X$ should be known. However, for discriminative models, we directly deal with the unknown $C$, based on a given $X$. Discriminative models are generally simpler, faster, and have less parameters to adjust. Accordingly, while the family of conventional statistic approaches may be good enough for handling moderate-dimensional situations, traditional ML and modern DL based approaches (with discriminative models) are needed for processing big marine data problems (i.e. image, video, and other sensory information generated within the IoUT infrastructure).
		
		The concept of DL is built around the idea that artificial neurons are capable of automatically extracting features and learn a pattern, provided that there are enough hidden layers and unweighted neurons in their networks. Based on this concept, deep networks have evolved from the traditional NNs by invoking more than one hidden layer (technically, dozens of layers). Using these extra layers, deep networks become capable of extracting features and reduce the learning dimension as it will be introduced later in Section~\ref{SubSection_Feature}.
		
\begin{figure}[!t]
			\centering
			\includegraphics[width=.85\columnwidth, clip, trim={0 1 0 27}]{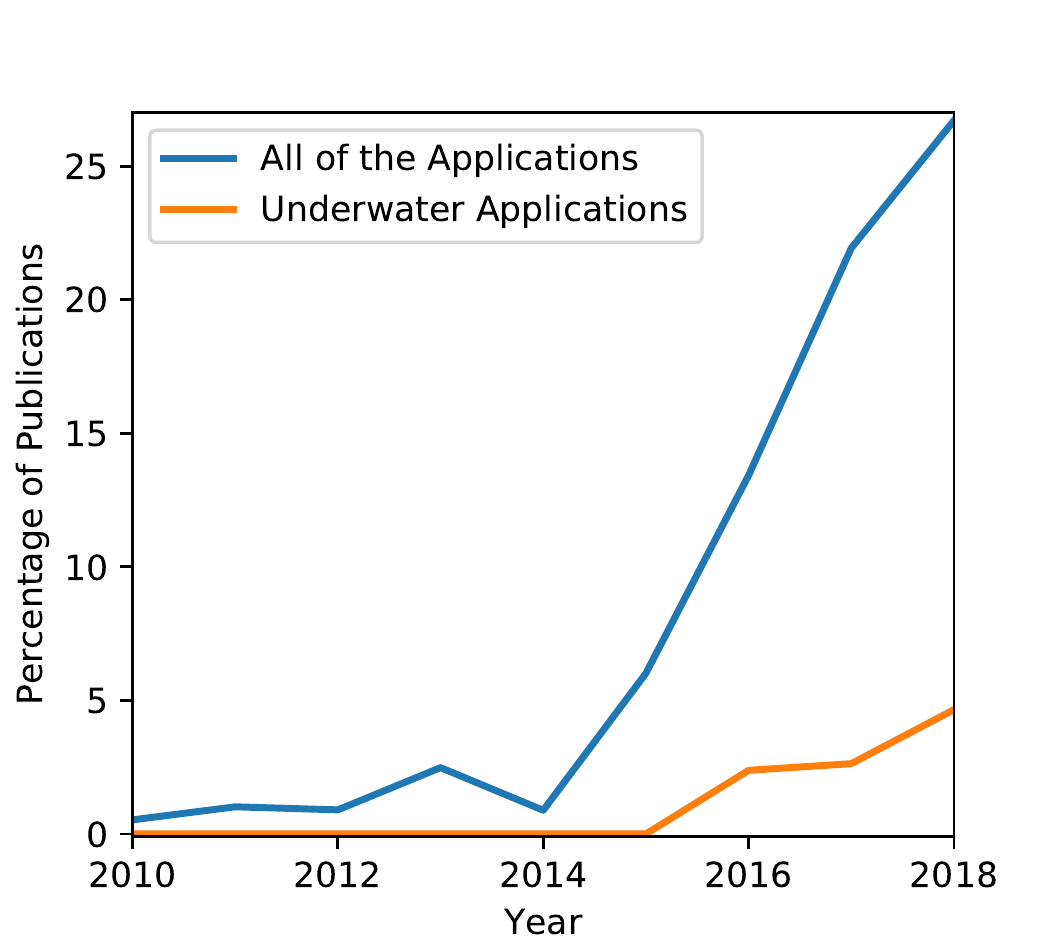}
			\caption{Percentage of deep learning usage in all machine learning publications, as searched between IEEE Explore as well as Web of Science databases.}
			\label{Fig_Plot_DL_Cmpred_Other_ML}
		\end{figure}

%
		The growth rate of DL usage in all ML publications in the post-2010 era is demonstrated in Fig.~\ref{Fig_Plot_DL_Cmpred_Other_ML}. This figure shows an astonishing factor 25 increase in the percentage of publications in as few as four years. These publications have substantially advanced the field by proposing new algorithms, networks, and strategies for improving the performance of deep networks. 
		Table~\ref{Table_DL_Cl_Networks} presents the most recognized supervised and unsupervised deep learning networks. All the tandem arrows and circular nodes in this table ($\rightarrow$\!$\bigcirc$) are representative of an artificial neuron processing the sum of weighted inputs and a subsequent activation function (e.g. Sigmoid, $\tanh$, etc.). Please note that every NN has a bias input. These inputs are deliberately eliminated in the shown diagrams, for presentation simplicity.
		
		\begin{table*}[!t]
			\renewcommand{\arraystretch}{1.2}
			\caption{Deep Networks with \colorbox{WordFillGreen}{Input}, \colorbox{WordLightBlue}{Hidden}, \colorbox{WordFillGray}{Mid-output}, and \colorbox{WordFillRed}{Output} Layer Neural Nodes to Evaluate Static, Dynamic, Sequential, or Hierarchical Input Data Types \cite{Hinton2006, Lecun1995, KimK2018, Socher2013, Hochreiter1997}}
			\label{Table_DL_Cl_Networks}
			\centering
			\includegraphics[width=\textwidth, clip, trim={37 60 40 60}]{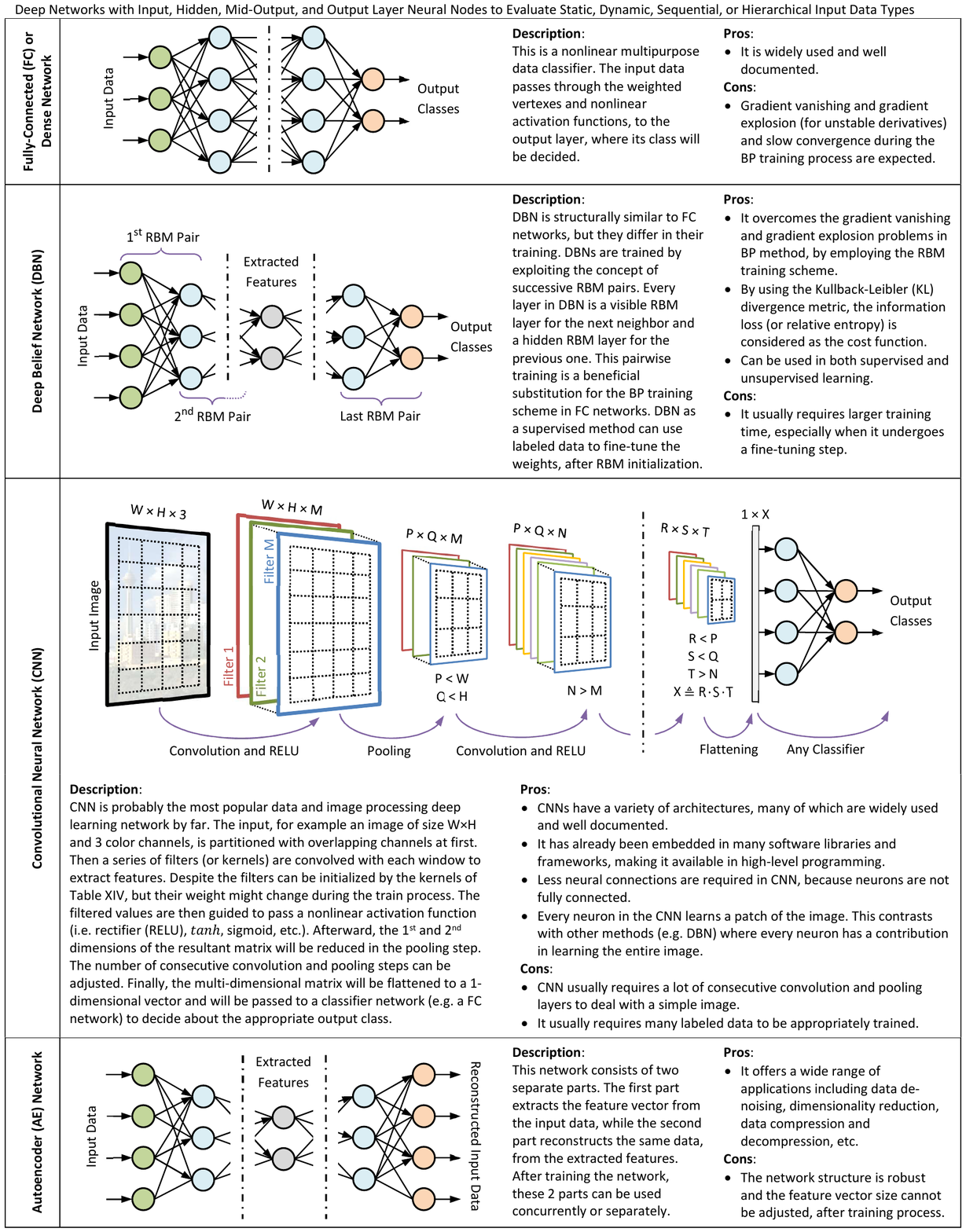}
		\end{table*}
		\addtocounter{table}{-1}
		\begin{table*}[!t]
			\renewcommand{\arraystretch}{1.2}
			\caption{(Continued)}
			\centering
			\includegraphics[width=\textwidth, clip, trim={37 487 40 60}]{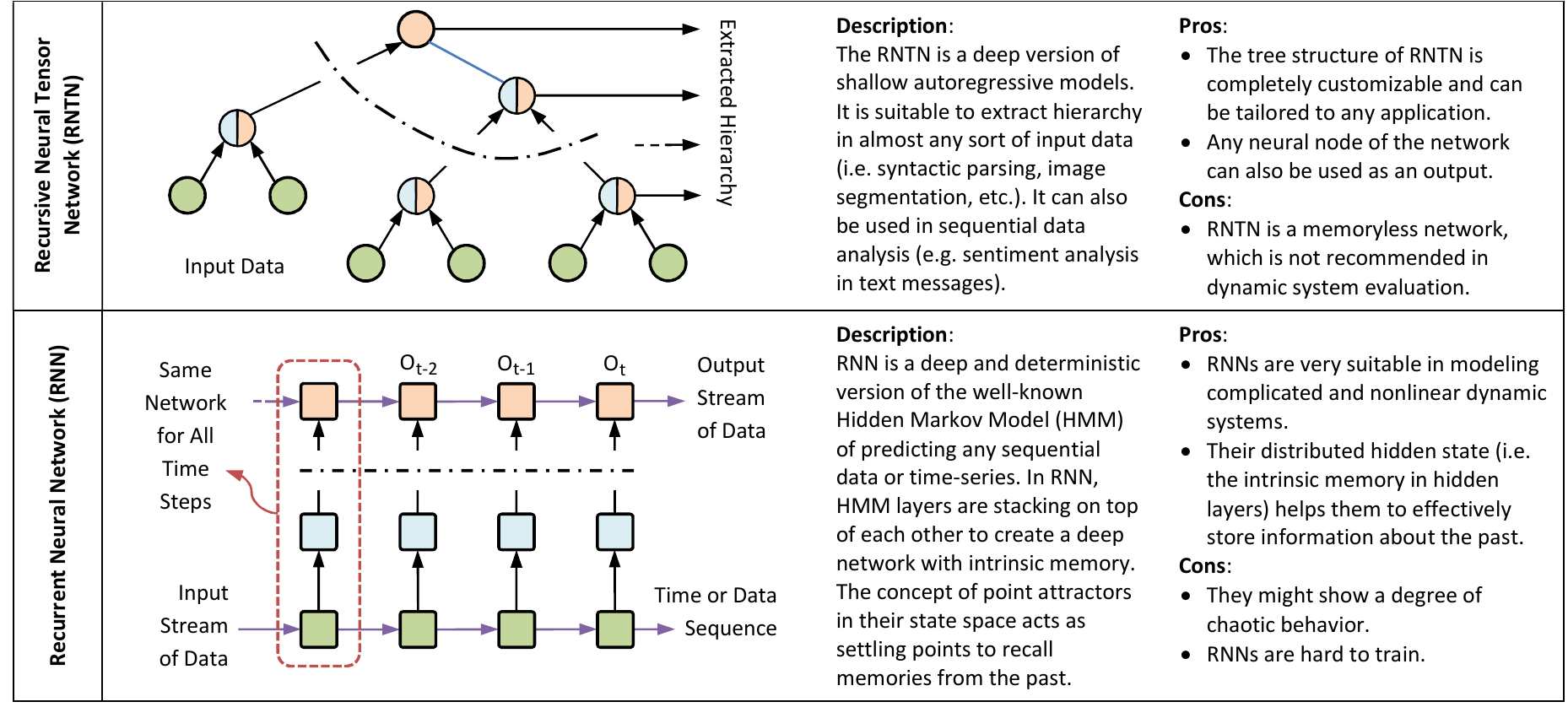}
		\end{table*}
		
		Again, deep networks have several hidden layers. This is shown in Table~\ref{Table_DL_Cl_Networks} by utilizing a dash-dot line ({$-$\,\!$\boldsymbol{\cdot}$\,\!$-$\,\!$\boldsymbol{\cdot}$}), whenever the network is capable of incorporating multiple hidden layers. However, it should also be considered that, as these networks grow deeper, they also require more training data. Further details on various types of networks and on their advantages/disadvantages are provided in Table~\ref{Table_DL_Cl_Networks} and will be discussed throughout this section. However, we study these networks mainly from an IoUT as well as BMD perspective and discuss what specific network types and algorithms would be particularly suitable both for BMD analysis and for IoUT applications.  
		In doing so, we appraise many reported use cases of ML in underwater applications in Section~\ref{SubSection_ML_in_BMD}, but before that, let us discuss the various software frameworks for architectural deep network design, suitable for big marine data types.
		
	\subsection{Deep Learning Frameworks and Libraries}\label{SubSection_DL_Frameworks}	
		To facilitate the development of various DL architectures, such as those listed in Tables~\ref{Table_DL_Cl_Networks} and \ref{Table_CNN_Architectures} for different applications, numerous software frameworks and libraries have been developed for open access by the ever-growing deep learning community \cite{Ravi2017}. Table~\ref{Table_DL_Frameworks} lists and compares a range of open-source DL tools and frameworks. Some of the abbreviations used in this table are Deep Learning for Java (DL4j), Microsoft Cognitive Toolkit (previously known as CNTK), and TensorFlow (TFlow).
		
		\begin{table*}[!t]
			\renewcommand{\arraystretch}{1.5}
			\caption{Comparing the Best Open-source Software Frameworks and Libraries for Deep Learning}
			\label{Table_DL_Frameworks}
			\centering
			\includegraphics[width=\textwidth, clip, trim={40 450 40 60}]{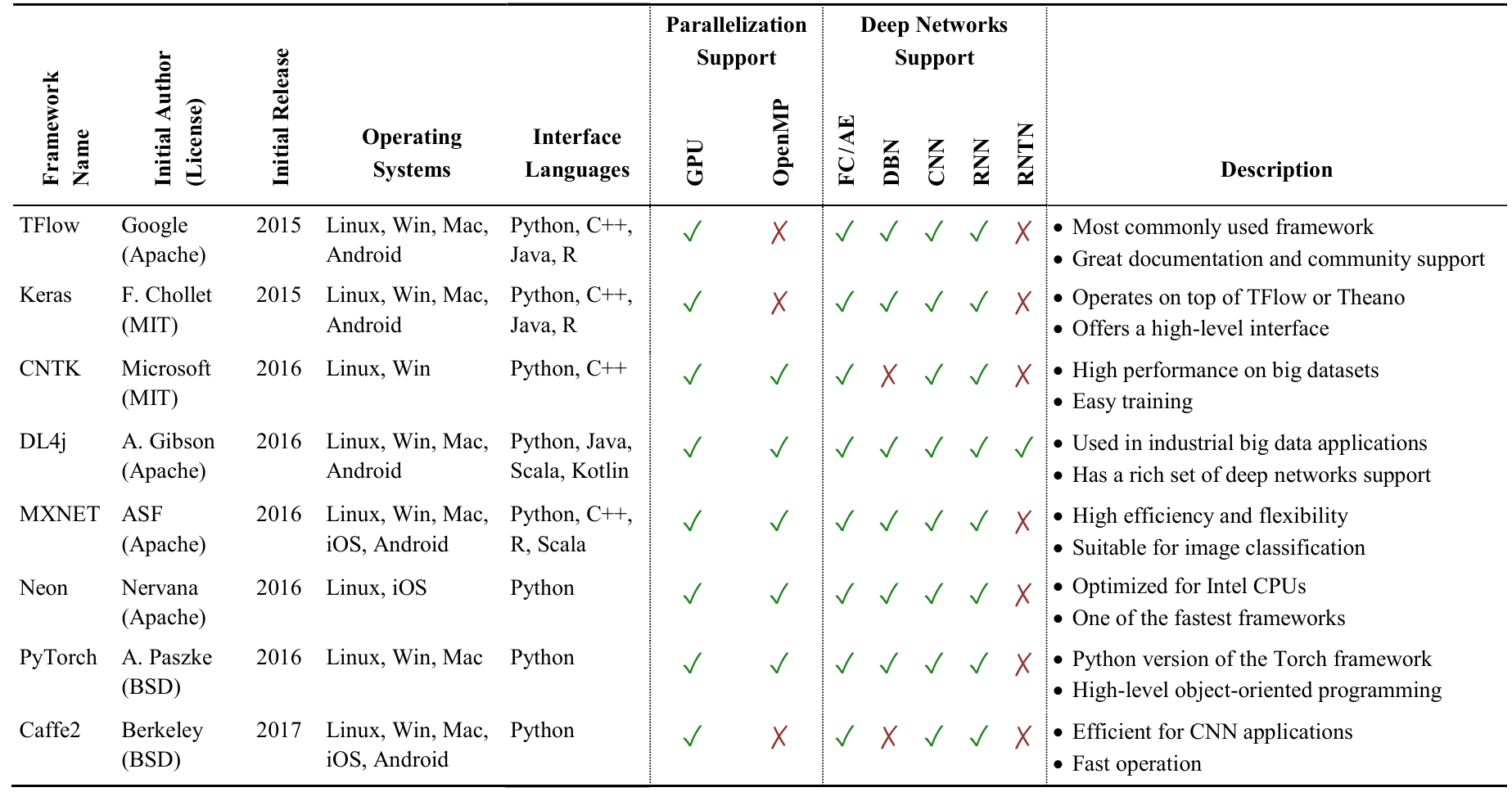}
		\end{table*}
		
		All the frameworks seen in Table~\ref{Table_DL_Frameworks} are capable of operating on NVidia\textsuperscript{\textregistered} CUDA-supported GPUs relying on parallel processing. By contrast, OpenMP is another shared-memory based multiprocessing programming interface, which is only supported by some of the frameworks. As it will be discussed in Section~\ref{SubSection_HW_Realization}, shared-memory based methods are critically important for speeding up the deep network's training process. 
		
		So far we have discussed various architectures capable of processing BMD and marine data. Once the data has been collected, some pre-processing steps have to be performed for ensuring that the data used to train the ML algorithm is clean and does not result in future learning problems. In the next section, this important issue is discussed in more details.
	
	\subsection{Marine Data Cleaning for Machine Learning}
	Data preparation and cleansing is a critical step in almost every ML project and it usually takes over half of the entire project duration to collect and clean the data. When dealing with the unaccessible harsh underwater environment of the IoUT, data cleansing becomes even more important. This is because the underwater environment is affected by many environmental factors, as shown in Fig.~\ref{Fig_InfoG_Phys_Layer_Roughness}, which make it almost impossible to acquire clean data. Additionally, data sources tend to be quite vulnerable to the hostile underwater environment. Therefore, data preprocessing and cleaning, before training is crucial for the success of any ML aided BMD processing in IoUT. Below, various IoUT data types and their main preparatory considerations are discussed.
	
	\subsubsection{Underwater sensor data cleaning}\label{SubSection_Sensor_Preparation}
	Sensors are inevitable parts of every IoUT subsystem and are used in nearly all underwater exploration and surveillance applications \cite{Xiao2019}. They continuously measure physical, chemical, as well as biological parameters and generate a huge volume of data. Examples of these sensor and data are listed in Table~\ref{Table_Primary_Data_Source_Sensor}.
	
	Data cleansing for sensory data sources in underwater BMD applications is typically concerned with missing values, contaminated measurements, and detecting outliers. There are a variety of techniques for these preparatory steps. To deal with a missing value in any sensor data, one can \cite{Aggarwal2015},
	
	\begin{itemize}
		\item Delete the entire record;
		\item Use a global constant;
		\item Use a statistical value (e.g. mean or median);
		\item Use an inter-class statistical value;
		\item Use the most likelihood (i.e. more probable) value;
		\item Use the most likelihood inter-class value.
	\end{itemize}
	
	Noise is another phenomenon affecting almost any sensory data. The process of de-noising a sensory measurement is termed as \textit{noise cancellation}. Although no universally applicable de-noising methods exist, one should find the one meeting the SNR criteria. The SNR is a very common parameter of characterizing the performance of different noise cancellation methods. Some of these methods include:
	
	\begin{itemize}
		\item Using a low pass filter \cite{Gupta2019};
		\item Using regression models \cite{James2017};
		\item Applying a data binning method \cite{WangM2017};
		\item Using wavelet methods in time-series \cite{Aggarwal2015}.
	\end{itemize}
	
	Among these methods, low pass filter is the most common one and wavelet-based methods are the most complex ones to implement. Additionally, data binning is not particularly popular for sensory data, but it is widely used for images as it will be discussed in the next section.
	
	The final step in underwater sensory data cleansing is the outlier detection. This step can be readily carried out by considering three well-known criteria, named the IQR, T\textsuperscript{2}, and Q criteria \cite{Aggarwal2015}. There are also some more advanced methodologies like, probabilistic models, clustering methods, distance-based detection, density-based detection, etc. \cite{Aggarwal2015}.
	
	\subsubsection{Underwater image and video data cleaning and quality enhancement}\label{SubSection_Image_Preparation}			
	As we have already discussed in Section~\ref{SubSection_UW_Comm}, underwater image acquisition suffers from strong absorption and scattering. Additionally, there are two other undersea signal degradation factors, namely chromatic aberrations \cite{LiC2018, PengY2017, WangN2017} (i.e. color distortion or chromatic distortion) and noise imposed by both natural and artificial light sources \cite{LiM2018}. These destructive factors significantly affect the quality of the captured video or image and should be mitigated by a data cleansing algorithm.
	
	Any image and video preparation algorithm has to assess the image quality first. Automatic image quality evaluation algorithms assign an objective metric, which is a weighted sum of the image colorfulness, contrast, sharpness, etc. \cite{WangYan2018, Panetta2016}. This metric is then used as an objective function to be maximized by other image enhancement procedures. Therefore, underwater image and video quality assessment constitutes an essential step before image retrieval, image quality optimization, video compression, and other visual signal processing steps. It can also act as a guide for determining the data bandwidth required by the underwater communication algorithms, as well as by other decision-making processes \cite{Yang2015}.
	
	After the automatic quality assessment of the underwater images and videos, software-based image preparation algorithms are used for enhancing the image quality. As a general rule, it is always cheaper to improve the image quality in software, instead of implementing bespoke high-cost imaging devices.
	
	In a comprehensive review paper by Han \textit{et al.} \cite{HanM2018}, the image preparation methods are divided into image dehazing and color enhancement. The authors then introduce and compare numerous methods for each category. However, they have missed the innovative method proposed by Ancuti \textit{et al.} \cite{Ancuti2018} that addresses both image dehazing and image enhancement at the same time. This method improves the global contrast that was degraded by light absorption, increases the edge sharpness impaired by light scattering, and exposes the dark regions with the aid of color balancing.
	The proposed method relies on a single-image camera-independent technique that can be applied to both photos and full-motion videos. Furthermore, it does not require any prior knowledge about the environmental conditions.
	
	\subsubsection{Underwater image data binning}\label{SubSection_Image_Data_Binning}
	In addition to image enhancement, image data binning is another data cleansing method that can be used in IoUT, which relies on grouping pixels into distinct partitions based on the similarity of their characteristics \cite{WangM2017}.
	The goal of constructing these partitions is both to reduce the amount of noise and to resolve data complexity. After binning, every pixel has a label to explicitly indicate its parent partition, based on its membership index. Although image binning is not an essential preparation step, it is recommended before invoking feature extraction for reducing the computational complexity and to speed up the ML process \cite{WangM2017}.
	
	
	Image clustering methods, such as image binning, were developed much earlier than the state-of-the-art ML and DL algorithms and date back to the age of statistical mathematics. However, none of these methods are universally accepted and it is still a challenging task to select the most appropriate image partitioning method for a given application. A list of image clustering methods, that are beneficial in underwater applications is provided in Table~\ref{Table_Image_Clustering_Methods}. These methods either tend to exploit image discontinuities (like edges) or similar regions to partition a given image.
	
	\begin{table*}[!t]
		\renewcommand{\arraystretch}{1.2}
		\caption{Image Clustering Methods and Algorithms in Underwater Applications}
		\label{Table_Image_Clustering_Methods}
		\centering
		\begin{tabular}{@{}m{0.45\columnwidth}@{} @{}m{0.07\columnwidth}@{} @{}m{0.7\columnwidth}@{} @{}m{0.07\columnwidth}@{} @{}m{0.7\columnwidth}@{}}
			\toprule
			\textbf{Method (basis)} && \multicolumn{1}{c}{\textbf{Algorithms}} && \multicolumn{1}{c}{\textbf{Reported Underwater Applications}}\\
			\midrule
			Edge-detection \newline (based on the 1\textsuperscript{st} and 2\textsuperscript{nd} \newline order image derivatives)\newline\newline&&
			$\bullet$ Watershed \newline
			$\bullet$ Snakes (Active Contour Model) \newline
			$\bullet$ Blob detection (i.e. Laplacian of the Gaussian,\newline \textcolor{white}{..}
			Determinant of Hessian, Difference of\newline \textcolor{white}{..}
			Gaussians, etc.)&&
			$\bullet$ Jellyfish Detection by Watershed and Snakes \cite{Rife2003} \newline
			$\bullet$ Plankton Recognition by Snakes \cite{Kocak1999} \newline				
			$\bullet$ Deep Sea Megafauna Recognition by blob \cite{Schoening2015}\newline\newline\\
			\midrule
			Thresholding \newline (based on a threshold value \newline (clip-level) in feature space) \newline&&
			$\bullet$ Otsu's Method (Maximum Variance) \newline
			$\bullet$ Expectation-Maximization \newline
			$\bullet$ Maximum Entropy Method \newline
			$\bullet$ Color or Intensity Histogram Thresholding&&
			$\bullet$ Fish Detection \cite{Chuang2011} \newline
			$\bullet$ Jellyfish Detection by Expect.-Maximization \cite{Rife2003} \newline
			$\bullet$ Deep Sea Megafauna Recognition \cite{Schoening2012} \newline
			$\bullet$ Plankton Recognition \cite{Faillettaz2016, Tong2004}\\
			\midrule
			Distance- or Region-based \newline (based on the similarity in \newline attributes) \newline \newline \newline \newline \newline \newline \newline \newline \newline&&
			$\bullet$ Region Growing (RG) \newline
			$\bullet$ Simple Linear Iterative Clustering (SLIC) \newline
			$\bullet$ Region Splitting and Merging \newline
			$\bullet$ Morphological Spatial Pattern Analysis (MSPA) \newline
			$\bullet$ Statistical Clustering Methods (K-Means, \newline \textcolor{white}{..} Subtractive Clustering, AGNES, DBSCAN) \newline
			$\bullet$ Neural Clustering Methods (Self-Organizing \newline \textcolor{white}{..} Map (SOM), Hierarchical SOM, Hierarchically \newline \textcolor{white}{..} Growing Hyperbolic SOM) \newline
			$\bullet$ Fuzzy Clustering Method (FCM) \newline
			$\bullet$ Wavelet Transform \newline
			$\bullet$ Compression (Texture and Boundary)&&
			$\bullet$ Shrimp Detection by SLIC \cite{Osterloff2016} \newline
			$\bullet$ Plant Detection by Gabor Wavelet \cite{Bonin-Font2017} \newline
			$\bullet$ Plant Recognition by Gabor Wavelet \cite{Johnson-Roberson2006} \newline
			$\bullet$ Plankton Recognition by MSPA \cite{Schmid2016, Hu2006} \newline
			$\bullet$ Coral Detection by Clustering \cite{Purser2009} \newline
			$\bullet$ Object Detection by SLIC \cite{Jian2018} \newline
			$\bullet$ Object Detection by Gabor Wavelet \cite{Reggiannini2017} \newline
			$\bullet$ Object Detection by Clustering \cite{Schoening2016} \newline
			$\bullet$ General Underwater Clustering by FCM \cite{LiX2016} \newline
			$\bullet$ General Underwater Clustering by RG \cite{Peizhe2011} \newline
			$\bullet$ Coral Recognition by MSPA \cite{Shihavuddin2013} \newline
			$\bullet$ Coral Recognition by RG \cite{Benfield2007} \\
			\midrule
			Math Calculus \newline (based on solving a Differential or Integral Equation)&&
			$\bullet$ Level Set Model \newline
			$\bullet$ Mumford Shah Model \newline
			$\bullet$ Chan-Vese Model&&
			To the best of authors' knowledge,  this method has never been used in underwater applications.\newline \\
			\midrule
			Graph Cut \newline (based on the undirected graph partitioning (i.e. Markov Random Fields) to model the impact of pixel neighborhoods) \newline&&
			$\bullet$ Normalized Cuts \newline
			$\bullet$ Lazy Random Walks \newline
			$\bullet$ Isoperimetric Partitioning \newline
			$\bullet$ Minimum Spanning Tree-based \newline
			$\bullet$ GrabCut \newline
			$\bullet$ Model Based&&
			$\bullet$ Fish Recognition by GrabCut \cite{Chuang2016, HuangP2015} \newline \newline \newline \newline \newline \\
			\midrule
			Video Motion \newline (based on the object movement) \newline&&
			$\bullet$ Subtracting a Pair of Images \newline \newline&&
			$\bullet$ Fish Detection \cite{Chambah2004, Zhang2016} \newline
			$\bullet$ Fish Recognition \cite{Spampinato2010} \newline
			$\bullet$ Lobster Detection \cite{Pons2010} \\
			\bottomrule
		\end{tabular}
	\end{table*}
	
	In addition to the algorithms introduced in Table~\ref{Table_Image_Clustering_Methods}, there are many others that have however been more rarely used in the literature. These methods are reviewed by Flake \textit{et al.} \cite{Flake2004} as well as Wang \textit{et al.} \cite{WangM2017} and might be worth investigating in future studies. Some of them are the Minimum Cut, Mean-Shift, Turbo-Pixels, Lattice Cut, Compact Super-Pixels, Constant Intensity Super-Pixels, Entropy Rate Super-Pixels, Homogeneous Super-Pixels, Topology Preserved Regular Super-Pixel, SEEDS, VCells, Depth-Adaptive Super-Pixels, Voxel Cloud Connectivity Segmentation, Structure Sensitive Super-Pixels, Saliency-based Super-Pixel, and Linear Spectral Clustering.
	
	In addition to data cleansing, feature extraction is another essential step for any ML applications, which rely on statistical methods and traditional NNs. This substantial step in ML is discussed in the next section. 
	
	\subsection{Feature Extraction for Marine Data Analytics}\label{SubSection_Feature}
	Feature extraction is a mathematical step, in which raw data is replaced by its numerical descriptors. This step is responsible for transforming large vectors of sensory data and large matrices of image data into their low-dimensional vector-based representatives.
	It is usually the most intricate part of almost any ML-aided computer vision problem and the solution should be tailored for the particular needs of the specific project at hand. Although all of the modern deep learning algorithms extract their own feature vectors automatically, these deep networks are hard to train, when relying on limited training datasets.
	
	Finding a series of useful features is even harder in underwater image processing applications in high-dynamic environments subject to non-uniform light illumination, variable scene brightness, and degraded colors. Table~\ref{Table_Feature_Sets} lists a number of salient descriptor routinely used in underwater applications. These features are categorized into four classes, including color, texture, shape (boundary), and other descriptors.
	The papers listed in Table~\ref{Table_Feature_Sets} tend to use a series of descriptors, depending on their specific target applications. In one case, the number of descriptors used has been as high as 66 features in the Fish Recognition project carried out by Huang \textit{et al.} \cite{HuangP2015}.
	
	\begin{table}[!t]
		\renewcommand{\arraystretch}{1.2}
		\caption{Data Processing and Machine Learning Feature-sets in Published Underwater Applications}
		\label{Table_Feature_Sets}
		\centering
		\begin{tabular}{@{}m{0.6\columnwidth}@{} @{}m{0.02\columnwidth}@{} @{}m{0.38\columnwidth}@{}}
			\toprule
			\multicolumn{1}{c}{\textbf{Color Descriptors}} && \multicolumn{1}{c}{\textbf{Underwater Applications}}\\
			\cmidrule{1-3}
			$\bullet$ Image statistics (e.g. Hu's 7 invariant \newline \textcolor{white}{..} moments, affine moment invariants, \newline \textcolor{white}{..} Skewness, kurtosis, mean) && \hspace{-0.1pt}\cite{Osterloff2016, Schmid2016, Faillettaz2016, Benfield2007, HuangP2015, Spampinato2010, Hu2006, Ludtke2012, Westling2014} \newline\\
			
			$\bullet$ Segment contrast && \hspace{-0.1pt}\cite{Osterloff2016}\\
			
			$\bullet$ Histogram descriptor \newline && \hspace{-0.1pt}\cite{Schoening2016, Osterloff2016, Shihavuddin2013, Schoening2015, Chuang2016, HuangP2015, Pugh2014}\\
			
			$\bullet$ Five MPEG7 color features && \hspace{-0.1pt}\cite{Schoening2015, Schoening2012}\\
			
			$\bullet$ Transparency ratio && \hspace{-0.1pt}\cite{Tong2004}\\
			
			\cmidrule{1-1}
			\multicolumn{1}{c}{\textbf{Texture Descriptors}} &&\\ 
			\cmidrule{1-1}
			$\bullet$ Three MPEG7 texture features && \hspace{-0.1pt}\cite{Schoening2015, Schoening2012}\\
			
			$\bullet$ Wavelet transform && \hspace{-0.1pt}\cite{Purser2009}\\
			
			$\bullet$ Gabor filter \newline && \hspace{-0.1pt}\cite{Bonin-Font2017, Shihavuddin2013, Schoening2015, Johnson-Roberson2006, HuangP2015, Spampinato2010, Pugh2014}\\
			
			$\bullet$ Filter banks (e.g. Schmid, maximum \newline \textcolor{white}{..} response, Leung and Malik, root \newline \textcolor{white}{..} filter set) \newline && \hspace{-0.1pt}To the best of the authors' knowledge, these descriptors have never been used in underwater applications.\\
			
			$\bullet$ Grey level co-occurrence matrices \newline \textcolor{white}{..} properties (contrast, correlation, energy, \newline \textcolor{white}{..} and homogeneity) && \hspace{-0.1pt}\cite{Lau2012, Schmid2016, Faillettaz2016, Shihavuddin2013, HuangP2015, Spampinato2010, Hu2006, Ludtke2012, Spampinato2008} \newline\\
			
			\cmidrule{1-1}
			\multicolumn{1}{c}{\textbf{Shape Descriptors and Key Points}} &&\\ 
			\cmidrule{1-1}
			$\bullet$ Hough transform && \hspace{-0.1pt}\cite{Foresti2001}\\
			
			$\bullet$ HOG (Histograms of Oriented Gradients) \newline \textcolor{white}{..} based (e.g. SIFT\textsuperscript{\textcopyright}, SURF\textsuperscript{\textcopyright}, GLOH) && \hspace{-0.1pt}\cite{Reggiannini2017, Schoening2015, Chuang2016, HuangP2015, Fouad2013} \newline\\
			
			$\bullet$ Binary descriptors && \hspace{-0.1pt}\cite{Sun2018, AniBrownMary2017, Shihavuddin2013, Bonin-Font2016, Chuang2014} \\
			
			$\bullet$ Weber local descriptor && \hspace{-0.1pt}\cite{Elawady2015}\\
			
			$\bullet$ Convexity && \hspace{-0.1pt}\cite{Schmid2016, Tong2004, Westling2014}\\
			
			
			$\bullet$ Fourier descriptors && \hspace{-0.1pt}\cite{Schmid2016, Faillettaz2016, HuangP2015, Spampinato2010, Hu2006}\\
			
			$\bullet$ Frequency domain descriptors like phase \newline \textcolor{white}{..} congruency && \hspace{-0.1pt}\cite{Elawady2015} \newline\\
			
			$\bullet$ Curvature scale space transform && \hspace{-0.1pt}\cite{Spampinato2010}\\
			
			\cmidrule{1-1}
			\multicolumn{1}{c}{\textbf{Other  Descriptors}} &&\\ 
			\cmidrule{1-1}
			$\bullet$ Granulometrics (size, area, and orientation) \newline \textcolor{white}{..} to recognize an already detected object && \hspace{-0.1pt}\cite{Lau2012, Schmid2016, Faillettaz2016, Chuang2015, Tong2004, HuangP2015, Hu2006}\\
			
			$\bullet$ Eigenvalues and covariance matrix \newline \textcolor{white}{..} to recognize an already detected object && \hspace{-0.1pt}\cite{Tong2004} \newline \\
			
			$\bullet$ Motion related && \hspace{-0.1pt}\cite{Lau2012}\\
			
			$\bullet$ 3D shape, surfaces, and texture descriptors && \hspace{-0.1pt}\cite{Johnson-Roberson2006}\\
			\bottomrule
		\end{tabular}
	\end{table}
	
	After selecting a number of features, it is recommended to mathematically evaluate their correlation and then reduce the number of features to the number of truly independent descriptors. This process is termed as \textit{data dimensionality reduction} and can be performed by feature reduction techniques, feature selection, and feature aggregation \cite{Ambika2019}.
	
	Some of the popular dimensionality reduction techniques found in literature are Principal Component Analysis (PCA), Linear Discriminant Analysis (LDA),
	independent component analysis, non-negative matrix factorization, Self-Organizing Map (SOM), sequential forward search, sequential backward search, bag of words, etc. \cite{Ambika2019, Gepshtein2019, Mahdianpari2018, Tang2019, Zhao2018}.
	
	So far in this section, the software components suitable for ML and DL aided IoUT have been studied. However, these ML-aided solutions also rely on appropriate high-performance hardware platforms.

	\subsection{Hardware Platforms for ML in IoUT}\label{SubSection_HW_Realization}
	The implementation of any ML solution, from its data cleansing and algorithm development to its final deployment, may rely on a variety of hardware platforms. These hardware platforms typically boost the overall throughput by parallel processing. These processing methods can be broadly divided into the following two main categories:
	
	\begin{itemize}
		\item \textit{Shared-memory multi-processors}: They rely on multiple processors that all share a memory unit \cite{Ceze2011} and the ML algorithm resides within this memory unit. Some of the more popular shared-memory methods are based on Application-Specific Integrated Circuits (ASIC) \cite{LeeJ2019}, Field-Programmable Gate Arrays (FPGA) \cite{Venieris2019}, multi-purpose and multi-core CPUs, and Graphics Processing Units (GPU) \cite{Manderson2018}.
		
		\item \textit{System of networked processors}: As already introduced in Section~\ref{SubSection_Data_Processing_Platforms}, a Distributed Computing System (DCS) is a system of networked processors, which coordinates the processors' actions by passing messages to each other \cite{Coulouris2011, Xingang2019}. They can be used to parallelize ML algorithms in three different ways, including:
		
		\begin{itemize}
			\item \textit{Data parallelization}: This is performed by running the same ML algorithm in all of the distributed computers and then dividing the data between them. Every computer estimates all parameters based on a separate dataset, before they exchange their estimates for formulating a final estimate. MapReduce is one of the most popular data parallelization methods \cite{XiaD2016, XuZ2019}.
			
			\item \textit{Model parallelization}: In this paradigm, the ML algorithm will have to be decomposed into different functions and operands. The algorithmic parts are then shared among multiple computers and every part has its own specific set of parameters. The input data, however, will be simultaneously fed to all of them, where every computer is responsible for estimating the set of parameters assigned. During the parameter training process, those parallelized computers exchange their partial error vectors back and forth to meet the convergence criteria 
			and to come up with the final parameter estimations \cite{MingY2018, Garcia2012}.
			
			\item \textit{Pipelined parallelization}: In this method, the algorithm is shared among distributed computers, similar to the above-mentioned model parallelization method. These parallelized algorithm parts are chained together from one input block to one (or more) output blocks, termed as \textit{pipelines}. In contrast to model parallelization, data will also split into a series of records. At the beginning, the first record is passed to the first computer in the DCS, to carry out its own task (i.e. local parameter adjustment). The output of this computer is then relayed to the next computer to carry out its own task over the first data record. Meanwhile, the first computer starts processing the second data record. This pipe-lined process continues until the last record in the database exits the DCS and consequently, the local parameters of the last block in the pipeline are updated \cite{Schwartz2019}.
		\end{itemize}
		
	\end{itemize}
	
	A rudimentary qualitative comparison of different parallel processing hardware platforms used for ML-aided and BMD processing is offered in Table~\ref{Table_Hardware_Platforms}.
	Here, the strength of each platform in terms of a specific criterion is indicated by the $+$ and $-$ signs, where more $+$s represents better performance in the context of that specific benchmark.
	The last two rows of this table summarize the entire table with respect to the associated research or industrial applications. For instance, as the Table shows, the CPUs require the lowest design time and impose the lowest design complexity, but they suffer from the lowest operational speed in all design and application phases, i.e. data cleansing, NN training, and final inference. Therefore, they may be used for research, but they are less suitable for industrial applications.

	\begin{table}[!t]
		\renewcommand{\arraystretch}{1.3}
		\caption{Comparing Different Parallel Processing Hardware Platforms to Carry out a Typical Machine Learning Algorithm}
		\label{Table_Hardware_Platforms}
		\centering
		\begin{tabular}{@{}m{0.29\columnwidth}@{} @{}m{0.02\columnwidth}@{} @{}m{0.1\columnwidth}@{} @{}m{0.02\columnwidth}@{} @{}m{0.1\columnwidth}@{} @{}m{0.02\columnwidth}@{} @{}m{0.1\columnwidth}@{} @{}m{0.02\columnwidth}@{} @{}m{0.1\columnwidth}@{} @{}m{0.02\columnwidth}@{} @{}m{0.1\columnwidth}@{}}
			\toprule
			\multicolumn{1}{c}{\textbf{Criteria}} && \multicolumn{1}{c}{\textbf{ASIC}} && \multicolumn{1}{c}{\textbf{FPGA}} && \multicolumn{1}{c}{\textbf{CPU}} && \multicolumn{1}{c}{\textbf{GPU}} && \multicolumn{1}{c}{\textbf{DCS}}\\
			\midrule
			
			Inference Speed && \multicolumn{1}{c}{$++$} && \multicolumn{1}{c}{$++$} && \multicolumn{1}{c}{$--$} && \multicolumn{1}{c}{$+$} && \multicolumn{1}{c}{$-$}\\
			
			Product Unit Cost && \multicolumn{1}{c}{$++$} && \multicolumn{1}{c}{$+$} && \multicolumn{1}{c}{$+$} && \multicolumn{1}{c}{$-$} && \multicolumn{1}{c}{$--$}\\
			
			Design Complexity && \multicolumn{1}{c}{$--$} && \multicolumn{1}{c}{$-$} && \multicolumn{1}{c}{$++$} && \multicolumn{1}{c}{$+$} && \multicolumn{1}{c}{$-$}\\
			
			Occupied Space && \multicolumn{1}{c}{$++$} && \multicolumn{1}{c}{$++$} && \multicolumn{1}{c}{$-$} && \multicolumn{1}{c}{$+^a$} && \multicolumn{1}{c}{$--$}\\
			
			\midrule
			Research Friendly && \multicolumn{1}{c}{$--$} && \multicolumn{1}{c}{$-$} && \multicolumn{1}{c}{$+$} && \multicolumn{1}{c}{{\cellcolor{WordLightGreen}$++$}} && \multicolumn{1}{c}{$+^b$}\\
			
			Market Friendly && \multicolumn{1}{c}{{\cellcolor{WordLightGreen}$++$}} && \multicolumn{1}{c}{$+$} && \multicolumn{1}{c}{$--$} && \multicolumn{1}{c}{$-$} && \multicolumn{1}{c}{$+^c$}\\
			\bottomrule
		\end{tabular}
		\begin{flushleft}
			$^{a}$\,New GPU platforms such as NVIDIA Jetson occupy small spaces.
			
			$^{b,\,c}$\,Cloud-based DCS platforms such as AWS EMR are readily available.
		\end{flushleft}
	\end{table}
	
	Perhaps the most prominent choice for typical scientific application would be the GPU \cite{Modasshir2018}. The hardware and software resources required for GPUs are affordable and their speed is high enough to cover almost any application. Both the research-based as well as the industrial-scale employment of GPUs in underwater data processing and ML applications have been frequently reported in the literature. These applications range from obstacle detection and collision avoidance, to image-based SLAM localization, and even further to underwater object detection (e.g., coral detection) \cite{KimH2019, Manderson2018}.
	
	However, the major problem with GPUs is their dependence on a bulky host computer. This has limited their implementation in low-power and lightweight IoUT platforms as well as underwater vehicles. To address this problem, new compact GPU designs have emerged to offer AI-ready computing resources. For example, \href{https://developer.nvidia.com/embedded/develop/hardware}{NVIDIA Jetson} is a standalone GPU-accelerated embedded system, which has a small volume. The high performance, low power, and compact form-factor of the Jetson family makes them ideal for example for deep learning aided computer vision applications \cite{KimH2019}.
	
	Since the Jetson embedded systems are empowered by the Linux Ubuntu operating system, they are eminently suitable for deep NN frameworks of Table~\ref{Table_DL_Frameworks}, e.g., TensorFlow. Thus, the low-latency inference capability of machine learning algorithms can be readily realized in underwater vehicles and platforms \cite{Manderson2018}.
	
	Additionally, NVIDIA's Compute Unified Device Architecture (\href{http://developer.download.nvidia.com/compute/cuda/2_0/docs/CudaReferenceManual_2.0.pdf}{CUDA}) library unleashes the GPU's parallel processing capabilities for applications other than machine vision. As previously mentioned in Section~\ref{SubSection_DL_Frameworks}, CUDA is an API model that allows engineers to use a CUDA-enabled GPU for general purpose processing, this is also referred to in parlance as a \textit{General Purpose GPU} (GPGPU). For instance, Pallayil \textit{et al.} \cite{Pallayil2016} have used a GPGPU for performing high-complexity real-time beamforming in their underwater acoustic phased array. They simply used the popular C Language in the Linux environment to harness the fast Fourier transform in the CUDA library, to implement their underwater frequency-domain beamformer.
	
	ASICs and FPGAs are listed in Table~\ref{Table_Hardware_Platforms}, which have the lowest form-factor. They also offer a high throughput and high power efficiency, which makes them eminently suitable for industry-scale IoUT projects \cite{Karabchevsky2011}. Given these compelling attributes, FPGAs are repeatedly featured in real-time and low-power underwater applications. For example, Karabchevsky \textit{et al.} \cite{Karabchevsky2011} have implemented a standalone FPGA architecture for noise suppression in underwater sonars. Their proposed signal processing implementation is claimed to overcome the sonar-based underwater visibility problems.
	
	Following the above survey of various data processing methods, platforms, and their hardware realizations, diverse ML algorithms used for underwater applications will be discussed in the next section.
	
	\subsection{ML Techniques in BMD Applications}\label{SubSection_ML_in_BMD}
	The advantages of ML techniques make them eminently suitable for most underwater applications. The categorical investigation of these techniques as well as their implementation in BMD applications will be carried out in this section.
	
		\subsubsection{Deep NNs for static IoUT data}
		Static data is exemplified by still images in contrast to full-motion video clips. In the context of Table~\ref{Table_DL_Cl_Networks}, the fully-connected, Deep Belief Network (DBN) \cite{Hinton2006}, CNN \cite{Lecun1995}, Autoencoder \cite{KimK2018}, and Recursive Neural Tensor Network (RNTN) \cite{Socher2013} networks are eminently suitable for static marine data processing.
		In the case of RNTNs, they are usually fed by a single static underwater image in every single data entry step. This image consists of multiple segments, and RNTN is supposed to determine the hierarchy of every segment inside the given image (e.g. background, coral, plant, fish, etc.). The double-colored nodes of RNTN in Table~\ref{Table_DL_Cl_Networks} are hidden neurons that can also be designated by the network designer to be an output neuron, representing those expected hierarchical segments.
		
		Among the static data processing networks seen in Table~\ref{Table_DL_Cl_Networks}, Autoencoders are the only unsupervised clustering NNs. Every Autoencoder consists of two parts, which can be used either separately or simultaneously. The first part processes a static input data such as an underwater image and extracts its features. The second part on the other hand, takes the feature vector and tries to reconstruct the input image again. If the inherent image features were adequately captured, the reconstructed input data appearing at the output will be similar to the input data itself. From an application-oriented perspective, one can use the first part of the Autoencoder for dimensionality reduction and data compression, while the second part is suitable for data decompression. Meanwhile, by using both parts simultaneously together, Autoencoder will act like a noise reducing NN.
		
		Another network architecture in Table~\ref{Table_DL_Cl_Networks} is the family of DBNs. These networks consist of consecutive shallow learning Restricted Boltzmann Machine (RBM) pairs, which gradually reduces the dimension of input data from the top-level of the entire search/classification towards the final unique classifier. Any mid-layer in DBN may act as output layer of the feature extraction. Again, by continuously decreasing the feature vector size, the procedure culminates by unambiguously classifying the input data. Therefore, DBN can act as an unsupervised clustering NN, provided that it is terminated somewhere at a mid-layer. By contrast, it can also act as a supervised classifier, if the number of nodes is reduced to the specific number of classes and if labeled data is used to train the network.
		
		The final deep network for static data that will be discussed here is the convolutional neural network. This network relies on multiple stages of convolution and pooling, as seen in Table~\ref{Table_DL_Cl_Networks}. It is considered to be the best deep classification method, especially when dealing with static images or previously recorded outputs of sensors and of hydrophones. Several architectures of this popular network have been designed for scientific use, some of which are also supported by Goggle and Microsoft (as exemplified by Inception \cite{Szegedy2017} and ResNet \cite{HeK2016}). Several of these CNN architectures are compared in Table~\ref{Table_CNN_Architectures}, based on the excellent review article by Canziani \textit{et al.} \cite{Canziani2017}. The Top-1 accuracy\footnote{Classifiers usually assign a probability value to all of their output classes. Thereafter, the class with the highest probability (top-1) will be considered as the final answer, which is not always true. Sometimes the correct answer is among the top-N classes. Using top-5 accuracy is common, when comparing different classifiers.} listed in this table is measured with the aid of the single central-crop sampling technique\footnote{A given image may have four corner crops and one central crop.} of \cite{ZagoruykoGIT2016} for all of the networks.

		\begin{table}[!t]
			\renewcommand{\arraystretch}{1.5}
			\caption{Comparison of a selected number of CNN architectures submitted to the Annual ImageNet Challenge \cite{Canziani2017, ZagoruykoGIT2016}}
			\label{Table_CNN_Architectures}
			\centering
			\begin{tabular}{@{}m{0.27\columnwidth}@{} @{}m{0.0\columnwidth}@{} @{}m{0.17\columnwidth}@{} @{}m{0.0\columnwidth}@{} @{}m{0.17\columnwidth}@{} @{}m{0.0\columnwidth}@{} @{}m{0.17\columnwidth}@{}}
				\toprule
				\multicolumn{1}{c}{\textbf{\shortstack[c]{Architecture \\ Name}}} && \multicolumn{1}{c}{\textbf{\shortstack[c]{Top-1 \\ Accuracy}}} && \multicolumn{1}{c}{\textbf{\shortstack[c]{No. of \\ Parameters}}} && \multicolumn{1}{c}{\textbf{\shortstack[c]{No. of \\ Operations}}}\\
				\midrule
				AlexNet \cite{Krizhevsky2012} && \multicolumn{1}{l}{\;\,$>56\%$} && \multicolumn{1}{l}{\;\;\;$\approx$ 50\,M} && \multicolumn{1}{l}{\;\;\;\;\textbf{$\approx$ 2\,G}}\\
				
				NIN \cite{LinM2013} && \multicolumn{1}{l}{\;\,$>62\%$} && \multicolumn{1}{l}{\;\;\;\textbf{$\approx$ 6\,M}} && \multicolumn{1}{l}{\;\;\;\;$\approx$ 3\,G}\\
				
				VGGNet-19 \cite{Simonyan2014} && \multicolumn{1}{l}{\;\,$>70\%$} && \multicolumn{1}{l}{\;\;\;$\approx$ 150\,M} && \multicolumn{1}{l}{\;\;\;\;$\approx$ 39\,G}\\
				
				Inception-v4 \cite{Szegedy2017} && \multicolumn{1}{l}{\;\,\textbf{$>$ 80\%}} && \multicolumn{1}{l}{\;\;\;$\approx$ 35\,M} && \multicolumn{1}{l}{\;\;\;\;$\approx$ 18\,G}\\
				
				ResNet-152 \cite{HeK2016} && \multicolumn{1}{l}{\;\,\textbf{$>$ 77\%}} && \multicolumn{1}{l}{\;\;\;$\approx$ 50\,M} && \multicolumn{1}{l}{\;\;\;\;$\approx$ 23\,G}\\
				
				ENet \cite{PaszkeGIT2016} && \multicolumn{1}{l}{\;\,$>67\%$} && \multicolumn{1}{l}{\;\;\;\textbf{$\approx$ 5\,M}} && \multicolumn{1}{l}{\;\;\;\;\textbf{$\approx$ 2\,G}}\\
				
				\bottomrule
			\end{tabular}
		\end{table}
		
		The number of network parameters in Table~\ref{Table_CNN_Architectures} is proportional both to their memory footprint and to their required training time. On the other hand, the number of operations required for a single forward pass, as shown in the table is capable of indicating the overall inference speed of the network. The lower the number of operations, the higher the inference speed. Here, ENet is not a CNN, but it is directly inspired by CNNs. By the same token, Network in Network (NIN) is not a CNN, but it relies on convolution operations. Some of these architectures are also included in the deep learning libraries of Section~\ref{SubSection_DL_Frameworks}
		and can be readily involved for any project.
		Let us now extend our discussions to cover deep NNs in dynamic systems.
		
		\subsubsection{Deep NNs for dynamic IoUT data}
		Recurrent Neural Network (RNN) \cite{Rumelhart1986, Hochreiter1997} and its variants (i.e. long short-term memory, gated recurrent unit, etc.) constitute the only deep NN architecture in Table~\ref{Table_DL_Cl_Networks} that can be used for nonlinear dynamic systems, as exemplified by continuous underwater sensor outputs. RNNs are capable of using both time-series and sequential data streams to construct supervised classifiers.
		
		RNNs constitute a deep version of Hidden Markov Models (HMMs) \cite{Rumelhart1986}, which represent a simple form of the broader family of dynamic Bayesian networks. Markov models (i.e. Markov chains) are stochastic models, in which the output of a NN in response to the current state, only depends on the output of certain selected neurons in the previous state. The values of all neurons in all states are visible and are considered as the outputs.
		
		Hidden Markov models are similar to Markov models, but they rely on non-observable or hidden neurons. In these networks, the visible output is directly calculated by applying a nonlinear function to the output of the hidden neurons.
		As illustrated in Table~\ref{Table_DL_Cl_Networks}, an RNN is constructed by stacking HMMs on top of each other. The rectangular shape of the nodes in this picture indicates that in contrast to the circles of the other networks, these nodes are not constituted by a single neuron, but rather they are a combination of neurons in the form of a HMM.
		After studying the DNNs suitable for static and dynamic data analysis, let us study their capabilities to solve real-life problems in marine environment.
		
		\subsubsection{ML solutions in underwater image applications}
		The concepts of image and video quality assessment as well as image restoration were studied previously in Section~\ref{SubSection_Image_Preparation}. We have also stated the fact that underwater imaging applications heavily rely on image enhancement algorithms to cope with the destructive effects of visible-light environments. To address this problem, deep learning techniques have been employed for enhancing images suffering from noise, absorption, scattering, and color distortion effects \cite{LiC2018, Adler2017, LiY2016}. Furthermore, the benefits of DNNs in underwater image quality assessment can be assessed in future studies with reference to their overwater counterparts \cite{ZhangW2020}. These applications, while in their infancy, are rapidly expanding.
		
		Other applications that can benefit from deep learning are underwater image clustering and binning. We briefly reviewed these concepts in Section~\ref{SubSection_Image_Data_Binning}, noting that some underwater clustering algorithms are also introduced in Table~\ref{Table_Image_Clustering_Methods}. By the way, using novel deep learning techniques in underwater image clustering has not as yet attracted the attention it deserves. For instance, one can beneficially exploit the embedded image clustering technique of \cite{YanY2020} as an unsupervised DNN methodology in underwater image segmentation.
		
		Additionally, using ML techniques in underwater object recognition have been previously used for various underwater applications and have shown different performance. Here, we provide a summary of these techniques applied to some common underwater object of interest recognition and compare their performance.
		In order to have a fair comparison, the Average Precision (AP) parameter, which is a widely accepted performance measure, can be used.
		The average precision in any statistical or ML-based classifier can be calculated as \cite{Meng2018},
		
		\begin{equation}\label{Eq_AP}
		AP\overset{{\Delta}}{=}\dfrac{1}{N_C} \sum_{i=1}^{N_C} \frac{{TP}_i}{{TP}_i+{FP}_i} \,,
		\end{equation}
		where the True Positive (TP) and False Positive (FP) values are calculated for $N_C$ number of classes. The AP parameter of many ML techniques published in the underwater plant, fish, and coral recognition literature are compared in Table~\ref{Table_Compare_ML_Recognition}. All the algorithms referenced in this table are based on the well-known Fish4Knowledge and EILAT datasets introduced in Table~\ref{Table_Image_Data_Sources} and are also based on the video footages recorded by AUV and ROV vehicles.
		
		\begin{table*}[!t]
			\renewcommand{\arraystretch}{1.3}
			\caption{Comparing the Precision of Multiple Machine Learning Techniques to Recognize Three Different Underwater Species}
			\label{Table_Compare_ML_Recognition}
			\centering
			\begin{tabular}{!{\vrule width 1pt} @{}m{0.23\columnwidth}@{} | @{}m{0.1\columnwidth}@{} | @{}m{0.1\columnwidth}@{} | @{}m{0.1\columnwidth}@{} | @{}m{0.1\columnwidth}@{} | @{}m{0.1\columnwidth}@{} | @{}m{0.1\columnwidth}@{} | @{}m{0.1\columnwidth}@{} | @{}m{0.1\columnwidth}@{} | @{}m{0.1\columnwidth}@{} | @{}m{0.1\columnwidth}@{} | @{}m{0.1\columnwidth}@{} | @{}m{0.1\columnwidth}@{} | @{}m{0.1\columnwidth}@{} @{}m{0.0\columnwidth}@{}!{\vrule width 1pt}}
				\thickhline
				\centering \textbf{Application} & \multicolumn{13}{c}{{\cellcolor[gray]{0.95}Fish Recognition}} &\\
				\hline
				\centering \textbf{Dataset} & \multicolumn{13}{c}{Fish4Knowledge Dataset \cite{Meng2018, Sun2018, Chuang2016, HuangP2015, Qin2016}}&\\
				\hline
				\centering \textbf{Methods} & \centering\rotatebox[origin=c]{90}{\parbox[c]{2.6cm}{\textcolor{white}{..}Regression Tree}} & \centering\rotatebox[origin=c]{90}{\parbox[c]{2.6cm}{\textcolor{white}{..}LeNet}} & \centering\rotatebox[origin=c]{90}{\parbox[c]{2.6cm}{\textcolor{white}{..}Linear Discriminant\newline \textcolor{white}{..}Analysis and SVM}} & \centering\rotatebox[origin=c]{90}{\parbox[c]{2.6cm}{\textcolor{white}{..}Flat SVM}} & \centering\rotatebox[origin=c]{90}{\parbox[c]{2.6cm}{\textcolor{white}{..}BEOTR}} & \centering\rotatebox[origin=c]{90}{\parbox[c]{2.6cm}{\textcolor{white}{..}Decision Tree \newline \textcolor{white}{..}and SVM}} & \centering\rotatebox[origin=c]{90}{\parbox[c]{2.6cm}{\textcolor{white}{..}Softmax}} & \centering\rotatebox[origin=c]{90}{\parbox[c]{2.6cm}{\textcolor{white}{..}K-Nearest \newline \textcolor{white}{..}Neighbors}} & \centering\rotatebox[origin=c]{90}{\parbox[c]{2.6cm}{\textcolor{white}{..}DeepFish}} & \centering\rotatebox[origin=c]{90}{\parbox[c]{2.6cm}{\textcolor{white}{..}GoogLeNet}} & \centering\rotatebox[origin=c]{90}{\parbox[c]{2.6cm}{\textcolor{white}{..}VGGNet}} & \centering\rotatebox[origin=c]{90}{\parbox[c]{2.6cm}{\textcolor{white}{..}DeepFish and SVM}} & \centering\rotatebox[origin=c]{90}{\parbox[c]{2.6cm}{\textcolor{white}{..}AlexNet}}&\\
				\hline
				\centering \textbf{Average Precision (\%)} & \centering 75.4 & \centering 76.8 & \centering 80.1 & \centering 82.9 & \centering 84.5 & \centering 84.6 & \centering 87.6 & \centering 89.8 & \centering 90.1 & \centering 93.9 & \centering 96.6 & \centering 98.2 & \centering 99.7&\\
				\thickhline
			\end{tabular}
			\begin{tabular}{!{\vrule width 1pt} @{}m{0.227\columnwidth}@{} | @{}m{0.1\columnwidth}@{} | @{}m{0.1\columnwidth}@{} | @{}m{0.1\columnwidth}@{} | @{}m{0.1\columnwidth}@{} | @{}m{0.1\columnwidth}@{} | @{}m{0.1\columnwidth}@{} | @{}m{0.1\columnwidth}@{} @{}m{0.0\columnwidth}@{}!{\vrule width 1pt} @{}m{0.1\columnwidth}@{} | @{}m{0.1\columnwidth}@{} | @{}m{0.1\columnwidth}@{} | @{}m{0.1\columnwidth}@{} | @{}m{0.1\columnwidth}@{} | @{}m{0.1\columnwidth}@{} @{}m{0.0\columnwidth}@{}!{\vrule width 1pt}}
				\centering \textbf{Application} & \multicolumn{7}{c}{{\cellcolor[gray]{0.95}Plant Recognition}} && \multicolumn{6}{c}{{\cellcolor[gray]{0.95}Coral Recognition}} &\\
				\hline
				\centering \textbf{Dataset} & \multicolumn{7}{c}{ROV and AUV Video Footages \cite{Ludtke2012, Bonin-Font2016}} && \multicolumn{6}{c}{EILAT Dataset \cite{Shihavuddin2013, Wahidin2015}} &\\
				\hline
				\centering \textbf{Methods} & \centering\rotatebox[origin=c]{90}{\parbox[c]{2.4cm}{\textcolor{white}{..}Self-Organizing \newline \textcolor{white}{..}Map (SOM)}} & \centering\rotatebox[origin=c]{90}{\parbox[c]{2.4cm}{\textcolor{white}{..}Learning Vector \newline \textcolor{white}{..}Quantization}} & \centering\rotatebox[origin=c]{90}{\parbox[c]{2.4cm}{\textcolor{white}{..}Decision Tree}} & \centering\rotatebox[origin=c]{90}{\parbox[c]{2.4cm}{\textcolor{white}{..}K-Nearest \newline \textcolor{white}{..}Neighbors}} & \centering\rotatebox[origin=c]{90}{\parbox[c]{2.4cm}{\textcolor{white}{..}Random Forest}} & \centering\rotatebox[origin=c]{90}{\parbox[c]{2.4cm}{\textcolor{white}{..}Fully-Connected}} & \centering\rotatebox[origin=c]{90}{\parbox[c]{2.4cm}{\textcolor{white}{..}SVM}} && \centering\rotatebox[origin=c]{90}{\parbox[c]{2.4cm}{\textcolor{white}{..}Decision Tree}} & \centering\rotatebox[origin=c]{90}{\parbox[c]{2.4cm}{\textcolor{white}{..}Bayes}} & \centering\rotatebox[origin=c]{90}{\parbox[c]{2.4cm}{\textcolor{white}{..}K-Nearest \newline \textcolor{white}{..}Neighbors}} & \centering\rotatebox[origin=c]{90}{\parbox[c]{2.4cm}{\textcolor{white}{..}Random Forest}} & \centering\rotatebox[origin=c]{90}{\parbox[c]{2.4cm}{\textcolor{white}{..}Fully-Connected}} & \centering\rotatebox[origin=c]{90}{\parbox[c]{2.4cm}{\textcolor{white}{..}SVM}}&\\
				\hline
				\centering \textbf{Average Precision (\%)} & \centering 93.8 & \centering 94.1 & \centering 95.1 & \centering 95.3 & \centering 95.8 & \centering 96.1 & \centering 96.1 && \centering 69.6 & \centering 82.1 & \centering 83.3 & \centering 84.6 & \centering 85.2 & \centering 90.8&\\
				\thickhline
			\end{tabular}
		\end{table*}
		
		It is quite common in ML to cascade different algorithms into a single method to attain an improved performance (e.g. decision tree with SVM, and CNN with fully-connected). As an instance of algorithm merging in underwater applications, Faillettaz \textit{et al.} \cite{Faillettaz2016} as well as Hu and Davis \cite{Hu2006} cascaded a fully-connected classifier over shape-based feature-sets with an SVM classifier over texture-based feature-sets. By combining the results from these two classifiers, they claimed to have achieved an improved average precision, as defined in \eqref{Eq_AP}.
		
		Another noteworthy cascaded solution was conceived by Schoening \textit{et al.} \cite{Schoening2012}, where they employ multiple cascaded SVM binary detectors to construct a deep sea megafauna recognizer. The binary SVM detectors have object-specific operation, which makes them more accurate. Accordingly, by combining these binary detectors, high precision multi-object recognition was achieved. However, combining those object-specific binary detectors requires more hardware resources than a single classifier.
		
		Finally, it is worth mentioning that underwater imaging applications are not limited at all to the visible light domain. For example, undersea sonar imagery can provide high-resolution images of the seabed, even in turbid water with low visibility. Some of the reported applications of sonar images include ocean mapping, mine-countermeasures, oil prospecting, and underwater search and rescue (e.g. finding the drowned corpses, wrecks, and airplanes) \cite{HuoG2020}. In this regard, relying on human operators in sonar-based underwater object recognition applications is not recommended, since they will experience fatigue by staring at the display screen, and they might consequently miss the object of interest. Therefore, intelligent image recognition methods can be trained to replace human operators, especially in long-duration search scenarios.
		
		To highlight the accuracy of machine learning techniques in underwater sonar-based object recognition, the AP metric of \eqref{Eq_AP} is used here in Table~\ref{Table_Compare_ML_Recognition_Sonar}. This table compares the performance of deep NNs to that of several statistical methods as well as to that of traditional NNs. While the statistical and traditional networks seen in this table are trained using HOG features from Table~\ref{Table_Feature_Sets}, CNN deep networks do not need any feature extraction based preprocessing. Additionally, as seen in this table, appropriately designed and well-trained DNNs may outperform traditional models.
		
		\begin{table}[!t]
			\renewcommand{\arraystretch}{1.3}
			\caption{Comparing the Precision of Multiple Machine Learning Techniques to Recognize Objects in Sonar Imagery}
			\label{Table_Compare_ML_Recognition_Sonar}
			\centering
			\begin{tabular}{!{\vrule width 1pt} @{}m{0.35\columnwidth}@{} | @{}m{0.13\columnwidth}@{} | @{}m{0.13\columnwidth}@{} | @{}m{0.13\columnwidth}@{} | @{}m{0.13\columnwidth}@{} | @{}m{0.13\columnwidth}@{} @{}m{0.0\columnwidth}@{}!{\vrule width 1pt}}
				\thickhline
				\centering \textbf{Application} & \multicolumn{5}{c}{{\cellcolor[gray]{0.95}Object Recognition}}&\\
				\hline
				\centering \textbf{Dataset} & \multicolumn{5}{c}{Echoscope Sonar Image Dataset \cite{JinL2019}}&\\
				\hline
				\centering \textbf{Methods} & \centering\rotatebox[origin=c]{90}{\parbox[c]{1.6cm}{\textcolor{white}{..}K-Nearest \newline \textcolor{white}{..}Neighbors}} & \centering\rotatebox[origin=c]{90}{\parbox[c]{1.6cm}{\textcolor{white}{..}Multilayer\newline \textcolor{white}{..}Perceptron}} & \centering\rotatebox[origin=c]{90}{\parbox[c]{1.6cm}{\textcolor{white}{..}SVM}} & \centering\rotatebox[origin=c]{90}{\parbox[c]{1.6cm}{\textcolor{white}{..}CNN\newline \textcolor{white}{..}(AlexNet)}} & \centering\rotatebox[origin=c]{90}{\parbox[c]{1.6cm}{\textcolor{white}{..}CNN\newline \textcolor{white}{..}(LeNet)}}&\\
				\hline
				\centering \textbf{Average Precision (\%)} & \centering 72.0 & \centering 89.3 & \centering 92.7 & \centering 94.1 & \centering 97.0&\\
				\thickhline
			\end{tabular}
		\end{table}

		\subsubsection{ML solutions in underwater video applications}
		As already discussed in Section~\ref{SubSection_HW_Realization}, the capability of hardware platforms to train and analyze DNNs was impressively improved recently. This improvement has attracted increasing attention to the subject area of real-time IoUT video applications. It was also pointed out in Section~\ref{SubSection_UW_Video_Data} that the rapid growth in underwater video data volumes will require the development of automatic video processing, which can be carried out by machine learning techniques. These automatic solutions will be used in a variety of underwater video data processing applications such as:
		
		\begin{itemize}
			\item \textit{Visible light video tracking}: This application is designed for scanning video sequences to follow a specific element of interest \cite{Trucco2006}. A pair of basic problems in a video object tracking solution is how to predict the location of a moving element in the next frame and how to detect the element within this predicted region. Both of these be handled by deep NNs \cite{Chuang2015}. This application was also discussed in more detail in Section~\ref{SubSection_Ocean_Data_Application}.
			
			\item \textit{Sonar video tracking}: Sonar videos are recorded with the aid of acoustic waves, instead of visible light. Within turbid undersea environments, sonar systems offer a significant advantage by supporting long-range and low data-rate imaging \cite{Negahdaripour2013}. Similar to visible light video object tracking, sonar-based video object tracking can also be carried out with the aid of deep NNs \cite{Perry2004}.
			
			\item \textit{Underwater photo mosaicing}: This is the act of combining separate visible light or sonar video frames, for capturing a wider perspective of the region of interest. Machine learning based photo mosaicing of underwater images is now routinely performed by state-of-the-art underwater vehicles both for exploration and for navigation \cite{Bonin-Font2017, Paull2014, Singh2004}. Automatic localization and positioning of submersible vehicles with the aid of photo mosaicing is termed as SLAM \cite{Norgren2018} and it is studied in Section~\ref{SubSection_Ocean_Data_Application}.
			
			\item \textit{Marine life studies}: The analysis of underwater species is an indispensable part of any observatory video system. Thanks to the wide availability of machine learning based data processing toolsets, marine biologists are now capable of analyzing the high-volume video data captured for extracting the desired information. Scientific studies based on marine life video data applications have been published in different areas, including underwater species behavior understanding (ethology) \cite{Spampinato2010, Pons2010, Nian2014}, abundance and counting \cite{Westling2014, Kocak1999, Lau2012, LuoS2015}, size measurement \cite{Westling2014}, detection and recognition \cite{Sun2018, Kocak1999, HuangP2015, Meng2018, Zhang2016, LuoS2015}, and tracking \cite{ Rife2003, Kocak1999, Chuang2015, Chuang2017}. We have covered some of these aspects of marine life analysis in Section~\ref{SubSection_Ocean_Data_Application}.
		\end{itemize}
		
		\subsubsection{ML solutions in underwater sensor applications}
		Despite the unique benefits of applying RNNs to the processing of sensor outputs and in nonlinear dynamic systems control, there is a paucity of reported use cases in real-world IoUT applications. Their underwater applications are limited to a few scenarios, such as IoUT sensor data forecasting \cite{ZhangK2020}, underwater vehicle sensor read and fault diagnosis \cite{WangJ2009}, and the dynamic control of underwater movements \cite{Shan2013}.
		
		Among these applications, RNN-based predictive models conceived for IoUT sensory data forecasting as well as for missed sensory data implantation are better investigated \cite{ZhangY2019}. More specifically, sea water temperature and salinity predictions are claimed to be important, because:
		\begin{itemize}
			\item Water temperature and salinity have a direct effect on the acoustic communications between IoUT nodes \cite{GouY2020},
			\item Oceanic temperature has a substantial impact on both the land and the marine ecosystems by regulating the global climate \cite{ZhangK2020}.
		\end{itemize}
		
		Hence, the employment of machine learning techniques to design accurate predictive models is promising in BMD. For example, the influence of Long Short-Term Memory (LSTM) \cite{YangY2018} and Gated Recurrent Unit (GRU) \cite{XieJSST2019}, as a pair of common variants of RNNs on highly accurate water temperature prediction is presented in Table~\ref{Table_MSE_PHL}. Even though sea surface temperature forecasting is challenging due to the influence of numerous complex and nonlinear thermodynamic factors, data-driven DNNs are capable of learning these dynamic behaviors. By comparing RNN-powered networks to traditional machine learning solutions in Table~\ref{Table_MSE_PHL}, the efficiency benefit of DNNs in IoUT sensor data prediction becomes explicit.
		
		\begin{table}[!t]
			\renewcommand{\arraystretch}{1.3}
			\caption{Comparing the Mean Squared Error of ML Techniques in Sea Surface Temperature Prediction at Philippine Sea}
			\label{Table_MSE_PHL}
			\centering
			\begin{tabular}{!{\vrule width 1pt} @{}m{0.38\columnwidth}@{} | @{}m{0.12\columnwidth}@{} | @{}m{0.12\columnwidth}@{} | @{}m{0.12\columnwidth}@{} | @{}m{0.12\columnwidth}@{} | @{}m{0.12\columnwidth}@{}
			@{}m{0.0\columnwidth}@{}!{\vrule width 1pt}}
				\thickhline
				\centering \textbf{Application} & \multicolumn{5}{c}{{\cellcolor[gray]{0.95}Sensor Timeseries Forecasting}}&\\
				\hline
				\centering \textbf{Dataset} & \multicolumn{5}{c}{\href{https://www.noaa.gov/}{NOAA} Dataset in Table~\ref{Table_Primary_Data_Sources}}&\\
				\hline
				
				\centering \textbf{Methods} & \centering SVM \cite{YangY2018} & \centering CFCC-LSTM \cite{YangY2018} & \centering SVR \cite{XieJSST2019} & \centering FC-LSTM \cite{XieJSST2019} & \centering GRU-ED \cite{XieJSST2019} &
				\\
				\hline

				\centering \textbf{One Day Forecasting} & \centering 0.434 & \centering 0.166 & \centering 0.095 & \centering 0.061 & \centering 0.063 &
				\\
				\hline

				\centering \textbf{One Week Forecasting} & \centering N/A & \centering N/A & \centering 0.214 & \centering 0.168 & \centering 0.162 &
				\\
				\hline

				\centering \textbf{One Month Forecasting} & \centering 1.478 & \centering 1.145 & \centering 0.212 & \centering 0.343 & \centering 0.207 &
				\\
				\thickhline
			\end{tabular}
		\end{table}

		To recap, the current usage of RNN and its variants in IoUT applications are rare. Nonetheless, their versatile applications in IoT in smart homes \cite{KimH2017}, smart cities \cite{Zyner2018}, weather forecasting \cite{Marzano2007}, and other areas promise a similar growth for IoUT applications in the near future.
	
	\subsection{Section Summary}
	In this section, the employment of ML-based techniques in BMD processing was studied. We started this section by the definition of machine learning, classifying it into: classic statistical methods, traditional neural networks, and modern (i.e. deep) neural networks. The associated chronological perspective was also presented. Then, we discussed NNs by surveying the major deep networks in the literature and their potential applications both in static and dynamic underwater data processing. Despite the rapid development of sophisticated, but complex deep NNs, light-weight traditional NNs might remain typical in limited-complexity underwater applications. Therefore, more detailed discussions were provided in this section to cover data cleaning and feature extraction techniques in BMD. We then surveyed the available software frameworks and hardware platforms, including a collection of freely available libraries and frameworks. We also provided a comparison of the hardware infrastructures suitable for the software products discussed. Finally, the average precision and accuracy of diverse machine learning approaches suitable for underwater applications was studied.
	
	\section{Challenges and Future Directions in IoUT and BMD}\label{Section_Challenges}
	Having reviewed the state-of-the-art research in the areas of IoUT, BMD, and machine intelligence, the challenges and opportunities in these growing fields will be discussed. We also propose solutions and future research directions to address the challenges and to pursue the opportunities.
	
	One of the main obstacles that has been hindering further advances in the IoUT domain is that well-known terrestrial technologies, which perform well in the IoT domain, tend to be unsuitable in underwater applications. Many issues in the oversea application domain can be readily solved, easily, while they pose a significant challenge in underwater scenarios.
	Below we will continue with a list of challenges, opportunities, and future trends in the IoUT, BMD, and ML fields. Some of these challenges may be mitigated with the aid of big data processing and analytics, while some others require research efforts from the broader engineering community, hardware vendors, and policy makers.
	
	\subsection{Underwater Network Management System}
	Due to the significant growth in the number of Internet-connected underwater devices, the IoUT infrastructure tends to exhibit increased complexity. Consequently, improved Network Management Systems (NMS) are required, which represent the process of monitoring and controlling every aspects of the underlying network, for ensuring its seamless operation \cite{PrasA2007}. The monitoring must be automatic and prompt in locating, measuring, and reporting faults. Additionally, their control should be capable of efficient and reliable resource allocation or troubleshooting \cite{RajD2020}.
	
	While the concepts of network management in IoUT are somewhat similar to those in IoT, the methodologies are different, as discussed in Section~\ref{SubSection_Channel_Model} owing to the differences between overwater and underwater communication channels. Hence, the extension of NMS in IoT to Underwater NMS (U-NMS) in IoUT requires further research \cite{UrunovH2017} for each of the six aspects of U-NMS in the FCAPSC model, namely the Fault-, Configuration-, Account-, Performance-, Security-, and Constraint-managements \cite{UrunovH2017}. It is worth mentioning that FCAPSC of IoUT was derived from the original five elements in FCAPS of IoT, which was introduced by the International Organization for Standardization (ISO) in the 1990s. The constraint-management element requires U-NMS to deliver continuous connectivity even in the face of the hostile underwater channel, node mobility, device fragility, environmental dynamics, and technological heterogeneity.
	
	To facilitate the implementation of the FCAPSC model, U-NMS protocols tend to divide their influence domains into the family of network functionalities (e.g. routing management, protocol assignment, security checks, etc.) and of device operations (e.g. UWSN maintenance, energy conservation, device positioning, time synchronization, etc.) \cite{RajD2020}. The relevant studies around these domains were comprehensively surveyed in Section~\ref{Section_IoUT}. To the best of our knowledge, the Underwater Simple Network Management Protocol (U-SNMP) is the only U-NMS protocol, which covers the first domain of influence (i.e. the network functionalities). U-SNMP in IoUT is again, an extension of SNMP in IoT, and it is a manager–agent-based protocol, which is used in communication between devices \cite{UrunovH2017}. On the other hand, Lightweight Machine to Machine (LWM2M) is an IoT protocol that covers the second domain of influence (i.e. the device operations). LWM2M is applicable to IoUT as well, subject to some modest adjustments \cite{RajD2020}.
	
	Despite its popularity, U-SNMP lacks facilities for network configuration, and LWM2M suffers from challenges owing to the associated heterogeneous network support \cite{RajD2020}. By contrast, there are other network management protocols in IoT that offer better performance. Future studies might consider appropriately adopting those overwater protocols to underwater applications. For example, the Common Management Information Protocol (CMIP) offers better security features and it is suitable for wide area networks \cite{WongA2004}. Another alternative might be the LoWPAN Network Management Protocol (LNMP) in IoT, which has low data rates, low power consumption, low cost, and supports flexible topologies \cite{MukhtarH2008}.
	
	While the above protocols are better suited for distributed network managements, the centralized SDN management techniques of Section~\ref{SubSection_SDN} are also in need for underwater management systems. Here, the Open vSwitch Database (OVSDB) protocol at overwater SDNs constitutes a promising base for designing its underwater counterpart. However, any attempt to adopt OVSDB to underwater SDNs requires further research for addressing its current security issues \cite{SharmaS2014}.
	
	Even after dividing the U-NMS responsibilities into two major domains of influence and limiting the U-NMS protocols correspondingly, the number of challenges in each domain will remain significant \cite{RajD2020}. This is a direct consequence of the broad nature of the U-NMS topic itself. To better understand these challenges and study their future directions, we have discussed each challenge in a dedicated subsection. The following subsections will cover multiple aspects of U-NMS, ranging from energy conservation to device maintenance, to its security issues and communications. Some of these subsections discuss using BMD analytics as well as powerful deep learning techniques.
	
	\subsection{Energy Conservation and Harvesting in IoUT Devices}
	Energy conservation and harvesting in IoUT devices are of prime concern in almost all underwater applications, while they can be readily addressed in overwater scenarios, where energy can be harvested from the sun and preserved in the system \cite{Bito2017}. Some innovative methods of gleaning energy in underwater environments include:
	
	\begin{itemize}
		\item \textit{Overwater solar energy}: As already discussed in Section~\ref{SubSection_Edge_Computing}, by performing computations in above-water edge-devices, such as surface-floated buoys and vehicles, solar energy can be harvested. Indeed, solar powered buoys are amongst the oldest methods of environmental energy harvesting techniques \cite{Alippi2011, Joshi2011}.
		
		\item \textit{Underwater solar energy}: It was stated in Section~\ref{SubSection_UW_Comm} that light is strongly absorbed in water and it additionally suffers from color distortion. However, as mentioned in Section~\ref{SubSection_UW_Image_Data}, blue light will penetrate water deeper than other visible light frequencies. Some previous studies suggested the use of solar cells to harvest the blue frequency band of the solar energy underwater. These photovoltaics are claimed to provide useful power at the depth of more than $9\,m$ \cite{Jenkins2014}. However, many underwater applications are invoked at depths well beyond the light penetration domain.  
		
		\item \textit{Tidal-wave energy}: Harvesting kinetic energy from waves in the littoral tidal basin can be readily achieved by using today's technology. This can also offer a source of energy in the underwater benthic zone. These systems are based on piezoelectric elements and the energy generated is high enough to power UWSNs and their devices \cite{Toma2015}. Similarly, some contributions report the employment of the same technique to harvest energy from fluid-flow in pipelines \cite{Qureshi2017}.
		
		\item \textit{Wireless energy transfer}: This method was introduced for RFIDs and acoustic tags in Section~\ref{SubSection_Marine_Life_Tracking}. Additionally, a remotely powered acoustic UWSN was reported in \cite{Bereketli2012} as another energy acquisition alternative for the IoUT. Thereby, sensor nodes harvest the mechanical wave power supplied by an external acoustic source. Another use case of wireless energy transfer may rely on in-situ magnetic charging stations, where underwater vehicles (e.g. AUV, ROV, etc.) can use these stations to recharge their batteries \cite{KanT2018}.
		
		\item \textit{Wired energy transfer}: In a clear contrast to the previous items, every close-to-shore IoUT application may rely on energy transmission through a cabled network. In this method, the energy arrive from a solar, wind, or urban power network and it is directly transferred to the UWSN. These systems, however, are costly because of the cabled infrastructures on both land and at sea \cite{Perez2017}.
	\end{itemize}
	
	The above methods have the potential to be used in marine type projects. They can extend the lifetime of IoUT networks and boost their QoS. However, except for the \textit{wired energy transfer} method, they tend to be unpredictable, hence none of them guarantees the uninterrupted delivery of energy. To address this issue, the following pair of solutions may assist:
	
	\begin{itemize}
		\item \textit{Using rechargeable batteries}: All the aforementioned methods can be accompanied by tandem batteries to store energy. These batteries are recharged during the instances of energy acquisition and deliver their stored energy afterwards. This solution obviously prolongs the sensor network's lifetime, but requires maintenance and increases the cost of the system \cite{Perez2017}.
		
		\item \textit{Managing energy consumption}: Just like any other electrical grid worldwide, the IoUT has to manage its energy demands by optimizing the power allocation to guarantee uninterrupted data collection and transmission \cite{Jing2017}.
        Indeed, all components of the IoUT should be energy-aware. In this context, an innovative energy-aware robot was proposed by Wu \textit{et al.}~\cite{WuZ2020}, which had the shape of a killer-whale. This robot has had reduced energy consumption for its propulsion as a benefit of its excellent lift-to-drag ratio, which is important for effortless gliding in water. Compared to the Seagliders introduced in Table~\ref{Table_DA_Tools_Usage}, the controllable flukes of this robot offer substantial energy savings, better maneuverability as well as enhanced endurance.
	\end{itemize}
	
	\subsection{Development of Low-cost and Affordable Sensors}\label{SubSection_Challenge_Low_Cost_sensors}
	The underwater sensors listed in Table~\ref{Table_Primary_Data_Source_Sensor} and multi-sensor buoys are usually very costly compared to their overwater counterparts. To address this multi-disciplinary challenge, the following approaches can be considered:
	
	\begin{itemize}
		\item \textit{Quality vs. cost trade-off}: Low-cost sensing devices having lower precision measurements could be purchased to strike a quality vs. cost trade-ff.
		
		\item  \textit{Transferring specimen for inland assessment}: The establishment and operation of an underwater in-situ sensing unit is generally more expensive than a laboratory-based experiment. Therefore, whenever possible, samples could be transferred to an inland lab, to avoid the need for costly in-situ processing and evaluation. 
		
		\item \textit{Inferential measurements}: It is a common technique in industrial instrumentation to estimate a parameter from the values of other parameters, which are easier to measure. For example, calculating the total amount of ions dissolved in water is always easier by measuring its electrical resistance instead of utilizing costly electrochemical sensors \cite{Hem2012}.
	\end{itemize}
	
	In addition to the above-mentioned general recommendations, designing specific cost-effective sensors can help the evolution of IoUT technology. In order to design such sensors, scientists from different research backgrounds have to cooperate. The result of this cooperation will be \textit{ad hoc} solutions, which are tailored to the predefined need of any project. For instance, here we list several contributions involving low-cost underwater sensors.
	
	\begin{itemize}
		\item Islam \textit{et al.} \cite{Islam2018} proposed a low-profile and low-cost microstrip patch antenna to measure the salinity of water. They found that the antenna's reflection coefficient is proportional to the amount of salt or sugar dissolved in water.
		
		\item Vorathin \textit{et al.} \cite{Vorathin2019} constructed a high-resolution hydrostatic pressure and depth sensor by attaching a fiber Bragg grating on a rubber diaphragm. Their sensor is claimed to enhance the sensitivity and to compensate the temperature effects.
		
		\item  Wang \textit{et al.} \cite{WangY2018} conceived a low-cost turbidity sensor, based on the $90^{\circ}$ scattered light detection principles. To elaborate, they used off-the-shelf infrared LEDs having controlled light emission to construct a low-cost, yet accurate product.
		
		\item Kirkey \textit{et al.} \cite{Kirkey2018} proposed an inexpensive fluorometer based on an optical backscatter transducer. Explicitly, their idea is to use low-frequency circuitry for modulating the light source. Using this technique, their product will be very cost-effective.
		
		
		
		\item
		By relying on the fundamental concepts of Time-Domain Reflectometry (TDR), Time-Domain Transmissometry (TDT), or Fiber Bragg Grating (FBG) in optical physics, optical fibers can act as a sensor to detect a wide range of underwater physical parameters. Examples include leak detection in pipelines and estimating its location \cite{LiuC2019}, stress response of the offshore platforms (i.e. legs of the jacket structure) and detecting its deformation \cite{WangP2018}, bending moments of the flexible risers in a hang-off position to avoid exceeding its absolute maximum ratings \cite{WangP2018}, temperature and pressure measurement as well as eliminating the temperature-pressure cross-sensitivity \cite{ZhouX2019}, etc.
		
	\end{itemize}
	
	In addition to the above techniques of reducing the sensors costs, reducing the physical size of underwater sensors also tends to reduce their production cost. Examples of low-cost smaller sensors include miniature underwater robots \cite{Berlinger2018} and low-power nano-sensors \cite{Zulkifli2018}.
	
	Furthermore, using the edge computing capabilities discussed in Section~\ref{SubSection_Edge_Computing}, it is possible to establish a laboratory on board of an oceanic exploration platform \cite{QiuT2020}. This limited-capability on-board lab may conduct preliminary experiments on samples, before transmitting the numerical results through the web. Additionally, by using a video-empowered command and control system, there would be no need for an expert to be present in-situ. This laboratory on the edge will also eliminate the need for sending the specimen to an inland lab for evaluation, which would be both expensive and time-consuming.
	
	\subsection{Large-scale IoUT Underwater Communications}
	In terrestrial telecommunication, electromagnetic waves, copper cables, and optical fibers are the mainstream transmission media. However, as studied in Sections~\ref{SubSection_UW_Comm} and \ref{SubSection_UW_Physical_Layer}, fiber-optics are expensive to deploy and maintain under water, hence typically acoustic, electromagnetic, and optical technologies are deployed, which do not propagate well, hence making IoUT telecommunication challenging.
	
	Electromagnetic and optical technologies only cover short communication distances and are therefore unsuitable for long-range IoUT communications. Acoustic technologies tend to be more amenable to long-range IoUT networks, but they have a narrow frequency bandwidth and are prone to cross-talk with other local acoustic applications \cite{Makled2018}. Hence they are also unsuitable for large-scale networks. 
	
	A promising, but costly technique of addressing the communication challenge in IoUT is to combine heterogeneous communication technologies \cite{CelikA2020}. However, the design of multi-technology multi-mode gateways for undersea applications is a challenging task, especially when considering the energy harvesting difficulties. These gateways may be combined with the SDN and cognitive radio concepts for efficiently sharing the limited spectrum undersea \cite{CelikA2020}. Future research should conceive energy-aware software and low-power hardware solutions for these gateways to improve the quality of communication in IoUT networks.   
	
	Another promising solution to underwater communication problems, especially in the case of long-range inter-continental telecommunications, is the inter-connection of the IoUT and {aeronautical technology}. In this technique and according to Table~\ref{Table_Network_Archit}, the data in the application layer of underwater networks that is produced by sensors and imagery equipment can be transferred to the overwater buoys and to the floating ships. Therefrom, data will be handed to the low-earth orbit or medium earth orbit communication satellites, or to the aerial vehicles. A similar example of this technique is provided by \href{https://www.ncdc.noaa.gov/data-access/satellite-data}{NCEI}, Landsat, \href{https://aquarius.nasa.gov/}{Aquarius}, SARAL, \href{https://earth.esa.int/web/guest/data-access/browse-data-products}{CryoSat}, \href{https://www.eumetsat.int/website/home/Data/DataDelivery/index.html}{Jason}, \href{https://www.eumetsat.int/website/home/Data/DataDelivery/index.html}{HY2-A}, and IRS satellites in Table~\ref{Table_Primary_Data_Sources} for remote sensing of ocean surface parameters. Another example is offered by \cite{WangQ2020} for the connectivity of unmanned aerial vehicles and IoUT.
	
	Additionally, reducing the data volume to make it suitable for transmission using narrow-bandwidth acoustic technology could be used for mitigating the communication challenges in large-scale IoUT networks. This was discussed under the concept of MCC and MEC edge computing paradigms in Section~\ref{SubSection_Edge_Computing}. According to this method, edge-processing is capable of reducing the volume of raw data \cite{WangT2020}. This consequently reduces the bandwidth requirement.
	
	In this regard, the combination of the MEC paradigm with smart unmanned vehicles can offer an alternative solution to the problem of long-distance underwater communications. In this solution, autonomous underwater vehicles \cite{CaiS2019}, unmanned aerial vehicles \cite{WangQ2020}, etc. can be used for IoUT data collection. Here, edge computing can undertake some essential computations, so that big data collection will be mitigated, while latency-sensitive situations are handled promptly \cite{WangQ2020}. However, the lack of energy in undersea environments, makes the employment of MEC a challenge. In closing we note that MEC has a similar architecture to the previously suggested inter-connection between the IoT and IoUT.
	
	\subsection{Dynamic IoUT Signal Routing and Traffic Control}\label{SubSection_Challenge_Routing_and_Traffic}
	The underwater propagation environment is quite hostile; hence it is of low channel capacity, which makes even point-to-point single-link data transmission challenging in the IoUT. This becomes even more challenging when a network of concurrently communicating nodes is considered. For example, in the presence of a network supporting multiple transmitters, the tele-traffic escalates, and traffic control becomes a challenging task, when aiming for a reasonable QoS \cite{Al-Fuqaha2015}.
	
	To avoid any tele-traffic congestion, efficient routing management is crucial. 
	To elaborate a little further, in \textit{ad hoc} networks the design-dilemma is whether to use more short hops at the cost of an increased delay or fewer longer hops. It is beneficial to use low-complexity non-coherent transceiver techniques and take into account the battery-charge during routing.
	
	Additionally, intelligent traffic control systems, using both the deep learning approaches \cite{Fadlullah2018} and the SDNs \cite{XieJ2019, Mao2018} may be devised in underwater applications. These systems are capable of efficiently handling concurrent data transfers to avoid congestion.
	
	To employ deep learning in traffic control systems, we first have to define the action space. To do so, consider a heterogeneous IoUT network constructed of both wired and wireless connections. As detailed in Section~\ref{SubSection_Hop_Count_optim}, all the nodes of such a network may connect with one another, using multiple hops. As a consequent of both the limited underwater transmission range \cite{SuY2020} or the large-scale infrastructural size of the network \cite{CelikA2020}, the number of possible hops escalates, leading to numerous potential paths for a data packet to travel between a pair of nodes. In this context, every path can be considered as an action in the action space. After taking an action, the system's feedback (a.k.a. reward) can be quantified by the traffic load level of the nodes in the following timeslot, which is formatted as an award value matrix. The combination of the actions and their consequent rewards provides the required training data for semi-supervised deep reinforcement learning, as advocated in \cite{Fadlullah2018}.
	To the best of the authors' knowledge, such datasets for IoT, do not exist for IoUT. Once collected, these datasets of IoUT routing management and traffic control can then be used to train ML-based models.
	After successfully passing the training phase, this deep model can be used to infer the best overall path for a given data packet.
	
	Although the above DNN can be conveniently used in most of the IoUT structures, it might falter as the IoUT architecture gets wider. By expanding the network scale, the number of possible paths will increase exponentially, and the deep NN can no longer learn the patterns in the data flow. To address this problem, Fadlullah \textit{et al.} \cite{Fadlullah2018} have proposed a solution, which can also be employed in underwater applications. To elaborate a little further, they suggested to:
	\begin{enumerate}
		\item Change the action space from the entire set of all path combinations to simply the next hop destination;
		\item Replace the semi-supervised reinforcement learning by a cascaded combination of supervised CNN and DBN;
		\item 3)	Predict the award value matrix (by CNN) before deciding on the best action (by DBN).
	\end{enumerate}
	
	These recommendations for using DNNs for wide-scale IoUT traffic control will reduce the packet loss rate as well as improve the network throughput. As a result, the routing performance will improve compared to that of the conventional methods.
	
	The other technique of improving the network throughput is that of adhering to the SDN methodology. This type of network is described in Section~\ref{SubSection_SDN}, which relies on a centralized control system, for monitoring the network's traffic flow. According to Xie \textit{et al.} \cite{XieJ2019}, the centralized management of SDNs substantially benefits from using DNNs. For example, the functions of routing optimization, traffic prediction, path load prediction, node deployment optimization, delay prediction, QoS prediction, content delivery optimization, resource allocation, SDN reconfiguration, optimized spectrum sharing, number of active nodes estimation, and intrusion detection would all benefit from using DNNs in overwater SDNs. Hence it is promising to critically appraise their synergies also in underwater applications.
	
	\subsection{Securing Underwater Networks}
	As discussed in Section~\ref{SubSection_Network_Security}, network security has to be considered throughout the entire IoUT architectural model, from the physical layer to the application layer. However, the highly unreliable underwater communication channels having high propagation delays and low energy resources, make network security a challenging topic \cite{Lal2017}.
	
	In this context, and according to a review article by Jiang \textit{et al.} \cite{JiangS2019}, two major concerns in underwater communication security, which require further investigations and therefore offer research opportunities are:
	
	\begin{itemize}
		\item According to both Section~\ref{SubSection_Network_Architect} and Table~\ref{Table_Network_Archit}, underwater sensors in IoUT are considered as network endpoints, located at the highest TCP/IP layer (i.e., the application layer). These endpoints are vulnerable to potential intrusion attacks and therefore, have to be secured. But the main focus in UWSN security researches was on the lower physical, data link, and network layers, missing out the higher transport and application layers. It is difficult to guarantee a secure connection in IoUT by relying on the lower three layers alone.
		On the other hand, the corresponding application layer security protocols of the conventional IoT are often computationally complex and hence power-thirsty. Those protocols are not readily applicable to underwater endpoints.
		Therefore, upper-layer security should be further investigated to provide a reliable underwater network.
		
		\item Considering the resource-constrained nature of the IoUT, compared to the terrestrial IoT, a comprehensive framework is required for optimally distributing the security-related functionalities across all the layers. This optimal distribution of security tasks will minimize the overall resource consumption of the UWSN. To the best of our knowledge, no comprehensive framework is available for energy-effective security enforcement in IoUT at the time of writing.
	\end{itemize}
	
	To address these challenges, taking a similar approach to the IoT can be helpful. Earlier in Section~\ref{SubSection_Network_Security}, we introduced several layer-based security protocols to be employed in the IoUT infrastructure. We also suggested ECC as a cryptographic primitive in the application layer of the IoUT. 
	However, due to the large-scale nature of IoUT systems and their complex environment, traditional cyber security techniques cannot be readily adapted. Therefore, new research directions relying on DL may be adopted. To address this issue, large datasets of different IoUT security attacks have to be collected first. To the best of our knowledge, such datasets do not exist for underwater applications. However, datasets of various security attacks are already available for download for the conventional IoT (e.g. CICIDS2017) \cite{Roopak2019}. Once collected, these datasets of IoUT security attacks can then be used to train ML-based models for protecting IoUT networks against a wide range of security threats.
	
	Having said that, if a large and clean dataset is not collectable for a given IoUT attack, semi-supervised reinforcement learning might be considered. As already discussed in Section~\ref{SubSection_Challenge_Routing_and_Traffic}, the deep reinforcement learning models can be trained by the combination of actions and their consequent rewards. As a result, these models can be trained during a practical experiment, leaving out the data gathering step. Such an experiment has been conducted by Xiao \textit{et al.} \cite{XiaoL2020} in a pool with underwater transducers. They have used this setup to train their proposed reinforcement learning model against sophisticated jamming attacks in UWSNs. With no prior data gathering, the node has inferred, where the heavily jammed locations are and avoided them. It has also learned to adjust its power-level for achieving the required bit error rate.

	The IoUT architectural model of Table~\ref{Table_Network_Archit} is based on the similar 5-layer TCP/IP model of IoT. As a result, the layer-wise security threats of IoUT are of similar nature to the common layer-wise attacks experienced in IoT. Some examples of these common IoT/IoUT security issues are related to the recognition and segregation of malicious requests (e.g. distributed denial-of-service protection), prevention of policy violations (e.g. anomaly as well as intrusion detection), provision of authentication (e.g. man-in-the-middle attack prevention and data-injection protection), etc. \cite{JiangS2019, Lal2017}.
	
	When choosing an ML-based model for satisfying certain security requirements in IoT, CNN as well as RNN deep NNs are proven to have a better single-attack detection performance, than to other traditional NNs. Furthermore, if a sole ML model has to detect a collection of multiple attacks, DBNs might be considered as a better choice \cite{Roopak2019, ZhanhY2019}. However, the application of these ML models in IoUT requires further investigations. We anticipate that the security domain in IoUT can significantly benefit from the techniques already available in IoT. This offers a great research opportunity for the developers to adapt these methods to the challenging underwater environment.
	
	\subsection{Deleterious Effects of Imprecise Channel Modeling}
	Naturally, the channel plays an essential role in designing the underwater deployment of endpoint nodes, relay nodes, and sinks. The strategic deployment of nodes is capable of increasing the entire system's battery life and improving the QoS. Almost all of the underwater acoustic, electromagnetic, and optical channel models as described in Sections~\ref{SubSection_UW_Comm} and \ref{SubSection_Channel_Model} and also the underwater magnetic induction channel model of \cite{LiY2019} rely on approximations to simplify the overall model.
	For example, many channel models assume straight signal propagation undersea, which simplifies numerical calculations \cite{LiY2018}. However, approximations and idealized simplifying assumptions will result in imprecise models and inaccurate communication.
	
	Imprecise channel modeling can also lead to inaccurate simulation-based modeling of underwater communications. Below, we provide a list of recent publications, which showed the impact of imperfect channel modelling in the simulations of underwater communication, and therefore used precise channel models for their simulations. 
	
	\begin{itemize}
		\item Using a preamble alerts the receiver about the reception of an incoming data burst and switches it from its low-power dormant mode to its high-power active mode. Therefore, both missing the detection of a preamble and declaring its reception, when it is actually absent reduces the receiver's battery life. A beneficial preamble detection method was proposed by Li \textit{et al.} \cite{LiW2019}, for an underwater digital communication system, which coexists with other deployed networks. Using an accurate channel model is critical in this identification method in order to prevent a receiver from being triggered by other systems and consequently extends the underwater battery life.
		
		\item IoUT communication is affected by numerous signal impairments, such as a high propagation delay and high signal attenuation, as shown in Fig.~\ref{Fig_InfoG_Phys_Layer_Roughness}. These reduce the link reliability, which can be mitigated by sophisticated Forward Error Correction (FEC) and Automatic Repeat reQuest (ARQ) techniques. However, at low SNRs the throughput may be reduced by excessive data retransmission. Liang \textit{et al.} \cite{Liang2018} optimized the overall transmission redundancy to be used in UWSNs by relying on an accurate channel model.
		
		\item
		In the simulation-based investigation of adaptive transmission in time varying underwater acoustic channels conducted by Wang \textit{et al.} \cite{WangC2018}, the transmitter's data queue length and the predicted channel conditions were relied upon for designing the adaptive transmitter parameter values. Although their method relies on reinforcement learning to yield considerable energy saving, it cannot correctly learn the accurate parameter values without a precise channel model.
	\end{itemize}
	
	To address the need for precise channel modeling, further research is required for devising precise yet computationally efficient models, for ensuring that the simulation of underwater communications in context of the emerging IoUT is as reliable as possible. This would mitigate the need for and the likelihood of future network upgrades.
	
	\subsection{Sparse and High-maintenance Sensing Devices in IoUT}
	In contrast to the IoT, the sensing devices of the IoUT are sparsely deployed and exposed to severe environmental effects \cite{Domingo2012}. The sparse configuration and the harsh environmental conditions make the maintenance of the IoUT, costly. Explicitly, maintenance should mitigate the effects of erosion, corrosion, sediments, pollutions, and other phenomena imposed by seawater.
	
	To address these issues and to reduce the maintenance cost of sparse high-maintenance nodes in the IoUT, a compelling solution is to incorporate self-management capability \cite{Atyabi2018}, including self-evaluation, self-configuration, and automatic reports to human operators. Therefore, developing intelligent ML-based hardware nodes for the IoUT, which have a self-management and decision-making capability and conceiving their required software are promising avenues for research in IoUT, which need further attention.
	
	\subsection{Poor Underwater Positioning and Navigation}
	As mentioned earlier, GPS signals do not penetrate the sea, hence other navigation techniques have to be used. A number of navigation methods such as blind positioning, acoustic transponders, ranging sonars, image-based positioning, and SLAM were introduced in Section~\ref{SubSection_Geo_Localization}. However, none of these stand-alone techniques offer a non-accumulating positioning error, therefore, none of them are adequate \cite{Liu2018}.
	
	Hence, further research is required for improving underwater navigation systems. The challenges to address are:
	
	\begin{itemize}
		\item Selecting and combining the large amount of data from the aforementioned stand-alone navigation techniques is a challenging task, while considering the system's cost and complexity as well as precision. A research opportunity to address this issue is to study different combinations of navigation techniques, while striking a trade-off between cost and accuracy. For instance, Bonin-Font \textit{et al.} \cite{Bonin-Font2017} have combined image-based positioning with SLAM to achieve improved navigation. Another study has combined blind positioning with a long-baseline acoustic transponder to reduce the positioning errors caused by acoustic ray bending and variable sound velocities \cite{ZhangT2018}.
		
		\item Preliminary environmental survey for acquiring offline data is a challenging mandatory step, which has to be conducted before scene analysis based localization techniques (i.e. image-based positioning systems as well as acoustic transponders). This step is required for extracting positional fingerprints (i.e. features), which will be subsequently used for accurately training a localization ML algorithm. However, gathering these big datasets to train ML algorithms is not trivial in underwater environments.
			
		Furthermore, traditional ML models can be easily mislead by any variation in the underlying high-dynamic underwater scenes. To address this, using deep NNs having automatic feature extraction capability is highly recommended. The benefits of diverse statistical methods, traditional NNs, and deep NNs in scene-based indoor positioning systems are reported in \cite{Zafari2019}. These quick lines can be adapted for underwater applications.
		
		\item Finding and implementing new natural phenomena for improving underwater positioning as well as navigation is another challenge. In this context, a gravity-aided navigation system was conceived in \cite{WangB2019}, which is based on exploiting the difference between the observed and the predicted gravity. These methods however, suffer from biases and error accumulation, which has to be addressed in future research.
		
		\item Adapting the existing Low-Power Wide Area Network (LPWAN) technologies, which has been discussed in Section~\ref{SubSection_UW_Physical_Layer}, to carry out IoUT localization is a potential opportunity. In a review article of Zafari \textit{et al.} \cite{Zafari2019}, a collection of these wireless technologies (i.e. SigFox, LoRaWAN, Weightless, etc.) was studied in IoT localization. Using the same approach would be beneficial for IoUT, as they all consume extremely low energy and operate in a wide reception range.
	\end{itemize}
	
	\subsection{Non-Destructive Testing in Underwater Applications}
	Non-destructive testing (NDT) is the process of inspecting a system, device, or component, without imposing any changes on its shape or material. NDT can be performed while the Device Under Test (DUT) continues its normal operation. For example, both the TDR and TDT methods discussed in Section~\ref{SubSection_Challenge_Low_Cost_sensors}, are variants of NDT methodologies relying on optical fibers, which are undertaken without interrupting the normal operation of their DUT~\cite{LiuC2019}.
	
	Apart from optical fibers, NDT can also be carried out by other equipment. Some of these NDT tools include visible light cameras, electromagnetic flux detectors, ultrasonic transceivers, and magnetic inductors (i.e. eddy current sensors). Using these equipment in underwater NDT are reported in many diverse applications, such as:
	
	\begin{itemize}
		\item Leakage detection \cite{LiuC2019}, vibration recognition \cite{WuH2019}, and non-stationary disturbances as well as strain sensing \cite{MaK2017} in pipelines, by fiber-optics;
		\item Bending and deformation inspection in flexible risers, by fiber-optics \cite{WangP2018};
		\item Temperature and pressure monitoring in downhole tools, by fiber-optics \cite{ZhouX2019};
		\item Determining water-level in unmanned water resource management systems, by visual cameras \cite{PanJ2018};
		\item Inspecting the outer surface of large ship hulls, by electromagnetic flux detectors \cite{Hedayati2010};
		\item High-sensitivity hydrophones for opto-acoustic imaging, by ultrasonic \cite{LiC2019};
		\item Welding inspection and defect characterization in offshore platforms, by magnetic induction \cite{Gros1995}.
	\end{itemize}
	
	Developing a classic NDT algorithm for a typical component is not an easy task, requiring knowledge about both the DUT and the underlying physics of the NDT itself (i.e. ultrasonic, electromagnetic, or optical wave scattering). Nevertheless, in the modern age of ML techniques, one can gather a big dataset from any DUT. These data samples can then be fed to a deep NN for training purposes. By implementing this deep learning approach, developing NDT algorithms can be carried out faster, with minimal knowledge about the underlying physical concepts \cite{WuH2019, PanJ2018, MaK2017}.
	
	Another opportunity in underwater NDT is to use the recent developments in the field of distributed and cloud-based BMD Processing tools, which has been discussed in Section~\ref{SubSection_Data_Processing_Platforms}. Relying on the frameworks listed in Table~\ref{Table_Apache_Platforms}, one can readily glean data from multiple independent NDT equipment, and then employ a data fusion technique from Section~\ref{SubSection_Data_Fusion}, to combine and process the gathered big data. It is proven by Bayes' theorem that the uncertainty in the final test result will dramatically decrease by fusing data from independent data sources \cite{Gros1995}.
	
	\subsection{Lack of Strong Data Leveraging Tools}
	The processing of BMD requires powerful hardware and software tools that can automatically extract knowledge from large databases. Some of these hardware tools were introduced throughout this paper, specially in Sections~\ref{SubSection_Data_Processing_Platforms} and \ref{SubSection_HW_Realization}. However, more advanced software techniques are required for automatic long-term data-gathering and data-monitoring applications.
	For instance, even though unmanned auto-annotation based industrial software is in very high demand for classification and labeling underwater objects, plants, or creatures, the existing marine image-annotation software packages are only semi-automated, at best \cite{Gomes-Pereira2016}.
	
	
	The shortages in automatic data leveraging techniques is partially due to the uncertainty in selecting feature-set. Ambiguity in feature selection and in the ensuing feature reduction is a consistent challenge in automated ML-aided projects.
	In a conventional neural network, there are a variety of features and descriptors, hence selecting the most useful ones is challenging. These diverse underwater feature-sets have previously been discussed in Section~\ref{SubSection_Feature} and summarized in Table~\ref{Table_Feature_Sets}. Recall that relying on a single feature is usually inadequate for accurate underwater classification and clustering \cite{HuangP2015}.
	
	Although modern deep learning approaches (e.g. fully-connected, CNN, Autoencoder, etc.) are promising in terms of overcoming this challenge, the advantages of classical ML methods may nonetheless provide better results. These benefits of classical ML methods include having fewer parameters, more rapid convergence during their training, better insights into the tangible physical interpretation of their operation, and much easier debugging as well as tunning the network.
	Further research is required for developing techniques and algorithms to infer useful features from a dataset, or even to automate the feature selection process. This will be invaluable for the ML community, because feature selection has a direct impact on the performance of ML-based solutions.
	
	\subsection{Training Deep Networks}
	When using deep networks, the feature extraction will be automatically handled by the hidden layers of the network. Despite this automatic feature extraction, the deep networks have the disadvantage of requiring large amounts of data for tuning their weights and biases, during the training step. The deeper and wider these networks become, the more useful features can be extracted \cite{Pang2018}. Ultimately, we have to strike a trade-off between the training data volume required, the network size/power and its overall performance. 
	
	To address this trade-off, a pair of general solutions could offered. The first solution is to satisfy the deep network's hunger by more data. The extra data can be generated automatically (for example by various data augmentation techniques or by employing generative neural networks, such as the Generative Adversarial Network (GAN) \cite{LiN2018, LiJ2018}) or manually (by using web-based technologies to enlist the assistance of international experts and to produce a large volume of user-generated contents).
	
	The second solution is to enhance the deep network's efficiency by modifying their building blocks and the inter-connection of neurons (like the convolution, pooling, and activation operations of a typical CNN \cite{JafariA2019} or pruning a deep NN to ease their operation on mobile devices \cite{ChangJ2019}), in order to reduce the number of network parameters. The vibrant deep learning community will no doubt continue to improve the deep network training and tackle the challenges. This will definitely improve the IoUT applications and advance BMD analytics.
	
	\subsection{Degraded Underwater Images}
	Undersea photography is always affected by environmental factors. Although the contributions surveyed in Section~\ref{SubSection_Image_Preparation} have addressed some of these issues, there is a lot of room for improvement. For instance, developing an imaging system capable of both real-time forward- and back-scattering elimination is critically needed.
	
	Furthermore, the underwater image quality could be significantly improved by applying some well-known hardware methods and techniques, such as light polarization \cite{Tian2019}, 
	multi-spectral imaging \cite{Zawada2003}, or stereoscopic imaging \cite{Suresh2019}. Additional improvements may also be attained by software methods, such as wavelength compensation and color reconstruction \cite{HanM2018, Ancuti2018} both in active and passive underwater photography. Finally, the new concept of image reconstruction with DL \cite{LiC2018, Adler2017, LiY2016} has promise for future research.

	
	\section{Conclusion}\label{Section_conclusion}
	The recent advances in IoT technology and the extension of its influence both to coastal and open sea areas has led to the proliferation of the number of Internet-connected objects both in over- and underwater applications. This technological evolution inspired the new scientific concept of IoUT constituted by marine sensors, cameras, hydrophones, etc. This concept opens many new research directions for undersea data acquisition, data communication, BMD handling, and oceanic data processing.
	
	In this article, we commenced by surveying the state-of-the-art in underwater communications. Given the harsh underwater propagation environment, data communication is quite a challenging task in the IoUT. The families of advanced underwater communication models of acoustic, electromagnetic, and optical technologies were introduced and innovative solutions were proposed for increasing the overall link reliability by topology and routing optimization, security improvement, and protocol enhancements. Furthermore, the underwater channel modeling was studied along with software tools to simulate both those channel models as well as the communication protocols. Both SDN and edge computing techniques were also reviewed as a promising technique of improving underwater communications. We also reviewed the IoUT network architecture, based on the well-known 5-layer TCP/IP standard model.
	
	Naturally, the IoUT leads to BMD generation and the associated challenges include data storage, transportation, preparation, and analysis. Because manual and semi-automatic data processing methods are no longer appropriate in the new era of the IoUT, the five system components of BMD solutions were discussed and the most recent frontier-research and a range of practical solutions were discussed for each component. These solutions covered the areas of sensor, image, and video data sources, marine geographic data, localization and tracking, open access databases, distributed data processing, and cloud-based services. A complete section was also dedicated to machine intelligence (i.e. ML and DL) and its applications to marine data processing. The most recent research articles in both the hardware and software aspects of the IoUT and BMD processing were also surveyed, along with the critical appraisal of these works.
	Finally, numerous open research issues and future study directions were presented to provide an insight into the prospective applications, trends, and challenges. Do join this vibrant interdisciplinary research community, valued colleagues.

	\ifCLASSOPTIONcaptionsoff
	\newpage
	\fi
	
	\bibliographystyle{IEEEtran}
	\bibliography{./BibDoNotUpdateHere/IoUTAbbr}

\begin{thebibliography}{100}
\providecommand{\url}[1]{#1}
\csname url@samestyle\endcsname
\providecommand{\newblock}{\relax}
\providecommand{\bibinfo}[2]{#2}
\providecommand{\BIBentrySTDinterwordspacing}{\spaceskip=0pt\relax}
\providecommand{\BIBentryALTinterwordstretchfactor}{4}
\providecommand{\BIBentryALTinterwordspacing}{\spaceskip=\fontdimen2\font plus
\BIBentryALTinterwordstretchfactor\fontdimen3\font minus
  \fontdimen4\font\relax}
\providecommand{\BIBforeignlanguage}[2]{{%
\expandafter\ifx\csname l@#1\endcsname\relax
\typeout{** WARNING: IEEEtran.bst: No hyphenation pattern has been}%
\typeout{** loaded for the language `#1'. Using the pattern for}%
\typeout{** the default language instead.}%
\else
\language=\csname l@#1\endcsname
\fi
#2}}
\providecommand{\BIBdecl}{\relax}
\BIBdecl

\bibitem{Al-Fuqaha2015}
A.~Al-Fuqaha, M.~Guizani, M.~Mohammadi, M.~Aledhari, and M.~Ayyash, ``Internet
  of things: a survey on enabling technologies, protocols, and applications,''
  \emph{IEEE Commun. Surveys Tuts.}, vol.~17, no.~4, pp. 2347--2376, Jun. 2015.

\bibitem{Eldefrawy2019}
M.~H. {Eldefrawy}, N.~{Pereira}, and M.~{Gidlund}, ``Key distribution protocol
  for industrial internet of things without implicit certificates,'' \emph{IEEE
  Internet Things J.}, vol.~6, no.~1, pp. 906--917, Feb. 2019.

\bibitem{Kao2017}
C.-C. Kao, Y.-S. Lin, G.-D. Wu, and C.-J. Huang, ``A comprehensive study on the
  internet of underwater things: applications, challenges, and channel
  models,'' \emph{Sensors}, vol.~17, no.~7, Jul. 2017.

\bibitem{Domingo2012}
M.~C. Domingo, ``An overview of the internet of underwater things,'' \emph{J.
  Netw. Comput. Appl.}, vol.~35, no.~6, pp. 1879--1890, Nov. 2012.

\bibitem{WHOIKnowYourOcean}
\BIBentryALTinterwordspacing
WHOI, ``Know your ocean,'' Feb. 2019. [Online]. Available:
  \url{http://www.whoi.edu/know-your-ocean/}
\BIBentrySTDinterwordspacing

\bibitem{LiY2018}
Y.~Li, Y.~Zhang, W.~Li, and T.~Jiang, ``Marine wireless big data: efficient
  transmission, related applications, and challenges,'' \emph{IEEE Wireless
  Commun.}, vol.~25, no.~1, pp. 19--25, Feb. 2018.

\bibitem{Xiao2017}
G.~Xiao, B.~Wang, Z.~Deng, M.~Fu, and Y.~Ling, ``An acoustic communication time
  delays compensation approach for master-slave {AUV} cooperative navigation,''
  \emph{IEEE Sensors J.}, vol.~17, no.~2, pp. 504--513, Jan. 2017.

\bibitem{Zeng2017}
Z.~Zeng, S.~Fu, H.~Zhang, Y.~Dong, and J.~Cheng, ``A survey of underwater
  optical wireless communications,'' \emph{IEEE Commun. Surveys Tuts.},
  vol.~19, no.~1, pp. 204--238, Jan. 2017.

\bibitem{Perez2017}
C.~A. Pérez, F.~S. Valles, R.~T. Sánchez, M.~J. Buendía,
  F.~López-Castejón, and J.~G. Cervera, ``Design and deployment of a wireless
  sensor network for the {Mar Menor} coastal observation system,'' \emph{IEEE
  J. Ocean. Eng.}, vol.~42, no.~4, pp. 966--976, Oct. 2017.

\bibitem{Javed2018}
F.~Javed, M.~K. Afzal, M.~Sharif, and B.~Kim, ``Internet of things ({IoT})
  operating systems support, networking technologies, applications, and
  challenges: a comparative review,'' \emph{IEEE Commun. Surveys Tuts.},
  vol.~20, no.~3, pp. 2062--2100, Mar. 2018.

\bibitem{Cao2018}
X.~Cao, L.~Liu, Y.~Cheng, and X.~S. Shen, ``Towards energy-efficient wireless
  networking in the big data era: a survey,'' \emph{IEEE Commun. Surveys
  Tuts.}, vol.~20, no.~1, pp. 303--332, Jan. 2018.

\bibitem{Verma2017}
S.~Verma, Y.~Kawamoto, Z.~M. Fadlullah, H.~Nishiyama, and N.~Kato, ``A survey
  on network methodologies for real-time analytics of massive {IoT} data and
  open research issues,'' \emph{IEEE Commun. Surveys Tuts.}, vol.~19, no.~3,
  pp. 1457--1477, Apr. 2017.

\bibitem{Maimo2018}
L.~F. Maimo, A.~L.~P. Gomez, F.~J.~G. Clemente, M.~G. Perez, and G.~M. Perez,
  ``A self-adaptive deep learning-based system for anomaly detection in {5G}
  networks,'' \emph{{IEEE} Access}, vol.~6, pp. 7700--7712, Feb. 2018.

\bibitem{Meng2018}
L.~Meng, T.~Hirayama, and S.~Oyanagi, ``Underwater-drone with panoramic camera
  for automatic fish recognition based on deep learning,'' \emph{{IEEE}
  Access}, vol.~6, pp. 17\,880--17\,886, Mar. 2018.

\bibitem{Sun2018}
X.~Sun, J.~Shi, L.~Liu, J.~Dong, C.~Plant, X.~Wang, and H.~Zhou, ``Transferring
  deep knowledge for object recognition in low-quality underwater videos,''
  \emph{Neurocomp.}, vol. 275, pp. 897--908, Jan. 2018.

\bibitem{AniBrownMary2017}
N.~Ani Brown~Mary and D.~Dharma, ``Coral reef image classification employing
  improved {LDP} for feature extraction,'' \emph{J. Vis. Commun. Image
  Repres.}, vol.~49, pp. 225--242, Nov. 2017.

\bibitem{Reggiannini2017}
M.~Reggiannini and O.~Salvetti, ``Seafloor analysis and understanding for
  underwater archeology,'' \emph{J. Cultur. Heritage}, vol.~24, pp. 147--156,
  Apr. 2017.

\bibitem{Jian2018}
M.~Jian, Q.~Qi, J.~Dong, Y.~Yin, and K.-M. Lam, ``Integrating {QDWD} with
  pattern distinctness and local contrast for underwater saliency detection,''
  \emph{J. Vis. Commun. Image Repres.}, vol.~53, pp. 31--41, May 2018.

\bibitem{Schoening2016}
T.~Schoening, T.~Kuhn, D.~O.~B. Jones, E.~Simon-Lledo, and T.~W. Nattkemper,
  ``Fully automated image segmentation for benthic resource assessment of
  poly-metallic nodules,'' \emph{Methods Ocean.}, vol. 15-16, pp. 78--89, Jul.
  2016.

\bibitem{Bonin-Font2017}
F.~Bonin-Font, A.~Burguera, and J.~L. Lisani, ``Visual discrimination and large
  area mapping of {Posidonia Oceanica} using a lightweight {AUV},''
  \emph{{IEEE} Access}, vol.~5, pp. 24\,479--24\,494, Oct. 2017.

\bibitem{Chuang2017}
M.~C. Chuang, J.~N. Hwang, J.~H. Ye, S.~C. Huang, and K.~Williams, ``Underwater
  fish tracking for moving cameras based on deformable multiple kernels,''
  \emph{IEEE Trans. Syst., Man, Cybern.: Syst.}, vol.~47, no.~9, pp.
  2467--2477, Sep. 2017.

\bibitem{Lau2012}
P.~Y. Lau, P.~L. Correia, P.~Fonseca, and A.~Campos, ``Estimating {Norway}
  lobster abundance from deep-water videos: an automatic approach,'' \emph{IET
  Image Process.}, vol.~6, no.~1, pp. 22--30, Feb. 2012.

\bibitem{Osterloff2016}
J.~Osterloff, I.~Nilssen, and T.~W. Nattkemper, ``A computer vision approach
  for monitoring the spatial and temporal shrimp distribution at the {LoVe}
  observatory,'' \emph{Methods Ocean.}, vol. 15-16, pp. 114--128, Jul. 2016.

\bibitem{Schmid2016}
M.~S. Schmid, C.~Aubry, J.~Grigor, and L.~Fortier, ``The {LOKI} underwater
  imaging system and an automatic identification model for the detection of
  zooplankton taxa in the {Arctic Ocean},'' \emph{Methods Ocean.}, vol. 15-16,
  pp. 129--160, Jul. 2016.

\bibitem{Faillettaz2016}
R.~Faillettaz, M.~Picheral, J.~Y. Luo, C.~Guigand, R.~K. Cowen, and J.-O.
  Irisson, ``Imperfect automatic image classification successfully describes
  plankton distribution patterns,'' \emph{Methods Ocean.}, vol. 15-16, pp.
  60--77, Jul. 2016.

\bibitem{Pearlman2016}
J.~Pearlman, S.~Jirka, J.~d. Rio, E.~Delory, L.~Frommhold, S.~Martinez, and
  T.~O. Reilly, ``{Oceans of Tomorrow} sensor interoperability for in-situ
  ocean monitoring,'' in \emph{Proc. Ocean.}\hskip 1em plus 0.5em minus
  0.4em\relax Monterey, USA: IEEE, Sep. 2016, pp. 1--8.

\bibitem{Foresti2001}
G.~L. Foresti, ``Visual inspection of sea bottom structures by an autonomous
  underwater vehicle,'' \emph{IEEE Trans. Syst., Man, Cybern., Part B
  (Cybern.)}, vol.~31, no.~5, pp. 691--705, Oct. 2001.

\bibitem{Khalifeh2016}
R.~Khalifeh, M.~S. Yasri, B.~Lescop, F.~Gallée, E.~Diler, D.~Thierry, and
  S.~Rioual, ``Development of wireless and passive corrosion sensors for
  material degradation monitoring in coastal zones and immersed environment,''
  \emph{IEEE J. Ocean. Eng.}, vol.~41, no.~4, pp. 776--782, Oct. 2016.

\bibitem{Paull2014}
L.~Paull, S.~Saeedi, M.~Seto, and H.~Li, ``{AUV} navigation and localization: a
  review,'' \emph{IEEE J. Ocean. Eng.}, vol.~39, no.~1, pp. 131--149, Jan.
  2014.

\bibitem{Miyoshi2016}
{T. Miyoshi \textit{et al.}}, ``“big data assimilation” toward
  post-petascale severe weather prediction: an overview and progress,''
  \emph{Proc. IEEE}, vol. 104, no.~11, pp. 2155--2179, Nov. 2016.

\bibitem{Kalyvas2017}
C.~Kalyvas, A.~Kokkos, and T.~Tzouramanis, ``A survey of official online
  sources of high-quality free-of-charge geospatial data for maritime
  geographic information systems applications,'' \emph{Inf. Syst.}, vol.~65,
  pp. 36--51, Apr. 2017.

\bibitem{Ahmed2017}
E.~Ahmed, I.~Yaqoob, I.~A.~T. Hashem, I.~Khan, A.~I.~A. Ahmed, M.~Imran, and
  A.~V. Vasilakos, ``The role of big data analytics in internet of things,''
  \emph{Comput. Netw.}, vol. 129, pp. 459--471, Dec. 2017.

\bibitem{Mohammadi2018}
M.~Mohammadi, A.~Al-Fuqaha, S.~Sorour, and M.~Guizani, ``Deep learning for
  {IoT} big data and streaming analytics: a survey,'' \emph{IEEE Commun.
  Surveys Tuts.}, vol.~20, no.~4, pp. 2923--2960, Jun. 2018.

\bibitem{HuangD2015}
D.~Huang, D.~Zhao, L.~Wei, Z.~Wang, and Y.~Du, ``Modeling and analysis in
  marine big data: advances and challenges,'' \emph{Math. Probl. Eng.}, vol.
  2015, pp. 1--13, Sep. 2015.

\bibitem{XuG2019}
G.~Xu, Y.~Shi, X.~Sun, and W.~Shen, ``Internet of things in marine environment
  monitoring: a review,'' \emph{Sensors}, vol.~19, no.~7, p. 1711, Apr. 2019.

\bibitem{Jouhari2019}
M.~{Jouhari}, K.~{Ibrahimi}, H.~{Tembine}, and J.~{Ben-Othman}, ``Underwater
  wireless sensor networks: a survey on enabling technologies, localization
  protocols, and internet of underwater things,'' \emph{{IEEE} Access}, vol.~7,
  pp. 96\,879--96\,899, Jul. 2019.

\bibitem{QiuT2020}
T.~{Qiu}, Z.~{Zhao}, T.~{Zhang}, C.~{Chen}, and C.~L.~P. {Chen}, ``Underwater
  internet of things in smart ocean: system architecture and open issues,''
  \emph{IEEE Trans. Ind. Inform.}, vol.~16, no.~7, pp. 4297--4307, Jul. 2020.

\bibitem{RajD2020}
D.~Raj, J.~Lee, E.~Ko, S.~Shin, J.-I. Namgung, S.-H. Yum, and S.-H. Park,
  ``Underwater network management system in internet of underwater things: open
  challenges, benefits, and feasible solution,'' \emph{Electron.}, vol.~9, p.
  1142, Jul. 2020.

\bibitem{KhalilR2020}
R.~{Khalil}, M.~{Babar}, T.~{Jan}, and N.~{Saeed}, ``Towards the internet of
  underwater things: recent developments and future challenges,'' \emph{IEEE
  Consum. Electron. Mag.}, p.~1, May 2020.

\bibitem{Zhou2016}
Z.~Zhou, B.~Yao, R.~Xing, L.~Shu, and S.~Bu, ``{E-CARP}: an energy efficient
  routing protocol for {UWSNs} in the internet of underwater things,''
  \emph{IEEE Sensors J.}, vol.~16, no.~11, pp. 4072--4082, Jun. 2016.

\bibitem{Adamo2015}
F.~Adamo, F.~Attivissimo, C.~G.~C. Carducci, and A.~M.~L. Lanzolla, ``A smart
  sensor network for sea water quality monitoring,'' \emph{IEEE Sensors J.},
  vol.~15, no.~5, pp. 2514--2522, May 2015.

\bibitem{Al-Zaidi2018}
R.~Al-Zaidi, J.~C. Woods, M.~Al-Khalidi, and H.~Hu, ``Building novel
  {VHF}-based wireless sensor networks for the internet of marine things,''
  \emph{IEEE Sensors J.}, vol.~18, no.~5, pp. 2131--2144, Mar. 2018.

\bibitem{Hughes2017}
{T. P. Hughes \textit{et al.}}, ``Global warming and recurrent mass bleaching
  of corals,'' \emph{{Nature}}, vol. 543, pp. 373--377, Mar. 2017.

\bibitem{Fuentes-Perez2018}
J.~F. Fuentes-Pérez, C.~Meurer, J.~A. Tuhtan, and M.~Kruusmaa, ``Differential
  pressure sensors for underwater speedometry in variable velocity and
  acceleration conditions,'' \emph{IEEE J. Ocean. Eng.}, vol.~43, no.~2, pp.
  418--426, Apr. 2018.

\bibitem{BalanisAnt2016}
C.~A. Balanis, \emph{Antenna theory: analysis and design}.\hskip 1em plus 0.5em
  minus 0.4em\relax John Wiley \& Sons, 2016.

\bibitem{Ainslie1998}
M.~A. Ainslie and J.~G. McColm, ``A simplified formula for viscous and chemical
  absorption in sea water,'' \emph{J. Acoust. Soc. America}, vol. 103, no.~3,
  pp. 1671--1672, Jun. 1998.

\bibitem{Gabriel2013}
C.~Gabriel, M.~A. Khalighi, S.~Bourennane, P.~Leon, and V.~Rigaud,
  ``{Monte-Carlo-based} channel characterization for underwater optical
  communication systems,'' \emph{IEEE/OSA J. Opt. Commun. Netw.}, vol.~5,
  no.~1, pp. 1--12, Jan. 2013.

\bibitem{IlliE2019}
E.~{Illi}, F.~E. {Bouanani}, K.~{Park}, F.~{Ayoub}, and M.~{Alouini}, ``An
  improved accurate solver for the time-dependent {RTE} in underwater optical
  wireless communications,'' \emph{{IEEE} Access}, vol.~7, pp.
  96\,478--96\,494, Jul. 2019.

\bibitem{Jaffe2015}
J.~S. Jaffe, ``Underwater optical imaging: the past, the present, and the
  prospects,'' \emph{IEEE J. Ocean. Eng.}, vol.~40, no.~3, pp. 683--700, Jul.
  2015.

\bibitem{WangJ2018}
J.~{Wang}, H.~{Zhou}, Y.~{Li}, Q.~{Sun}, Y.~{Wu}, S.~{Jin}, T.~Q.~S. {Quek},
  and C.~{Xu}, ``Wireless channel models for maritime communications,''
  \emph{{IEEE} Access}, vol.~6, pp. 68\,070--68\,088, Dec. 2018.

\bibitem{SunW2018}
W.~{Sun}, X.~{Yuan}, J.~{Wang}, Q.~{Li}, L.~{Chen}, and D.~{Mu}, ``End-to-end
  data delivery reliability model for estimating and optimizing the link
  quality of industrial {WSNs},'' \emph{IEEE Trans. Autom. Sci. Eng.}, vol.~15,
  no.~3, pp. 1127--1137, Jul. 2018.

\bibitem{ZhangH2017}
H.~{Zhang}, X.~{Liu}, C.~{Li}, Y.~{Chen}, X.~{Che}, L.~Y. {Wang}, F.~{Lin}, and
  G.~{Yin}, ``Scheduling with predictable link reliability for wireless
  networked control,'' \emph{IEEE Trans. Wireless Commun.}, vol.~16, no.~9, pp.
  6135--6150, Sep. 2017.

\bibitem{Gomes2019}
R.~D. Gomes, C.~Benavente-Peces, I.~E. Fonseca, and M.~S. Alencar, ``Adaptive
  and beacon-based multi-channel protocol for industrial wireless sensor
  networks,'' \emph{J. Netw. Comput. Appl.}, vol. 132, pp. 22--39, Apr. 2019.

\bibitem{Murali2019}
S.~{Murali} and A.~{Jamalipour}, ``Mobility-aware energy-efficient parent
  selection algorithm for low power and lossy networks,'' \emph{IEEE Internet
  Things J.}, vol.~6, no.~2, pp. 2593--2601, Apr. 2019.

\bibitem{Fink2005}
J.~C. Fink, K.~Fermanich, and T.~Ehlinger, \emph{The effects of urbanization on
  {Baird Creek}, {Green Bay}, {Wisconsin}}.\hskip 1em plus 0.5em minus
  0.4em\relax University of Wisconsin-Green Bay, 2005.

\bibitem{Kaushal2016}
H.~{Kaushal} and G.~{Kaddoum}, ``Underwater optical wireless communication,''
  \emph{{IEEE} Access}, vol.~4, pp. 1518--1547, Apr. 2016.

\bibitem{LiZ2017}
Z.~{Li} and Y.~{Wu}, ``Smooth mobility and link reliability-based optimized
  link state routing scheme for {MANETs},'' \emph{IEEE Commun. Lett.}, vol.~21,
  no.~7, pp. 1529--1532, Jul. 2017.

\bibitem{Rani2017}
S.~Rani, S.~H. Ahmed, J.~Malhotra, and R.~Talwar, ``Energy efficient chain
  based routing protocol for underwater wireless sensor networks,'' \emph{J.
  Netw. Comput. Appl.}, vol.~92, pp. 42--50, Aug. 2017.

\bibitem{TranDang2019}
H.~{Tran-Dang} and D.~{Kim}, ``Channel-aware cooperative routing in underwater
  acoustic sensor networks,'' \emph{J. Commun. Netw.}, vol.~21, no.~1, pp.
  33--44, Feb. 2019.

\bibitem{ZhaoR2019}
R.~{Zhao}, H.~{Long}, O.~A. {Dobre}, X.~{Shen}, T.~M.~N. {Ngatched}, and
  H.~{Mei}, ``Time reversal based {MAC} for multi-hop underwater acoustic
  networks,'' \emph{IEEE Syst. J.}, vol.~13, no.~3, pp. 2531--2542, Sep. 2019.

\bibitem{Fu2014}
B.~{Fu}, Y.~{Xiao}, H.~{Deng}, and H.~{Zeng}, ``A survey of cross-layer designs
  in wireless networks,'' \emph{IEEE Commun. Surveys Tuts.}, vol.~16, no.~1,
  pp. 110--126, Jan. 2014.

\bibitem{Dye2007}
M.~Dye, R.~McDonald, and A.~Rufi, \emph{Network fundamentals, {CCNA}
  exploration companion guide}, 1st~ed.\hskip 1em plus 0.5em minus 0.4em\relax
  Cisco press, Oct. 2007.

\bibitem{Cui2006}
J.-H. Cui, J.~Kong, M.~Gerla, and S.~Zhou, ``The challenges of building mobile
  underwater wireless networks for aquatic applications,'' \emph{IEEE Netw.},
  vol.~20, no.~3, pp. 12--18, May 2006.

\bibitem{MusilovaS2020}
S.~Musilová, ``Profiling and detection of {IoT} attacks in {Telnet} traffic,''
  Master's Thesis, Czech Technical University in Prague, Department of Computer
  Science, Jan. 2020.

\bibitem{Xu2017}
M.~Xu and L.~Liu, ``Sender-receiver role-based energy-aware scheduling for
  internet of underwater things,'' \emph{IEEE Trans. Emerg. Topics Comput.},
  p.~1, Nov. 2016.

\bibitem{Koseoglu2017}
M.~Koseoglu, E.~Karasan, and L.~Chen, ``Cross-layer energy minimization for
  underwater {ALOHA} networks,'' \emph{IEEE Syst. J.}, vol.~11, no.~2, pp.
  551--561, Jun. 2017.

\bibitem{AlfouzanF2019}
F.~A. {Alfouzan}, A.~{Shahrabi}, S.~M. {Ghoreyshi}, and T.~{Boutaleb}, ``A
  collision-free graph coloring {MAC} protocol for underwater sensor
  networks,'' \emph{{IEEE} Access}, vol.~7, pp. 39\,862--39\,878, Mar. 2019.

\bibitem{Feng2018}
X.~Feng, Z.~Wang, G.~Han, W.~Qu, and A.~Chen, ``Distributed receiver-oriented
  adaptive multichannel {MAC} for underwater sensor networks,'' \emph{{IEEE}
  Access}, vol.~6, pp. 11\,666--11\,675, Feb. 2018.

\bibitem{ChenYJ2007}
Y.~{Chen} and H.~{Wang}, ``Ordered {CSMA}: a collision-free {MAC} protocol for
  underwater acoustic networks,'' in \emph{Proc. Ocean.}\hskip 1em plus 0.5em
  minus 0.4em\relax Vancouver, Canada: IEEE, Oct. 2007, pp. 1--6.

\bibitem{Jiang2018}
S.~Jiang, ``State-of-the-art {Medium Access Control (MAC)} protocols for
  underwater acoustic networks: a survey based on a {MAC} reference model,''
  \emph{IEEE Commun. Surveys Tuts.}, vol.~20, no.~1, pp. 96--131, Jan. 2018.

\bibitem{Chen2014}
M.~{Chen}, Y.~{Shen}, J.~{Luis}, and C.~{Chou}, ``Energy-efficient {OR-based}
  {MAC} protocol for underwater sensor networks,'' in \emph{Proc.
  Sensors}.\hskip 1em plus 0.5em minus 0.4em\relax Valencia, Spain: IEEE, Nov.
  2014, pp. 118--121.

\bibitem{LiC2016}
C.~Li, Y.~Xu, C.~Xu, Z.~An, B.~Diao, and X.~Li, ``{DTMAC}: a delay tolerant
  {MAC} protocol for underwater wireless sensor networks,'' \emph{IEEE Sensors
  J.}, vol.~16, no.~11, pp. 4137--4146, Jun. 2016.

\bibitem{Fujihashi2018}
T.~Fujihashi, S.~Saruwatari, and T.~Watanabe, ``Multiview video transmission
  over underwater acoustic path,'' \emph{IEEE Trans. Multimedia}, vol.~20,
  no.~8, pp. 2166--2181, Aug. 2018.

\bibitem{Mohamed2010}
N.~{Mohamed}, I.~{Jawhar}, J.~{Al-Jaroodi}, and L.~{Zhang}, ``Monitoring
  underwater pipelines using sensor networks,'' in \emph{Proc. 12th Int. Conf.
  High Perform. Comput. Commun.}\hskip 1em plus 0.5em minus 0.4em\relax
  Melbourne, Australia: IEEE, Sep. 2010, pp. 346--353.

\bibitem{Carpenter2013}
B.~E. Carpenter, \emph{Network geeks: how they built the internet}.\hskip 1em
  plus 0.5em minus 0.4em\relax Springer, Apr. 2013.

\bibitem{Aalsalem2018}
M.~Y. Aalsalem, W.~Z. Khan, W.~Gharibi, M.~K. Khan, and Q.~Arshad, ``Wireless
  sensor networks in oil and gas industry: recent advances, taxonomy,
  requirements, and open challenges,'' \emph{J. Netw. Comput. Appl.}, vol. 113,
  pp. 87--97, Jul. 2018.

\bibitem{BergmannN2014}
N.~W. {Bergmann}, J.~{Juergens}, L.~{Hou}, Y.~{Wang}, and J.~{Trevathan},
  ``Wireless underwater power and data transfer,'' in \emph{Proc. 38th Annual
  IEEE Conf. Local Comput. Netw.}\hskip 1em plus 0.5em minus 0.4em\relax
  Sydney, Australia: IEEE, Mar. 2014, pp. 104--107.

\bibitem{Lentz2007}
S.~T. {Lentz}, ``The {NEPTUNE} canada communications network,'' in \emph{Proc.
  Ocean.}\hskip 1em plus 0.5em minus 0.4em\relax Vancouver, Canada: IEEE, Sep.
  2007, pp. 1--5.

\bibitem{Marchetti2016}
L.~Marchetti and R.~Reggiannini, ``An efficient receiver structure for
  sweep-spread-carrier underwater acoustic links,'' \emph{IEEE J. Ocean. Eng.},
  vol.~41, no.~2, pp. 440--449, Apr. 2016.

\bibitem{Qarabaqi2013}
P.~{Qarabaqi} and M.~{Stojanovic}, ``Statistical characterization and
  computationally efficient modeling of a class of underwater acoustic
  communication channels,'' \emph{IEEE J. Ocean. Eng.}, vol.~38, no.~4, pp.
  701--717, Oct. 2013.

\bibitem{ZhouJ2019}
J.~{Zhou}, H.~{Jiang}, P.~{Wu}, and Q.~{Chen}, ``Study of propagation channel
  characteristics for underwater acoustic communication environments,''
  \emph{{IEEE} Access}, vol.~7, pp. 79\,438--79\,445, Jun. 2019.

\bibitem{Naderi2018}
M.~{Naderi}, A.~G. {Zajić}, and M.~{Pätzold}, ``A nonisovelocity
  geometry-based underwater acoustic channel model,'' \emph{IEEE Trans. Veh
  Technol.}, vol.~67, no.~4, pp. 2864--2879, Apr. 2018.

\bibitem{Naderi2017}
M.~{Naderi}, M.~{Pätzold}, R.~{Hicheri}, and N.~{Youssef}, ``A geometry-based
  underwater acoustic channel model allowing for sloped ocean bottom
  conditions,'' \emph{IEEE Trans. Wireless Commun.}, vol.~16, no.~4, pp.
  2394--2408, Apr. 2017.

\bibitem{LiW2016}
W.~{Li}, Z.~{Huang}, M.~{Gong}, and Y.~{Ji}, ``Land-to-underwater optical
  communication system based on underwater visible light communication network
  units and fiber links,'' in \emph{Proc. Asia Commun. Photon. Conf.}\hskip 1em
  plus 0.5em minus 0.4em\relax Wuhan, China: IEEE, Nov. 2016, pp. 1--3.

\bibitem{LiW2019}
W.~Li, S.~Zhou, P.~Willett, and Q.~Zhang, ``Preamble detection for underwater
  acoustic communications based on sparse channel identification,'' \emph{IEEE
  J. Ocean. Eng.}, vol.~44, no.~1, pp. 256--268, Jan. 2019.

\bibitem{Liang2018}
M.~Liang, J.~Duan, and D.~Zhao, ``Optimal redundancy control strategy for
  fountain code-based underwater acoustic communication,'' \emph{{IEEE}
  Access}, vol.~6, pp. 69\,321--69\,334, Nov. 2018.

\bibitem{WangC2018}
C.~Wang, Z.~Wang, W.~Sun, and D.~R. Fuhrmann, ``Reinforcement learning-based
  adaptive transmission in time-varying underwater acoustic channels,''
  \emph{{IEEE} Access}, vol.~6, pp. 2541--2558, Dec. 2017.

\bibitem{Atyabi2018}
A.~Atyabi, S.~MahmoudZadeh, and S.~Nefti-Meziani, ``Current advancements on
  autonomous mission planning and management systems: an {AUV} and {UAV}
  perspective,'' \emph{Annual Rev. Control}, vol.~46, pp. 196--215, Aug. 2018.

\bibitem{Naumann2019}
R.~{Naumann}, S.~{Dietzel}, and B.~{Scheuermann}, ``Push the barrier: discrete
  event protocol emulation,'' \emph{IEEE/ACM Trans. Netw.}, vol.~27, no.~2, pp.
  635--648, Apr. 2019.

\bibitem{Petrioli2015}
C.~Petrioli, R.~Petroccia, J.~R. Potter, and D.~Spaccini, ``The {SUNSET}
  framework for simulation, emulation and at-sea testing of underwater wireless
  sensor networks,'' \emph{{Ad Hoc} Netw.}, vol.~34, pp. 224--238, Nov. 2015.

\bibitem{Dorathy2018}
I.~Dorathy and M.~Chandrasekaran, ``Simulation tools for mobile ad hoc
  networks: a survey,'' \emph{J. Appl. Research Technol.}, vol.~16, no.~5, pp.
  437--445, Oct. 2018.

\bibitem{Veltri2019}
L.~Veltri, L.~Davoli, R.~Pecori, A.~Vannucci, and F.~Zanichelli, ``{NEMO}: a
  flexible and highly scalable network emulator,'' \emph{Softwarex}, vol.~10,
  p. 100248, Dec. 2019.

\bibitem{Ghaleb2017}
M.~{Ghaleb}, E.~{Felemban}, S.~{Subramaniam}, A.~A. {Sheikh}, and S.~B.
  {Qaisar}, ``A performance simulation tool for the analysis of data gathering
  in both terrestrial and underwater sensor networks,'' \emph{{IEEE} Access},
  vol.~5, pp. 4190--4208, Mar. 2017.

\bibitem{KisacikR2019}
{R. Kisacik \textit{et al.}}, ``Distance and power based experimental
  verification of channel model in visible light communication,'' in
  \emph{Proc. 27th Signal Process. Commun. Appl. Conf.}\hskip 1em plus 0.5em
  minus 0.4em\relax Sivas, Turkey: IEEE, Apr. 2019, pp. 1--4.

\bibitem{YangG2019}
G.~Yang, L.~Dai, G.~Si, S.~Wang, and S.~Wang, ``Challenges and security issues
  in underwater wireless sensor networks,'' \emph{Proc. Comput. Sci.}, vol.
  147, pp. 210--216, Nov. 2019.

\bibitem{Editorial2019}
Editorial, ``Underwater acoustic communications: where we stand and what is
  next?'' \emph{IEEE J. Ocean. Eng.}, vol.~44, no.~1, pp. 1--6, Jan. 2019.

\bibitem{JiangS2019}
S.~Jiang, ``On securing underwater acoustic networks: a survey,'' \emph{IEEE
  Commun. Surveys Tuts.}, vol.~21, no.~1, pp. 729--752, Jan. 2019.

\bibitem{XuM2020}
M.~{Xu}, Y.~{Fan}, and L.~{Liu}, ``Multi-party secret key generation over
  underwater acoustic channels,'' \emph{IEEE Wireless Commun. Lett.}, vol.~9,
  no.~7, pp. 1075--1079, Jul. 2020.

\bibitem{YangT2008}
T.~C. Yang and W.~Yang, ``Low probability of detection underwater acoustic
  communications for mobile platforms,'' in \emph{Proc. Ocean.}\hskip 1em plus
  0.5em minus 0.4em\relax Quebec, Canada: IEEE, Sep. 2008, pp. 1--6.

\bibitem{JiajiaJ2020}
J.~{Jiajia}, W.~{Xianquan}, D.~{Fajie}, F.~{Xiao}, L.~{Chunyue}, and
  S.~{Zhongbo}, ``A basic bio-inspired camouflage communication frame design
  and applications for secure underwater communication among military
  underwater platforms,'' \emph{{IEEE} Access}, vol.~8, pp. 24\,927--24\,940,
  Jan. 2020.

\bibitem{HuangS2020}
S.~{Huang}, X.~{Hou}, W.~{Liu}, G.~{Liu}, Y.~{Dai}, and W.~{Tian}, ``Mimicking
  ship-radiated noise with chaos signal for covert underwater acoustic
  communication,'' \emph{{IEEE} Access}, vol.~8, pp. 180\,341--180\,351, Sep.
  2020.

\bibitem{Dini2011}
G.~{Dini} and A.~L. {Duca}, ``{SeFLOOD}: a secure network discovery protocol
  for underwater acoustic networks,'' in \emph{Proc. IEEE Symp. Comput.
  Commun.}\hskip 1em plus 0.5em minus 0.4em\relax Kerkyra, Greece: IEEE, Jun.
  2011, pp. 636--638.

\bibitem{Zuba2014}
M.~{Zuba}, M.~{Fagan}, Z.~{Shi}, and J.~{Cui}, ``A resilient pressure routing
  scheme for underwater acoustic networks,'' in \emph{Proc. IEEE Glob. Commun.
  Conf.}\hskip 1em plus 0.5em minus 0.4em\relax Austin, USA: IEEE, Dec. 2014,
  pp. 637--642.

\bibitem{Capossele2015}
A.~Capossele, G.~De~Cicco, and C.~Petrioli, ``{R-CARP}: a reputation based
  channel aware routing protocol for underwater acoustic sensor networks,'' in
  \emph{Proc. 10th Int. Conf. Underwater Netw. Syst.}\hskip 1em plus 0.5em
  minus 0.4em\relax New York, USA: ACM, Oct. 2015, pp. 1--6.

\bibitem{LuoH2018}
H.~Luo, K.~Wu, R.~Ruby, Y.~Liang, Z.~Guo, and L.~M. Ni, ``Software-defined
  architectures and technologies for underwater wireless sensor networks: a
  survey,'' \emph{IEEE Commun. Surveys Tuts.}, vol.~20, no.~4, pp. 2855--2888,
  May 2018.

\bibitem{Dol2017}
H.~S. Dol, P.~Casari, T.~van~der Zwan, and R.~Otnes, ``Software-defined
  underwater acoustic modems: historical review and the {NILUS} approach,''
  \emph{IEEE J. Ocean. Eng.}, vol.~42, no.~3, pp. 722--737, Jul. 2017.

\bibitem{LuoY2014}
Y.~Luo, L.~Pu, M.~Zuba, Z.~Peng, and J.~Cui, ``Challenges and opportunities of
  underwater cognitive acoustic networks,'' \emph{IEEE Trans. Emerg. Topics
  Comput.}, vol.~2, no.~2, pp. 198--211, Jun. 2014.

\bibitem{Akyildiz2016}
I.~F. Akyildiz, P.~Wang, and S.-C. Lin, ``{SoftWater}: software-defined
  networking for next-generation underwater communication systems,'' \emph{{Ad
  Hoc} Netw.}, vol.~46, pp. 1--11, Aug. 2016.

\bibitem{Srimathi2013}
C.~Srimathi, S.-H. Park, and N.~Rajesh, ``Proposed framework for underwater
  sensor cloud for environmental monitoring,'' in \emph{Proc. 5th Int. Conf.
  Ubiquitous Future Netw.}\hskip 1em plus 0.5em minus 0.4em\relax Da Nang,
  Vietnam: IEEE, Jul. 2013, pp. 104--109.

\bibitem{CelikA2020}
A.~{Celik}, N.~{Saeed}, B.~{Shihada}, T.~Y. {Al-Naffouri}, and M.~{Alouini},
  ``A software-defined opto-acoustic network architecture for internet of
  underwater things,'' \emph{IEEE Commun. Mag.}, vol.~58, no.~4, pp. 88--94,
  Apr. 2020.

\bibitem{Lal2017}
C.~{Lal}, R.~{Petroccia}, K.~{Pelekanakis}, M.~{Conti}, and J.~{Alves},
  ``Toward the development of secure underwater acoustic networks,'' \emph{IEEE
  J. Ocean. Eng.}, vol.~42, no.~4, pp. 1075--1087, Oct. 2017.

\bibitem{LiuJ2018}
J.~Liu, G.~Shou, Y.~Liu, Y.~Hu, and Z.~Guo, ``Performance evaluation of
  integrated multi-access edge computing and fiber-wireless access networks,''
  \emph{{IEEE} Access}, vol.~6, pp. 30\,269--30\,279, May 2018.

\bibitem{WangT2020}
T.~{Wang}, D.~{Zhao}, S.~{Cai}, W.~{Jia}, and A.~{Liu}, ``Bidirectional
  prediction-based underwater data collection protocol for end-edge-cloud
  orchestrated system,'' \emph{IEEE Trans. Ind. Inform.}, vol.~16, no.~7, pp.
  4791--4799, Jul. 2020.

\bibitem{CaiS2019}
S.~{Cai}, Y.~{Zhu}, T.~{Wang}, G.~{Xu}, A.~{Liu}, and X.~{Liu}, ``Data
  collection in underwater sensor networks based on mobile edge computing,''
  \emph{{IEEE} Access}, vol.~7, pp. 65\,357--65\,367, May 2019.

\bibitem{WangQ2020}
Q.~{Wang}, H.~{Dai}, Q.~{Wang}, M.~K. {Shukla}, W.~{Zhang}, and C.~G. {Soares},
  ``On connectivity of {UAV}-assisted data acquisition for underwater internet
  of things,'' \emph{IEEE Internet Things J.}, vol.~7, no.~6, pp. 5371--5385,
  Mar. 2020.

\bibitem{Porambage2018}
P.~Porambage, J.~Okwuibe, M.~Liyanage, M.~Ylianttila, and T.~Taleb, ``Survey on
  multi-access edge computing for internet of things realization,'' \emph{IEEE
  Commun. Surveys Tuts.}, vol.~20, no.~4, pp. 2961--2991, Jun. 2018.

\bibitem{DiX2018}
X.~Di, Y.~Zhang, T.~Liu, S.~Kang, and Y.~Zhao, ``Mobile fog computing-assisted
  resource allocation for two-hop {SWIPT} {OFDM} networks,'' \emph{Wireless
  Commun. Mob. Comput.}, vol. 2018, Sep. 2018.

\bibitem{Safavat2019}
S.~Safavat, N.~N. Sapavath, and D.~B. Rawat, ``Recent advances in mobile edge
  computing and content caching,'' \emph{Digit. Commun. Netw.}, Sep. 2019.

\bibitem{ChiangM2016}
M.~{Chiang} and T.~{Zhang}, ``Fog and {IoT}: an overview of research
  opportunities,'' \emph{IEEE Internet Things J.}, vol.~3, no.~6, pp. 854--864,
  Dec. 2016.

\bibitem{Brown2014}
M.~S. Brown, \emph{Data mining for dummies}, 1st~ed.\hskip 1em plus 0.5em minus
  0.4em\relax John Wiley \& Sons, Sep. 2014.

\bibitem{Chhabra2015}
S.~Chhabra and D.~Singh, ``Data fusion and data aggregation/summarization
  techniques in {WSNs}: a review,'' \emph{Int. J. Comput. Appl.}, vol. 121,
  no.~19, Jul. 2015.

\bibitem{Misra2017}
S.~{Misra} and R.~{Kumar}, ``An analytical study of {LEACH} and {PEGASIS}
  protocol in wireless sensor network,'' in \emph{Proc. Int. Conf. Innov. Inf.,
  Embed. Commun. Syst.}\hskip 1em plus 0.5em minus 0.4em\relax Coimbatore,
  India: IEEE, Mar. 2017, pp. 1--5.

\bibitem{Chand2014}
S.~Chand, S.~Singh, and B.~Kumar, ``Heterogeneous {HEED} protocol for wireless
  sensor networks,'' \emph{Wireless Personal Commun.}, vol.~77, no.~3, pp.
  2117--2139, Feb. 2014.

\bibitem{Ma2018}
J.~Ma, S.~Wang, C.~Meng, Y.~Ge, and J.~Du, ``Hybrid energy-efficient {APTEEN}
  protocol based on ant colony algorithm in wireless sensor network,''
  \emph{Wireless Commun. Netw.}, vol. 2018, no.~1, p. 102, Dec. 2018.

\bibitem{Qiu2019}
J.~{Qiu}, Z.~{Xing}, C.~{Zhu}, K.~{Lu}, J.~{He}, Y.~{Sun}, and L.~{Yin},
  ``Centralized fusion based on interacting multiple model and adaptive
  {Kalman} filter for target tracking in underwater acoustic sensor networks,''
  \emph{{IEEE} Access}, vol.~7, pp. 25\,948--25\,958, Feb. 2019.

\bibitem{Castanedo2013}
F.~Castanedo, ``A review of data fusion techniques,'' \emph{Sci. World J.},
  vol. 2013, Sep. 2013.

\bibitem{Nomura2018}
K.~{Nomura}, D.~{Sugimura}, and T.~{Hamamoto}, ``Underwater image color
  correction using exposure-bracketing imaging,'' \emph{IEEE Signal Process.
  Lett.}, vol.~25, no.~6, pp. 893--897, Jun. 2018.

\bibitem{Zulkifli2018}
S.~N. Zulkifli, H.~A. Rahim, and W.-J. Lau, ``Detection of contaminants in
  water supply: a review on state-of-the-art monitoring technologies and their
  applications,'' \emph{Sensors Actuat. B: Chemic.}, vol. 255, pp. 2657--2689,
  Feb. 2018.

\bibitem{Kassal2018}
P.~Kassal, M.~D. Steinberg, and I.~M. Steinberg, ``Wireless chemical sensors
  and biosensors: a review,'' \emph{Sensors Actuat. B: Chemic.}, vol. 266, pp.
  228--245, Aug. 2018.

\bibitem{Hewitt2019}
J.~E. Hewitt and S.~F. Thrush, ``Monitoring for tipping points in the marine
  environment,'' \emph{J. Env. Manag.}, vol. 234, pp. 131--137, Mar. 2019.

\bibitem{Wright2016}
N.~G. Wright and H.~K. Chan, ``Low-cost internet of things ocean observation,''
  in \emph{Proc. Ocean.}\hskip 1em plus 0.5em minus 0.4em\relax Monterey, USA:
  IEEE, Sep. 2016, pp. 1--5.

\bibitem{Palaniswami2017}
M.~Palaniswami, A.~S. Rao, and S.~Bainbridge, ``Real-time monitoring of the
  {Great Barrier Reef} using internet of things with big data analytics,''
  \emph{ITU J.: ICT Discov.}, vol.~1, no.~13, pp. 1--10, Oct. 2017.

\bibitem{Godo2014}
O.~Godø, S.~Johnsen, and T.~Torkelsen, ``The {LoVe} ocean observatory is in
  operation,'' \emph{Marine Technol. Soc. J.}, vol.~48, Mar. 2014.

\bibitem{Shihavuddin2013}
A.~S.~M. Shihavuddin, N.~Gracias, R.~Garcia, C.~A. Gleason, and B.~Gintert,
  ``Image-based coral reef classification and thematic mapping,'' \emph{Remote
  Sens.}, vol.~5, no.~4, Apr. 2013.

\bibitem{HanM2018}
M.~Han, Z.~Lyu, T.~Qiu, and M.~Xu, ``A review on intelligence dehazing and
  color restoration for underwater images,'' \emph{IEEE Trans. Syst., Man,
  Cybern.: Syst.}, vol.~PP, no.~99, pp. 1--13, Jan. 2018.

\bibitem{Duntley1963}
S.~Q. Duntley, ``Light in the sea,'' \emph{J. Opt. Soc. America}, vol.~53,
  no.~2, pp. 214--233, 1963.

\bibitem{Zaneveld2003}
J.~R.~V. Zaneveld and W.~S. Pegau, ``Robust underwater visibility parameter,''
  \emph{Opt. Exp.}, vol.~11, no.~23, pp. 2997--3009, Nov. 2003.

\bibitem{Salih2018}
A.~A.~M. Salih, K.~Hasikin, and N.~A.~M. Isa, ``Adaptive fuzzy exposure local
  contrast enhancement,'' \emph{{IEEE} Access}, vol.~6, pp. 58\,794--58\,806,
  Sep. 2018.

\bibitem{Spampinato2010}
C.~Spampinato, D.~Giordano, R.~Di~Salvo, Y.-H.~J. Chen-Burger, R.~B. Fisher,
  and G.~Nadarajan, ``Automatic fish classification for underwater species
  behavior understanding,'' in \emph{Proc. 1st Int. Worksh. Anal. Retriev.
  Tracked Events Motion Imagery Streams}.\hskip 1em plus 0.5em minus
  0.4em\relax Firenze, Italy: ACM, Oct. 2010, pp. 45--50.

\bibitem{Chuang2015}
M.~C. Chuang, J.~N. Hwang, K.~Williams, and R.~Towler, ``Tracking live fish
  from low-contrast and low-frame-rate stereo videos,'' \emph{IEEE Trans.
  Circuits Syst. Video Technol.}, vol.~25, no.~1, pp. 167--179, Jan. 2015.

\bibitem{Negahdaripour2013}
S.~Negahdaripour, ``On {3-D} motion estimation from feature tracks in {2-D}
  {FS} sonar video,'' \emph{IEEE Trans. Robot.}, vol.~29, no.~4, pp.
  1016--1030, Aug. 2013.

\bibitem{ChengCh2018}
C.~Cheng and H.~Yang, ``Realization of {3D} stereo imaging for underwater
  applications,'' in \emph{Proc. Int. Conf. Intell. Robot. Control Eng.}\hskip
  1em plus 0.5em minus 0.4em\relax Lanzhou, China: IEEE, Aug. 2018, pp.
  242--246.

\bibitem{Mazzei2015}
L.~Mazzei, L.~Corgnati, S.~Marini, E.~Ottaviani, and B.~Isoppo, ``Low cost
  stereo system for imaging and {3D} reconstruction of underwater organisms,''
  in \emph{Proc. Ocean.}\hskip 1em plus 0.5em minus 0.4em\relax Genoa, Italy:
  IEEE, May 2015, pp. 1--4.

\bibitem{Assalih2009}
H.~Assalih, Y.~Petillot, and J.~Bell, ``Acoustic stereo imaging ({ASI})
  system,'' in \emph{Proc. Ocean.}\hskip 1em plus 0.5em minus 0.4em\relax
  Bremen, Germany: IEEE, May 2009, pp. 1--7.

\bibitem{Anwer2016}
A.~Anwer, S.~S.~A. Ali, A.~Khan, and F.~Mériaudeau, ``Real-time underwater
  {3D} scene reconstruction using commercial depth sensor,'' in \emph{Proc.
  Int. Conf. Underwater Syst. Technol.: Theory Appl.}\hskip 1em plus 0.5em
  minus 0.4em\relax Penang, Malaysia: IEEE, Dec. 2016, pp. 67--70.

\bibitem{WangCh2000}
C.-C. Wang, S.-W. Shyue, and S.-H. Cheng, ``Underwater structure inspection
  with laser light stripes,'' in \emph{Proc. Int. Symp. Underwater
  Technol.}\hskip 1em plus 0.5em minus 0.4em\relax Tokyo, Japan: IEEE, May
  2000, pp. 201--205.

\bibitem{Ishibashi2009}
S.~Ishibashi, ``The stereo vision system for an underwater vehicle,'' in
  \emph{Proc. Ocean.}\hskip 1em plus 0.5em minus 0.4em\relax Bremen, Germany:
  IEEE, May 2009, pp. 1--6.

\bibitem{Negahdaripour2009}
S.~Negahdaripour, H.~Sekkati, and H.~Pirsiavash, ``Opti-acoustic stereo
  imaging: on system calibration and {3-D} target reconstruction,'' \emph{IEEE
  Trans. Image Process.}, vol.~18, no.~6, pp. 1203--1214, Jun. 2009.

\bibitem{Marouchos2017}
A.~Marouchos, M.~Sherlock, A.~Filisetti, and A.~Williams, ``Underwater imaging
  on self-contained tethered systems,'' in \emph{Proc. Ocean.}\hskip 1em plus
  0.5em minus 0.4em\relax Anchorage, USA: IEEE, Sep. 2017, pp. 1--5.

\bibitem{Westling2014}
F.~Westling, C.~Sun, and D.~Wang, ``A modular learning approach for fish
  counting and measurement using stereo baited remote underwater video,'' in
  \emph{Proc. Int. Conf. Digit. Image Comput.: Techn. Appl.}\hskip 1em plus
  0.5em minus 0.4em\relax Wollongong, Australia: IEEE, Nov. 2014, pp. 1--7.

\bibitem{Mulsow2014}
C.~Mulsow and H.~Maas, ``A universal approach for geometric modelling in
  underwater stereo image processing,'' in \emph{Proc. ICPR Worksh. Comput.
  Vis. Anal. Underwater Imagery}.\hskip 1em plus 0.5em minus 0.4em\relax
  Stockholm, Sweden: IEEE, Aug. 2014, pp. 49--56.

\bibitem{Trucco2006}
E.~Trucco and K.~Plakas, ``Video tracking: a concise survey,'' \emph{IEEE J.
  Ocean. Eng.}, vol.~31, no.~2, pp. 520--529, Apr. 2006.

\bibitem{Johnson-Roberson2006}
M.~Johnson-Roberson, S.~Kumar, O.~Pizarro, and S.~Willams, ``Stereoscopic
  imaging for coral segmentation and classification,'' in \emph{Proc.
  Ocean.}\hskip 1em plus 0.5em minus 0.4em\relax Boston, USA: IEEE, Sep. 2006,
  pp. 1--6.

\bibitem{Xingang2019}
W.~Xingang, ``A research review of distributed computing system,'' in
  \emph{Recent Develop. Intell. Comput., Commun. Devices}.\hskip 1em plus 0.5em
  minus 0.4em\relax Springer, Aug. 2018, pp. 357--368.

\bibitem{ASFProjectDirectory}
\BIBentryALTinterwordspacing
ASF, ``Apache project directory,'' Feb. 2019. [Online]. Available:
  \url{https://projects.apache.org/}
\BIBentrySTDinterwordspacing

\bibitem{Power2018}
B.~Power and J.~Weinman, ``Revenue growth is the primary benefit of the
  cloud,'' \emph{IEEE Cloud Comput.}, vol.~5, no.~4, pp. 89--94, Jul. 2018.

\bibitem{QuboleBDTrends2018}
\BIBentryALTinterwordspacing
Qubole, ``Big data trends and challenges,'' Oct. 2018. [Online]. Available:
  \url{https://go.qubole.com/CA---RP---2018-Big-Data-Survey_-LP.html}
\BIBentrySTDinterwordspacing

\bibitem{Zhelev2017}
S.~Zhelev and A.~Rozeva, ``Big data processing in the cloud - challenges and
  platforms,'' in \emph{Proc. 43rd Int. Conf. Appl. Math. Eng. Econo.}, vol.
  1910, no.~1.\hskip 1em plus 0.5em minus 0.4em\relax Sozopol, Bulgaria: AIP,
  Dec. 2017.

\bibitem{Gomes-Pereira2016}
{J. N. Gomes-Pereira \textit{et al.}}, ``Current and future trends in marine
  image annotation software,'' \emph{Prog. Ocean.}, vol. 149, pp. 106--120,
  Dec. 2016.

\bibitem{Barale2018}
V.~Barale, ``A supporting marine information system for maritime spatial
  planning: the {European} atlas of the seas,'' \emph{Ocean Coast. Manag.},
  Dec. 2018.

\bibitem{Liu2018}
P.~Liu, B.~Wang, Z.~Deng, and M.~Fu, ``{INS}/{DVL}/{PS} tightly coupled
  underwater navigation method with limited {DVL} measurements,'' \emph{IEEE
  Sensors J.}, vol.~18, no.~7, pp. 2994--3002, Apr. 2018.

\bibitem{ZhangT2018}
T.~Zhang, L.~Chen, and Y.~Yan, ``Underwater positioning algorithm based on
  {SINS}/{LBL} integrated system,'' \emph{{IEEE} Access}, vol.~6, pp.
  7157--7163, Jan. 2018.

\bibitem{WangB2019}
B.~Wang, J.~Zhu, Z.~Deng, and M.~Fu, ``A characteristic parameter matching
  algorithm for gravity-aided navigation of underwater vehicles,'' \emph{IEEE
  Trans. Ind. Electron.}, vol.~66, no.~2, pp. 1203--1212, Feb. 2019.

\bibitem{Chen2016}
Y.~Chen, D.~Zheng, P.~A. Miller, and J.~A. Farrell, ``Underwater inertial
  navigation with long baseline transceivers: a near-real-time approach,''
  \emph{IEEE Trans. Control Syst. Technol.}, vol.~24, no.~1, pp. 240--251, Jan.
  2016.

\bibitem{Song2018}
Z.~Song and K.~Mohseni, ``Long-term inertial navigation aided by dynamics of
  flow field features,'' \emph{IEEE J. Ocean. Eng.}, vol.~43, no.~4, pp.
  940--954, Oct. 2018.

\bibitem{Batista2015}
P.~Batista, ``{GES} long baseline navigation with unknown sound velocity and
  discrete-time range measurements,'' \emph{IEEE Trans. Control Syst.
  Technol.}, vol.~23, no.~1, pp. 219--230, Jan. 2015.

\bibitem{HanY2018}
Y.~Han, C.~Shi, D.~Sun, and J.~Zhang, ``Research on integrated navigation
  algorithm based on ranging information of single beacon,'' \emph{Appl.
  Acoust.}, vol. 131, pp. 203--209, Feb. 2018.

\bibitem{Renner2017}
C.~Renner, ``Packet-based ranging with a low-power, low-cost acoustic modem for
  micro {AUVs},'' in \emph{Proc. 11th ITG Int. Conf. Syst., Commun.
  Coding}.\hskip 1em plus 0.5em minus 0.4em\relax Hamburg, Germany: VDE, Feb.
  2017, pp. 1--6.

\bibitem{Sarmiento2018}
T.~A. Sarmiento and R.~R. Murphy, ``Insights on obstacle avoidance for small
  unmanned aerial systems from a study of flying animal behavior,''
  \emph{Robot. Auton. Syst.}, vol.~99, pp. 17--29, Jan. 2018.

\bibitem{Menna2018}
F.~Menna, P.~Agrafiotis, and A.~Georgopoulos, ``State of the art and
  applications in archaeological underwater {3D} recording and mapping,''
  \emph{J. Cultur. Heritage}, Oct. 2018.

\bibitem{Al-Kaff2018}
A.~Al-Kaff, D.~Martín, F.~García, A.~d.~l. Escalera, and J.~María~Armingol,
  ``Survey of computer vision algorithms and applications for unmanned aerial
  vehicles,'' \emph{Expert Syst. Appl.}, vol.~92, pp. 447--463, Feb. 2018.

\bibitem{Gao2016}
J.~Gao, A.~A. Proctor, Y.~Shi, and C.~Bradley, ``Hierarchical model predictive
  image-based visual servoing of underwater vehicles with adaptive neural
  network dynamic control,'' \emph{IEEE Trans. Cybern.}, vol.~46, no.~10, pp.
  2323--2334, Oct. 2016.

\bibitem{Palomeras2018}
N.~Palomeras, N.~Hurtós, M.~Carreras, and P.~Ridao, ``Autonomous mapping of
  underwater {3-D} structures: from view planning to execution,'' \emph{IEEE
  Robot. Autom. Lett.}, vol.~3, no.~3, pp. 1965--1971, Jul. 2018.

\bibitem{Norgren2018}
P.~Norgren and R.~Skjetne, ``A multibeam-based {SLAM} algorithm for iceberg
  mapping using {AUVs},'' \emph{{IEEE} Access}, vol.~6, pp. 26\,318--26\,337,
  Apr. 2018.

\bibitem{Fore2017}
M.~Føre, K.~Frank, T.~Dempster, J.~A. Alfredsen, and E.~Høy, ``Biomonitoring
  using tagged sentinel fish and acoustic telemetry in commercial salmon
  aquaculture: a feasibility study,'' \emph{Aquacult. Eng.}, vol.~78, pp.
  163--172, Aug. 2017.

\bibitem{Schmidhuber2015}
J.~Schmidhuber, ``Deep learning in neural networks: an overview,'' \emph{Neural
  Netw.}, vol.~61, pp. 85--117, Jan. 2015.

\bibitem{NohY2018}
Y.~{Noh}, B.~{Zhang}, and D.~D. {Lee}, ``Generative local metric learning for
  nearest neighbor classification,'' \emph{IEEE Trans. Pattern Anal. Mach.
  Intell.}, vol.~40, no.~1, pp. 106--118, Jan. 2018.

\bibitem{AdeliE2019}
E.~{Adeli}, K.~{Thung}, L.~{An}, G.~{Wu}, F.~{Shi}, T.~{Wang}, and D.~{Shen},
  ``Semi-supervised discriminative classification robust to sample-outliers and
  feature-noises,'' \emph{IEEE Trans. Pattern Anal. Mach. Intell.}, vol.~41,
  no.~2, pp. 515--522, Feb. 2019.

\bibitem{Hinton2006}
G.~E. Hinton, S.~Osindero, and Y.-W. Teh, ``A fast learning algorithm for deep
  belief nets,'' \emph{Neural Comput.}, vol.~18, no.~7, pp. 1527--1554, Jul.
  2006.

\bibitem{Lecun1995}
Y.~LeCun, Y.~Bengio \emph{et~al.}, ``Convolutional networks for images, speech,
  and time series,'' \emph{Handbook Brain Theory Neural Netw.}, vol. 3361,
  no.~10, pp. 255--258, Apr. 1995.

\bibitem{KimK2018}
K.~Kim and H.~Myung, ``Autoencoder-combined generative adversarial networks for
  synthetic image data generation and detection of jellyfish swarm,''
  \emph{{IEEE} Access}, vol.~6, pp. 54\,207--54\,214, Sep. 2018.

\bibitem{Socher2013}
R.~Socher, A.~Perelygin, J.~Wu, J.~Chuang, C.~D. Manning, A.~Ng, and C.~Potts,
  ``Recursive deep models for semantic compositionality over a sentiment
  treebank,'' in \emph{Proc. Empirical Methods Natural Language Process.}\hskip
  1em plus 0.5em minus 0.4em\relax Seattle, USA: Association for Computational
  Linguistics, Oct. 2013, pp. 1631--1642.

\bibitem{Hochreiter1997}
S.~Hochreiter and J.~Schmidhuber, ``Long short-term memory,'' \emph{Neural
  Comput.}, vol.~9, no.~8, pp. 1735--1780, Nov. 1997.

\bibitem{Ravi2017}
D.~Ravì, C.~Wong, F.~Deligianni, M.~Berthelot, J.~Andreu-Perez, B.~Lo, and
  G.~Yang, ``Deep learning for health informatics,'' \emph{IEEE J. Biomed.
  Health Inform.}, vol.~21, no.~1, pp. 4--21, Jan. 2017.

\bibitem{Xiao2019}
L.~Xiao, G.~Sheng, X.~Wan, W.~Su, and P.~Cheng, ``Learning-based {PHY}-layer
  authentication for underwater sensor networks,'' \emph{IEEE Commun. Lett.},
  vol.~23, no.~1, pp. 60--63, Jan. 2019.

\bibitem{Aggarwal2015}
C.~C. Aggarwal, \emph{Data mining: the textbook}.\hskip 1em plus 0.5em minus
  0.4em\relax Springer, 2015.

\bibitem{Gupta2019}
S.~Gupta and S.~Roy, ``Medav filter---filter for removal of image noise with
  the combination of median and average filters,'' in \emph{Proc. Recent Trends
  Signal Image Process.}\hskip 1em plus 0.5em minus 0.4em\relax Singapore:
  Springer, May 2018, pp. 11--19.

\bibitem{James2017}
G.~James, D.~Witten, T.~Hastie, and R.~Tibshirani, \emph{An introduction to
  statistical learning}, 7th~ed.\hskip 1em plus 0.5em minus 0.4em\relax
  Springer, Jun. 2017.

\bibitem{WangM2017}
M.~Wang, X.~Liu, Y.~Gao, X.~Ma, and N.~Q. Soomro, ``Superpixel segmentation: a
  benchmark,'' \emph{Signal Process.: Image Commun.}, vol.~56, pp. 28--39, Aug.
  2017.

\bibitem{LiC2018}
C.~Li, J.~Guo, and C.~Guo, ``Emerging from water: underwater image color
  correction based on weakly supervised color transfer,'' \emph{IEEE Signal
  Process. Lett.}, vol.~25, no.~3, pp. 323--327, Mar. 2018.

\bibitem{PengY2017}
Y.~Peng and P.~C. Cosman, ``Underwater image restoration based on image
  blurriness and light absorption,'' \emph{IEEE Trans. Image Process.},
  vol.~26, no.~4, pp. 1579--1594, Apr. 2017.

\bibitem{WangN2017}
N.~Wang, H.~Zheng, and B.~Zheng, ``Underwater image restoration via maximum
  attenuation identification,'' \emph{{IEEE} Access}, vol.~5, pp.
  18\,941--18\,952, Sep. 2017.

\bibitem{LiM2018}
M.~Li, J.~Liu, W.~Yang, X.~Sun, and Z.~Guo, ``Structure-revealing low-light
  image enhancement via robust retinex model,'' \emph{IEEE Trans. Image
  Process.}, vol.~27, no.~6, pp. 2828--2841, Jun. 2018.

\bibitem{WangYan2018}
Y.~Wang, N.~Li, Z.~Li, Z.~Gu, H.~Zheng, B.~Zheng, and M.~Sun, ``An
  imaging-inspired no-reference underwater color image quality assessment
  metric,'' \emph{Comput. Electr. Eng.}, vol.~70, pp. 904--913, Aug. 2018.

\bibitem{Panetta2016}
K.~Panetta, C.~Gao, and S.~Agaian, ``Human-visual-system-inspired underwater
  image quality measures,'' \emph{IEEE J. Ocean. Eng.}, vol.~41, no.~3, pp.
  541--551, Jul. 2016.

\bibitem{Yang2015}
M.~Yang and A.~Sowmya, ``An underwater color image quality evaluation metric,''
  \emph{IEEE Trans. Image Process.}, vol.~24, no.~12, pp. 6062--6071, Dec.
  2015.

\bibitem{Ancuti2018}
C.~O. Ancuti, C.~Ancuti, C.~D. Vleeschouwer, and P.~Bekaert, ``Color balance
  and fusion for underwater image enhancement,'' \emph{IEEE Trans. Image
  Process.}, vol.~27, no.~1, pp. 379--393, Jan. 2018.

\bibitem{Rife2003}
J.~Rife and S.~M. Rock, ``Segmentation methods for visual tracking of
  deep-ocean jellyfish using a conventional camera,'' \emph{IEEE J. Ocean.
  Eng.}, vol.~28, no.~4, pp. 595--608, Oct. 2003.

\bibitem{Kocak1999}
D.~M. Kocak, N.~d.~V. Lobo, and E.~A. Widder, ``Computer vision techniques for
  quantifying, tracking, and identifying bioluminescent plankton,'' \emph{IEEE
  J. Ocean. Eng.}, vol.~24, no.~1, pp. 81--95, Jan. 1999.

\bibitem{Schoening2015}
T.~Schoening, ``Automated detection in benthic images for megafauna
  classification and marine resource exploration: supervised and unsupervised
  methods for classification and regression tasks in benthic images with
  efficient integration of expert knowledge,'' Ph.D. dissertation, Der
  Technischen Fakultät der Universität Bielefeld, 2015.

\bibitem{Chuang2011}
M.~C. Chuang, J.~N. Hwang, K.~Williams, and R.~Towler, ``Automatic fish
  segmentation via double local thresholding for trawl-based underwater camera
  systems,'' in \emph{Proc. 18th Int. Conf. Image Process.}\hskip 1em plus
  0.5em minus 0.4em\relax Brussels, Belgium: IEEE, Sep. 2011, pp. 3145--3148.

\bibitem{Schoening2012}
T.~Schoening, M.~Bergmann, J.~Ontrup, J.~Taylor, J.~Dannheim, J.~Gutt,
  A.~Purser, and T.~W. Nattkemper, ``Semi-automated image analysis for the
  assessment of megafaunal densities at the arctic deep-sea observatory
  {HAUSGARTEN},'' \emph{{PLoS} One}, vol.~7, no.~6, p. 38179, Jun. 2012.

\bibitem{Tong2004}
L.~Tong, K.~Kramer, D.~B. Goldgof, L.~O. Hall, S.~Samson, A.~Remsen, and
  T.~Hopkins, ``Recognizing plankton images from the shadow image particle
  profiling evaluation recorder,'' \emph{IEEE Trans. Syst., Man, Cybern., Part
  B (Cybern.)}, vol.~34, no.~4, pp. 1753--1762, Aug. 2004.

\bibitem{Hu2006}
Q.~Hu and C.~Davis, ``Accurate automatic quantification of taxa-specific
  plankton abundance using dual classification with correction,'' \emph{Marine
  Ecolog. Prog. Series}, vol. 306, pp. 51--61, Jan. 2006.

\bibitem{Purser2009}
A.~Purser, M.~Bergmann, T.~Lundälv, and T.~Nattkemper, ``Use of
  machine-learning algorithms for the automated detection of cold-water coral
  habitats: a pilot study,'' \emph{Marine Ecolog. Prog. Series}, vol. 397, p.
  241, Dec. 2009.

\bibitem{LiX2016}
X.~Li, J.~Song, F.~Zhang, X.~Ouyang, and S.~U. Khan, ``{MapReduce-based} fast
  fuzzy c-means algorithm for large-scale underwater image segmentation,''
  \emph{Future Generat. Comput. Syst.}, vol.~65, pp. 90--101, Dec. 2016.

\bibitem{Peizhe2011}
Z.~Peizhe and L.~Canping, ``Region-based color image segmentation of fishes
  with complex background in water,'' in \emph{Proc. Int. Conf. Comput. Sci.
  Autom. Eng.}, vol.~1.\hskip 1em plus 0.5em minus 0.4em\relax Shanghai, China:
  IEEE, Jun. 2011, pp. 596--600.

\bibitem{Benfield2007}
S.~L. Benfield, H.~M. Guzman, J.~M. Mair, and J.~A.~T. Young, ``Mapping the
  distribution of coral reefs and associated sublittoral habitats in {Pacific
  Panama}: a comparison of optical satellite sensors and classification
  methodologies,'' \emph{Int. J. Remote Sens.}, vol.~28, no.~22, pp.
  5047--5070, Oct. 2007.

\bibitem{Chuang2016}
M.~C. Chuang, J.~N. Hwang, and K.~Williams, ``A feature learning and object
  recognition framework for underwater fish images,'' \emph{IEEE Trans. Image
  Process.}, vol.~25, no.~4, pp. 1862--1872, Apr. 2016.

\bibitem{HuangP2015}
P.~X. Huang, B.~J. Boom, and R.~B. Fisher, ``Hierarchical classification with
  reject option for live fish recognition,'' \emph{Mach. Vis. Appl.}, vol.~26,
  no.~1, pp. 89--102, Jan. 2015.

\bibitem{Chambah2004}
M.~Chambah, D.~Semani, A.~Renouf, P.~Coutellemont, and A.~Rizzi, ``Underwater
  color constancy: enhancement of automatic live fish recognition,'' in
  \emph{Proc. 16th Annual Symp. Electron. Imaging}.\hskip 1em plus 0.5em minus
  0.4em\relax San Jose, USA: International Society for Optics and Photonics,
  Dec. 2003, pp. 157--168.

\bibitem{Zhang2016}
D.~Zhang, G.~Kopanas, C.~Desai, S.~Chai, and M.~Piacentino, ``Unsupervised
  underwater fish detection fusing flow and objectiveness,'' in \emph{Proc.
  IEEE Winter Appl. Comput. Vis. Worksh.}\hskip 1em plus 0.5em minus
  0.4em\relax Lake Placid, USA: IEEE, Mar. 2016, pp. 1--7.

\bibitem{Pons2010}
S.~Pons, J.~Piera, and J.~Aguzzi, ``Video-image processing applied to the
  analysis of the behaviour of deep-water lobsters ({Nephrops Norvegicus}),''
  in \emph{Proc. Ocean.}\hskip 1em plus 0.5em minus 0.4em\relax Sydney,
  Australia: IEEE, May 2010, pp. 1--4.

\bibitem{Flake2004}
G.~W. Flake, R.~E. Tarjan, and K.~Tsioutsiouliklis, ``Graph clustering and
  minimum cut trees,'' \emph{Internet Math.}, vol.~1, no.~4, pp. 385--408,
  2004.

\bibitem{Ludtke2012}
A.~Lüdtke, K.~Jerosch, O.~Herzog, and M.~Schlüter, ``Development of a machine
  learning technique for automatic analysis of seafloor image data: case
  example, {Pogonophora} coverage at mud volcanoes,'' \emph{Comput. Geosci.},
  vol.~39, pp. 120--128, Feb. 2012.

\bibitem{Pugh2014}
M.~Pugh, B.~Tiddeman, H.~Dee, and P.~Hughes, ``Towards automated classification
  of seabed substrates in underwater video,'' in \emph{Proc. ICPR Worksh.
  Comput. Vis. Anal. Underwater Imagery}.\hskip 1em plus 0.5em minus
  0.4em\relax Stockholm, Sweden: IEEE, Aug. 2014, pp. 9--16.

\bibitem{Spampinato2008}
C.~Spampinato, Y.-H. Chen-Burger, G.~Nadarajan, and R.~Fisher, ``Detecting,
  tracking and counting fish in low quality unconstrained underwater videos,''
  in \emph{Proc. 3rd Int. Conf. Comput. Vis. Theory Appl.}, vol.~2, Madeira,
  Portugal, 2008, pp. 514--519.

\bibitem{Fouad2013}
M.~M.~M. Fouad, H.~M. Zawbaa, N.~El-Bendary, and A.~E. Hassanien, ``Automatic
  {Nile} {Tilapia} fish classification approach using machine learning
  techniques,'' in \emph{Proc. 13th Int. Conf. Hybrid Intell. Syst.}\hskip 1em
  plus 0.5em minus 0.4em\relax Gammarth, Tunisia: IEEE, Dec. 2013, pp.
  173--178.

\bibitem{Bonin-Font2016}
F.~Bonin-Font, M.~M. Campos, and G.~O. Codina, ``Towards visual detection,
  mapping and quantification of posidonia oceanica using a lightweight auv,''
  \emph{IFAC-PapersOnLine}, vol.~49, no.~23, pp. 500--505, 2016.

\bibitem{Chuang2014}
M.~C. Chuang, J.~N. Hwang, F.~F. Kuo, M.~K. Shan, and K.~Williams,
  ``Recognizing live fish species by hierarchical partial classification based
  on the exponential benefit,'' in \emph{Proc. Int. Conf. Image Process.}\hskip
  1em plus 0.5em minus 0.4em\relax Paris, France: IEEE, Oct. 2014, pp.
  5232--5236.

\bibitem{Elawady2015}
M.~Elawady, ``Sparse coral classification using deep convolutional neural
  networks,'' \emph{Comput. Research Repos.}, Nov. 2015.

\bibitem{Ambika2019}
P.~R. Ambika and A.~B. Malakreddy, ``Review of existing research contribution
  toward dimensional reduction methods in high-dimensional data,'' in
  \emph{Proc. Int. Conf. Comput. Netw. Commun. Technol.}, Seoul, South Korea,
  Jun. 2018, pp. 21--23.

\bibitem{Gepshtein2019}
S.~Gepshtein and Y.~Keller, ``Iterative spectral independent component
  analysis,'' \emph{Signal Process.}, vol. 155, pp. 368--376, Feb. 2019.

\bibitem{Mahdianpari2018}
M.~Mahdianpari, B.~Salehi, F.~Mohammadimanesh, B.~Brisco, S.~Mahdavi, M.~Amani,
  and J.~E. Granger, ``Fisher linear discriminant analysis of coherency matrix
  for wetland classification using {PolSAR} imagery,'' \emph{Remote Sens.
  Env.}, vol. 206, pp. 300--317, Mar. 2018.

\bibitem{Tang2019}
X.~Tang, Y.~Dai, and Y.~Xiang, ``Feature selection based on feature
  interactions with application to text categorization,'' \emph{Expert Syst.
  Appl.}, vol. 120, pp. 207--216, Apr. 2019.

\bibitem{Zhao2018}
R.~Zhao and K.~Mao, ``Fuzzy bag-of-words model for document representation,''
  \emph{IEEE Trans. Fuzzy Syst.}, vol.~26, no.~2, pp. 794--804, Apr. 2018.

\bibitem{Ceze2011}
L.~H. Ceze, \emph{Encyclopedia of parallel computing}.\hskip 1em plus 0.5em
  minus 0.4em\relax Springer, 2011, ch. Shared-Memory Multiprocessors, pp.
  1810--1812.

\bibitem{LeeJ2019}
J.~Lee, C.~Kim, S.~Kang, D.~Shin, S.~Kim, and H.~Yoo, ``{UNPU}: an
  energy-efficient deep neural network accelerator with fully variable weight
  bit precision,'' \emph{IEEE J. Solid-State Circuits}, vol.~54, no.~1, pp.
  173--185, Jan. 2019.

\bibitem{Venieris2019}
S.~I. Venieris and C.~Bouganis, ``{fpgaConvNet}: mapping regular and irregular
  convolutional neural networks on {FPGAs},'' \emph{IEEE Trans. Neural Netw.
  Learn. Syst.}, vol.~30, no.~2, pp. 326--342, Feb. 2019.

\bibitem{Manderson2018}
T.~Manderson and G.~Dudek, ``{GPU}-assisted learning on an autonomous marine
  robot for vision-based navigation and image understanding,'' in \emph{Proc.
  Ocean.}\hskip 1em plus 0.5em minus 0.4em\relax Charleston, USA: IEEE, Oct.
  2018, pp. 1--6.

\bibitem{Coulouris2011}
G.~F. Coulouris, J.~Dollimore, and T.~Kindberg, \emph{Distributed systems:
  concepts and design}, 5th~ed.\hskip 1em plus 0.5em minus 0.4em\relax Pearson
  Education, 2011.

\bibitem{XiaD2016}
D.~Xia, H.~Li, B.~Wang, Y.~Li, and Z.~Zhang, ``A map reduce-based nearest
  neighbor approach for big-data-driven traffic flow prediction,'' \emph{{IEEE}
  Access}, vol.~4, pp. 2920--2934, Jun. 2016.

\bibitem{XuZ2019}
Z.~Xu, B.~Liu, S.~Zhe, H.~Bai, Z.~Wang, and J.~Neville, ``Variational random
  function model for network modeling,'' \emph{IEEE Trans. Neural Netw. Learn.
  Syst.}, vol.~30, no.~1, pp. 318--324, Jan. 2019.

\bibitem{MingY2018}
Y.~Ming, E.~Zhu, M.~Wang, Y.~Ye, X.~Liu, and J.~Yin, ``{DMP-ELMs}: data and
  model parallel extreme learning machines for large-scale learning tasks,''
  \emph{Neurocomp.}, vol. 320, pp. 85--97, Dec. 2018.

\bibitem{Garcia2012}
S.~Garcia, D.~Jeon, C.~Louie, and M.~B. Taylor, ``The {Kremlin} oracle for
  sequential code parallelization,'' \emph{{IEEE} Micro}, vol.~32, no.~4, pp.
  42--53, Jul. 2012.

\bibitem{Schwartz2019}
E.~Schwartz, R.~Giryes, and A.~M. Bronstein, ``{DeepISP}: toward learning an
  end-to-end image processing pipeline,'' \emph{IEEE Trans. Image Process.},
  vol.~28, no.~2, pp. 912--923, Feb. 2019.

\bibitem{Modasshir2018}
M.~Modasshir, A.~Q. Li, and I.~Rekleitis, ``Deep neural networks: a comparison
  on different computing platforms,'' in \emph{Proc. 15th Conf. Comput. Robot
  Vis.}\hskip 1em plus 0.5em minus 0.4em\relax Toronto, Canada: IEEE, May 2018,
  pp. 383--389.

\bibitem{KimH2019}
H.~{Kim}, J.~{Koo}, D.~{Kim}, B.~{Park}, Y.~{Jo}, H.~{Myung}, and D.~{Lee},
  ``Vision-based real-time obstacle segmentation algorithm for autonomous
  surface vehicle,'' \emph{{IEEE} Access}, vol.~7, pp. 179\,420--179\,428, Dec.
  2019.

\bibitem{Pallayil2016}
V.~Pallayil, M.~Chitre, S.~Kuselan, A.~Raichur, M.~Ignatius, and J.~R. Potter,
  ``Development of a second-generation underwater acoustic ambient noise
  imaging camera,'' \emph{IEEE J. Ocean. Eng.}, vol.~41, no.~1, pp. 175--189,
  Jan. 2016.

\bibitem{Karabchevsky2011}
S.~Karabchevsky, D.~Kahana, O.~Ben-Harush, and H.~Guterman, ``{FPGA}-based
  adaptive speckle suppression filter for underwater imaging sonar,''
  \emph{IEEE J. Ocean. Eng.}, vol.~36, no.~4, pp. 646--657, Oct. 2011.

\bibitem{Szegedy2017}
C.~Szegedy, S.~Loffe, V.~Vanhoucke, and A.~A. Alemi, ``Inception-v4,
  inception-resnet and the impact of residual connections on learning,'' in
  \emph{Proc. 31st AAAI Conf. Artif. Intell.}\hskip 1em plus 0.5em minus
  0.4em\relax San Francisco, USA: AAAI, Feb. 2017.

\bibitem{HeK2016}
K.~He, X.~Zhang, S.~Ren, and J.~Sun, ``Deep residual learning for image
  recognition,'' in \emph{Proc. IEEE Conf. Comput. Vis. Pattern Recogn.}\hskip
  1em plus 0.5em minus 0.4em\relax Las Vegas, USA: IEEE, 2016, pp. 770--778.

\bibitem{Canziani2017}
A.~Canziani, A.~Paszke, and E.~Culurciello, ``An analysis of deep neural
  network models for practical applications,'' \emph{Comput. Research Repos.},
  Apr. 2017.

\bibitem{ZagoruykoGIT2016}
\BIBentryALTinterwordspacing
S.~Zagoruyko, ``{ImageNet-validation.torch},'' Jun. 2016. [Online]. Available:
  \url{https://github.com/szagoruyko/imagenet-validation.torch}
\BIBentrySTDinterwordspacing

\bibitem{Krizhevsky2012}
A.~Krizhevsky, I.~Sutskever, and G.~E. Hinton, ``{ImageNet} classification with
  deep convolutional neural networks,'' in \emph{Adv. Neural Inf. Process.
  Syst.}\hskip 1em plus 0.5em minus 0.4em\relax Curran Associates, Inc., 2012,
  pp. 1097--1105.

\bibitem{LinM2013}
M.~Lin, Q.~Chen, and S.~Yan, ``Network in network,'' \emph{ArXiv}, Dec. 2013.

\bibitem{Simonyan2014}
K.~Simonyan and A.~Zisserman, ``Very deep convolutional networks for
  large-scale image recognition,'' \emph{ArXiv}, Apr. 2015.

\bibitem{PaszkeGIT2016}
\BIBentryALTinterwordspacing
A.~Paszke and A.~Jackson, ``{Torch-opCounter},'' 2016. [Online]. Available:
  \url{https://github.com/apaszke/torch-opCounter}
\BIBentrySTDinterwordspacing

\bibitem{Rumelhart1986}
D.~E. Rumelhart, G.~E. Hinton, and R.~J. Williams, ``Learning representations
  by back-propagating errors,'' \emph{{Nature}}, vol. 323, p. 533, Oct. 1986.

\bibitem{Adler2017}
J.~Adler and O.~Öktem, ``Solving ill-posed inverse problems using iterative
  deep neural networks,'' \emph{Inver. Probl.}, vol.~33, no.~12, p. 124007, May
  2017.

\bibitem{LiY2016}
Y.~Li, H.~Lu, J.~Li, X.~Li, Y.~Li, and S.~Serikawa, ``Underwater image
  de-scattering and classification by deep neural network,'' \emph{Comput.
  Electr. Eng.}, vol.~54, pp. 68--77, Aug. 2016.

\bibitem{ZhangW2020}
W.~{Zhang}, K.~{Ma}, J.~{Yan}, D.~{Deng}, and Z.~{Wang}, ``Blind image quality
  assessment using a deep bilinear convolutional neural network,'' \emph{IEEE
  Trans. Circuits Syst. Video Technol.}, vol.~30, no.~1, pp. 36--47, Jan. 2020.

\bibitem{YanY2020}
Y.~{Yan}, H.~{Hao}, B.~{Xu}, J.~{Zhao}, and F.~{Shen}, ``Image clustering via
  deep embedded dimensionality reduction and probability-based triplet loss,''
  \emph{IEEE Trans. Image Process.}, vol.~29, pp. 5652--5661, Apr. 2020.

\bibitem{Qin2016}
H.~Qin, X.~Li, J.~Liang, Y.~Peng, and C.~Zhang, ``{DeepFish}: accurate
  underwater live fish recognition with a deep architecture,''
  \emph{Neurocomp.}, vol. 187, pp. 49--58, Apr. 2016.

\bibitem{Wahidin2015}
N.~Wahidin, V.~P. Siregar, B.~Nababan, I.~Jaya, and S.~Wouthuyzen,
  ``Object-based image analysis for coral reef benthic habitat mapping with
  several classification algorithms,'' \emph{Proc. Env. Sci.}, vol.~24, pp.
  222--227, Dec. 2015.

\bibitem{HuoG2020}
G.~{Huo}, Z.~{Wu}, and J.~{Li}, ``Underwater object classification in sidescan
  sonar images using deep transfer learning and semisynthetic training data,''
  \emph{{IEEE} Access}, vol.~8, pp. 47\,407--47\,418, Mar. 2020.

\bibitem{JinL2019}
L.~{Jin}, H.~{Liang}, and C.~{Yang}, ``Accurate underwater {ATR} in
  forward-looking sonar imagery using deep convolutional neural networks,''
  \emph{{IEEE} Access}, vol.~7, pp. 125\,522--125\,531, Sep. 2019.

\bibitem{Perry2004}
S.~W. Perry and L.~Guan, ``A recurrent neural network for detecting objects in
  sequences of sector-scan sonar images,'' \emph{IEEE J. Ocean. Eng.}, vol.~29,
  no.~3, pp. 857--871, Jul. 2004.

\bibitem{Singh2004}
H.~Singh, J.~Howland, and O.~Pizarro, ``Advances in large-area photomosaicking
  underwater,'' \emph{IEEE J. Ocean. Eng.}, vol.~29, no.~3, pp. 872--886, Jul.
  2004.

\bibitem{Nian2014}
R.~Nian, B.~He, B.~Zheng, M.~van Heeswijk, Q.~Yu, Y.~Miche, and A.~Lendasse,
  ``Extreme learning machine towards dynamic model hypothesis in fish ethology
  research,'' \emph{Neurocomp.}, vol. 128, pp. 273--284, Mar. 2014.

\bibitem{LuoS2015}
S.~Luo, X.~Li, D.~Wang, J.~Li, and C.~Sun, ``Automatic fish recognition and
  counting in video footage of fishery operations,'' in \emph{Proc. Int. Conf.
  Comput. Intell. Commun. Netw.}\hskip 1em plus 0.5em minus 0.4em\relax
  Jabalpur, India: IEEE, Dec. 2015, pp. 296--299.

\bibitem{ZhangK2020}
K.~{Zhang}, X.~{Geng}, and X.~{Yan}, ``Prediction of {3-D} ocean temperature by
  multilayer convolutional {LSTM},'' \emph{IEEE Geosci. Remote Sens. Lett.},
  pp. 1--5, Jan. 2020.

\bibitem{WangJ2009}
J.~Wang, G.~Wu, L.~Wan, Y.~Sun, and D.~Jiang, ``Recurrent neural network
  applied to fault diagnosis of underwater robots,'' in \emph{Proc. Int. Conf.
  Intell. Comput. Intell. Syst.}, vol.~1.\hskip 1em plus 0.5em minus
  0.4em\relax Shanghai, China: IEEE, Nov. 2009, pp. 593--598.

\bibitem{Shan2013}
Y.~Shan, Z.~Yan, and J.~Wang, ``Model predictive control of underwater gliders
  based on a one-layer recurrent neural network,'' in \emph{Proc. 6th Int.
  Conf. Adv. Comput. Intell.}\hskip 1em plus 0.5em minus 0.4em\relax Hangzhou,
  China: IEEE, Oct. 2013, pp. 328--333.

\bibitem{ZhangY2019}
Y.~{Zhang}, P.~J. {Thorburn}, W.~{Xiang}, and P.~{Fitch}, ``{SSIM}—a deep
  learning approach for recovering missing time series sensor data,''
  \emph{IEEE Internet Things J.}, vol.~6, no.~4, pp. 6618--6628, Aug. 2019.

\bibitem{GouY2020}
Y.~{Gou}, T.~{Zhang}, J.~{Liu}, L.~{Wei}, and J.~{Cui}, ``{DeepOcean}: a
  general deep learning framework for spatio-temporal ocean sensing data
  prediction,'' \emph{{IEEE} Access}, vol.~8, pp. 79\,192--79\,202, Apr. 2020.

\bibitem{YangY2018}
Y.~{Yang}, J.~{Dong}, X.~{Sun}, E.~{Lima}, Q.~{Mu}, and X.~{Wang}, ``A
  {CFCC-LSTM} model for sea surface temperature prediction,'' \emph{IEEE
  Geosci. Remote Sens. Lett.}, vol.~15, no.~2, pp. 207--211, Dec. 2018.

\bibitem{XieJSST2019}
J.~{Xie}, J.~{Zhang}, J.~{Yu}, and L.~{Xu}, ``An adaptive scale sea surface
  temperature predicting method based on deep learning with attention
  mechanism,'' \emph{IEEE Geosci. Remote Sens. Lett.}, vol.~17, no.~5, pp.
  740--744, Aug. 2020.

\bibitem{KimH2017}
H.~Kim and J.~Y. Kim, ``Environmental sound event detection in wireless
  acoustic sensor networks for home telemonitoring,'' \emph{China Commun.},
  vol.~14, no.~9, pp. 1--10, Sep. 2017.

\bibitem{Zyner2018}
A.~Zyner, S.~Worrall, and E.~Nebot, ``A recurrent neural network solution for
  predicting driver intention at unsignalized intersections,'' \emph{IEEE
  Robot. Autom. Lett.}, vol.~3, no.~3, pp. 1759--1764, Jul. 2018.

\bibitem{Marzano2007}
F.~S. Marzano, G.~Rivolta, E.~Coppola, B.~Tomassetti, and M.~Verdecchia,
  ``Rainfall nowcasting from multisatellite passive-sensor images using a
  recurrent neural network,'' \emph{IEEE Trans. Geosci. Remote Sens.}, vol.~45,
  no.~11, pp. 3800--3812, Nov. 2007.

\bibitem{PrasA2007}
A.~{Pras}, J.~{Schonwalder}, M.~{Burgess}, O.~{Festor}, G.~M. {Perez},
  R.~{Stadler}, and B.~{Stiller}, ``Key research challenges in network
  management,'' \emph{IEEE Commun. Mag.}, vol.~45, no.~10, pp. 104--110, Oct.
  2007.

\bibitem{UrunovH2017}
H.~Urunov, S.~Y. Shin, S.~H. Park, and K.~Yi, ``{U-SNMP} for the internet of
  underwater things,'' \emph{Int. J. Control Autom.}, vol.~10, pp. 199--216,
  Jan. 2017.

\bibitem{WongA2004}
A.~K.~Y. {Wong}, {An Chi Chen}, N.~{Paramesh}, and P.~{Rav}, ``Ontology mapping
  for network management systems,'' in \emph{Proc. Netw. Operat. Manag. Symp.},
  vol.~1.\hskip 1em plus 0.5em minus 0.4em\relax Seoul, South Korea: IEEE/IFIP,
  Apr. 2004, pp. 885--886.

\bibitem{MukhtarH2008}
H.~Mukhtar, K.~Kang-Myo, S.~A. Chaudhry, A.~H. Akbar, K.~Ki-Hyung, and S.-W.
  Yoo, ``{LNMP}- management architecture for {IPv6} based low-power wireless
  personal area networks ({6LoWPAN}),'' in \emph{Proc. Netw. Operat. Manag.
  Symp.}\hskip 1em plus 0.5em minus 0.4em\relax Salvador, Brazil: IEEE, Aug.
  2008, pp. 417--424.

\bibitem{SharmaS2014}
S.~{Sharma}, D.~{Staessens}, D.~{Colle}, D.~{Palma}, J.~{Gonçalves},
  R.~{Figueiredo}, D.~{Morris}, M.~{Pickavet}, and P.~{Demeester},
  ``Implementing quality of service for the software defined networking enabled
  future internet,'' in \emph{Proc. 3rd Europ. Worksh. Softw. Defin.
  Netw.}\hskip 1em plus 0.5em minus 0.4em\relax London, UK: IEEE, Sep. 2014,
  pp. 49--54.

\bibitem{Bito2017}
J.~Bito, R.~Bahr, J.~G. Hester, S.~A. Nauroze, A.~Georgiadis, and M.~M.
  Tentzeris, ``A novel solar and electromagnetic energy harvesting system with
  a {3-D} printed package for energy efficient internet-of-things wireless
  sensors,'' \emph{IEEE Trans. Microw. Theory Techn.}, vol.~65, no.~5, pp.
  1831--1842, May 2017.

\bibitem{Alippi2011}
C.~Alippi, R.~Camplani, C.~Galperti, and M.~Roveri, ``A robust, adaptive,
  solar-powered {WSN} framework for aquatic environmental monitoring,''
  \emph{IEEE Sensors J.}, vol.~11, no.~1, pp. 45--55, Jan. 2011.

\bibitem{Joshi2011}
K.~B. Joshi, J.~H. Costello, and S.~Priya, ``Estimation of solar energy
  harvested for autonomous jellyfish vehicles ({AJVs}),'' \emph{IEEE J. Ocean.
  Eng.}, vol.~36, no.~4, pp. 539--551, Oct. 2011.

\bibitem{Jenkins2014}
P.~P. Jenkins, S.~Messenger, K.~M. Trautz, S.~I. Maximenko, D.~Goldstein,
  D.~Scheiman, R.~Hoheisel, and R.~J. Walters, ``High-bandgap solar cells for
  underwater photovoltaic applications,'' \emph{IEEE J. Photovolt.}, vol.~4,
  no.~1, pp. 202--207, Jan. 2014.

\bibitem{Toma2015}
D.~M. Toma, J.~del Rio, M.~Carbonell-Ventura, and J.~M. Masalles, ``Underwater
  energy harvesting system based on plucked-driven piezoelectrics,'' in
  \emph{Proc. Ocean.}\hskip 1em plus 0.5em minus 0.4em\relax Genova, Italy:
  IEEE, May 2015, pp. 1--5.

\bibitem{Qureshi2017}
F.~U. Qureshi, A.~Muhtaroğlu, and K.~Tuncay, ``Near-optimal design of scalable
  energy harvester for underwater pipeline monitoring applications with
  consideration of impact to pipeline performance,'' \emph{IEEE Sensors J.},
  vol.~17, no.~7, pp. 1981--1991, Apr. 2017.

\bibitem{Bereketli2012}
A.~Bereketli and S.~Bilgen, ``Remotely powered underwater acoustic sensor
  networks,'' \emph{IEEE Sensors J.}, vol.~12, no.~12, pp. 3467--3472, Dec.
  2012.

\bibitem{KanT2018}
T.~Kan, Y.~Zhang, Z.~Yan, P.~P. Mercier, and C.~C. Mi, ``A rotation-resilient
  wireless charging system for lightweight autonomous underwater vehicles,''
  \emph{IEEE Trans. Veh Technol.}, vol.~67, no.~8, pp. 6935--6942, Aug. 2018.

\bibitem{Jing2017}
L.~Jing, C.~He, J.~Huang, and Z.~Ding, ``Energy management and power allocation
  for underwater acoustic sensor network,'' \emph{IEEE Sensors J.}, vol.~17,
  no.~19, pp. 6451--6462, Oct. 2017.

\bibitem{WuZ2020}
Z.~{Wu}, J.~{Yu}, J.~{Yuan}, M.~{Tan}, and S.~{Qi}, ``Gliding motion regulation
  of a robotic dolphin based on a controllable fluke,'' \emph{IEEE Trans. Ind.
  Electron.}, vol.~67, no.~4, pp. 2945--2953, Apr. 2020.

\bibitem{Hem2012}
J.~D. Hem, \emph{Water analysis. inorganic species}.\hskip 1em plus 0.5em minus
  0.4em\relax Academic Press, 2012, vol.~1, no.~1, ch. Conductance: a
  collective measure of dissolved ions, pp. 137--161.

\bibitem{Islam2018}
M.~T. Islam, M.~N. Rahman, M.~S.~J. Singh, and M.~Samsuzzaman, ``Detection of
  salt and sugar contents in water on the basis of dielectric properties using
  microstrip antenna-based sensor,'' \emph{{IEEE} Access}, vol.~6, pp.
  4118--4126, Jan. 2018.

\bibitem{Vorathin2019}
E.~Vorathin, Z.~M. Hafizi, A.~M. Aizzuddin, M.~K.~A. Zaini, and K.~S. Lim, ``A
  novel temperature-insensitive hydrostatic liquid-level sensor using chirped
  {FBG},'' \emph{IEEE Sensors J.}, vol.~19, no.~1, pp. 157--162, Jan. 2019.

\bibitem{WangY2018}
Y.~Wang, S.~M. S.~M. Rajib, C.~Collins, and B.~Grieve, ``Low-cost turbidity
  sensor for low-power wireless monitoring of fresh-water courses,'' \emph{IEEE
  Sensors J.}, vol.~18, no.~11, pp. 4689--4696, Jun. 2018.

\bibitem{Kirkey2018}
W.~D. Kirkey, J.~S. Bonner, and C.~B. Fuller, ``Low-cost submersible turbidity
  sensors using low-frequency source light modulation,'' \emph{IEEE Sensors
  J.}, vol.~18, no.~22, pp. 9151--9162, Nov. 2018.

\bibitem{LiuC2019}
C.~Liu, Y.~Li, and M.~Xu, ``An integrated detection and location model for
  leakages in liquid pipelines,'' \emph{J. Petrol. Sci. Eng.}, vol. 175, pp.
  852--867, Apr. 2019.

\bibitem{WangP2018}
P.~Wang, X.~Tian, T.~Peng, and Y.~Luo, ``A review of the state-of-the-art
  developments in the field monitoring of offshore structures,'' \emph{Ocean
  Eng.}, vol. 147, pp. 148--164, Jan. 2018.

\bibitem{ZhouX2019}
X.~Zhou, Q.~Yu, and W.~Peng, ``Fiber-optic {Fabry–Perot} pressure sensor for
  down-hole application,'' \emph{Opt. Lasers Eng.}, vol. 121, pp. 289--299,
  Oct. 2019.

\bibitem{Berlinger2018}
F.~Berlinger, J.~Dusek, M.~Gauci, and R.~Nagpal, ``Robust maneuverability of a
  miniature, low-cost underwater robot using multiple fin actuation,''
  \emph{IEEE Robot. Autom. Lett.}, vol.~3, no.~1, pp. 140--147, Jan. 2018.

\bibitem{Makled2018}
E.~A. Makled, A.~Yadav, O.~A. Dobre, and R.~D. Haynes, ``Hierarchical
  full-duplex underwater acoustic network: a {NOMA} approach,'' in \emph{Proc.
  Ocean.}\hskip 1em plus 0.5em minus 0.4em\relax Charleston, USA: IEEE, Oct.
  2018, pp. 1--6.

\bibitem{Fadlullah2018}
Z.~M. Fadlullah, F.~Tang, B.~Mao, J.~Liu, and N.~Kato, ``On intelligent traffic
  control for large-scale heterogeneous networks: a value matrix-based deep
  learning approach,'' \emph{IEEE Commun. Lett.}, vol.~22, no.~12, pp.
  2479--2482, Dec. 2018.

\bibitem{XieJ2019}
J.~{Xie}, F.~R. {Yu}, T.~{Huang}, R.~{Xie}, J.~{Liu}, C.~{Wang}, and Y.~{Liu},
  ``A survey of machine learning techniques applied to software defined
  networking ({SDN}): research issues and challenges,'' \emph{IEEE Commun.
  Surveys Tuts.}, vol.~21, no.~1, pp. 393--430, Jan. 2019.

\bibitem{Mao2018}
B.~Mao, F.~Tang, Z.~M. Fadlullah, N.~Kato, O.~Akashi, T.~Inoue, and
  K.~Mizutani, ``A novel non-supervised deep-learning-based network traffic
  control method for software defined wireless networks,'' \emph{IEEE Wireless
  Commun.}, vol.~25, no.~4, pp. 74--81, Aug. 2018.

\bibitem{SuY2020}
Y.~{Su}, L.~{Guo}, Z.~{Jin}, and X.~{Fu}, ``A mobile-beacon based iterative
  localization mechanism in large-scale underwater acoustic sensor networks,''
  \emph{IEEE Internet Things J.}, Sep. 2020.

\bibitem{Roopak2019}
M.~{Roopak}, G.~{Yun Tian}, and J.~{Chambers}, ``Deep learning models for cyber
  security in {IoT} networks,'' in \emph{Proc. 9th Annual Comput. Commun.
  Worksh. Conf.}\hskip 1em plus 0.5em minus 0.4em\relax Las Vegas, USA: IEEE,
  Jan. 2019, pp. 0452--0457.

\bibitem{XiaoL2020}
L.~{Xiao}, D.~{Jiang}, Y.~{Chen}, W.~{Su}, and Y.~{Tang},
  ``Reinforcement-learning-based relay mobility and power allocation for
  underwater sensor networks against jamming,'' \emph{IEEE J. Ocean. Eng.},
  vol.~45, no.~3, pp. 1148--1156, Jul. 2020.

\bibitem{ZhanhY2019}
Y.~{Zhang}, P.~{Li}, and X.~{Wang}, ``Intrusion detection for {IoT} based on
  improved genetic algorithm and deep belief network,'' \emph{{IEEE} Access},
  vol.~7, pp. 31\,711--31\,722, Mar. 2019.

\bibitem{LiY2019}
Y.~{Li}, S.~{Wang}, C.~{Jin}, Y.~{Zhang}, and T.~{Jiang}, ``A survey of
  underwater magnetic induction communications: fundamental issues, recent
  advances, and challenges,'' \emph{IEEE Commun. Surveys Tuts.}, vol.~21,
  no.~3, pp. 2466--2487, Feb. 2019.

\bibitem{Zafari2019}
F.~{Zafari}, A.~{Gkelias}, and K.~K. {Leung}, ``A survey of indoor localization
  systems and technologies,'' \emph{IEEE Commun. Surveys Tuts.}, vol.~21,
  no.~3, pp. 2568--2599, Apr. 2019.

\bibitem{WuH2019}
H.~{Wu}, J.~{Chen}, X.~{Liu}, Y.~{Xiao}, M.~{Wang}, Y.~{Zheng}, and Y.~{Rao},
  ``One-dimensional {CNN}-based intelligent recognition of vibrations in
  pipeline monitoring with {DAS},'' \emph{J. Lightw. Technol.}, vol.~37,
  no.~17, pp. 4359--4366, Sep. 2019.

\bibitem{MaK2017}
K.~{Ma}, H.~{Leung}, E.~{Jalilian}, and D.~{Huang}, ``Fiber-optic
  acoustic-based disturbance prediction in pipelines using deep learning,''
  \emph{IEEE Sensors Lett.}, vol.~1, no.~6, pp. 1--4, Dec. 2017.

\bibitem{PanJ2018}
J.~Pan, Y.~Yin, J.~Xiong, W.~Luo, G.~Gui, and H.~Sari, ``Deep learning-based
  unmanned surveillance systems for observing water levels,'' \emph{{IEEE}
  Access}, vol.~6, pp. 73\,561--73\,571, Nov. 2018.

\bibitem{Hedayati2010}
M.~R. {Hedayati}, A.~A. {Amidian}, S.~A. {Sadr}, and A.~{Razazan},
  ``Intelligent ship hull inspection and {NDT} using {ROV} based flux leakage
  expert system,'' in \emph{Proc. 2nd Int. Conf. Comput. Intell., Model.
  Simul.}\hskip 1em plus 0.5em minus 0.4em\relax Tuban, Indonesia: IEEE, Sep.
  2010, pp. 412--415.

\bibitem{LiC2019}
C.~{Li}, X.~{Peng}, J.~{Liu}, C.~{Wang}, S.~{Fan}, and S.~{Cao}, ``D-shaped
  fiber bragg grating ultrasonic hydrophone with enhanced sensitivity and
  bandwidth,'' \emph{J. Lightw. Technol.}, vol.~37, no.~9, pp. 2100--2108, May
  2019.

\bibitem{Gros1995}
X.~E. {Gros}, P.~{Strachan}, and D.~{Lowden}, ``Fusion of multiprobe {NDT} data
  for {ROV} inspection,'' in \emph{Proc. Ocean. Chall. Our Chang. Glob. Env.},
  vol.~3.\hskip 1em plus 0.5em minus 0.4em\relax San Diego, USA: IEEE, Oct.
  1995, pp. 2046--2050.

\bibitem{Pang2018}
Y.~Pang, M.~Sun, X.~Jiang, and X.~Li, ``Convolution in convolution for network
  in network,'' \emph{IEEE Trans. Neural Netw. Learn. Syst.}, vol.~29, no.~5,
  pp. 1587--1597, May 2018.

\bibitem{LiN2018}
N.~Li, Z.~Zheng, S.~Zhang, Z.~Yu, H.~Zheng, and B.~Zheng, ``The synthesis of
  unpaired underwater images using a multistyle generative adversarial
  network,'' \emph{{IEEE} Access}, vol.~6, pp. 54\,241--54\,257, Sep. 2018.

\bibitem{LiJ2018}
J.~Li, K.~A. Skinner, R.~M. Eustice, and M.~Johnson-Roberson, ``{WaterGAN}:
  unsupervised generative network to enable real-time color correction of
  monocular underwater images,'' \emph{IEEE Robot. Autom. Lett.}, vol.~3,
  no.~1, pp. 387--394, Jan. 2018.

\bibitem{JafariA2019}
A.~Jafari, A.~Ganesan, C.~S.~K. Thalisetty, V.~Sivasubramanian, T.~Oates, and
  T.~Mohsenin, ``{SensorNet}: a scalable and low-power deep convolutional
  neural network for multimodal data classification,'' \emph{IEEE Trans.
  Circuits Syst. I: Reg. Papers}, vol.~66, no.~1, pp. 274--287, Jan. 2019.

\bibitem{ChangJ2019}
J.~Chang and J.~Sha, ``Prune deep neural networks with the modified ${L}_{1/2}$
  penalty,'' \emph{{IEEE} Access}, vol.~7, pp. 2273--2280, Dec. 2018.

\bibitem{Tian2019}
Y.~{Tian}, B.~{Liu}, X.~{Su}, L.~{Wang}, and K.~{Li}, ``Underwater imaging
  based on {LF} and polarization,'' \emph{IEEE Photon. J.}, vol.~11, no.~1, pp.
  1--9, Feb. 2019.

\bibitem{Zawada2003}
D.~G. {Zawada}, ``Image processing of underwater multispectral imagery,''
  \emph{IEEE J. Ocean. Eng.}, vol.~28, no.~4, pp. 583--594, Oct. 2003.

\bibitem{Suresh2019}
S.~{Suresh}, E.~{Westman}, and M.~{Kaess}, ``Through-water stereo {SLAM} with
  refraction correction for {AUV} localization,'' \emph{IEEE Robot. Autom.
  Lett.}, vol.~4, no.~2, pp. 692--699, Apr. 2019.

\end{thebibliography}
	
	%
	
	\begin{IEEEbiography}[{\includegraphics[width=1in,height=1.25in,clip,keepaspectratio]{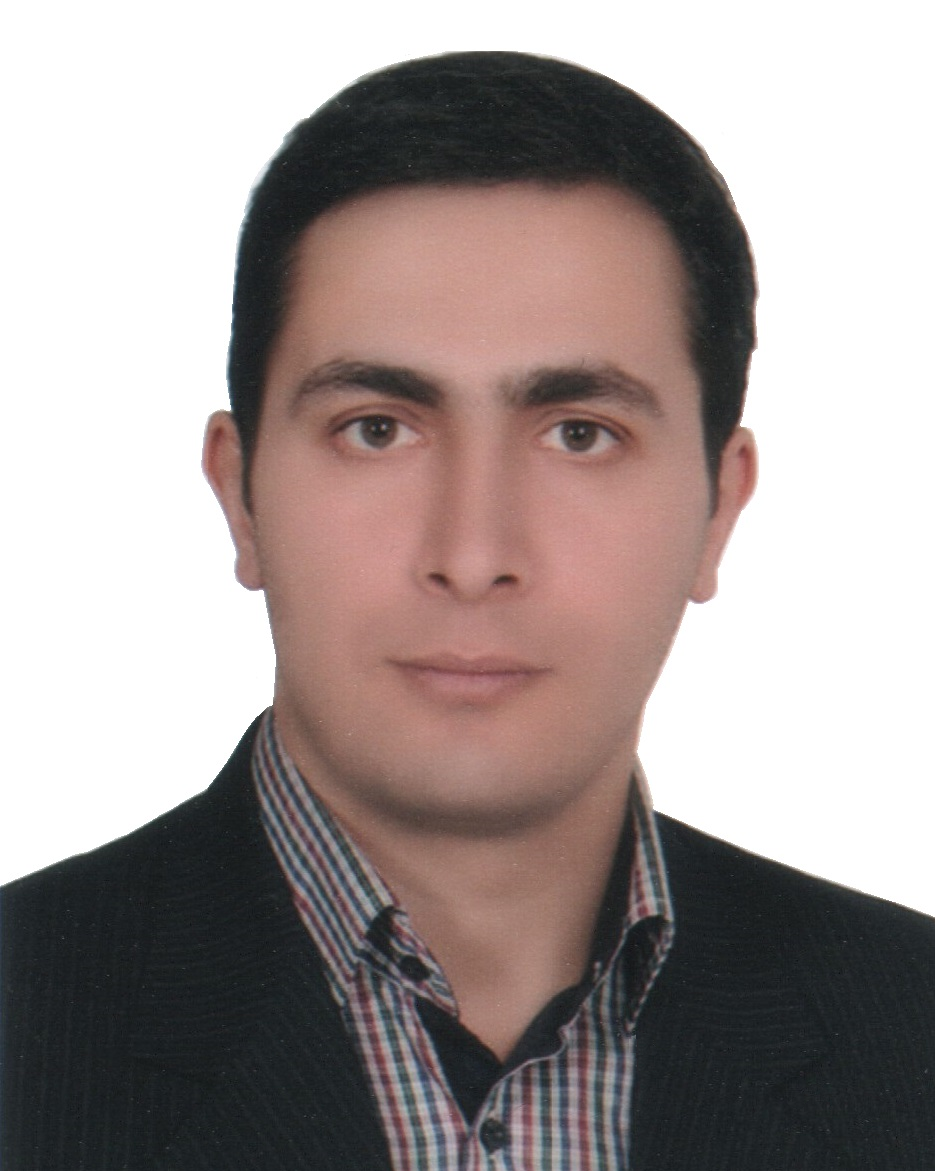}}]{Mohammad Jahanbakht}
		(S'18) received the B.Eng., M.Eng., and Ph.D. degrees, all in telecommunication engineering, from the Islamic Azad University of Iran, in 2003, 2005, 2009, respectively. He currently holds an adjunct faculty position at the Department of Electronic Engineering, Shahr-e-Qods Branch, Islamic Azad University, Tehran, Iran. In March 2018, he was awarded a Research Training Program (RTP) scholarship by the Australian Government to start a new project on machine intelligence for marine data at the College of Science and Engineering, James Cook University, Australia. His research interests are data science, machine learning, RF and microwave engineering, numerical methods, and signal processing.
	\end{IEEEbiography}

	\begin{IEEEbiography}[{\includegraphics[width=1in,height=1.25in,clip,keepaspectratio]{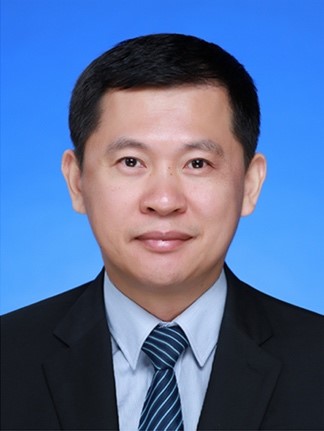}}]{Wei Xiang}
		(S’00–M’04–SM’10) received the B.Eng. and M.Eng. degrees, both in electronic engineering, from the University of Electronic Science and Technology of China, Chengdu, China, in 1997 and 2000, respectively, and the Ph.D. degree in telecommunications engineering from the University of South Australia, Adelaide, Australia, in 2004.
		
		Professor Wei Xiang is Cisco Research Chair of AI and IoT and Director Cisco-La Trobe Centre for AI and IoT at La Trobe University. Previously, he was Foundation Chair and Head of Discipline of IoT Engineering at James Cook University, Cairns, Australia. Due to his instrumental leadership in establishing Australia’s first accredited Internet of Things Engineering degree program, he was inducted into Pearcy Foundation’s Hall of Fame in October 2018. He is an elected Fellow of the IET in UK and Engineers Australia. He received the TNQ Innovation Award in 2016, and Pearcey Entrepreneurship Award in 2017, and Engineers Australia Cairns Engineer of the Year in 2017. He was a co-recipient of four Best Paper Awards at WiSATS’2019, WCSP’2015, IEEE WCNC’2011, and ICWMC’2009. He has been awarded several prestigious fellowship titles. He was named a Queensland International Fellow (2010-2011) by the Queensland Government of Australia, an Endeavour Research Fellow (2012-2013) by the Commonwealth Government of Australia, a Smart Futures Fellow (2012-2015) by the Queensland Government of Australia, and a JSPS Invitational Fellow jointly by the Australian Academy of Science and Japanese Society for Promotion of Science (2014-2015). He is the Vice Chair of the IEEE Northern Australia Section. He was an Editor for IEEE Communications Letters (2015-2017), and is currently an Associate Editor for IEEE Internet of Things Journal and IEEE Access. He has published over 250 peer-reviewed papers including 3 books and 180 journal articles. He has severed in a large number of international conferences in the capacity of General Co-Chair, TPC Co-Chair, Symposium Chair, etc. His research interest includes the Internet of Things, wireless communications, machine learning for IoT data analytics, and computer vision.

	\end{IEEEbiography}

	\begin{IEEEbiography}[{\includegraphics[width=1in,height=1.25in,clip,keepaspectratio]{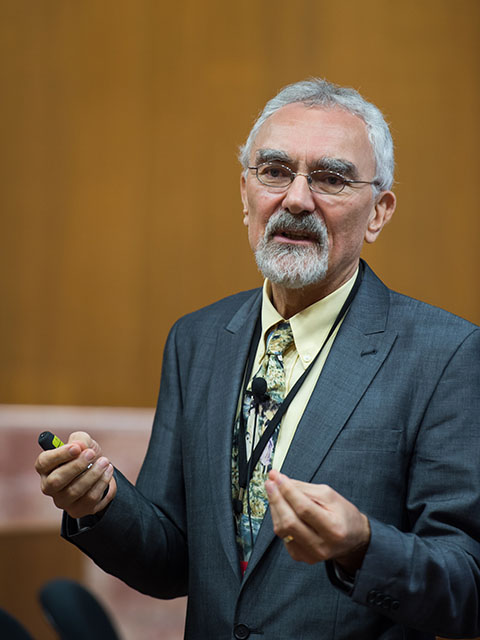}}]{Lajos Hanzo}
	(\href{http://www-mobile.ecs.soton.ac.uk}{http://www-mobile.ecs.soton.ac.uk}, \href{https://en.wikipedia.org/wiki/Lajos_Hanzo}{https://en.wikipedia.org/wiki/Lajos\_Hanzo}) (FIEEE'04) received his Master degree and Doctorate in 1976 and 1983, respectively from the Technical University (TU) of Budapest. He was also awarded the Doctor of Sciences (DSc) degree by the University of Southampton (2004) and Honorary Doctorates by the TU of Budapest (2009) and by the University of Edinburgh (2015).  He is a Foreign Member of the Hungarian Academy of Sciences and a former Editor-in-Chief of the IEEE Press.  He has served several terms as Governor of both IEEE ComSoc and of VTS.  He has published 2000+ contributions at IEEE Xplore, 19 Wiley-IEEE Press books and has helped the fast-track career of 123 PhD students. Over 40 of them are Professors at various stages of their careers in academia and many of them are leading scientists in the wireless industry. He is also a Fellow of the Royal Academy of Engineering (FREng), of the IET and of EURASIP.
	\end{IEEEbiography}

	\begin{IEEEbiography}[{\includegraphics[width=1in,height=1.25in,clip,keepaspectratio]{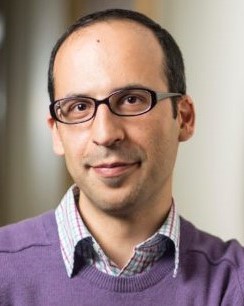}}]{Mostafa Rahimi Azghadi}
		(S’07–M’14–SM’19) completed his PhD in Electrical \& Electronic Engineering at The University of Adelaide, Australia, earning the Doctoral Research Medal, and the 2015 Adelaide University Alumni Medal. He is currently a senior lecturer at the College of Science and Engineering, James Cook University, Australia, where he researches Machine Learning software and hardware design for a variety of applications including automation, precision agriculture, aquaculture, marine sciences, mining, and medical imaging. His research has attracted over \$0.7 Million in funding from national and international resources.
		
		Dr. Rahimi Azghadi was the recipient of several accolades including a 2015 South Australia Science Excellence award, a 2016 Endeavour Research Fellowship, a 2017 Queensland Young Tall Poppy Science Award, a 2018 Rising Star ECR Leader Fellowship, a 2019 Fresh Science Queensland finalist, and a 2020 Award for Excellence in Innovation and Change.
%
	\end{IEEEbiography}


	
	
\end{document}